\documentclass[12pt]{article}

\usepackage[dvips]{graphicx}
\usepackage{indentfirst,hyperref,color,psfrag,verbatim}
\usepackage{amsmath,amssymb,mathrsfs,graphicx,dsfont,caption,
subcaption,mathtools,mathabx}
\usepackage{scalerel}
\usepackage{feynmp-auto}

\numberwithin{equation}{section}

\newcommand{\dfn}{\vcentcolon=}
\newcommand{\n}{\varstar}

\newcommand{\dm}{\,\,
\begin{gathered}
	\begin{fmffile}{dm}
		\begin{fmfgraph}(40,40)
			\fmfset{dash_len}{1.2mm}
			\fmfset{wiggly_len}{1.1mm} \fmfset{dot_len}{0.5mm}
			\fmfpen{0.25mm}
			\fmfvn{decor.shape=circle,decor.filled=shaded, decor.size=5thin}{u}{1}
			\fmfleft{i}
			\fmfright{o}
			\fmf{phantom,fore=black,tension=5}{i,u1,o}
			\fmffreeze
			\fmfforce{(0w,0.35h)}{i}
			\fmfforce{(0w,0.35h)}{u1}
			\fmfforce{(0w,0.35h)}{o}
		\end{fmfgraph}
	\end{fmffile}
\end{gathered}\!\!\!\!
}

\newcommand{\wf}{\!
\begin{gathered}
	\begin{fmffile}{wf}
		\begin{fmfgraph}(40,40)
			\fmfset{dash_len}{1.2mm}
			\fmfset{wiggly_len}{1.1mm} \fmfset{dot_len}{0.5mm}
			\fmfpen{0.25mm}
			\fmfvn{decor.shape=square,decor.filled=shaded, decor.size=5thin}{u}{1}
			\fmfleft{i}
			\fmfright{o}
			\fmf{phantom,fore=black,tension=5}{i,u1,o}
			\fmffreeze
			\fmfforce{(0.5w,0.35h)}{i}
			\fmfforce{(0.5w,0.35h)}{u1}
			\fmfforce{(0.5w,0.35h)}{o}
		\end{fmfgraph}
	\end{fmffile}
\end{gathered}
}

\newcommand{\vanilla}{
\begin{gathered}
	\begin{fmffile}{vanilla}
		\begin{fmfgraph}(40,40)
			\fmfset{dash_len}{1.2mm}
			\fmfset{wiggly_len}{1.1mm} \fmfset{dot_len}{0.5mm}
			\fmfpen{0.25mm}
			\fmfleft{i}
			\fmfright{o}
			\fmf{plain,fore=black}{i,o}
			\fmffreeze
			\fmfforce{(-1w,0.35h)}{i}
			\fmfforce{(1w,0.35h)}{o}
		\end{fmfgraph}
	\end{fmffile}
\end{gathered}\,\,\,
}

\newcommand{\wiggly}{\hspace{0.5cm}
\begin{gathered}
	\begin{fmffile}{wiggly}
		\begin{fmfgraph}(40,40)
			\fmfset{dash_len}{1.2mm}
			\fmfset{wiggly_len}{1.1mm} \fmfset{dot_len}{0.5mm}
			\fmfpen{0.25mm}
			\fmfleft{i}
			\fmfright{o}
			\fmf{wiggly,fore=black}{i,o}
			\fmffreeze
			\fmfforce{(-1w,0.35h)}{i}
			\fmfforce{(1w,0.35h)}{o}
		\end{fmfgraph}
	\end{fmffile}
\end{gathered}\,\,\,}

\newcommand{\dashes}{
\begin{gathered}
	\begin{fmffile}{dashes}
		\begin{fmfgraph}(40,40)
			\fmfset{dash_len}{1.2mm}
			\fmfset{wiggly_len}{1.1mm} \fmfset{dot_len}{0.5mm}
			\fmfpen{0.25mm}
			\fmfleft{i}
			\fmfright{o}
			\fmf{dashes,fore=black}{i,o}
			\fmffreeze
			\fmfforce{(-1w,0.35h)}{i}
			\fmfforce{(1w,0.35h)}{o}
		\end{fmfgraph}
	\end{fmffile}
\end{gathered}\,\,\,
}

\fmfcmd{%
vardef cross_bar (expr p, len, ang) = 
((-len/2,0)--(len/2,0)) rotated (ang + angle direction length(p)/2 of p) 
shifted point length(p)/2 of p
enddef;
style_def crossed expr p =
cdraw p;
ccutdraw cross_bar (p, 5mm,  45);
ccutdraw cross_bar (p, 5mm, -45)
enddef;}

\begin{document}

\title{\Large \bf Complete Normal Ordering 1: Foundations} 

\date{}

\maketitle

\vspace{-5.5cm}
\rightline{\small KCL-PH-TH/2015-55, LCTS/2015-43, CERN-PH-TH/2015-295}
\vspace{5.5cm}

\vspace{-1.8cm}
\centerline{\author{\large \bf John~Ellis$^{(a,b)}$, Nick~E.~Mavromatos$^{(a,b)}$ and Dimitri~P.~Skliros$^{(a,c)}$}}

\vspace{0.3cm}

\begin{center}

{\sl (a) Theoretical Particle Physics and Cosmology Group\\
Department of Physics, King's~College~London\\
London WC2R 2LS, United Kingdom}

\vskip 0.08in
{ \sl (b) Theory Division, CERN, CH-1211 Geneva 23, Switzerland} 

\vskip 0.08in
{ \sl (c) School of Physics and Astronomy, University of Nottingham}\\
{\sl Nottingham, NG7 2RD, UK} 

\vskip 0.1in
{\tt \small john.ellis@cern.ch; nikolaos.mavromatos@kcl.ac.uk; d.p.skliros@gmail.com}

\end{center}

\vskip 0.1 in

\abstract{
\begin{center}
\begin{minipage}[c]{0.85\textwidth}
We introduce a new prescription for quantising scalar field theories (in generic spacetime dimension and background) perturbatively around a true minimum of the full quantum effective action, which is to `complete normal order' the bare action of interest. 
When the true vacuum of the theory is located at zero field value, the key property of this prescription is the automatic cancellation, to any finite order in perturbation theory, of all tadpole and, more generally, all `cephalopod' Feynman diagrams. The latter are connected diagrams that can be disconnected into two pieces by cutting one internal vertex, with either one or both pieces free from external lines. In addition, this procedure of `complete normal ordering' (which is an extension of the standard field theory definition of normal ordering) reduces by a substantial factor the number of Feynman diagrams to be calculated at any given loop order. We illustrate explicitly the complete normal ordering procedure and the cancellation of cephalopod diagrams in scalar field theories with non-derivative interactions, and by using a point splitting `trick' we extend this result to theories with derivative interactions, such as those appearing as non-linear $\sigma$-models in the world-sheet formulation of string theory. We focus here on theories with  trivial vacua, generalising the discussion to non-trivial vacua in a follow-up paper.
\end{minipage} 
\end{center}
}

\newpage
\tableofcontents

\section{Overview}

One of the important tools in quantum field theory is the procedure of `normal ordering' \cite{Coleman11}, $\mathcal{O}(\phi)\rightarrow \,\,:\!\mathcal{O}(\phi)\!:\,$, 
that is invaluable, e.g., when defining products of field operators at coincident points and when evaluating correlation functions using Wick's theorem. A crucial property is that expectation values of normal-ordered operators vanish in the \emph{free} theory: $\langle:\!\!\mathcal{O}(\phi)\!\!:\rangle_0=0$. 
There are various formulations of this notion~\cite{Polchinski_v1}, including creation-annihilation operator normal ordering, conformal normal ordering,
functional integral normal ordering, etc., which are often interrelated. The following is one concise definition of standard normal ordering,
\begin{equation}\label{eq:NO1}
: \!\mathcal{O}(\phi)\!:\,\,=\mathcal{O}(\delta_X)\,e^{-\frac{1}{2}\int_{z}\int_w\mathcal{G}(z,w)X(z)X(w)+\int X\phi}\big|_{X=0} \, ,
\end{equation}
where $\mathcal{G}(z,w)$ is the \emph{free} Feynman propagator of the theory, 
and $\delta_X$ is a functional derivative with respect to the (unphysical) source $X$. For example, for an interaction term $\mathcal{O}(\phi)=\frac{1}{N!}g_N\phi^N$ in the action of interest, upon normal ordering we obtain:
\begin{equation}\label{eq:Bell::phiN}
: \!\phi^N\!:\,\, =B_N(\phi,-\mathcal{G},0,\dots,0),
\end{equation}
with $B_N(a_1,\dots,a_N)$ a complete Bell polynomial\footnote{These satisfy $B_n(a_1,\dots,a_n)=\sum_{r=0}^n\binom{n}{r}B_{n-r}(0,a_2,\dots,a_{n-r})a_1^r$ and $B_n(a_1,\dots,a_n)z^n=B_n(z a_1,\dots,z^na_n)$ 
and are defined by the generating function: 
\begin{equation}\label{eq:Bn genfunc_foot}
\sum_{n=0}^{\infty}\frac{1}{n!}\,B_n(a_1,\dots,a_n)z^n\dfn\exp\Big(\sum_{n=1}^{\infty}\frac{1}{n!}\,a_nz^n\Big).
\end{equation}
} \cite{Riordan58,RomanRota78} and $\mathcal{G}$ the Feynman propagator at coincident points. 
In the interaction picture of quantum field theory this standard normal ordering of the action produces counterterms that eradicate Feynman diagrams with internal lines that begin and end on the same internal vertex \cite{Coleman75},
such as the one-loop two-point amplitude in $\phi^4$, as well as some (but not all) tadpole diagrams in $\phi^3$ scalar field theory.\footnote{Other diagrams that are cancelled by normal ordering are: $
\begin{gathered}
	\begin{fmffile}{wgtadpole}
		\fmfset{dash_len}{1.2mm}
		\begin{fmfgraph}(50,50)
			\fmfset{dash_len}{1.2mm}
			\fmfset{wiggly_len}{1.1mm} \fmfset{dot_len}{0.5mm}
			\fmfpen{0.25mm}
			\fmfleft{i}
			\fmfright{o}
			\fmf{phantom,tension=5}{i,v1}
			\fmf{wiggly,fore=black,tension=0.8}{v2,o}
			\fmf{wiggly,fore=black,left,tension=0.08}{v1,v2,v1}
			\fmf{phantom}{v1,v2}
		\end{fmfgraph}
	\end{fmffile}
\end{gathered}
$, 
$
\begin{gathered}
	\begin{fmffile}{wg4-3tadpole}
		\begin{fmfgraph}(35,35)
			\fmfset{dash_len}{1.2mm}
			\fmfset{wiggly_len}{1.1mm} \fmfset{dot_len}{0.5mm}
			\fmfpen{0.25mm}
			\fmfsurround{u1,u2,u3}
			\fmf{wiggly,fore=black,tension=1}{u1,v}
			\fmf{wiggly,fore=black,tension=1}{u2,v}
			\fmf{wiggly,fore=black,tension=1}{u3,v}
			\fmf{wiggly,fore=black,tension=1,left}{u1,u1}
			\fmf{wiggly,fore=black,tension=1,left}{u2,u2}
			\fmf{wiggly,fore=black,tension=1,left}{u3,u3}
		\end{fmfgraph}
	\end{fmffile}
\end{gathered}\,\,
$, \!\!\!
$
\begin{gathered}
	\begin{fmffile}{wkappa-tadpole}
		\begin{fmfgraph}(60,60)
			\fmfset{dash_len}{1.2mm}
			\fmfset{wiggly_len}{1.1mm} \fmfset{dot_len}{0.5mm}
			\fmfpen{0.25mm}
			\fmftop{t1,t2,t3}
			\fmfbottom{b1,b2,b3}
			\fmf{phantom}{t1,v1,b1}
			\fmf{phantom}{t2,v2,b2}
			\fmf{phantom}{t3,v3,b3}
			\fmf{wiggly,fore=black,right,tension=1}{v1,v2,v1}
			\fmf{wiggly,fore=black,right,tension=1}{v2,v3,v2}
			\fmf{wiggly,fore=black,tension=1}{v2,b2}
		\end{fmfgraph}
	\end{fmffile}
\end{gathered}\!\!
$, 
$
\begin{gathered}
	\begin{fmffile}{wglambda-tadpole}
		\begin{fmfgraph}(45,45)
			\fmfset{dash_len}{1.2mm}
			\fmfset{wiggly_len}{1.1mm} \fmfset{dot_len}{0.5mm}
			\fmfpen{0.25mm}
			\fmftop{t1,t2,t3,t4}
        			\fmfbottom{b1,b2,b3,b4}
        			\fmf{phantom}{t1,v1,b1}
        			\fmf{phantom}{t2,v2,b2}
			\fmf{phantom}{t3,v3,b3}
			\fmf{phantom}{t4,v4,b4}
        			\fmffreeze
			\fmf{wiggly,fore=black,right,tension=0.7}{v1,v2,v1}
        			\fmf{wiggly,fore=black,right,tension=0.7}{v2,v3,v2}
        			\fmf{wiggly,fore=black,tension=3}{v3,v4}
			\fmffreeze
			\fmfforce{(1.1w,0.5h)}{v4}
		\end{fmfgraph}
	\end{fmffile}
\end{gathered}, \,\,
\begin{gathered}
	\begin{fmffile}{wgamma-vacuum-1PI}
		\begin{fmfgraph}(50,50)
			\fmfset{dash_len}{1.2mm}
			\fmfset{wiggly_len}{1.1mm} \fmfset{dot_len}{0.5mm}
			\fmfpen{0.25mm}
			\fmfsurroundn{x}{3}
			\fmf{phantom,fore=black}{x1,v}
			\fmf{phantom,fore=black}{x2,v}
			\fmf{phantom,fore=black}{x3,v}
			\fmf{wiggly,fore=black,tension=0.7}{v,v}
			\fmf{wiggly,fore=black,tension=0.7,right}{v,v}
			\fmf{wiggly,fore=black,tension=0.7,left}{v,v}
		\end{fmfgraph}
	\end{fmffile}
\end{gathered}\!$, 
$
\begin{gathered}
	\begin{fmffile}{wgamma-2pt1loop-1PI}
		\begin{fmfgraph}(45,45)
			\fmfset{dash_len}{1.2mm}
			\fmfset{wiggly_len}{1.1mm} \fmfset{dot_len}{0.5mm}
			\fmfpen{0.25mm}
			\fmfleft{i}
			\fmfright{o}
			\fmf{wiggly,fore=black,tension=0.7}{i,v,v,o}
			\fmf{wiggly,fore=black,left=90,tension=0.7}{v,v}
		\end{fmfgraph}
	\end{fmffile}
\end{gathered}\!,
$ 
$
\begin{gathered}
	\begin{fmffile}{wgamma-4pt1loop-1PI}
		\begin{fmfgraph}(45,45)
			\fmfset{dash_len}{1.2mm}
			\fmfset{wiggly_len}{1.1mm} \fmfset{dot_len}{0.5mm}
			\fmfpen{0.25mm}
			\fmfsurroundn{x}{8}
			\fmf{phantom,fore=black}{x1,c,x5}
			\fmf{phantom,fore=black}{x2,c,x6}
			\fmf{phantom,fore=black}{x3,c,x7}
			\fmf{phantom,fore=black}{x4,c,x8}
			\fmf{wiggly,fore=black}{x1,c}
			\fmf{wiggly,fore=black}{x8,c}
			\fmf{wiggly,fore=black}{x7,c}
			\fmf{wiggly,fore=black}{x6,c}
			\fmfi{wiggly,fore=black}{fullcircle scaled .38w shifted (0.46w,.58h)}	
		\end{fmfgraph}
	\end{fmffile}
\end{gathered}$, etc.} These counterterms, call them $\delta_{::} g_n$, are read off from the right-hand side of (\ref{eq:Bell::phiN}), and in particular $\frac{1}{N!}g_N: \!\phi^N\!:\,\,$ equals $\frac{1}{N!}g_N\phi^N$ plus  a polynomial $\sum_{n=0}^{N-1}\frac{1}{n!}\delta_{::} g_n\phi^n$ with:
\begin{equation}\label{eq:deltagn}
\delta_{::} g_n=\frac{g_N}{(N-n)!}B_{N-n}(0,-\mathcal{G},0,\dots,0).
\end{equation}
This quick exercise yields one of the terms contributing to $\delta_{::} g_n$, and there may be various such contributions if we also  sum over $N$ (as we typically should to preserve renormalisability).  Substituting these counterterms into the generating function of interest is clearly equivalent to having started from a path integral with a normal-ordered action in the first place.

In 2 dimensions and in the absence of derivative interactions these counterterms  make all correlation functions (of elementary fields) UV-finite. But this still does not guarantee that we are quantising the theory around a true minimum of the full quantum effective action, given that there will generically be an infinite number of (physical and unphysical) tadpole diagrams. In addition, there will generically also be an infinite number of \emph{cephalopod} diagrams in the resulting amplitudes, and these can also be removed by an appropriate choice of counterterms. 

Cephalopods  \cite{Skliros15} are generalisations of the familiar `tadpole' diagrams: the 1PI version of a `cephalopod' diagram has an arbitrary number (0,1,2,..) of
external legs and an arbitrary number (1,2,3,..) of bubbles or `heads'.  There is no restriction on the number of loops in the head,
but the `neck' joining the head(s) and leg(s) is formed by a single vertex. The following are some examples of cephalopod diagrams:
$$\label{cephalopod diagrams}
\begin{gathered}\label{eq:cephalopods-examples}
	\begin{fmffile}{wg2L-3loop2pt-1PR-xx}
		\begin{fmfgraph}(46,46)
			\fmfset{dash_len}{1.2mm}
			\fmfset{wiggly_len}{1.1mm} \fmfset{dot_len}{0.5mm}
			\fmfpen{0.25mm}
			\fmfleft{i}
			\fmfright{o}
			\fmf{phantom,tension=5}{i,v1}
			\fmf{wiggly,fore=black,tension=2.5}{v2,o}
			\fmf{wiggly,fore=black,left,tension=0.5}{v1,v2,v1}
			\fmf{wiggly,fore=black}{v1,v2}
			\fmffreeze
			\fmfforce{(1.1w,0.5h)}{o}
			\fmffreeze
			\fmfright{n,m}
			\fmf{wiggly,fore=black,tension=1}{o,n}
			\fmf{wiggly,fore=black,tension=1}{o,m}
			\fmfforce{(1.35w,0.9h)}{n}
			\fmfforce{(1.35w,0.1h)}{m}
		\end{fmfgraph}
	\end{fmffile}
\end{gathered}
\,\,\,\,\,\,, 
\!\! \!\!\!\!\!\!
\quad\,\,
\begin{gathered}
	\begin{fmffile}{wg4-3tadpole-xx}
		\begin{fmfgraph}(40,40)
			\fmfset{dash_len}{1.2mm}
			\fmfset{wiggly_len}{1.1mm} \fmfset{dot_len}{0.5mm}
			\fmfpen{0.25mm}
			\fmfsurround{u1,u2,u3}
			\fmf{wiggly,fore=black,tension=1}{u1,v}
			\fmf{wiggly,fore=black,tension=1}{u2,v}
			\fmf{wiggly,fore=black,tension=1}{u3,v}
			\fmf{wiggly,fore=black,tension=1,left}{u1,u1}
			\fmf{wiggly,fore=black,tension=1,left}{u2,u2}
			\fmf{wiggly,fore=black,tension=1,left}{u3,u3}
		\end{fmfgraph}
	\end{fmffile}
\end{gathered}\,\,\,\,,\,\,
\,
\begin{gathered}
	\begin{fmffile}{wgkappa-4ptself1PR-xx}
		\begin{fmfgraph}(40,40)
			\fmfset{dash_len}{1.2mm}
			\fmfset{wiggly_len}{1.1mm} \fmfset{dot_len}{0.5mm}
			\fmfpen{0.25mm}
			\fmftop{t}
			\fmfbottom{a,b,c}
			\fmf{wiggly,fore=black,tension=1}{a,v}
			\fmf{wiggly,fore=black,tension=1}{b,v}
			\fmf{wiggly,fore=black,tension=1}{c,v}
			\fmf{wiggly,fore=black,tension=1.4,left}{v,t,v}
			\fmffreeze
			\fmfbottom{x,z}
			\fmf{wiggly,fore=black}{c,x}
			\fmf{wiggly,fore=black}{c,z}
			\fmfforce{(w,-.4h)}{x}
			\fmfforce{(1.45w,.4h)}{z}
		\end{fmfgraph}
	\end{fmffile}
\end{gathered}\,\,
\,\,\,, \,\,\,
\begin{gathered}
	\begin{fmffile}{wlambdaself-xx}
		\begin{fmfgraph}(40,40)
			\fmfset{dash_len}{1.2mm}
			\fmfset{wiggly_len}{1.1mm} \fmfset{dot_len}{0.5mm}
			\fmfpen{0.25mm}
			\fmftop{s}
			\fmfleft{a}
			\fmfright{b}
			\fmf{wiggly,fore=black}{a,v}
			\fmf{wiggly,fore=black}{b,v}
			\fmf{wiggly,fore=black,right,tension=.7}{v,v}
			\fmffreeze
			\fmfforce{(0w,0.2h)}{a}
			\fmfforce{(w,0.2h)}{b}
		\end{fmfgraph}
	\end{fmffile}
\end{gathered}
\,, 
\,\,\,
\begin{gathered}
	\begin{fmffile}{wgamma-vacuum-1PI-xx}
		\begin{fmfgraph}(52,52)
			\fmfset{dash_len}{1.2mm}
			\fmfset{wiggly_len}{1.1mm} \fmfset{dot_len}{0.5mm}
			\fmfpen{0.25mm}
			\fmfsurroundn{x}{3}
			\fmf{phantom,fore=black}{x1,v}
			\fmf{phantom,fore=black}{x2,v}
			\fmf{phantom,fore=black}{x3,v}
			\fmf{wiggly,fore=black,tension=0.7}{v,v}
			\fmf{wiggly,fore=black,tension=0.7,right}{v,v}
			\fmf{wiggly,fore=black,tension=0.7,left}{v,v}
		\end{fmfgraph}
	\end{fmffile}
\end{gathered}, 
\,\,\,
\begin{gathered}
	\begin{fmffile}{wgamma-2pt1loop-1PI-xx}
		\begin{fmfgraph}(51,51)
			\fmfset{dash_len}{1.2mm}
			\fmfset{wiggly_len}{1.1mm} \fmfset{dot_len}{0.5mm}
			\fmfpen{0.25mm}
			\fmfleft{i}
			\fmfright{o}
			\fmf{wiggly,fore=black,tension=0.7}{i,v,v,o}
			\fmf{wiggly,fore=black,left=90,tension=0.7}{v,v}
		\end{fmfgraph}
	\end{fmffile}
\end{gathered}\!, 
\,\,\,
\begin{gathered}
	\begin{fmffile}{wgamma-4pt1loop-1PI-xx}
		\begin{fmfgraph}(53,53)
			\fmfset{dash_len}{1.2mm}
			\fmfset{wiggly_len}{1.1mm} \fmfset{dot_len}{0.5mm}
			\fmfpen{0.25mm}
			\fmfsurroundn{x}{8}
			\fmf{phantom,fore=black}{x1,c,x5}
			\fmf{phantom,fore=black}{x2,c,x6}
			\fmf{phantom,fore=black}{x3,c,x7}
			\fmf{phantom,fore=black}{x4,c,x8}
			\fmf{wiggly,fore=black}{x1,c}
			\fmf{wiggly,fore=black}{x8,c}
			\fmf{wiggly,fore=black}{x7,c}
			\fmf{wiggly,fore=black}{x6,c}
			\fmfi{wiggly,fore=black}{fullcircle scaled .38w shifted (0.46w,.58h)}	
		\end{fmfgraph}
	\end{fmffile}
\end{gathered},
\,\,
\begin{gathered}
	\begin{fmffile}{wgkappa-bubble-2pt-1PI-xx}
		\begin{fmfgraph}(40,40)
			\fmfset{dash_len}{1.2mm}
			\fmfset{wiggly_len}{1.1mm} \fmfset{dot_len}{0.5mm}
			\fmfpen{0.25mm}
			\fmfleft{i}
			\fmfright{o}
			\fmf{phantom,tension=5}{i,v1}
			\fmf{phantom,tension=5}{v2,o}
			\fmf{wiggly,fore=black,left,tension=0.4}{v1,v2,v1}
			\fmf{wiggly,fore=black}{v1,v2}
			\fmffreeze
			\fmfright{o1,o2}
			\fmf{wiggly,fore=black,tension=1}{v2,o1}
			\fmf{wiggly,fore=black,tension=1}{v2,o2}
			\fmfforce{(1.1w,0.9h)}{o1}
			\fmfforce{(1.1w,0.1h)}{o2}
		\end{fmfgraph}
	\end{fmffile}
\end{gathered}\,
\,, \,\,\,\,\,\,
\!\!\!\begin{gathered}
	\begin{fmffile}{wg2L-bubble2ptz-xx}
		\begin{fmfgraph}(50,50)
			\fmfset{dash_len}{1.2mm}
			\fmfset{wiggly_len}{1.1mm} \fmfset{dot_len}{0.5mm}
			\fmfpen{0.25mm}
			\fmftop{t1,t2,t3}
			\fmfbottom{b1,b2,b3}
			\fmf{phantom}{t1,v1,b1}
			\fmf{phantom}{t2,v2,b2}
			\fmf{phantom}{t3,v3,b3}
			\fmffreeze
			\fmf{wiggly,fore=black,right}{v1,v2,v1}
			\fmf{wiggly,fore=black}{v2,t3}
			\fmf{wiggly,fore=black}{v2,b3}
			\fmf{wiggly,fore=black,tension=1}{t1,b1}
			\fmfforce{(0.25w,0.7h)}{t1}
			\fmfforce{(0.25w,0.3h)}{b1}
			\fmfforce{(0.9w,0.9h)}{t3}
			\fmfforce{(0.9w,0.1h)}{b3}
		\end{fmfgraph}
	\end{fmffile}
\end{gathered}, 
\,\,\,
\begin{gathered}
	\begin{fmffile}{wg2L-bubble-xx}
		\begin{fmfgraph}(43,43)
			\fmfset{dash_len}{1.2mm}
			\fmfset{wiggly_len}{1.1mm} \fmfset{dot_len}{0.5mm}
			\fmfpen{0.25mm}
			\fmftop{t1,t2,t3}
			\fmfbottom{b1,b2,b3}
			\fmf{phantom}{t1,v1,b1}
			\fmf{phantom}{t2,v2,b2}
			\fmf{phantom}{t3,v3,b3}
			\fmffreeze
			\fmf{wiggly,fore=black,right}{v1,v2,v1}
			\fmf{wiggly,fore=black,right}{v2,v3,v2}
			\fmf{wiggly,fore=black,tension=1}{t1,b1}
			\fmfforce{(0.25w,0.7h)}{t1}
			\fmfforce{(0.25w,0.3h)}{b1}
		\end{fmfgraph}
	\end{fmffile}
\end{gathered},\,\,\,
\begin{gathered}
	\begin{fmffile}{wgtadpole-xx}
		\fmfset{dash_len}{1.2mm}
		\begin{fmfgraph}(57,57)
			\fmfset{dash_len}{1.2mm}
			\fmfset{wiggly_len}{1.1mm} \fmfset{dot_len}{0.5mm}
			\fmfpen{0.25mm}
			\fmfleft{i}
			\fmfright{o}
			\fmf{phantom,tension=5}{i,v1}
			\fmf{wiggly,fore=black,tension=0.8}{v2,o}
			\fmf{wiggly,fore=black,left,tension=0.08}{v1,v2,v1}
			\fmf{phantom}{v1,v2}
		\end{fmfgraph}
	\end{fmffile}
\end{gathered}
,  
\!\!\!\!
\,\,
\begin{gathered}
	\begin{fmffile}{wg3tadpole1-xx}
		\begin{fmfgraph}(70,70)
			\fmfset{dash_len}{1.2mm}
			\fmfset{wiggly_len}{1.1mm} \fmfset{dot_len}{0.5mm}
			\fmfpen{0.25mm}
			\fmftop{t}
			\fmfbottom{b}
			\fmfleft{l}
			\fmfright{r}
			\fmf{phantom,fore=black,tension=9}{t,u,v,b}
			\fmf{phantom,fore=black,tension=9}{l,s,x,r}
			\fmf{wiggly,fore=black,tension=.01,left}{u,v,u}
			\fmf{phantom,fore=black,tension=0.01}{s,x,s}
			\fmf{wiggly,fore=black,tension=1}{u,v}
			\fmf{wiggly,fore=black,tension=1}{x,r}
		\end{fmfgraph}
	\end{fmffile}
\end{gathered}
,
\dots
$$ 
{\it ~~~~~~~~~~~~~~~~~~~~~~~~~~~~~Some examples of `cephalopod' diagrams.}
~\\

\vspace{0cm}
$\!\!\!\!\!\!\!\!\!\!\!$Examples of the way in which these diagrams can be disconnected into two pieces by cutting one internal vertex with either one or both resulting pieces free from external lines are:
$$
\begin{gathered}
	\begin{fmffile}{wgkappa-bubble-2pt-1PI-xx}
		\begin{fmfgraph}(40,40)
			\fmfset{dash_len}{1.2mm}
			\fmfset{wiggly_len}{1.1mm} \fmfset{dot_len}{0.5mm}
			\fmfpen{0.25mm}
			\fmfleft{i}
			\fmfright{o}
			\fmf{phantom,tension=5}{i,v1}
			\fmf{phantom,tension=5}{v2,o}
			\fmf{wiggly,fore=black,left,tension=0.4}{v1,v2,v1}
			\fmf{wiggly,fore=black}{v1,v2}
			\fmffreeze
			\fmfright{o1,o2}
			\fmf{wiggly,fore=black,tension=1}{v2,o1}
			\fmf{wiggly,fore=black,tension=1}{v2,o2}
			\fmfforce{(1.1w,0.9h)}{o1}
			\fmfforce{(1.1w,0.1h)}{o2}
		\end{fmfgraph}
	\end{fmffile}
\end{gathered}\,
\,\qquad \rightarrow\qquad  \begin{gathered}
	\begin{fmffile}{wgkappa-bubble-2pt-1PI-xxvv}
		\begin{fmfgraph}(40,40)
			\fmfset{dash_len}{1.2mm}
			\fmfset{wiggly_len}{1.1mm} \fmfset{dot_len}{0.5mm}
			\fmfpen{0.25mm}
			\fmfleft{i}
			\fmfright{o}
			\fmf{phantom,tension=5}{i,v1}
			\fmf{phantom,tension=5}{v2,o}
			\fmf{wiggly,fore=black,left,tension=0.4}{v1,v2,v1}
			\fmf{wiggly,fore=black}{v1,v2}
			\fmffreeze
			\fmfright{o1,o2}
			\fmf{phantom,fore=black,tension=1}{v2,o1}
			\fmf{phantom,fore=black,tension=1}{v2,o2}
			\fmfforce{(1.1w,0.9h)}{o1}
			\fmfforce{(1.1w,0.1h)}{o2}
			\fmfv{decor.shape=hexagram,decor.filled=full, decor.size=3thin}{v2}
		\end{fmfgraph}
	\end{fmffile}
\end{gathered}\!\!
\begin{gathered}
	\begin{fmffile}{wgkappa-bubble-2pt-1PI-xxv}
		\begin{fmfgraph}(40,40)
			\fmfset{dash_len}{1.2mm}
			\fmfset{wiggly_len}{1.1mm} \fmfset{dot_len}{0.5mm}
			\fmfpen{0.25mm}
			\fmfleft{i}
			\fmfright{o}
			\fmf{phantom,tension=5}{i,v1}
			\fmf{phantom,tension=5}{v2,o}
			\fmf{phantom,fore=black,left,tension=0.4}{v1,v2,v1}
			\fmf{phantom,fore=black}{v1,v2}
			\fmffreeze
			\fmfright{o1,o2}
			\fmf{wiggly,fore=black,tension=1}{v2,o1}
			\fmf{wiggly,fore=black,tension=1}{v2,o2}
			\fmfforce{(1.1w,0.9h)}{o1}
			\fmfforce{(1.1w,0.1h)}{o2}
			\fmfv{decor.shape=hexagram,decor.filled=full, decor.size=3thin}{v2}
		\end{fmfgraph}
	\end{fmffile}
\end{gathered}
$$ 
$$
\begin{gathered}\label{eq:cephalopods-examples}
	\begin{fmffile}{wg2L-3loop2pt-1PR-x3x}
		\begin{fmfgraph}(56,56)
			\fmfset{dash_len}{1.2mm}
			\fmfset{wiggly_len}{1.1mm} \fmfset{dot_len}{0.5mm}
			\fmfpen{0.25mm}
			\fmfleft{i}
			\fmfright{o}
			\fmf{phantom,tension=5}{i,v1}
			\fmf{wiggly,fore=black,tension=2.5}{v2,o}
			\fmf{wiggly,fore=black,left,tension=0.5}{v1,v2,v1}
			\fmf{wiggly,fore=black}{v1,v2}
			\fmffreeze
			\fmfforce{(1.1w,0.5h)}{o}
			\fmffreeze
			\fmfright{n,m}
			\fmf{wiggly,fore=black,tension=1}{o,n}
			\fmf{wiggly,fore=black,tension=1}{o,m}
			\fmfforce{(1.35w,0.9h)}{n}
			\fmfforce{(1.35w,0.1h)}{m}
		\end{fmfgraph}
	\end{fmffile}
\end{gathered}\qquad\,\,\,\ \rightarrow \!\qquad
\begin{gathered}\label{eq:cephalopods-examples}
	\begin{fmffile}{wg2L-3loop2pt-1PR-xxvv}
		\begin{fmfgraph}(56,56)
			\fmfset{dash_len}{1.2mm}
			\fmfset{wiggly_len}{1.1mm} \fmfset{dot_len}{0.5mm}
			\fmfpen{0.25mm}
			\fmfleft{i}
			\fmfright{o}
			\fmf{phantom,tension=5}{i,v1}
			\fmf{phantom,fore=black,tension=2.5}{v2,o}
			\fmf{wiggly,fore=black,left,tension=0.5}{v1,v2,v1}
			\fmf{wiggly,fore=black}{v1,v2}
			\fmffreeze
			\fmfforce{(1.1w,0.5h)}{o}
			\fmffreeze
			\fmfright{n,m}
			\fmf{phantom,fore=black,tension=1}{o,n}
			\fmf{phantom,fore=black,tension=1}{o,m}
			\fmfforce{(1.35w,0.9h)}{n}
			\fmfforce{(1.35w,0.1h)}{m}
			\fmfv{decor.shape=hexagram,decor.filled=full, decor.size=3thin}{v2}
		\end{fmfgraph}
	\end{fmffile}
\end{gathered}\!\!\!
\begin{gathered}\label{eq:cephalopods-examples}
	\begin{fmffile}{wg2L-3loop2pt-1PR-xxv}
		\begin{fmfgraph}(56,56)
			\fmfset{dash_len}{1.2mm}
			\fmfset{wiggly_len}{1.1mm} \fmfset{dot_len}{0.5mm}
			\fmfpen{0.25mm}
			\fmfleft{i}
			\fmfright{o}
			\fmf{phantom,tension=5}{i,v1}
			\fmf{wiggly,fore=black,tension=2.5}{v2,o}
			\fmf{phantom,fore=black,left,tension=0.5}{v1,v2,v1}
			\fmf{phantom,fore=black}{v1,v2}
			\fmffreeze
			\fmfforce{(1.1w,0.5h)}{o}
			\fmffreeze
			\fmfright{n,m}
			\fmf{wiggly,fore=black,tension=1}{o,n}
			\fmf{wiggly,fore=black,tension=1}{o,m}
			\fmfforce{(1.35w,0.9h)}{n}
			\fmfforce{(1.35w,0.1h)}{m}
			\fmfv{decor.shape=hexagram,decor.filled=full, decor.size=3thin}{v2}
		\end{fmfgraph}
	\end{fmffile}
\end{gathered}
$$
$$
\begin{gathered}
	\begin{fmffile}{wg2L-bubble-3xx}
		\begin{fmfgraph}(55,55)
			\fmfset{dash_len}{1.2mm}
			\fmfset{wiggly_len}{1.1mm} \fmfset{dot_len}{0.5mm}
			\fmfpen{0.25mm}
			\fmftop{t1,t2,t3}
			\fmfbottom{b1,b2,b3}
			\fmf{phantom}{t1,v1,b1}
			\fmf{phantom}{t2,v2,b2}
			\fmf{phantom}{t3,v3,b3}
			\fmffreeze
			\fmf{wiggly,fore=black,right}{v1,v2,v1}
			\fmf{wiggly,fore=black,right}{v2,v3,v2}
			\fmf{wiggly,fore=black,tension=1}{t1,b1}
			\fmfforce{(0.25w,0.7h)}{t1}
			\fmfforce{(0.25w,0.3h)}{b1}
		\end{fmfgraph}
	\end{fmffile}
\end{gathered}\qquad \rightarrow \qquad 
\begin{gathered}
	\begin{fmffile}{wg2L-bubble-xxvv}
		\begin{fmfgraph}(55,55)
			\fmfset{dash_len}{1.2mm}
			\fmfset{wiggly_len}{1.1mm} \fmfset{dot_len}{0.5mm}
			\fmfpen{0.25mm}
			\fmftop{t1,t2,t3}
			\fmfbottom{b1,b2,b3}
			\fmf{phantom}{t1,v1,b1}
			\fmf{phantom}{t2,v2,b2}
			\fmf{phantom}{t3,v3,b3}
			\fmffreeze
			\fmf{wiggly,fore=black,right}{v1,v2,v1}
			\fmf{phantom,fore=black,right}{v2,v3,v2}
			\fmf{wiggly,fore=black,tension=1}{t1,b1}
			\fmfforce{(0.25w,0.7h)}{t1}
			\fmfforce{(0.25w,0.3h)}{b1}
			\fmfv{decor.shape=hexagram,decor.filled=full, decor.size=3thin}{v2}
		\end{fmfgraph}
	\end{fmffile}
\end{gathered}\!\!
\begin{gathered}
	\begin{fmffile}{wg2L-bubble-xxv}
		\begin{fmfgraph}(55,55)
			\fmfset{dash_len}{1.2mm}
			\fmfset{wiggly_len}{1.1mm} \fmfset{dot_len}{0.5mm}
			\fmfpen{0.25mm}
			\fmftop{t1,t2,t3}
			\fmfbottom{b1,b2,b3}
			\fmf{phantom}{t1,v1,b1}
			\fmf{phantom}{t2,v2,b2}
			\fmf{phantom}{t3,v3,b3}
			\fmffreeze
			\fmf{phantom,fore=black,right}{v1,v2,v1}
			\fmf{wiggly,fore=black,right}{v2,v3,v2}
			\fmf{phantom,fore=black,tension=1}{t1,b1}
			\fmfforce{(0.25w,0.7h)}{t1}
			\fmfforce{(0.25w,0.3h)}{b1}
			\fmfv{decor.shape=hexagram,decor.filled=full, decor.size=3thin}{v2}
		\end{fmfgraph}
	\end{fmffile}
\end{gathered}
$$
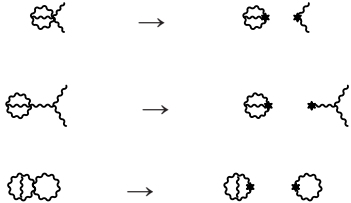
{\it ~~~~~~~~~~~~~~~~Some examples of ``cutting'' internal vertices in `cephalopod' diagrams.}
~\\


Since the cancellation induced by standard normal ordering the action is only partial, often with an infinite number of tadpole diagrams and cephalopods remaining, 
and since the expectation values of normal-ordered operators, $\langle:\!\mathcal{O}(\phi)\!\!:\rangle$, vanish only in the \emph{free} theory, 
one may hope to improve the definition of normal ordering. In particular, it would be extremely valuable to find a new form of normal ordering,
called here `\emph{complete normal ordering}', which ensures that one is doing perturbation theory around a true minimum of the full quantum effective action. When that minimum is at zero (corresponding to a trivial vacuum), such a `complete normal ordering' would cancel all tadpole diagrams to any order in perturbation theory and would ensure 
that the expectation values of `completely normal-ordered operators' computed in the full interacting theory vanish identically~\footnote{This is to be contrasted 
with `normal products' or `composite operators' where one requires that correlation functions involving generic insertions of such operators and elementary fields 
are well-defined \cite{Zimmermann73a,Zimmermann73b}.}. 

In this article we propose such a `complete normal ordering' procedure in the form
of a map $\mathcal{O}(\phi)\rightarrow \,\n\,\mathcal{O}(\phi)\,\n$ (by direct analogy to standard normal ordering) that ensures the cancellation of all `cephalopod' diagrams to any finite order in perturbation theory. By complete normal ordering the action of interest, the cancellation of \emph{all} cephalopods will be automatic and amounts to absorbing the `free from external line' pieces (of the above diagrams) into counterterms for the various couplings of the theory.\footnote{It may be useful to note that the only other approach (that the authors are aware of) that automatically cancels tadpoles in some contexts is the 2PI (or 3PI, etc.) quantum effective action approach \cite{CornwallJackiwTomboulis74}, see also \cite{PilaftsisTeresi15} and references therein. However, even though tadpoles are cancelled in, e.g., $\phi^4$ theory, in this approach tadpoles are \emph{not} cancelled for \emph{generic} interaction vertices, and one needs to consider higher and higher $n$PI formalisms to cancel all tadpoles -- a task that quickly becomes unwieldy in the $n$PI formalism. However, in the approach we propose here (based on `complete normal ordering'), all unwanted tadpoles will be cancelled automatically, to any loop order, and for \emph{generic} interactions.} We will phrase complete normal ordering in terms of generalisations of relations such as (\ref{eq:NO1}), (\ref{eq:Bell::phiN}) and (\ref{eq:deltagn}). These generalisations are the key equations for complete normal ordering and are displayed in ``boxed'' form in (\ref{eq:nmathcalFn2}), (\ref{eq:nphiNn}) and (\ref{eq:deltagn cno}) respectively.

We were led to make this proposal by a wish to simplify certain background field computations in string theories in non-trivial backgrounds. As is very well known, 
these may be formulated in terms of two-dimensional (super)conformal field theories, and
critical string theories may be thought of~\cite{DijkgraafVerlindeVerlinde88,Polchinski_v2} as (super)conformal fixed points in the general space of all (supersymmetric)
two-dimensional quantum field theories. The latter also include non-critical string theories, which are of interest in their own right.
These considerations motivate the study of generic two-dimensional quantum field theories and the renormalisation group (RG) flows that connect them. 
String theories in non-trivial backgrounds can be described as non-linear $\sigma$ models on the world-sheet, which are two-dimensional non-linear field theories with
second-order derivative interactions. Their local couplings correspond to background quantities such as the space-time metric, the dilaton field, etc..
Clearly, in order to study RG flow in the space of these backgrounds and couplings, one must first understand the renormalisation of the theory. 

This topic has been discussed in a very large literature, see e.g., Ref.~\!\cite{Ketov} 
and references therein. Nevertheless, it involves subtle issues such as the non-linear renormalisation of quantum fields~\cite{HowePapadopoulosStelle88}, 
the r\^ole of string loops~\cite{FischlerSusskind86a,FischlerSusskind86b,Polchinski88,CallanLovelaceNappiYost87,Tseytlin90}, 
and the treatment of moduli or zero modes that parametrise the classical background, which should presumably
be integrated out in deriving the effective world-sheet action. Having derived consistency conditions on string backgrounds, 
one would like to quantise strings in such backgrounds, constructing vertex operators~ \cite{CallanGan86} and correlation functions, which would
be an important step in studies of strings in the contexts of early-Universe cosmology and black holes. Carrying out this programme encounters
technical problems. In the words of~\cite{Zams}, in the context of a study of Liouville field theory on the Lobachevskiy plane: 
\begin{quote}
`\emph{Here we will not develop further the loop perturbation theory .... To go at higher loop diagrammatic calculations it is worth first to improve the technique to better handle the tadpole diagrams (which become rather numerous at higher orders)}\dots'
\end{quote}
Complete normal ordering takes care automatically of all tadpole diagrams of the theory of interest, and ensures
that the expectation value of the renormalised quantum field is identified with a solution to the equations of motion of the full quantum effective action. (Of course, in the context of Liouville theory exact results for the correlators are known \cite{Zams}, but this is clearly not the case for generic theories, and it is often of interest to carry out a perturbative expansion instead.)

A novel approach to the computation of the quantum effective action was recently proposed~\cite{GarbrechtMillington15}, where it was stressed that one should evaluate the relevant path integral around a saddle point of the full quantum effective action, rather than around a saddle point of the classical action. This idea has a long history and ultimately\footnote{The authors would like to thank Apostolos Pilaftsis for stressing this point.} goes back to the classic work of Coleman and Weinberg \cite{ColemanWeinberg73}. This suggestion is particularly natural, and is relevant, e.g., to problems where the quantum and classical trajectories are non-perturbatively far away from each other, see \cite{GarbrechtMillington15} and references therein. In the current document and in a follow-up paper \cite{EllisMavromatosSkliros15b} we propose that doing perturbation theory around a saddle-point of the full quantum effective action can be simply accomplished  by complete normal ordering the bare action of interest, as this ensures that the expectation value of the full renormalised quantum field is identified with a solution of the equations of motion of the full renormalised quantum effective action. When this solution (equivalently, vacuum configuration) is a trivial vacuum, this statement is equivalent to the requirement that all tadpole diagrams vanish, and complete normal ordering automatically cancels all tadpoles and (as a byproduct) all cephalopod Feynman diagrams from the corresponding renormalised generating function of connected Green functions.

In the current article we provide evidence for this conjecture in the case where the full quantum effective action has a minimum at vanishing field value. This (combined with the requirement that the quantum effective action be convex) is in turn equivalent to the statement that complete normal ordering cancels tadpoles to all orders in perturbation theory, and we demonstrate this explicitly for the case of a single scalar field with arbitrary non-derivative interaction vertices $\phi^3+\phi^4+\phi^5+\phi^6$, in generic spacetime backgrounds and with arbitrary (perturbative) local couplings. We also show (using a point-splitting ``trick") that this cancellation also holds for theories with generic derivative interactions. 

In Sec.~\ref{sec:GF} we introduce some preliminary material, we introduce the generating function of connected Green functions of interest, and determine its explicit form with generic bare couplings. We also recall the traditional tadpole cancellation procedure and apply it to the class of theories of interest.

In Sec.~\ref{sec:CNO} we introduce the notion of `complete normal ordering'. Starting from the definition in subsection \ref{sec:CNO-dfn}, we discuss a useful combinatorial interpretation in subsection \ref{sec:CI-CNO}, and a counterterm interpretation in subsection \ref{sec:counterterms}.


In Sec.~\ref{sec:CC} we present explicit perturbative evidence (up to three loops) that `complete normal ordering' the bare theory of interest results in tadpole- and cephalopod-free Feynman diagrams, focusing on the case where the vacuum of the full quantum effective action is at zero field value. In particular, in subsection \ref{sec:TCT} we show that the source counterterm that results from complete normal ordering is precisely identical to what was computed by the traditional (brute force) method of tadpole cancellation of Sec.~\ref{sec:GF}. In subsection \ref{sec:RCT} we display explicitly all the counterterms induced by complete normal ordering up to three loops. In subsection \ref{sec:VC}  we compute the vacuum contribution induced by complete normal ordering, and in subsection \ref{sec:CNOGF} we derive the tadpole- and cephalopod-free renormalised generating function of connected Green functions. Finally, in subsection \ref{sec:PS-CNOn} we generalise the results to arbitrary derivative interaction local scalar field theories.

In Sec.~\ref{sec:D} we end with a discussion of our results.
\vspace{0.1cm}

A concise overview of the results presented in this article can be found in \cite{Skliros15}.



\section{Preliminaries}\label{sec:GF}
Let us suppose that the quantum theory of interest has a generating function of the form:
\begin{equation}\label{eq:W(J)01}
e^{\tfrac{1}{\hbar}W(J)}=\int \mathcal{D}\phi \,e^{-\int \tfrac{1}{2}(\hbar\nabla \phi)^2+\frac{1}{2}m^2\phi^2-\frac{1}{2}\phi(\!\wf\!+\dm \!\!) \phi +\hat{\Lambda}+ \tfrac{1}{3!}\hat{g}\phi^3+ \tfrac{1}{4!}
\hat{\lambda}\phi^4+ \tfrac{1}{5!}\hat{\kappa}\phi^5+ \tfrac{1}{6!}\hat{\gamma}\phi^6+\mathcal{O}(\ell^5)}e^{\int (J+Y)\phi},
\end{equation}
(with $\int_z\dfn \int d^dz\sqrt{g}$) from which renormalised connected Green functions can be extracted via:
\begin{equation}\label{eq:W(J) GN}
\frac{1}{\hbar}W(J)=\sum\limits_{N=0}^{\infty}\frac{1}{N!}\int_{z_1}\dots\int_{z_N}G_N(z_1,\dots,z_N)J(z_1)\dots J(z_N).
\end{equation} 
The various local couplings appearing in (\ref{eq:W(J)01}) can take different forms depending on the theory of interest,\footnote{E.g., 
$
m^2(z)=M^2+\hbar^2\xi R_{(d)}+\dots,
$ 
with $R_{(d)}$ the $d$-dimensional Ricci scalar, $M^2$ a renormalised mass, and $\xi$ a renormalised  parameter that is required for the renormalisability of the theory.} and may also (with some caution) be interpreted as (strictly speaking non-linear) operators that generate derivative interactions --more about this later. 

If we denote the coupling associated to the $\phi^N$ interaction term by\footnote{So that $\hat{g}_0=\hat{\Lambda}$, $\hat{g}_1=-J-Y$, $\hat{g}_2=m^2-\dm $, $\hat{g}_3=\hat{g}$, $\hat{g}_4=\hat{\lambda}$, $\hat{g}_5=\hat{\kappa}$, and $\hat{g}_6=\hat{\gamma}$.\label{foot:couplings}} $\hat{g}_N$, then generically we can always decompose these as follows:
\begin{equation}\label{eq:ghatN}
\hat{g}_N\dfn \alpha^{N-2}(g_N+\delta_{\perp\n}g_N+\delta_{\n}g_N),\qquad {\rm with}\qquad g_N^B=Z^{-\frac{N}{2}}\hat{g}_N.
\end{equation}
$Z=1+\delta Z$ is a (possibly local) field renormalisation such that bare and renormalised fields are related via\footnote{We have found it convenient to define:
\begin{equation}\label{eq:deltam2 deltaZ}
\begin{aligned}
\begin{gathered}
	\begin{fmffile}{circle-shaded}
		\begin{fmfgraph}(40,40)
			\fmfset{dash_len}{1.2mm}
			\fmfset{wiggly_len}{1.1mm} \fmfset{dot_len}{0.5mm}
			\fmfpen{0.25mm}
			\fmfvn{decor.shape=circle,decor.filled=shaded, decor.size=5thin}{u}{1}
			\fmfleft{i}
			\fmfright{o}
			\fmf{phantom,fore=black,tension=5}{i,u1,o}
			\fmffreeze
			\fmfforce{(0w,0.35h)}{i}
			\fmfforce{(0w,0.35h)}{u1}
			\fmfforce{(0w,0.35h)}{o}
		\end{fmfgraph}
	\end{fmffile}
\end{gathered}\!\!\!
&\dfn\,-\delta_{\n} m^2 \, , \\
\begin{gathered}
	\begin{fmffile}{square-shaded}
		\begin{fmfgraph}(40,40)
			\fmfset{dash_len}{1.2mm}
			\fmfset{wiggly_len}{1.1mm} \fmfset{dot_len}{0.5mm}
			\fmfpen{0.25mm}
			\fmfvn{decor.shape=square,decor.filled=shaded, decor.size=5thin}{u}{1}
			\fmfleft{i}
			\fmfright{o}
			\fmf{phantom,fore=black,tension=5}{i,u1,o}
			\fmffreeze
			\fmfforce{(0.5w,0.35h)}{i}
			\fmfforce{(0.5w,0.35h)}{u1}
			\fmfforce{(0.5w,0.35h)}{o}
		\end{fmfgraph}
	\end{fmffile}
\end{gathered}&\dfn 
-\hbar^2\Big[\overleftarrow{\nabla}\delta Z\overrightarrow{\nabla}+\tfrac{1}{4}(\nabla \ln Z)^2+\tfrac{1}{2}(\nabla \delta Z)\overrightarrow{\nabla}+\tfrac{1}{2}\overleftarrow{\nabla} (\nabla \delta Z)\Big] \, , \\
\overleftarrow{\begin{gathered}
	\begin{fmffile}{square-shaded}
		\begin{fmfgraph}(40,40)
			\fmfset{dash_len}{1.2mm}
			\fmfset{wiggly_len}{1.1mm} \fmfset{dot_len}{0.5mm}
			\fmfpen{0.25mm}
			\fmfvn{decor.shape=square,decor.filled=shaded, decor.size=5thin}{u}{1}
			\fmfleft{i}
			\fmfright{o}
			\fmf{phantom,fore=black,tension=5}{i,u1,o}
			\fmffreeze
			\fmfforce{(0.5w,0.35h)}{i}
			\fmfforce{(0.5w,0.35h)}{u1}
			\fmfforce{(0.5w,0.35h)}{o}
		\end{fmfgraph}
	\end{fmffile}
\end{gathered}}&\dfn 
\hbar^2\Big[\overleftarrow{\nabla}\delta Z\overleftarrow{\nabla}-\tfrac{1}{4}(\nabla \ln Z)^2+\tfrac{1}{2}(\nabla^2 \delta Z)\Big] \, , \\
\overrightarrow{\begin{gathered}
	\begin{fmffile}{square-shaded}
		\begin{fmfgraph}(40,40)
			\fmfset{dash_len}{1.2mm}
			\fmfset{wiggly_len}{1.1mm} \fmfset{dot_len}{0.5mm}
			\fmfpen{0.25mm}
			\fmfvn{decor.shape=square,decor.filled=shaded, decor.size=5thin}{u}{1}
			\fmfleft{i}
			\fmfright{o}
			\fmf{phantom,fore=black,tension=5}{i,u1,o}
			\fmffreeze
			\fmfforce{(0.5w,0.35h)}{i}
			\fmfforce{(0.5w,0.35h)}{u1}
			\fmfforce{(0.5w,0.35h)}{o}
		\end{fmfgraph}
	\end{fmffile}
\end{gathered}}&\dfn 
\hbar^2\Big[\overrightarrow{\nabla}\delta Z\overrightarrow{\nabla}-\tfrac{1}{4}(\nabla \ln Z)^2+\tfrac{1}{2}(\nabla^2 \delta Z)\Big] \,  \\
\end{aligned}
\end{equation}} $\phi_B=Z^{\frac{1}{2}}\phi$. $g_N^B$ and $g_N$ are the corresponding bare and renormalised couplings respectively. The symbol $\alpha$ in (\ref{eq:ghatN}) contains all the $\hbar$ dependence of the couplings and in dimensional regularisation (where $d=n-2\epsilon$) also a scale  dependence $\mu$,
\begin{equation}\label{eq:alpha ell}
\alpha= \ell\big(\tfrac{\mu}{\hbar}\big)^{\epsilon}\qquad {\rm with}\qquad \ell=\hbar^{\frac{n}{2}}.
\end{equation}
(When these couplings are not operators, the beta functions, $\beta_N=\mu \frac{dg_N(\mu)}{d\mu}$, for the various renormalised couplings, $g_N(\mu)$, are defined by $dg_N^B/d\mu=0$.) 
We define the decomposition in (\ref{eq:ghatN}) such that the $\delta_{\n}g_N$ absorb all cephalopods (independently of whether or not they are naively infinite), and the $\delta_{\perp\n}g_N$ absorb all remaining divergences. The counterterm $\delta Z$ is fixed (e.g., by requiring the quantum effective action to have a canonical kinetic term) after all other divergences have been cancelled. 

Now, to $W(J)$ there corresponds a 1PI quantum effective action, $\Gamma(\varphi)$, with $\varphi(J)=\frac{1}{\hbar}\frac{1}{\sqrt{g}}\frac{\delta}{\delta J} W$, the two being related (under the assumtion that $J(\varphi)$ exists and is single-valued) via a Legendre transform, $\frac{1}{\hbar}W(J)=-\frac{1}{\hbar}\Gamma(\varphi)+\int J\varphi$. Suppose now that $\bar{\varphi}$ solves the full quantum equations of motion of $\Gamma(\varphi)$,
\begin{equation}\label{eq:dGdv=0}
\frac{1}{\hbar}\frac{1}{\sqrt{g}}\frac{\delta \Gamma(\varphi)}{\delta \varphi}\Big|_{\varphi=\bar{\varphi}}=0.
\end{equation}
Then from the Legendre transform it follows that the tadpole diagrams determine $\bar{\varphi}$:
\begin{equation}\label{eq:dWdJ=v}
\frac{1}{\hbar}\frac{\delta W(J)}{\delta J}\Big|_{J=0}=\bar{\varphi},
\end{equation}
and we want to set up the quantum field theory in such a way that the path integral computes quantum fluctuations around the minimum, $\bar{\varphi}$, of the quantum effective action. We will here consider theories for which the {\it true} vacuum is at $\bar\varphi=0$, and this amounts to choosing an appropriate source counterterm such that all tadpoles are absent. 

The main focus in this paper will be to show that `complete normal ordering' the bare action of interest determines uniquely the source counterterm that is required to cancel all tadpoles (to all orders in perturbation theory), while also cancelling all remaining cephalopods.  We will not have anything new to say about $\delta_{\perp\n}g_N$ (but it is useful to note that in the absence of derivative interactions and in $n=2$ dimensions all $\delta_{\perp\n}g_N$ can be set to zero). We will throughout absorb all $\delta_{\perp\n}g_N$'s into the $g_N$'s, and will usually not display the $\alpha$-dependence explicitly --one can always refer back to these definitions to restore either of these.\footnote{The mass and length dimensions of the various quantities appearing are:
\begin{equation*}\label{eq:dimensions}
\begin{aligned}
&[g_N] = M^{n-N(\frac{n}{2}-1)},\qquad [\phi]=[\phi_B] = \frac{1}{\sqrt{M L^{d-1}ML}},\qquad [\hbar]=ML,\\ 
&\qquad\qquad[\mu]=M,\qquad 
[\ell]=(ML)^{\frac{n}{2}},\qquad [\alpha]=(ML)^{\frac{n}{2}}L^{-\epsilon}
\end{aligned}
\end{equation*}}
That is, dropping the explicit $\alpha$-dependence we will write below:\label{coupling notation} $g_0=\Lambda$, $g_1=-J$, $g_2=m^2$, $g_3=g$, $g_4=\lambda$, $g_5=\kappa$, $g_6=\gamma$, and similarly for the counterterms, $\delta_{\n} g_0=\delta_{\n} \Lambda$, $\delta_{\n}g_1=-Y$, $\delta_{\n}g_2=-\dm$, $\delta_{\n} g_3=\delta_{\n}g$, $\delta_{\n}g_4=\delta_{\n}\lambda$, $\delta_{\n}g_5=\delta_{\n}\kappa$, and $\delta_{\n }g_6=\delta_{\n}\gamma$. 


Notice we are not assuming $W(0)=0$, because in curved space-time the $N=0$ terms (in particular the 1PI vacuum diagrams) will contribute to the vacuum energy. 

Evaluating (\ref{eq:W(J)01}) within perturbation theory (in $\ell$) (and for generic counterterms at this stage) is standard procedure and so we will be very brief. We will include contributions up to and including $\mathcal{O}(\ell^4$), and so we have  kept terms up to this order in the corresponding bare action -- recall the $\ell$-dependence of the couplings in (\ref{eq:ghatN}) and the comments below that. 
Using the standard trick \cite{BailinLove} of replacing the renormalised field, $\phi$, by a functional derivative with respect to the source, $\delta_{J}=\frac{1}{\hbar}\frac{1}{\sqrt{g}}\frac{\delta}{\delta J}$, in the interaction and counterterms, the first step is to write (\ref{eq:W(J)01}) in terms of functional derivatives of a (dressed) Gaussian:
\begin{equation}\label{eq:W(J)inter}
\begin{aligned}
e^{\tfrac{1}{\hbar}W(J)}
=&\, \,N\,e^{Q_1}e^{-\int \hat{\Lambda}}e^{\int Y\delta_J}e^{U(J)},
\end{aligned}
\end{equation}
with 
\begin{equation}\label{eq:U(J)expo defn}
e^{U(J)}\dfn e^{-\int \tfrac{1}{3!}\hat{g}\delta_J^3+ \tfrac{1}{4!}
\hat{\lambda}\delta_J^4+ \tfrac{1}{5!}\hat{\kappa}\delta_J^5+ \tfrac{1}{6!}\hat{\gamma}\delta_J^6+\mathcal{O}(\ell^5)}e^{\tfrac{1}{2}\wiggly}
\end{equation}
We have defined a determinant factor $N\dfn N'\,{\det}^{-\tfrac{1}{2}}\big(\Delta+m^2\big)$ (with $N'$ a normalisation constant and $\Delta$ the standard Laplacian\footnote{In particular, $\Delta\dfn -(\hbar\nabla)^2=-\tfrac{\hbar^2}{\sqrt{g}}\,\partial_a\big(\sqrt{g}g^{ab}\partial_{b}\big)$. }) associated to the free field Gaussian fluctuations, a set of vacuum bubble contributions, $Q_1$, associated to mass and field renormalisation counterterms,
\begin{equation}\label{eq:Q1}
Q_1\dfn \tfrac{1}{2}\Big(
\hspace{0.5cm}\begin{gathered}
	\begin{fmffile}{bubble1}
		\begin{fmfgraph}(40,40)
			\fmfset{dash_len}{1.2mm}
			\fmfset{wiggly_len}{1.1mm} \fmfset{dot_len}{0.5mm}
			\fmfpen{0.25mm}
			\fmfvn{decor.shape=circle,decor.filled=shaded, decor.size=5thin}{u}{1}
			\fmfleft{i}
			\fmfright{o}
			\fmf{dashes,fore=black,tension=5,left}{i,u1,i}
			\fmffreeze
			\fmfforce{(-w,0.35h)}{i}
			\fmfforce{(0w,0.35h)}{u1}
			\fmfforce{(1.1w,0.35h)}{o}
		\end{fmfgraph}\!\!\!\!
	\end{fmffile}
\end{gathered}\hspace{0cm}
+
\tfrac{1}{2}\hspace{0.6cm}
\begin{gathered}
	\begin{fmffile}{bubble2}
		\begin{fmfgraph}(40,40)
			\fmfset{dash_len}{1.2mm}
			\fmfset{wiggly_len}{1.1mm} \fmfset{dot_len}{0.5mm}
			\fmfpen{0.25mm}
			\fmfvn{decor.shape=circle,decor.filled=shaded, decor.size=5thin}{u}{2}
			\fmfleft{i}
			\fmfright{o}
			\fmf{dashes,fore=black,tension=5,left}{i,u1,u2,i}
			\fmffreeze
			\fmfforce{(-w,0.35h)}{i}
			\fmfforce{(0w,0.35h)}{u1}
			\fmfforce{(-1w,0.35h)}{u2}
			\fmfforce{(1.1w,0.35h)}{o}
		\end{fmfgraph}\!\!\!\!
	\end{fmffile}
\end{gathered}\hspace{0cm}
+
\tfrac{1}{3}\!\!
\begin{gathered}
	\begin{fmffile}{bubble3}
		\begin{fmfgraph}(80,80)
			\fmfset{dash_len}{1.2mm}
			\fmfset{wiggly_len}{1.1mm} \fmfset{dot_len}{0.5mm}
			\fmfpen{0.25mm}
			\fmfvn{decor.shape=circle,decor.filled=shaded, decor.size=5thin}{x}{3}
			\fmfsurroundn{u}{6}
			\fmf{phantom,fore=black,tension=1}{u1,x1,c,v,u4}
			\fmf{phantom,fore=black,tension=1}{u2,u,c,x3,u5}
			\fmf{phantom,fore=black,tension=1}{u3,x2,c,t,u6}
			\fmffreeze
			\fmf{dashes,fore=black,tension=1,right=.7}{x1,x2}
			\fmf{dashes,fore=black,tension=1,right=.7}{x2,x3}
			\fmf{dashes,fore=black,tension=1,right=.7}{x3,x1}
		\end{fmfgraph}\!\!
	\end{fmffile}
\end{gathered}
+\dots+
\hspace{0.5cm}
\begin{gathered}
	\begin{fmffile}{bubble1b}
		\begin{fmfgraph}(40,40)
			\fmfset{dash_len}{1.2mm}
			\fmfset{wiggly_len}{1.1mm} \fmfset{dot_len}{0.5mm}
			\fmfpen{0.25mm}
			\fmfvn{decor.shape=square,decor.filled=shaded, decor.size=5thin}{u}{1}
			\fmfleft{i}
			\fmfright{o}
			\fmf{plain,fore=black,tension=5,left}{i,u1,i}
			\fmffreeze
			\fmfforce{(-w,0.35h)}{i}
			\fmfforce{(0w,0.35h)}{u1}
			\fmfforce{(1.1w,0.35h)}{o}
		\end{fmfgraph}\!\!\!\!
	\end{fmffile}
\end{gathered}
+
\tfrac{1}{2}\hspace{0.6cm}
\begin{gathered}
	\begin{fmffile}{bubble2b}
		\begin{fmfgraph}(40,40)
			\fmfset{dash_len}{1.2mm}
			\fmfset{wiggly_len}{1.1mm} \fmfset{dot_len}{0.5mm}
			\fmfpen{0.25mm}
			\fmfvn{decor.shape=square,decor.filled=shaded, decor.size=5thin}{u}{2}
			\fmfleft{i}
			\fmfright{o}
			\fmf{plain,fore=black,tension=5,left}{i,u1,u2,i}
			\fmffreeze
			\fmfforce{(-w,0.35h)}{i}
			\fmfforce{(0w,0.35h)}{u1}
			\fmfforce{(-1w,0.35h)}{u2}
			\fmfforce{(1.1w,0.35h)}{o}
		\end{fmfgraph}\!\!\!\!
	\end{fmffile}
\end{gathered}
+
\tfrac{1}{3}\!\!
\begin{gathered}
	\begin{fmffile}{bubble3b}
		\begin{fmfgraph}(80,80)
			\fmfset{dash_len}{1.2mm}
			\fmfset{wiggly_len}{1.1mm} \fmfset{dot_len}{0.5mm}
			\fmfpen{0.25mm}
			\fmfvn{decor.shape=square,decor.filled=shaded, decor.size=5thin}{x}{3}
			\fmfsurroundn{u}{6}
			\fmf{phantom,fore=black,tension=1}{u1,x1,c,v,u4}
			\fmf{phantom,fore=black,tension=1}{u2,u,c,x3,u5}
			\fmf{phantom,fore=black,tension=1}{u3,x2,c,t,u6}
			\fmffreeze
			\fmf{plain,fore=black,tension=1,right=.7}{x1,x2}
			\fmf{plain,fore=black,tension=1,right=.7}{x2,x3}
			\fmf{plain,fore=black,tension=1,right=.7}{x3,x1}
		\end{fmfgraph}\!\!
	\end{fmffile}
\end{gathered}
+\dots\Big),
\end{equation}
and various (dressed) propagators \emph{with} sources:\footnote{It may also be useful to display a more explicit definition of the (wiggly) dressed propagator $\mathcal{G}(z,w)$:
\begin{equation*}\label{eq:W0(J)}
\begin{aligned}
N\,e^{Q_1}e^{\tfrac{1}{2}\wiggly}&\dfn \int \mathcal{D}\phi\,e^{-\int \tfrac{1}{2}(\hbar\nabla \phi)^2+\tfrac{1}{2}m^2\phi^2+\int\frac{1}{2}\phi(\dm \!\!+\!\wf\!) \phi}e^{\int J\phi}.
\end{aligned}
\end{equation*}
Clearly, when the counterterms $\delta Z$ and $\delta_{\n}  m^2$ vanish in the absence of interactions then the free renormalised propagator, $\mathcal{G}(z,w)$, in (\ref{eq:dressed_prop}) reduces to the plain propagator $G(z,w)$, and when only the $\delta_{\n}  m^2$ counterterm vanishes $\mathcal{G}(z,w)$ reduces to $\mathscr{G}(z,w)$; see also (\ref{eq:GGG}).}
\begin{equation}\label{eq:dressed_prop}
\begin{aligned}
&\begin{gathered}
	\begin{fmffile}{dressed-propagator}
		\begin{fmfgraph}(40,40)
			\fmfset{dash_len}{1.2mm}
			\fmfset{wiggly_len}{1.1mm} \fmfset{dot_len}{0.5mm}
			\fmfpen{0.25mm}
			\fmfleft{i}
			\fmfright{o}
			\fmf{wiggly,fore=black}{i,o}
			\fmffreeze
			\fmfforce{(-1w,0.35h)}{i}
			\fmfforce{(1w,0.35h)}{o}
		\end{fmfgraph}
	\end{fmffile}
\end{gathered}\,\,\,
=\hspace{0.6cm}
\begin{gathered}
	\begin{fmffile}{dressed-dashes0}
		\begin{fmfgraph}(40,40)
			\fmfset{dash_len}{1.2mm}
			\fmfset{wiggly_len}{1.1mm} \fmfset{dot_len}{0.5mm}
			\fmfpen{0.25mm}
			\fmfleft{i}
			\fmfright{o}
			\fmf{dashes,fore=black}{i,o}
			\fmffreeze
			\fmfforce{(-1w,0.35h)}{i}
			\fmfforce{(1w,0.35h)}{o}
		\end{fmfgraph}
	\end{fmffile}
\end{gathered}\,\,
+\hspace{0.6cm}
\begin{gathered}
	\begin{fmffile}{dressed-dashes1}
		\begin{fmfgraph}(40,40)
			\fmfset{dash_len}{1.2mm}
			\fmfset{wiggly_len}{1.1mm} \fmfset{dot_len}{0.5mm}
			\fmfpen{0.25mm}
			\fmfvn{decor.shape=circle,decor.filled=shaded, decor.size=5thin}{u}{1}
			\fmfleft{i}
			\fmfright{o}
			\fmf{dashes,fore=black,tension=5}{i,u1,o}
			\fmffreeze
			\fmfforce{(-w,0.35h)}{i}
			\fmfforce{(0w,0.35h)}{u1}
			\fmfforce{(1.1w,0.35h)}{o}
		\end{fmfgraph}
	\end{fmffile}
\end{gathered}\hspace{0.2cm}
+\hspace{0.6cm}
\begin{gathered}
	\begin{fmffile}{dressed-dashes2}
		\begin{fmfgraph}(40,40)
			\fmfset{dash_len}{1.2mm}
			\fmfset{wiggly_len}{1.1mm} \fmfset{dot_len}{0.5mm}
			\fmfpen{0.25mm}
			\fmfvn{decor.shape=circle,decor.filled=shaded, decor.size=5thin}{u}{2}
			\fmfleft{i}
			\fmfright{o}
			\fmf{dashes,fore=black,tension=5}{i,u1,u2,o}
			\fmffreeze
			\fmfforce{(-w,0.35h)}{i}
			\fmfforce{(0w,0.35h)}{u1}
			\fmfforce{(1w,0.35h)}{u2}
			\fmfforce{(2w,0.35h)}{o}
		\end{fmfgraph}
	\end{fmffile}
\end{gathered}\hspace{0.65cm}
+\hspace{0.65cm}
\begin{gathered}
	\begin{fmffile}{dressed-dashes3}
		\begin{fmfgraph}(40,40)
			\fmfset{dash_len}{1.2mm}
			\fmfset{wiggly_len}{1.1mm} \fmfset{dot_len}{0.5mm}
			\fmfpen{0.25mm}
			\fmfvn{decor.shape=circle,decor.filled=shaded, decor.size=5thin}{u}{3}
			\fmfleft{i}
			\fmfright{o}
			\fmf{dashes,fore=black,tension=5}{i,u1,u2,u3,o}
			\fmffreeze
			\fmfforce{(-w,0.35h)}{i}
			\fmfforce{(0w,0.35h)}{u1}
			\fmfforce{(1w,0.35h)}{u2}
			\fmfforce{(2w,0.35h)}{u3}
			\fmfforce{(3w,0.35h)}{o}
		\end{fmfgraph}
	\end{fmffile}
\end{gathered}\hspace{1.05cm}
+\hspace{0.5cm}\dots\\
&
\begin{gathered}
	\begin{fmffile}{dressed-propagator-wf}
		\begin{fmfgraph}(40,40)
			\fmfset{dash_len}{1.2mm}
			\fmfset{wiggly_len}{1.1mm} \fmfset{dot_len}{0.5mm}
			\fmfpen{0.25mm}
			\fmfleft{i}
			\fmfright{o}
			\fmf{dashes,fore=black}{i,o}
			\fmffreeze
			\fmfforce{(-1w,0.35h)}{i}
			\fmfforce{(1w,0.35h)}{o}
		\end{fmfgraph}
	\end{fmffile}
\end{gathered}\,\,\,
=\hspace{0.6cm}
\begin{gathered}
	\begin{fmffile}{dressed-plain0}
		\begin{fmfgraph}(40,40)
			\fmfset{dash_len}{1.2mm}
			\fmfset{wiggly_len}{1.1mm} \fmfset{dot_len}{0.5mm}
			\fmfpen{0.25mm}
			\fmfleft{i}
			\fmfright{o}
			\fmf{plain,fore=black}{i,o}
			\fmffreeze
			\fmfforce{(-1w,0.35h)}{i}
			\fmfforce{(1w,0.35h)}{o}
		\end{fmfgraph}
	\end{fmffile}
\end{gathered}\,\,
+\hspace{0.6cm}
\begin{gathered}
	\begin{fmffile}{dressed-plain1}
		\begin{fmfgraph}(40,40)
			\fmfset{dash_len}{1.2mm}
			\fmfset{wiggly_len}{1.1mm} \fmfset{dot_len}{0.5mm}
			\fmfpen{0.25mm}
			\fmfvn{decor.shape=square,decor.filled=shaded, decor.size=5thin}{u}{1}
			\fmfleft{i}
			\fmfright{o}
			\fmf{plain,fore=black,tension=5}{i,u1,o}
			\fmffreeze
			\fmfforce{(-w,0.35h)}{i}
			\fmfforce{(0w,0.35h)}{u1}
			\fmfforce{(1.1w,0.35h)}{o}
		\end{fmfgraph}
	\end{fmffile}
\end{gathered}\hspace{0.2cm}
+\hspace{0.6cm}
\begin{gathered}
	\begin{fmffile}{dressed-plain2}
		\begin{fmfgraph}(40,40)
			\fmfset{dash_len}{1.2mm}
			\fmfset{wiggly_len}{1.1mm} \fmfset{dot_len}{0.5mm}
			\fmfpen{0.25mm}
			\fmfvn{decor.shape=square,decor.filled=shaded, decor.size=5thin}{u}{2}
			\fmfleft{i}
			\fmfright{o}
			\fmf{plain,fore=black,tension=5}{i,u1,u2,o}
			\fmffreeze
			\fmfforce{(-w,0.35h)}{i}
			\fmfforce{(0w,0.35h)}{u1}
			\fmfforce{(1w,0.35h)}{u2}
			\fmfforce{(2w,0.35h)}{o}
		\end{fmfgraph}
	\end{fmffile}
\end{gathered}\hspace{0.65cm}
+\hspace{0.65cm}
\begin{gathered}
	\begin{fmffile}{dressed-plain3}
		\begin{fmfgraph}(40,40)
			\fmfset{dash_len}{1.2mm}
			\fmfset{wiggly_len}{1.1mm} \fmfset{dot_len}{0.5mm}
			\fmfpen{0.25mm}
			\fmfvn{decor.shape=square,decor.filled=shaded, decor.size=5thin}{u}{3}
			\fmfleft{i}
			\fmfright{o}
			\fmf{plain,fore=black,tension=5}{i,u1,u2,u3,o}
			\fmffreeze
			\fmfforce{(-w,0.35h)}{i}
			\fmfforce{(0w,0.35h)}{u1}
			\fmfforce{(1w,0.35h)}{u2}
			\fmfforce{(2w,0.35h)}{u3}
			\fmfforce{(3w,0.35h)}{o}
		\end{fmfgraph}
	\end{fmffile}
\end{gathered}\hspace{1.05cm}
+\hspace{0.5cm}\dots
\end{aligned}
\end{equation}
The propagators appearing will be denoted by:
\begin{equation}\label{eq:GGG}
\begin{aligned}
\mathcal{G}(z,w)\qquad &\leftrightarrow \qquad \wiggly \, , \\
\mathscr{G}(z,w)\qquad &\leftrightarrow \qquad \,\quad\dashes \, , \\
G(z,w)\qquad &\leftrightarrow \qquad\quad \,\vanilla \, ,
\end{aligned}
\end{equation}
The first propagator in (\ref{eq:GGG}) contains both mass and field renormalisation counterterm contributions, the second only field renormalisation counterterms, whereas the third is the ``plain'' vanilla propagator, defined by:
\begin{equation}\label{eq:vanillapropagator}
\vanilla \dfn \int_{z,w}J(z)G(z,w)J(w),\qquad \big(\Delta+m^2\big)G(z,w)\dfn \frac{1}{\sqrt{g}}\delta^d(z-w),
\end{equation}
The quantities appearing on the left-hand sides of (\ref{eq:dressed_prop}) are defined in a similar manner to the first equation in (\ref{eq:vanillapropagator}). 
A few examples will suffice to understand the notation on the right-hand sides of (\ref{eq:Q1}) and (\ref{eq:dressed_prop}):
$$
\hspace{0.6cm}
\begin{gathered}
	\begin{fmffile}{dressed-plainasd1}
		\begin{fmfgraph}(40,40)
			\fmfset{dash_len}{1.2mm}
			\fmfset{wiggly_len}{1.1mm} \fmfset{dot_len}{0.5mm}
			\fmfpen{0.25mm}
			\fmfvn{decor.shape=circle,decor.filled=shaded, decor.size=5thin}{u}{1}
			\fmfleft{i}
			\fmfright{o}
			\fmf{plain,fore=black,tension=5}{i,u1,o}
			\fmffreeze
			\fmfforce{(-w,0.35h)}{i}
			\fmfforce{(0w,0.35h)}{u1}
			\fmfforce{(1.1w,0.35h)}{o}
		\end{fmfgraph}
	\end{fmffile}
\end{gathered}\hspace{0.2cm}
= \int_w\Big(\int_zJ(z)G(z,w)\Big)(-\delta_{\n}  m^2) \Big(\int_{z'}J(z')G(z',w)\Big) \, ,
$$
\begin{equation*}
\begin{aligned}
\hspace{0.6cm}
\begin{gathered}
	\begin{fmffile}{dressed-dashes2asd}
		\begin{fmfgraph}(40,40)
			\fmfset{dash_len}{1.2mm}
			\fmfset{wiggly_len}{1.1mm} \fmfset{dot_len}{0.5mm}
			\fmfpen{0.25mm}
			\fmfvn{decor.shape=circle,decor.filled=shaded, decor.size=5thin}{u}{2}
			\fmfleft{i}
			\fmfright{o}
			\fmf{dashes,fore=black,tension=5}{i,u1,u2,o}
			\fmffreeze
			\fmfforce{(-w,0.35h)}{i}
			\fmfforce{(0w,0.35h)}{u1}
			\fmfforce{(1w,0.35h)}{u2}
			\fmfforce{(2w,0.35h)}{o}
		\end{fmfgraph}
	\end{fmffile}
\end{gathered}\hspace{0.65cm}
= \int_w&\int_{w'}\,\Big(\int_zJ(z)\mathscr{G}(z,w)\Big)(-\delta_{\n}  m^2) \mathscr{G}(w,w')(-\delta_{\n}  m^2) \Big(\int_{z'}J(z') \mathscr{G}(z',w')\Big) \, ,
\end{aligned}
\end{equation*}
$$
\hspace{0.6cm}
\begin{gathered}
	\begin{fmffile}{bubble2b}
		\begin{fmfgraph}(40,40)
			\fmfset{dash_len}{1.2mm}
			\fmfset{wiggly_len}{1.1mm} \fmfset{dot_len}{0.5mm}
			\fmfpen{0.25mm}
			\fmfvn{decor.shape=square,decor.filled=shaded, decor.size=5thin}{u}{2}
			\fmfleft{i}
			\fmfright{o}
			\fmf{plain,fore=black,tension=5,left}{i,u1,u2,i}
			\fmffreeze
			\fmfforce{(-w,0.35h)}{i}
			\fmfforce{(0w,0.35h)}{u1}
			\fmfforce{(-1w,0.35h)}{u2}
			\fmfforce{(1.1w,0.35h)}{o}
		\end{fmfgraph}\!\!\!\!
	\end{fmffile}
\end{gathered}=
\int_z\int_w\,G(z,w)\wf G(w,z)\wf ,
$$
where the right-most derivative appearing on the right-hand side of the last displayed relation, recall (\ref{eq:deltam2 deltaZ}), is understood to act on the left-most propagator, $G(z,w)$, as implied by the cyclicity of the diagram on the left-hand side.

Notice that the source counterterm, $\int Y\delta_J$, has not been included in the definition of $U(J)$, see (\ref{eq:U(J)expo defn}), and so $U(J)$ will contain \emph{all} connected Feynman diagram contributions (except for $Q_1$), \emph{including} tadpoles, one-particle reducible (1PR) diagrams and one-particle irreducible (1PI) diagrams. 

We next evaluate the Gaussian integrals in $U(J)$, see (\ref{eq:U(J)expo defn}), within perturbation theory, the natural expansion parameter being $\ell$. (We recall that 
the $\ell$-dependences of all couplings are explicit in (\ref{eq:ghatN}) (see also (\ref{eq:alpha ell}) and the footnote there). This is a standard procedure and we display the full result for all terms up to (and including) 
$\mathcal{O}(\ell^4)$:
\begin{equation}\label{eq:U(Jwiggly)}
\begin{aligned}
&U(J) = \tfrac{1}{2}
\begin{gathered}
	\begin{fmffile}{wfree}
		\begin{fmfgraph}(50,50)
			\fmfset{dash_len}{1.2mm}
			\fmfset{wiggly_len}{1.1mm} \fmfset{dot_len}{0.5mm}
			\fmfpen{0.25mm}
			\fmfleft{i}
			\fmfright{o}
			\fmf{wiggly,fore=black,tension=5}{i,o}
		\end{fmfgraph}
	\end{fmffile}
\end{gathered}
-\hat{g}
\Big(
\tfrac{1}{2}
\begin{gathered}
	\begin{fmffile}{wgtadpole}
		\fmfset{dash_len}{1.2mm}
		\begin{fmfgraph}(50,50)
			\fmfset{dash_len}{1.2mm}
			\fmfset{wiggly_len}{1.1mm} \fmfset{dot_len}{0.5mm}
			\fmfpen{0.25mm}
			\fmfleft{i}
			\fmfright{o}
			\fmf{phantom,tension=5}{i,v1}
			\fmf{wiggly,fore=black,tension=0.8}{v2,o}
			\fmf{wiggly,fore=black,left,tension=0.08}{v1,v2,v1}
			\fmf{phantom}{v1,v2}
		\end{fmfgraph}
	\end{fmffile}
\end{gathered}  
+\tfrac{1}{3!}
\begin{gathered}
	\begin{fmffile}{wg-3p}
		\begin{fmfgraph}(40,40)
			\fmfset{dash_len}{1.2mm}
			\fmfset{wiggly_len}{1.1mm} \fmfset{dot_len}{0.5mm}
			\fmfpen{0.25mm}
			\fmfleft{i}
			\fmfright{o1,o2}
			\fmf{wiggly,fore=black,tension=5}{i,v1}
			\fmf{wiggly,fore=black,tension=5}{v1,o1}
			\fmf{wiggly,fore=black,tension=5}{v1,o2}
		\end{fmfgraph}
	\end{fmffile}
\end{gathered}
\Big)
+\hat{g}^2
\Big(
\tfrac{1}{12}\!
\begin{gathered}
	\begin{fmffile}{wg2-2loopbubble-1PI}
		\begin{fmfgraph}(40,40)
			\fmfset{dash_len}{1.2mm}
			\fmfset{wiggly_len}{1.1mm} \fmfset{dot_len}{0.5mm}
			\fmfpen{0.25mm}
			\fmfleft{i}
			\fmfright{o}
			\fmf{phantom,tension=5}{i,v1}
			\fmf{phantom,tension=5}{v2,o}
			\fmf{wiggly,fore=black,left,tension=0.4}{v1,v2,v1}
			\fmf{wiggly,fore=black}{v1,v2}
		\end{fmfgraph}
	\end{fmffile}
\end{gathered}\!
+\tfrac{1}{8}
\begin{gathered}
	\begin{fmffile}{wbubble4}
		\begin{fmfgraph}(50,50)
			\fmfset{dash_len}{1.2mm}
			\fmfset{wiggly_len}{1.1mm} \fmfset{dot_len}{0.5mm}
			\fmfpen{0.25mm}
			\fmfleft{i}
			\fmfright{o}
			\fmf{phantom,tension=5}{i,v1}
			\fmf{phantom,tension=5}{v2,o}
			\fmf{wiggly,fore=black,left,tension=0.4}{v1,v3,v1}
			\fmf{wiggly,fore=black,right,tension=0.4}{v2,v4,v2}
			\fmf{wiggly,fore=black}{v3,v4}
		\end{fmfgraph}
	\end{fmffile}
\end{gathered} 
+\tfrac{1}{4}\,   
\begin{gathered}
	\begin{fmffile}{wg2-2pt}
		\begin{fmfgraph}(40,40)
			\fmfset{dash_len}{1.2mm}
			\fmfset{wiggly_len}{1.1mm} \fmfset{dot_len}{0.5mm}
			\fmfpen{0.25mm}
			\fmfleft{i}
			\fmfright{o}
			\fmf{wiggly,fore=black,tension=1}{i,v1}
			\fmf{wiggly,fore=black,tension=1}{v2,o}
			\fmf{wiggly,fore=black,left,tension=0.4}{v1,v2,v1}
			\fmffreeze
			\fmfforce{(-.12w,0.5h)}{i}
			\fmfforce{(1.1w,0.5h)}{o}
		\end{fmfgraph}
	\end{fmffile}
\end{gathered}\,
+\tfrac{1}{4}\,\,
\begin{gathered}
	\begin{fmffile}{wtadpole_g2b}
		\begin{fmfgraph}(40,40)
			\fmfset{dash_len}{1.2mm}
			\fmfset{wiggly_len}{1.1mm} \fmfset{dot_len}{0.5mm}
			\fmfpen{0.25mm}
			\fmfleft{i1,i2}
			\fmfright{o}
			\fmf{wiggly,fore=black,tension=1}{i1,v1}
			\fmf{wiggly,fore=black,tension=1}{i2,v1}
			\fmf{phantom,tension=1}{v3,o}
			\fmf{wiggly,fore=black,left,tension=0.2}{v2,v3,v2}
			\fmf{wiggly,fore=black,tension=0.5}{v1,v2}
			\fmffreeze
			\fmfforce{(-.3w,0.9h)}{i1}
			\fmfforce{(-.3w,0.1h)}{i2}
			\fmfforce{(-.05w,0.5h)}{v1}
		\end{fmfgraph}
	\end{fmffile}
\end{gathered}
+\tfrac{1}{8}
\begin{gathered}
	\begin{fmffile}{w2-2_g2}
		\begin{fmfgraph}(45,45)
			\fmfset{dash_len}{1.2mm}
			\fmfset{wiggly_len}{1.1mm} \fmfset{dot_len}{0.5mm}
			\fmfpen{0.25mm}
			\fmfsurroundn{i}{4}
			\fmf{wiggly,fore=black}{i1,n,m,i4}
			\fmf{wiggly,fore=black}{i2,n}
			\fmf{wiggly,fore=black}{m,i3}
		\end{fmfgraph}
	\end{fmffile}
\end{gathered}
\Big)\\
&-\hat{\lambda}
\Big(
\tfrac{1}{8}
\begin{gathered}
	\begin{fmffile}{wlambdabubble}
		\begin{fmfgraph}(35,35)
			\fmfset{dash_len}{1.2mm}
			\fmfset{wiggly_len}{1.1mm} \fmfset{dot_len}{0.5mm}
			\fmfpen{0.25mm}
			\fmftop{t1,t2,t3}
			\fmfbottom{b1,b2,b3}
			\fmf{phantom}{t1,v1,b1}
			\fmf{phantom}{t2,v2,b2}
			\fmf{phantom}{t3,v3,b3}
			\fmffreeze
			\fmf{wiggly,fore=black,right}{v1,v2,v1}
			\fmf{wiggly,fore=black,right}{v2,v3,v2}
		\end{fmfgraph}
	\end{fmffile}
\end{gathered}
+\tfrac{1}{4}
\begin{gathered}
	\begin{fmffile}{wlambdaself}
		\begin{fmfgraph}(33,33)
			\fmfset{dash_len}{1.2mm}
			\fmfset{wiggly_len}{1.1mm} \fmfset{dot_len}{0.5mm}
			\fmfpen{0.25mm}
			\fmftop{s}
			\fmfleft{a}
			\fmfright{b}
			\fmf{wiggly,fore=black}{a,v}
			\fmf{wiggly,fore=black}{b,v}
			\fmf{wiggly,fore=black,right,tension=.7}{v,v}
			\fmffreeze
			\fmfforce{(0w,0.2h)}{a}
			\fmfforce{(w,0.2h)}{b}
		\end{fmfgraph}
	\end{fmffile}
\end{gathered}
+\tfrac{1}{4!}
\begin{gathered}
	\begin{fmffile}{wlambdax}
		\begin{fmfgraph}(30,30)
			\fmfset{dash_len}{1.2mm}
			\fmfset{wiggly_len}{1.1mm} \fmfset{dot_len}{0.5mm}
			\fmfpen{0.25mm}
			\fmfleft{i1,i2}
			\fmfright{o1,o2}
			\fmf{wiggly,fore=black}{i1,v,o2}
			\fmf{wiggly,fore=black}{i2,v,o1}
		\end{fmfgraph}
	\end{fmffile}
\end{gathered}
\Big)
-\hat{g}^3
\Big(
\tfrac{1}{4}\!\!\!
\begin{gathered}
	\begin{fmffile}{wg3tadpole1}
		\begin{fmfgraph}(65,65)
			\fmfset{dash_len}{1.2mm}
			\fmfset{wiggly_len}{1.1mm} \fmfset{dot_len}{0.5mm}
			\fmfpen{0.25mm}
			\fmftop{t}
			\fmfbottom{b}
			\fmfleft{l}
			\fmfright{r}
			\fmf{phantom,fore=black,tension=9}{t,u,v,b}
			\fmf{phantom,fore=black,tension=9}{l,s,x,r}
			\fmf{wiggly,fore=black,tension=.01,left}{u,v,u}
			\fmf{phantom,fore=black,tension=0.01}{s,x,s}
			\fmf{wiggly,fore=black,tension=1}{u,v}
			\fmf{wiggly,fore=black,tension=1}{x,r}
		\end{fmfgraph}
	\end{fmffile}
\end{gathered}
+\tfrac{1}{4}
\begin{gathered}
	\begin{fmffile}{wg3-tadpole2}
		\begin{fmfgraph}(50,50)
			\fmfset{dash_len}{1.2mm}
			\fmfset{wiggly_len}{1.1mm} \fmfset{dot_len}{0.5mm}
			\fmfpen{0.25mm}
			\fmfleft{i}
			\fmfright{o}
			\fmf{phantom,tension=5}{i,v1}
			\fmf{wiggly,fore=black,tension=.04}{v2,o}
			\fmf{wiggly,fore=black,left,tension=0.01}{v1,v3,v1}
			\fmf{wiggly,fore=black,right,tension=0.01}{v2,v4,v2}
			\fmf{wiggly,fore=black,tension=0.03}{v3,v4}
			\fmffreeze
			\fmfforce{(1.2w,0.5h)}{o}
			\fmfforce{(.3w,0.5h)}{v3}
		\end{fmfgraph}
	\end{fmffile}
\end{gathered} \,
+\tfrac{1}{4}\,\,
\begin{gathered}
	\begin{fmffile}{wfish}
		\begin{fmfgraph}(40,40)
			\fmfset{dash_len}{1.2mm}
			\fmfset{wiggly_len}{1.1mm} \fmfset{dot_len}{0.5mm}
			\fmfpen{0.25mm}
			\fmfleft{i,j}
			\fmfright{o}
			\fmf{wiggly,fore=black,tension=1}{i,v1}
			\fmf{wiggly,fore=black,tension=1}{j,v1}
			\fmf{wiggly,fore=black}{v1,v2}
			\fmf{wiggly,fore=black,tension=1}{v3,o}
			\fmf{wiggly,fore=black,left,tension=0.3}{v2,v3,v2}
			\fmffreeze
			\fmfforce{(-.2w,0.8h)}{i}
			\fmfforce{(-.2w,0.2h)}{j}
			\fmfforce{(0.05w,0.5h)}{v1}
			\fmfforce{(1.2w,0.5h)}{o}
		\end{fmfgraph}
	\end{fmffile}
\end{gathered}
+\tfrac{1}{8}
\begin{gathered}
	\begin{fmffile}{wg3-2tadpoles}
		\begin{fmfgraph}(40,40)
			\fmfset{dash_len}{1.2mm}
			\fmfset{wiggly_len}{1.1mm} \fmfset{dot_len}{0.5mm}
			\fmfpen{0.25mm}
			\fmfleft{i}
			\fmfright{o1,o2}
			\fmf{wiggly,fore=black,tension=1}{i,v1}
			\fmf{phantom,tension=1}{v1,u1,01}
			\fmf{phantom,tension=1}{v1,u2,o2}
			\fmf{wiggly,fore=black,tension=0.8,left}{u1,o1,u1}
			\fmf{wiggly,fore=black,tension=0.4,right}{u2,o2,u2}
			\fmf{wiggly,fore=black,tension=1}{v1,u1}
			\fmf{wiggly,fore=black,tension=1}{v1,u2}
		\end{fmfgraph}
	\end{fmffile}
\end{gathered}
+\tfrac{1}{4}
\begin{gathered}
	\begin{fmffile}{wg3-cat}
		\begin{fmfgraph}(40,40)
			\fmfset{dash_len}{1.2mm}
			\fmfset{wiggly_len}{1.1mm} \fmfset{dot_len}{0.5mm}
			\fmfpen{0.25mm}
			\fmfleft{i1,i2}
			\fmfright{o1,o2}
			\fmf{phantom,tension=1}{i2,u,v1,v2,o1}
			\fmf{wiggly,fore=black,tension=3}{i1,v1,v2,o2}
			\fmf{wiggly,fore=black,tension=0.6,left}{i2,u,i2}
			\fmf{wiggly,fore=black,tension=1}{u,v1}
			\fmf{wiggly,fore=black,tension=1}{v2,o1}
			\fmffreeze
			\fmfforce{(1.1w,0.9h)}{o1}
			\fmfforce{(1.1w,0.1h)}{o2}
			\fmfforce{(0.8w,0.5h)}{v2}
			\fmfforce{(0.4w,0.5h)}{v1}
			\fmfforce{(0.08w,0.1h)}{i1}
		\end{fmfgraph}
	\end{fmffile}
\end{gathered}
+\tfrac{1}{8}
\begin{gathered}
	\begin{fmffile}{wg3-crystal}
		\begin{fmfgraph}(45,45)
			\fmfset{dash_len}{1.2mm}
			\fmfset{wiggly_len}{1.1mm} \fmfset{dot_len}{0.5mm}
			\fmfpen{0.25mm}
			\fmfsurround{i1,i2,i3,i4,i5,i6}
			\fmf{wiggly,fore=black,tension=1}{i6,v}
			\fmf{wiggly,fore=black,tension=1}{i1,v}
			\fmf{wiggly,fore=black,tension=1}{v,c}
		 	\fmf{wiggly,fore=black,tension=1}{c,u}
			\fmf{phantom,tension=1}{u,i2}
			\fmf{phantom,fore=black,tension=1}{u,i3}
			\fmf{wiggly,fore=black,tension=1}{c,s,i4}
			\fmf{wiggly,fore=black,tension=1}{s,i5}
		\end{fmfgraph}
	\end{fmffile}
\end{gathered}
+\tfrac{1}{6}
\begin{gathered}
	\begin{fmffile}{wg3-3ptlog}
		\begin{fmfgraph}(35,35)
			\fmfset{dash_len}{1.2mm}
			\fmfset{wiggly_len}{1.1mm} \fmfset{dot_len}{0.5mm}
			\fmfpen{0.25mm}
			\fmfsurroundn{i}{6}
			\fmf{phantom,fore=black}{i1,v,u,i4}
			\fmf{wiggly,fore=black}{v,i1}
			\fmf{phantom,fore=black}{i2,s,t,i5}
			\fmf{wiggly,fore=black}{i5,t}
			\fmf{phantom,fore=black}{i3,w,x,i6}
			\fmf{wiggly,fore=black}{i3,w}
			\fmfi{wiggly,fore=black}{fullcircle scaled .55w shifted (.51w,.5h)}
			\fmffreeze
			\fmfforce{(1.25w,0.5h)}{i1}
			\fmfforce{(0.16w,1.1h)}{i3}
			\fmfforce{(0.16w,-.1h)}{i5}
		\end{fmfgraph}
	\end{fmffile}
\end{gathered}\,\,
\Big)\\
&+\hat{g}\hat{\lambda}\Big(
\tfrac{1}{4}
\begin{gathered}
	\begin{fmffile}{wglambda-tadpole}
		\begin{fmfgraph}(45,45)
			\fmfset{dash_len}{1.2mm}
			\fmfset{wiggly_len}{1.1mm} \fmfset{dot_len}{0.5mm}
			\fmfpen{0.25mm}
			\fmftop{t1,t2,t3,t4}
        			\fmfbottom{b1,b2,b3,b4}
        			\fmf{phantom}{t1,v1,b1}
        			\fmf{phantom}{t2,v2,b2}
			\fmf{phantom}{t3,v3,b3}
			\fmf{phantom}{t4,v4,b4}
        			\fmffreeze
			\fmf{wiggly,fore=black,right,tension=0.7}{v1,v2,v1}
        			\fmf{wiggly,fore=black,right,tension=0.7}{v2,v3,v2}
        			\fmf{wiggly,fore=black,tension=3}{v3,v4}
			\fmffreeze
			\fmfforce{(1.1w,0.5h)}{v4}
		\end{fmfgraph}
	\end{fmffile}
\end{gathered}
+\tfrac{1}{4}
\begin{gathered}
	\begin{fmffile}{wglambda-tad}
		\begin{fmfgraph}(40,40)
			\fmfset{dash_len}{1.2mm}
			\fmfset{wiggly_len}{1.1mm} \fmfset{dot_len}{0.5mm}
			\fmfpen{0.25mm}
			\fmfsurround{a,b,c}
			\fmf{wiggly,fore=black,tension=1}{c,v,a}
			\fmf{wiggly,fore=black,left,tension=1.2}{v,b,v}
			\fmf{wiggly,fore=black,tension=1.3}{c,c}
		\end{fmfgraph}
	\end{fmffile}
\end{gathered}
+\tfrac{1}{4}\,
\begin{gathered}
	\begin{fmffile}{wglamba-swim}
		\begin{fmfgraph}(40,40)
			\fmfset{dash_len}{1.2mm}
			\fmfset{wiggly_len}{1.1mm} \fmfset{dot_len}{0.5mm}
			\fmfpen{0.25mm}
			\fmfleft{a}
			\fmfright{f,g}
			\fmf{wiggly,fore=black}{a,v,b}
			\fmf{wiggly,fore=black,tension=.6}{v,v}
			\fmf{wiggly,fore=black,tension=1}{b,f}
			\fmf{wiggly,fore=black,tension=1}{b,g}
			\fmffreeze
			\fmfforce{(1.1w,0.8h)}{f}
			\fmfforce{(1.1w,0.2h)}{g}
			\fmffreeze
			\fmfforce{(-.1w,0.5h)}{a}
		\end{fmfgraph}
	\end{fmffile}
\end{gathered}\,
+\tfrac{1}{3!}
\begin{gathered}
	\begin{fmffile}{wglambdatadpole}
		\begin{fmfgraph}(40,40)
			\fmfset{dash_len}{1.2mm}
			\fmfset{wiggly_len}{1.1mm} \fmfset{dot_len}{0.5mm}
			\fmfpen{0.25mm}
			\fmfleft{i}
			\fmfright{o}
			\fmf{phantom,tension=5}{i,v1}
			\fmf{wiggly,fore=black,tension=2.5}{v2,o}
			\fmf{wiggly,fore=black,left,tension=0.5}{v1,v2,v1}
			\fmf{wiggly,fore=black}{v1,v2}
			\fmffreeze
			\fmfforce{(1.1w,0.5h)}{o}
		\end{fmfgraph}
	\end{fmffile}
\end{gathered}
+\tfrac{1}{4}\,\,
\begin{gathered}
	\begin{fmffile}{wglambda-dart}
		\begin{fmfgraph}(40,40)
			\fmfset{dash_len}{1.2mm}
			\fmfset{wiggly_len}{1.1mm} \fmfset{dot_len}{0.5mm}
			\fmfpen{0.25mm}
			\fmfleft{i,j}
			\fmfright{o}
			\fmf{wiggly,fore=black,tension=5}{i,v1}
			\fmf{wiggly,fore=black,tension=5}{j,v1}
			\fmf{wiggly,fore=black,tension=0.8}{v2,o}
			\fmf{wiggly,fore=black,left,tension=0.08}{v1,v2,v1}
			\fmf{phantom}{v1,v2}
			\fmffreeze
			\fmfforce{(-.2w,0.8h)}{i}
			\fmfforce{(-.2w,0.2h)}{j}
		\end{fmfgraph}
	\end{fmffile}
\end{gathered}
+\tfrac{1}{12}
\begin{gathered}
	\begin{fmffile}{wglambda-3ptadpole}
		\begin{fmfgraph}(30,30)
			\fmfset{dash_len}{1.2mm}
			\fmfset{wiggly_len}{1.1mm} \fmfset{dot_len}{0.5mm}
			\fmfpen{0.25mm}
			\fmftop{t}
			\fmfbottom{a,b,c}
			\fmf{wiggly,fore=black,tension=1}{a,v}
			\fmf{wiggly,fore=black,tension=1}{b,v}
			\fmf{wiggly,fore=black,tension=1}{c,v}
			\fmf{wiggly,fore=black,tension=3}{v,t}
			\fmf{wiggly,fore=black,tension=0.7,left}{t,t}
		\end{fmfgraph}
	\end{fmffile}
\end{gathered}
+\tfrac{1}{12}
\begin{gathered}
	\begin{fmffile}{wglambda-5ptree}
		\begin{fmfgraph}(30,30)
			\fmfset{dash_len}{1.2mm}
			\fmfset{wiggly_len}{1.1mm} \fmfset{dot_len}{0.5mm}
			\fmfpen{0.25mm}
			\fmftop{t1,t2}
			\fmfbottom{a,b,c}
			\fmf{wiggly,fore=black,tension=1}{a,v}
			\fmf{wiggly,fore=black,tension=1}{b,v}
			\fmf{wiggly,fore=black,tension=1}{c,v}
			\fmf{wiggly,fore=black,tension=2}{v,t}
			\fmf{wiggly,fore=black,tension=0.7,left,straight}{t,t1}
			\fmf{wiggly,fore=black,tension=0.7,left,straight}{t,t2}
			\fmffreeze
			\fmfforce{(0.5w,-0.2h)}{b}
		\end{fmfgraph}
	\end{fmffile}
\end{gathered}
\Big)
-\hat{\kappa}
\Big(
\tfrac{1}{8}\!\!
\begin{gathered}
	\begin{fmffile}{wkappa-tadpole}
		\begin{fmfgraph}(60,60)
			\fmfset{dash_len}{1.2mm}
			\fmfset{wiggly_len}{1.1mm} \fmfset{dot_len}{0.5mm}
			\fmfpen{0.25mm}
			\fmftop{t1,t2,t3}
			\fmfbottom{b1,b2,b3}
			\fmf{phantom}{t1,v1,b1}
			\fmf{phantom}{t2,v2,b2}
			\fmf{phantom}{t3,v3,b3}
			\fmf{wiggly,fore=black,right,tension=1}{v1,v2,v1}
			\fmf{wiggly,fore=black,right,tension=1}{v2,v3,v2}
			\fmf{wiggly,fore=black,tension=1}{v2,b2}
		\end{fmfgraph}
	\end{fmffile}
\end{gathered}\!\!
+\tfrac{1}{12}
\begin{gathered}
	\begin{fmffile}{wkappa-3ptself}
		\begin{fmfgraph}(40,40)
			\fmfset{dash_len}{1.2mm}
			\fmfset{wiggly_len}{1.1mm} \fmfset{dot_len}{0.5mm}
			\fmfpen{0.25mm}
			\fmftop{t}
			\fmfbottom{a,b,c}
			\fmf{wiggly,fore=black,tension=1}{a,v}
			\fmf{wiggly,fore=black,tension=1}{b,v}
			\fmf{wiggly,fore=black,tension=1}{c,v}
			\fmf{wiggly,fore=black,tension=1.4,left}{v,t,v}
		\end{fmfgraph}
	\end{fmffile}
\end{gathered}
+\tfrac{1}{5!}
\begin{gathered}
	\begin{fmffile}{wkappa-5pt}
		\begin{fmfgraph}(40,40)
			\fmfset{dash_len}{1.2mm}
			\fmfset{wiggly_len}{1.1mm} \fmfset{dot_len}{0.5mm}
			\fmfpen{0.25mm}
			\fmfsurround{u1,u2,u3,u4,u5}
			\fmf{wiggly,fore=black,tension=1}{u1,v}
			\fmf{wiggly,fore=black,tension=1}{u2,v}
			\fmf{wiggly,fore=black,tension=1}{u3,v}
			\fmf{wiggly,fore=black,tension=1}{u4,v}
			\fmf{wiggly,fore=black,tension=1}{u5,v}
		\end{fmfgraph}
	\end{fmffile}
\end{gathered}
\Big)\\
&+
\hat{g}^4
\Big(
\tfrac{1}{48}\,
\begin{gathered}
	\begin{fmffile}{wg4-3tadpole}
		\begin{fmfgraph}(35,35)
			\fmfset{dash_len}{1.2mm}
			\fmfset{wiggly_len}{1.1mm} \fmfset{dot_len}{0.5mm}
			\fmfpen{0.25mm}
			\fmfsurround{u1,u2,u3}
			\fmf{wiggly,fore=black,tension=1}{u1,v}
			\fmf{wiggly,fore=black,tension=1}{u2,v}
			\fmf{wiggly,fore=black,tension=1}{u3,v}
			\fmf{wiggly,fore=black,tension=1,left}{u1,u1}
			\fmf{wiggly,fore=black,tension=1,left}{u2,u2}
			\fmf{wiggly,fore=black,tension=1,left}{u3,u3}
		\end{fmfgraph}
	\end{fmffile}
\end{gathered}\,\,
+\tfrac{1}{48}
\begin{gathered}
	\begin{fmffile}{wg4-6pttree}
		\begin{fmfgraph}(45,45)
			\fmfset{dash_len}{1.2mm}
			\fmfset{wiggly_len}{1.1mm} \fmfset{dot_len}{0.5mm}
			\fmfpen{0.25mm}
			\fmfsurround{u1,u2,u3,u4,u5,u6}
			\fmf{wiggly,fore=black,tension=1}{u1,v}
			\fmf{wiggly,fore=black,tension=1}{u2,v}
			\fmf{wiggly,fore=black,tension=1}{u3,u}
			\fmf{wiggly,fore=black,tension=1}{u4,u}
			\fmf{wiggly,fore=black,tension=1}{u5,s}
			\fmf{wiggly,fore=black,tension=1}{u6,s}
			\fmf{wiggly,fore=black,tension=1,left,straight}{s,c}
			\fmf{wiggly,fore=black,tension=1,left,straight}{u,c}
			\fmf{wiggly,fore=black,tension=1,left,straight}{v,c}
		\end{fmfgraph}
	\end{fmffile}
\end{gathered}
+\tfrac{1}{16}\,
\begin{gathered}
	\begin{fmffile}{wg4-2pt-2tadpole}
		\begin{fmfgraph}(24,24)
			\fmfset{dash_len}{1.2mm}
			\fmfset{wiggly_len}{1.1mm} \fmfset{dot_len}{0.5mm}
			\fmfpen{0.25mm}
			\fmfleft{i1,i2}
			\fmfright{o1,o2}
			\fmf{wiggly,fore=black,tension=1}{o1,v,u,i1}
			\fmf{wiggly,fore=black,tension=1}{o2,v}
			\fmf{wiggly,fore=black,tension=1}{u,i2}
			\fmf{wiggly,fore=black,tension=0.75}{i1,i1}
			\fmf{wiggly,fore=black,tension=0.75}{i2,i2}
			\fmffreeze
			\fmfforce{(1.3w,1.3h)}{o1}
			\fmfforce{(1.3w,-.3h)}{o2}
			\fmfforce{(0.9w,0.5h)}{v}
		\end{fmfgraph}
	\end{fmffile}
\end{gathered}
+\tfrac{1}{16}\,
\begin{gathered}
	\begin{fmffile}{wg4-4pt-1tadpole}
		\begin{fmfgraph}(40,40)
			\fmfset{dash_len}{1.2mm}
			\fmfset{wiggly_len}{1.1mm} \fmfset{dot_len}{0.5mm}
			\fmfpen{0.25mm}
			\fmfsurround{u1,u2,u3,u4,u5,u6}
			\fmf{wiggly,fore=black,tension=1}{u1,v}
			\fmf{wiggly,fore=black,tension=1}{u2,v}
			\fmf{wiggly,fore=black,tension=1}{u3,u}
			\fmf{wiggly,fore=black,tension=1}{u4,u}
			\fmf{phantom,fore=black,tension=1}{u5,s}
			\fmf{phantom,fore=black,tension=1}{u6,s}
			\fmf{wiggly,fore=black,tension=1,left,straight}{s,c}
			\fmf{wiggly,fore=black,tension=1,left,straight}{u,c}
			\fmf{wiggly,fore=black,tension=1,left,straight}{v,c}
			\fmf{wiggly,fore=black,tension=.6}{s,s}
			\fmffreeze
			\fmfforce{(0.5w,0.5h)}{u5}
			\fmfforce{(0.5w,0.5h)}{u6}
		\end{fmfgraph}
	\end{fmffile}
\end{gathered}
+\tfrac{1}{16}\,\,\,
\begin{gathered}
	\begin{fmffile}{wg4-0pt-2tadpole}
		\begin{fmfgraph}(40,40)
			\fmfset{dash_len}{1.2mm}
			\fmfset{wiggly_len}{1.1mm} \fmfset{dot_len}{0.5mm}
			\fmfpen{0.25mm}
			\fmfleft{i}
			\fmfright{o}
			\fmf{wiggly,fore=black,tension=1}{i,v1}
			\fmf{wiggly,fore=black,tension=1}{v2,o}
			\fmf{wiggly,fore=black,left,tension=0.4}{v1,v2,v1}
			\fmf{wiggly,fore=black,tension=0.55}{i,i}
			\fmf{wiggly,fore=black,tension=0.55}{o,o}
		\end{fmfgraph}
	\end{fmffile}
\end{gathered}\,\,
+\tfrac{1}{16}\,
\begin{gathered}
	\begin{fmffile}{wg4-0pt-3loop1PIa}
		\begin{fmfgraph}(27,27)
			\fmfset{dash_len}{1.2mm}
			\fmfset{wiggly_len}{1.1mm} \fmfset{dot_len}{0.5mm}
			\fmfpen{0.25mm}
			\fmfsurround{a,b,c,d}
			\fmf{phantom,fore=black,tension=1,curved}{a,b,c,d,a}
			\fmf{wiggly,fore=black,tension=1,right}{a,b}
			\fmf{wiggly,fore=black,tension=1,right}{c,d}
			\fmf{wiggly,fore=black,tension=1,right}{b,c}
			\fmf{wiggly,fore=black,tension=1,right}{d,a}
			\fmf{wiggly,fore=black,tension=1,straight}{a,b}
			\fmf{wiggly,fore=black,tension=1,straight}{c,d}
		\end{fmfgraph}
	\end{fmffile}
\end{gathered}\,
+\tfrac{1}{16}\,\,\,
\begin{gathered}
	\begin{fmffile}{wg4-4pt-1loop}
		\begin{fmfgraph}(40,40)
			\fmfset{dash_len}{1.2mm}
			\fmfset{wiggly_len}{1.1mm} \fmfset{dot_len}{0.5mm}
			\fmfpen{0.25mm}
			\fmfleft{a,b}
			\fmfright{c,d}
			\fmf{wiggly,fore=black,tension=1}{a,v}
			\fmf{wiggly,fore=black,tension=1}{b,v}
			\fmf{wiggly,fore=black,tension=1}{v,s}
			\fmf{wiggly,fore=black,tension=1}{t,u}
			\fmf{wiggly,fore=black,tension=0.3,left}{s,t,s}
			\fmf{wiggly,fore=black,tension=1}{u,c}
			\fmf{wiggly,fore=black,tension=1}{u,d}	
			\fmffreeze
			\fmfforce{(1.3w,0.8h)}{c}
			\fmfforce{(1.3w,0.2h)}{d}
			\fmfforce{(1w,0.5h)}{u}
			\fmfforce{(-0.06w,0.5h)}{v}
			\fmffreeze
			\fmfforce{(-0.3w,0.8h)}{a}
			\fmfforce{(-0.3w,0.2h)}{b}
		\end{fmfgraph}
	\end{fmffile}
\end{gathered}\,\,\,
+\tfrac{1}{8}\!\!\!
\begin{gathered}
	\begin{fmffile}{wg4tadpole1}
		\begin{fmfgraph}(65,65)
			\fmfset{dash_len}{1.2mm}
			\fmfset{wiggly_len}{1.1mm} \fmfset{dot_len}{0.5mm}
			\fmfpen{0.25mm}
			\fmftop{t}
			\fmfbottom{b}
			\fmfleft{l}
			\fmfright{r}
			\fmf{phantom,fore=black,tension=9}{t,u,v,b}
			\fmf{phantom,fore=black,tension=9}{l,s,x,r}
			\fmf{wiggly,fore=black,tension=.01,left}{u,v,u}
			\fmf{phantom,fore=black,tension=0.01}{s,x,s}
			\fmf{wiggly,fore=black,tension=1}{u,v}
			\fmf{wiggly,fore=black,tension=1}{x,r}
			\fmf{wiggly,fore=black,tension=0.7}{r,r}
		\end{fmfgraph}
	\end{fmffile}
\end{gathered}\,\,\,
+\tfrac{1}{8}\!\!\!
\begin{gathered}
	\begin{fmffile}{wg4-tadpole-2pt}
		\begin{fmfgraph}(65,65)
			\fmfset{dash_len}{1.2mm}
			\fmfset{wiggly_len}{1.1mm} \fmfset{dot_len}{0.5mm}
			\fmfpen{0.25mm}
			\fmftop{t}
			\fmfbottom{b}
			\fmfleft{l}
			\fmfright{r}
			\fmf{phantom,fore=black,tension=9}{t,u,v,b}
			\fmf{phantom,fore=black,tension=9}{l,s,x,r}
			\fmf{wiggly,fore=black,tension=.01,left}{u,v,u}
			\fmf{phantom,fore=black,tension=0.01}{s,x,s}
			\fmf{wiggly,fore=black,tension=1}{u,v}
			\fmf{wiggly,fore=black,tension=1}{x,r}
			\fmffreeze
			\fmfright{a,b}
			\fmf{wiggly,fore=black,tension=1}{r,a}
			\fmf{wiggly,fore=black,tension=1}{r,b}
			\fmffreeze
			\fmfforce{(1.2w,0.8h)}{a}
			\fmfforce{(1.2w,0.2h)}{b}
		\end{fmfgraph}
	\end{fmffile}
\end{gathered}\,\,\,\\
&+\tfrac{1}{8}\,\,
\begin{gathered}
	\begin{fmffile}{wg4-2pt-doubletadpole}
		\begin{fmfgraph}(40,40)
			\fmfset{dash_len}{1.2mm}
			\fmfset{wiggly_len}{1.1mm} \fmfset{dot_len}{0.5mm}
			\fmfpen{0.25mm}
			\fmfleft{i,j}
			\fmfright{o}
			\fmf{wiggly,fore=black,tension=1}{i,v1}
			\fmf{wiggly,fore=black,tension=1}{j,v1}
			\fmf{wiggly,fore=black}{v1,v2}
			\fmf{wiggly,fore=black,tension=1}{v3,o}
			\fmf{wiggly,fore=black,left,tension=0.3}{v2,v3,v2}
			\fmf{wiggly,fore=black,tension=0.75}{o,o}
			\fmffreeze
			\fmfforce{(-.2w,0.8h)}{i}
			\fmfforce{(-.2w,0.2h)}{j}
			\fmfforce{(0.05w,0.5h)}{v1}
			\fmfforce{(1.2w,0.5h)}{o}
		\end{fmfgraph}
	\end{fmffile}
\end{gathered}\,\,\,\,
+\tfrac{1}{8}\,\,
\begin{gathered}
	\begin{fmffile}{wg4-2pt-1PR-twoloop}
		\begin{fmfgraph}(50,50)
			\fmfset{dash_len}{1.2mm}
			\fmfset{wiggly_len}{1.1mm} \fmfset{dot_len}{0.5mm}
			\fmfpen{0.25mm}
			\fmfleft{i}
			\fmfright{o}
			\fmf{wiggly,fore=black,tension=5}{i,v1}
			\fmf{wiggly,fore=black,tension=5}{v2,o}
			\fmf{wiggly,fore=black,left,tension=0.4}{v1,v3,v1}
			\fmf{wiggly,fore=black,right,tension=0.4}{v2,v4,v2}
			\fmf{wiggly,fore=black}{v3,v4}
			\fmffreeze
			\fmfforce{(-.25w,0.5h)}{i}
			\fmfforce{(1.25w,0.5h)}{o}
			\end{fmfgraph}
	\end{fmffile}
\end{gathered}\,\,
+\tfrac{1}{8}\,\,\,\,
\begin{gathered}
	\begin{fmffile}{wg4-2pt-2tadpoles}
		\begin{fmfgraph}(30,30)
			\fmfset{dash_len}{1.2mm}
			\fmfset{wiggly_len}{1.1mm} \fmfset{dot_len}{0.5mm}
			\fmfpen{0.25mm}
			\fmfleft{i1,i2}
			\fmfright{o1,o2}
			\fmf{wiggly,fore=black}{i1,v1,v2,o1}
			\fmf{wiggly,fore=black}{i2,v1,v2,o2}
			\fmffreeze
			\fmfforce{(1.15w,0.9h)}{o1}
			\fmfforce{(-.2w,0.9h)}{i1}
			\fmfforce{(1.2w,-0.1h)}{o2}
			\fmfforce{(-.3w,-0.1h)}{i2}	
			\fmfforce{(0.75w,0.5h)}{v2}
			\fmfforce{(0.15w,0.5h)}{v1}
			\fmf{wiggly,fore=black,tension=0.75}{i1,i1}
			\fmf{wiggly,fore=black,tension=0.75}{o1,o1}
		\end{fmfgraph}
	\end{fmffile}
\end{gathered}\,\,\,\,
+\tfrac{1}{8}\,\,
\begin{gathered}
	\begin{fmffile}{wg4-4ptlog}
		\begin{fmfgraph}(40,40)
			\fmfset{dash_len}{1.2mm}
			\fmfset{wiggly_len}{1.1mm} \fmfset{dot_len}{0.5mm}
			\fmfpen{0.25mm}
			\fmfsurroundn{i}{4}
			\fmf{phantom,fore=black}{i1,v,u,i3}
			\fmf{wiggly,fore=black}{v,i1}
			\fmf{wiggly,fore=black}{i3,u}
			\fmf{phantom,fore=black}{i2,s,t,i4}
			\fmf{wiggly,fore=black}{i4,t}
			\fmf{wiggly,fore=black}{i2,s}
			\fmfi{wiggly,fore=black}{fullcircle scaled .5w shifted (.51w,.5h)}
			\fmffreeze
			\fmfforce{(1.15w,0.5h)}{i1}
			\fmfforce{(0.5w,1.1h)}{i2}
			\fmfforce{(-.1w,0.5h)}{i3}
			\fmfforce{(0.5w,-.1h)}{i4}
		\end{fmfgraph}
	\end{fmffile}
\end{gathered}\,
+\tfrac{1}{8}\begin{gathered}\,\,
	\begin{fmffile}{wg4-6pttreelong}
		\begin{fmfgraph}(55,55)
			\fmfset{dash_len}{1.2mm}
			\fmfset{wiggly_len}{1.1mm} \fmfset{dot_len}{0.5mm}
			\fmfpen{0.25mm}
			\fmfsurroundn{i}{6}
			\fmf{wiggly,fore=black}{i1,v,u,s,t,i4}
			\fmf{wiggly,fore=black}{i6,v}
			\fmf{wiggly,fore=black}{i4,t}
			\fmf{wiggly,fore=black}{i5,t}
			\fmf{wiggly,fore=black}{i2,u}
			\fmf{wiggly,fore=black}{i3,s}
			\fmffreeze
			\fmfforce{(1.12w,0.4h)}{i1}
			\fmfforce{(-.15w,0.423h)}{i4}
			\fmfforce{(.26w,0.16h)}{i5}
			\fmfforce{(0.67w,0.15h)}{i6}
		\end{fmfgraph}
	\end{fmffile}
\end{gathered}\,\,
+\tfrac{1}{4!}\!\!
\begin{gathered}
	\begin{fmffile}{wpeace}
		\begin{fmfgraph}(75,75)
			\fmfset{dash_len}{1.2mm}
			\fmfset{wiggly_len}{1.1mm} \fmfset{dot_len}{0.5mm}
			\fmfpen{0.25mm}
			\fmfsurroundn{i}{3}
			\fmf{phantom,fore=black}{i1,v,i2}
			\fmf{phantom,fore=black}{i2,u,i3}
			\fmf{phantom,fore=black}{i3,s,i1}
			\fmfi{wiggly,fore=black}{fullcircle scaled .4w shifted (.5w,.5h)}
			\fmf{wiggly,fore=black}{v,c}
			\fmf{wiggly,fore=black}{u,c}
			\fmf{wiggly,fore=black}{s,c}
		\end{fmfgraph}
	\end{fmffile}
\end{gathered}\!\!
+\tfrac{1}{4}\!
\begin{gathered}
	\begin{fmffile}{wg4-2ptlogtadpole}
		\begin{fmfgraph}(35,35)
			\fmfset{dash_len}{1.2mm}
			\fmfset{wiggly_len}{1.1mm} \fmfset{dot_len}{0.5mm}
			\fmfpen{0.25mm}
			\fmfsurroundn{i}{6}
			\fmf{phantom,fore=black}{i1,v,u,i4}
			\fmf{wiggly,fore=black}{v,i1}
			\fmf{phantom,fore=black}{i2,s,t,i5}
			\fmf{wiggly,fore=black}{i5,t}
			\fmf{phantom,fore=black}{i3,w,x,i6}
			\fmf{wiggly,fore=black}{i3,w}
			\fmfi{wiggly,fore=black}{fullcircle scaled .55w shifted (.51w,.5h)}
			\fmf{wiggly,fore=black,left,tension=0.59}{i1,i1}
			\fmffreeze
			\fmfforce{(1.3w,0.5h)}{i1}
			\fmfforce{(0.15w,1.1h)}{i3}
			\fmfforce{(0.15w,-.1h)}{i5}
		\end{fmfgraph}
	\end{fmffile}
\end{gathered}\,\,\,\,\,\,
+\tfrac{1}{4}
\begin{gathered}
	\begin{fmffile}{wg4-2pt2loop-1PIac}
		\begin{fmfgraph}(75,75)
			\fmfset{dash_len}{1.2mm}
			\fmfset{wiggly_len}{1.1mm} \fmfset{dot_len}{0.5mm}
			\fmfpen{0.25mm}
			\fmftop{t}
			\fmfbottom{b}
			\fmfleft{l}
			\fmfright{r}
			\fmf{phantom,fore=black,tension=9}{t,u,v,b}
			\fmf{phantom,fore=black,tension=9}{l,s,x,r}
			\fmf{wiggly,fore=black,tension=.01,left}{u,v,u}
			\fmf{phantom,fore=black,tension=0.01}{s,x,s}
			\fmf{wiggly,fore=black,tension=1}{u,v}
			\fmf{wiggly,fore=black,tension=1}{x,r}
			\fmf{wiggly,fore=black,tension=1}{l,s}
		\end{fmfgraph}
	\end{fmffile}
\end{gathered}
+\tfrac{1}{4}
\begin{gathered}
	\begin{fmffile}{wg4-2pt2loop-1PIbz}
		\begin{fmfgraph}(45,45)
			\fmfset{dash_len}{1.2mm}
			\fmfset{wiggly_len}{1.1mm} \fmfset{dot_len}{0.5mm}
			\fmfpen{0.25mm}
			\fmfleft{i}
			\fmfright{o}
			\fmf{phantom,tension=5}{i,v1}
			\fmf{phantom,tension=5}{v2,o}
			\fmf{wiggly,fore=black,left,tension=0.4}{v1,v2,v1}
			\fmf{wiggly,fore=black}{v1,v2}
			\fmfsurroundn{i}{6}
			\fmf{phantom}{i2,a,c,x1,i5}
			\fmf{phantom}{i3,b,c,x2,i6}
			\fmf{wiggly,fore=black}{x1,i5}
			\fmf{wiggly,fore=black}{x2,i6}
			\fmffreeze
			\fmfforce{(0.1w,-.2h)}{i5}
			\fmfforce{(0.9w,-.2h)}{i6}
			\end{fmfgraph}
	\end{fmffile}
\end{gathered}\,\\
&+\tfrac{1}{4}\,\,\,
\begin{gathered}
	\begin{fmffile}{wfishtadpole}
		\begin{fmfgraph}(40,40)
			\fmfset{dash_len}{1.2mm}
			\fmfset{wiggly_len}{1.1mm} \fmfset{dot_len}{0.5mm}
			\fmfpen{0.25mm}
			\fmfleft{i,j}
			\fmfright{o}
			\fmf{wiggly,fore=black,tension=1}{i,v1}
			\fmf{wiggly,fore=black,tension=1}{j,v1}
			\fmf{wiggly,fore=black}{v1,v2}
			\fmf{wiggly,fore=black,tension=1}{v3,o}
			\fmf{wiggly,fore=black,left,tension=0.3}{v2,v3,v2}
			\fmffreeze
			\fmfforce{(-.2w,0.8h)}{i}
			\fmfforce{(-.2w,0.2h)}{j}
			\fmfforce{(0.05w,0.5h)}{v1}
			\fmfforce{(1.2w,0.5h)}{o}
			\fmf{wiggly,fore=black,tension=0.75}{i,i}
			\end{fmfgraph}
	\end{fmffile}
\end{gathered}\,
+\tfrac{1}{4}
\begin{gathered}
	\begin{fmffile}{wg4-4ptlog1PR}
		\begin{fmfgraph}(35,35)
			\fmfset{dash_len}{1.2mm}
			\fmfset{wiggly_len}{1.1mm} \fmfset{dot_len}{0.5mm}
			\fmfpen{0.25mm}
			\fmfsurroundn{i}{6}
			\fmf{phantom,fore=black}{i1,v,u,i4}
			\fmf{wiggly,fore=black}{v,i1}
			\fmf{phantom,fore=black}{i2,s,t,i5}
			\fmf{wiggly,fore=black}{i5,t}
			\fmf{phantom,fore=black}{i3,w,x,i6}
			\fmf{wiggly,fore=black}{i3,w}
			\fmfi{wiggly,fore=black}{fullcircle scaled .55w shifted (.51w,.5h)}
			\fmffreeze
			\fmfforce{(1.25w,0.5h)}{i1}
			\fmfforce{(0.16w,1.1h)}{i3}
			\fmfforce{(0.16w,-.1h)}{i5}
			\fmffreeze
			\fmfright{x,z}
			\fmf{wiggly,fore=black,tension=1}{i1,x}
			\fmf{wiggly,fore=black,tension=1}{i1,z}
			\fmfforce{(1.5w,0.9h)}{z}
			\fmfforce{(1.5w,0.1h)}{x}
		\end{fmfgraph}
	\end{fmffile}
\end{gathered}\,\,\,\,
+\tfrac{1}{4}
\begin{gathered}\,\,
	\begin{fmffile}{wg4-5pttreelongtadpole}
		\begin{fmfgraph}(50,50)
			\fmfset{dash_len}{1.2mm}
			\fmfset{wiggly_len}{1.1mm} \fmfset{dot_len}{0.5mm}
			\fmfpen{0.25mm}
			\fmfsurroundn{i}{6}
			\fmf{wiggly,fore=black}{i1,v,u,s,t}
			\fmf{wiggly,fore=black}{i6,v}
			\fmf{phantom}{i4,t}
			\fmf{phantom}{i5,t}
			\fmf{wiggly,fore=black}{i2,u}
			\fmf{wiggly,fore=black}{i3,s}
			\fmffreeze
			\fmfforce{(1.12w,0.4h)}{i1}
			\fmfforce{(0.5w,h)}{i4}
			\fmfforce{(0.5w,h)}{i5}
			\fmfforce{(0.7w,0.1h)}{i6}
			\fmffreeze
			\fmf{wiggly,fore=black,tension=1.2}{t,t}
		\end{fmfgraph}
	\end{fmffile}
\end{gathered}\,\,
+\tfrac{1}{4}\,\,\,\,\,
\begin{gathered}
	\begin{fmffile}{wg4-4pt-1loopb}
		\begin{fmfgraph}(40,40)
			\fmfset{dash_len}{1.2mm}
			\fmfset{wiggly_len}{1.1mm} \fmfset{dot_len}{0.5mm}
			\fmfpen{0.25mm}
			\fmfsurroundn{i}{4}
			\fmf{wiggly,fore=black}{i1,n,m,i4}
			\fmf{wiggly,fore=black}{i2,n}
			\fmf{wiggly,fore=black}{m,i3}
			\fmffreeze
			\fmfleft{x}
			\fmf{wiggly,fore=black,tension=10,left}{i3,u,i3}
			\fmf{wiggly,fore=black}{u,x}
			\fmfforce{(-.4w,0.6h)}{u}
			\fmfforce{(-.75w,0.75h)}{x}
		\end{fmfgraph}
	\end{fmffile}
\end{gathered}
\Big)
-\hat{g}^2\hat{\lambda}
\Big(
\tfrac{1}{8}
\begin{gathered}
	\begin{fmffile}{wg2L-bubblez}
		\begin{fmfgraph}(45,45)
			\fmfset{dash_len}{1.2mm}
			\fmfset{wiggly_len}{1.1mm} \fmfset{dot_len}{0.5mm}
			\fmfpen{0.25mm}
			\fmftop{t1,t2,t3}
			\fmfbottom{b1,b2,b3}
			\fmf{phantom}{t1,v1,b1}
			\fmf{phantom}{t2,v2,b2}
			\fmf{phantom}{t3,v3,b3}
			\fmffreeze
			\fmf{wiggly,fore=black,right}{v1,v2,v1}
			\fmf{wiggly,fore=black,right}{v2,v3,v2}
			\fmf{wiggly,fore=black,tension=1}{t1,b1}
			\fmfforce{(0.25w,0.7h)}{t1}
			\fmfforce{(0.25w,0.3h)}{b1}
		\end{fmfgraph}
	\end{fmffile}
\end{gathered}
+\tfrac{1}{8}
\begin{gathered}
	\begin{fmffile}{wg2L-bubble2ptz}
		\begin{fmfgraph}(45,45)
			\fmfset{dash_len}{1.2mm}
			\fmfset{wiggly_len}{1.1mm} \fmfset{dot_len}{0.5mm}
			\fmfpen{0.25mm}
			\fmftop{t1,t2,t3}
			\fmfbottom{b1,b2,b3}
			\fmf{phantom}{t1,v1,b1}
			\fmf{phantom}{t2,v2,b2}
			\fmf{phantom}{t3,v3,b3}
			\fmffreeze
			\fmf{wiggly,fore=black,right}{v1,v2,v1}
			\fmf{wiggly,fore=black}{v2,t3}
			\fmf{wiggly,fore=black}{v2,b3}
			\fmf{wiggly,fore=black,tension=1}{t1,b1}
			\fmfforce{(0.25w,0.7h)}{t1}
			\fmfforce{(0.25w,0.3h)}{b1}
			\fmfforce{(0.9w,0.9h)}{t3}
			\fmfforce{(0.9w,0.1h)}{b3}
		\end{fmfgraph}
	\end{fmffile}
\end{gathered}
+\tfrac{1}{4}\,\,\,
\begin{gathered}
	\begin{fmffile}{wg2L-fishself}
		\begin{fmfgraph}(45,45)
			\fmfset{dash_len}{1.2mm}
			\fmfset{wiggly_len}{1.1mm} \fmfset{dot_len}{0.5mm}
			\fmfpen{0.25mm}
			\fmfleft{j}
			\fmfright{o}
			\fmf{wiggly,fore=black,tension=1}{j,v1}
			\fmf{wiggly,fore=black}{v1,v2}
			\fmf{wiggly,fore=black,tension=1}{v3,o}
			\fmf{wiggly,fore=black,left,tension=0.3}{v2,v3,v2}
			\fmf{wiggly,fore=black,tension=0.6}{v1,v1}
			\fmffreeze
			\fmfforce{(-.2w,0.2h)}{j}
			\fmfforce{(0.05w,0.5h)}{v1}
			\fmfforce{(1.2w,0.5h)}{o}
			\end{fmfgraph}
	\end{fmffile}
\end{gathered}\,
+\tfrac{1}{4}\,\,\,
\begin{gathered}
	\begin{fmffile}{wg2L-fishtadpole2}
		\begin{fmfgraph}(42,42)
			\fmfset{dash_len}{1.2mm}
			\fmfset{wiggly_len}{1.1mm} \fmfset{dot_len}{0.5mm}
			\fmfpen{0.25mm}
			\fmfleft{i,j}
			\fmfright{o}
			\fmf{wiggly,fore=black,tension=1}{i,v1}
			\fmf{wiggly,fore=black,tension=1}{j,v1}
			\fmf{wiggly,fore=black}{v1,v2}
			\fmf{wiggly,fore=black,tension=1}{v2,o}
			\fmffreeze
			\fmfforce{(-.2w,0.8h)}{i}
			\fmfforce{(-.2w,0.2h)}{j}
			\fmfforce{(0.05w,0.5h)}{v1}
			\fmfforce{(1.2w,0.5h)}{o}
			\fmf{wiggly,fore=black,tension=0.75}{v2,v2}
			\fmf{wiggly,fore=black,tension=0.75}{i,i}
			\end{fmfgraph}
	\end{fmffile}
\end{gathered}\,
+\tfrac{1}{12}\,\,\,
\begin{gathered}
	\begin{fmffile}{wg2L-4ptfish}
		\begin{fmfgraph}(40,40)
			\fmfset{dash_len}{1.2mm}
			\fmfset{wiggly_len}{1.1mm} \fmfset{dot_len}{0.5mm}
			\fmfpen{0.25mm}
			\fmfleftn{i}{3}
			\fmfright{o}
			\fmf{wiggly,fore=black,tension=1}{i1,v}
			\fmf{wiggly,fore=black,tension=1}{i2,v}
			\fmf{wiggly,fore=black,tension=1}{i3,v}
			\fmf{wiggly,fore=black,tension=0.5}{v,u}
			\fmf{wiggly,fore=black,tension=0.2,left}{u,x,u}
			\fmf{wiggly,fore=black,tension=1}{x,o}
			\fmffreeze
			\fmfforce{(-.1w,0.9h)}{i1}
			\fmfforce{(-.1w,0.1h)}{i3}
			\fmfforce{(-.35w,0.5h)}{i2}
			\fmfforce{(1.3w,0.5h)}{o}
		\end{fmfgraph}\,\,
	\end{fmffile}
\end{gathered}\\
&+\tfrac{1}{8}
\begin{gathered}
	\begin{fmffile}{wg2L-triplebubble}
		\begin{fmfgraph}(35,35)
			\fmfset{dash_len}{1.2mm}
			\fmfset{wiggly_len}{1.1mm} \fmfset{dot_len}{0.5mm}
			\fmfpen{0.25mm}
			\fmftop{t1,t2,t3}
			\fmfbottom{b1,b2,b3}
			\fmf{phantom}{t1,v1,b1}
			\fmf{phantom}{t2,v2,b2}
			\fmf{phantom}{t3,v3,b3}
			\fmffreeze
			\fmf{wiggly,fore=black,right}{v1,v2,v1}
			\fmf{wiggly,fore=black,right}{v2,v3,v2}
			\fmfright{o}
			\fmf{wiggly,fore=black,tension=1}{v3,o}
			\fmf{wiggly,fore=black,tension=0.64}{o,o}
			\fmffreeze
			\fmfforce{(1.5w,0.5h)}{o}
		\end{fmfgraph}
	\end{fmffile}
\end{gathered}\,\,\,\,\,\,\,
+\tfrac{1}{8}
\begin{gathered}
	\begin{fmffile}{wg2L-2pt-doublebubble}
		\begin{fmfgraph}(35,35)
			\fmfset{dash_len}{1.2mm}
			\fmfset{wiggly_len}{1.1mm} \fmfset{dot_len}{0.5mm}
			\fmfpen{0.25mm}
			\fmftop{t1,t2,t3}
			\fmfbottom{b1,b2,b3}
			\fmf{phantom}{t1,v1,b1}
			\fmf{phantom}{t2,v2,b2}
			\fmf{phantom}{t3,v3,b3}
			\fmffreeze
			\fmf{wiggly,fore=black}{t1,v2}
			\fmf{wiggly,fore=black}{b1,v2}
			\fmf{wiggly,fore=black,right}{v2,v3,v2}
			\fmfright{o}
			\fmf{wiggly,fore=black,tension=1}{v3,o}
			\fmf{wiggly,fore=black,tension=0.64}{o,o}
			\fmffreeze
			\fmfforce{(1.5w,0.5h)}{o}
			\fmfforce{(0w,0.9h)}{t1}
			\fmfforce{(0w,0.1h)}{b1}
		\end{fmfgraph}
	\end{fmffile}
\end{gathered}\,\,\,\,\,\,\,
+\tfrac{1}{12}\,\,
\begin{gathered}
	\begin{fmffile}{wg2L-4pttadpoleb}
		\begin{fmfgraph}(30,30)
			\fmfset{dash_len}{1.2mm}
			\fmfset{wiggly_len}{1.1mm} \fmfset{dot_len}{0.5mm}
			\fmfpen{0.25mm}
			\fmftop{t1,t2}
			\fmfbottom{a,b,c}
			\fmf{wiggly,fore=black,tension=1}{a,v}
			\fmf{wiggly,fore=black,tension=1}{b,v}
			\fmf{wiggly,fore=black,tension=1}{c,v}
			\fmf{wiggly,fore=black,tension=2}{v,t}
			\fmf{wiggly,fore=black,tension=0.7,left,straight}{t,t1}
			\fmf{wiggly,fore=black,tension=0.7,left,straight}{t,t2}
			\fmffreeze
			\fmfforce{(0.5w,-0.3h)}{b}
			\fmf{wiggly,fore=black,tension=0.65}{t1,t1}
			\fmfforce{(0.1w,0.8h)}{t1}
		\end{fmfgraph}
	\end{fmffile}
\end{gathered}\,
+\tfrac{1}{8}\,\,
\begin{gathered}
	\begin{fmffile}{wg2L-3loopbubble-1PI}
		\begin{fmfgraph}(27,27)
			\fmfset{dash_len}{1.2mm}
			\fmfset{wiggly_len}{1.1mm} \fmfset{dot_len}{0.5mm}
			\fmfpen{0.25mm}
			\fmfsurroundn{i}{3}
			\fmf{wiggly,fore=black,tension=1,right=1}{i1,i2}
			\fmf{wiggly,fore=black,tension=1,right=1}{i2,i3}
			\fmf{wiggly,fore=black,tension=1,right=0.8}{i3,i1}
			\fmf{wiggly,fore=black,tension=1}{i1,i2}
			\fmf{wiggly,fore=black,tension=1}{i2,i3}
		\end{fmfgraph}
	\end{fmffile}
\end{gathered}\,\,
+\tfrac{1}{8}
\begin{gathered}
	\begin{fmffile}{wg2L-2pt-doublebubbletadpole}
		\begin{fmfgraph}(35,35)
			\fmfset{dash_len}{1.2mm}
			\fmfset{wiggly_len}{1.1mm} \fmfset{dot_len}{0.5mm}
			\fmfpen{0.25mm}
			\fmftop{t1,t2,t3}
			\fmfbottom{b1,b2,b3}
			\fmf{phantom}{t1,v1,b1}
			\fmf{phantom}{t2,v2,b2}
			\fmf{phantom}{t3,v3,b3}
			\fmffreeze
			\fmf{wiggly,fore=black,right}{v1,v2,v1}
			\fmf{wiggly,fore=black,right}{v2,v3,v2}
			\fmfright{o}
			\fmf{wiggly,fore=black,tension=1}{v3,o}
			\fmf{wiggly,fore=black,tension=0.64}{o,t3}
			\fmf{wiggly,fore=black,tension=0.64}{o,b3}
			\fmffreeze
			\fmfforce{(1.5w,0.5h)}{o}
			\fmfforce{(1.85w,0.9h)}{t3}
			\fmfforce{(1.85w,0.1h)}{b3}
		\end{fmfgraph}
	\end{fmffile}
\end{gathered}\,\,\,\,\,\,\,
+\tfrac{1}{4}\,\,\,\,\,
\begin{gathered}
	\begin{fmffile}{wg2L-2ptdoublebubble-1PI}
		\begin{fmfgraph}(17,17)
			\fmfset{dash_len}{1.2mm}
			\fmfset{wiggly_len}{1.1mm} \fmfset{dot_len}{0.5mm}
			\fmfpen{0.25mm}
			\fmfright{o1,o2}
			\fmfleft{i}
			\fmf{wiggly,fore=black,tension=1,right=0.63}{i,o1}
			\fmf{wiggly,fore=black,tension=1,right=0.6}{o1,o2}
			\fmf{wiggly,fore=black,tension=1,right=0.63}{o2,i}
			\fmf{wiggly,fore=black,tension=5,left}{i,u,i}
			\fmffreeze
			\fmfforce{(-1.2w,0.5h)}{u}
			\fmffreeze
			\fmfright{x1,x2}
			\fmf{wiggly,fore=black,tension=1}{o1,x2}
			\fmf{wiggly,fore=black,tension=1}{o2,x1}
			\fmfforce{(1.5w,1.8h)}{x1}
			\fmfforce{(1.5w,-0.8h)}{x2}
		\end{fmfgraph}
	\end{fmffile}
\end{gathered}\,\,\,
+\tfrac{1}{4}\,\,\,\,\,
\begin{gathered}
	\begin{fmffile}{wg2L-4pt1loop-1PI}
		\begin{fmfgraph}(17,17)
			\fmfset{dash_len}{1.2mm}
			\fmfset{wiggly_len}{1.1mm} \fmfset{dot_len}{0.5mm}
			\fmfpen{0.25mm}
			\fmfright{o1,o2}
			\fmfleft{i}
			\fmf{wiggly,fore=black,tension=1,right=0.63}{i,o1}
			\fmf{wiggly,fore=black,tension=1,right=0.6}{o1,o2}
			\fmf{wiggly,fore=black,tension=1,right=0.63}{o2,i}
			\fmffreeze
			\fmfleft{s1,s2}
			\fmf{wiggly,fore=black,tension=1}{i,s2}
			\fmf{wiggly,fore=black,tension=1}{i,s1}
			\fmfforce{(-0.8w,1.6h)}{s1}
			\fmfforce{(-0.8w,-0.6h)}{s2}
			\fmffreeze
			\fmfright{x1,x2}
			\fmf{wiggly,fore=black,tension=1}{o1,x2}
			\fmf{wiggly,fore=black,tension=1}{o2,x1}
			\fmfforce{(1.4w,2h)}{x1}
			\fmfforce{(1.4w,-h)}{x2}
		\end{fmfgraph}
	\end{fmffile}
\end{gathered}\,\,\,
+\tfrac{1}{2}
\begin{gathered}
	\begin{fmffile}{wg2L-2pt2loop-1PI}
		\begin{fmfgraph}(40,40)
			\fmfset{dash_len}{1.2mm}
			\fmfset{wiggly_len}{1.1mm} \fmfset{dot_len}{0.5mm}
			\fmfpen{0.25mm}
			\fmfsurroundn{i}{4}
			\fmfi{wiggly,fore=black}{fullcircle scaled .63w shifted (0.5w,.5h)}
			\fmf{phantom,fore=black}{i1,u,v,i3}
			\fmf{wiggly,fore=black}{i1,v}
			\fmf{phantom,fore=black}{i2,s,t,i4}
			\fmf{wiggly,fore=black}{i2,s}
			\fmffreeze
			\fmfforce{(0.5w,1.2h)}{i2}
			\fmfforce{(0.2w,0.5h)}{v}
			\fmfforce{(1.3w,0.5h)}{i1}
		\end{fmfgraph}
	\end{fmffile}
\end{gathered}\,\,
+\tfrac{1}{4}\,\,\,\,\,
\begin{gathered}
	\begin{fmffile}{wg2L-4pt-1loopc}
		\begin{fmfgraph}(45,45)
			\fmfset{dash_len}{1.2mm}
			\fmfset{wiggly_len}{1.1mm} \fmfset{dot_len}{0.5mm}
			\fmfpen{0.25mm}
			\fmfsurroundn{i}{4}
			\fmf{wiggly,fore=black}{i1,n,m,i4}
			\fmf{wiggly,fore=black}{i2,n}
			\fmf{wiggly,fore=black}{m,i3}
			\fmffreeze
			\fmfleft{x}
			\fmf{wiggly,fore=black,tension=0.6,right}{i3,i3}
			\fmf{wiggly,fore=black}{i3,x}
			\fmfforce{(-.4w,0.63h)}{x}
		\end{fmfgraph}
	\end{fmffile}
\end{gathered}\\
&+\tfrac{1}{8}\,\,\,
\begin{gathered}
	\begin{fmffile}{wg2L-4pt1loopwiggly}
		\begin{fmfgraph}(35,35)
			\fmfset{dash_len}{1.2mm}
			\fmfset{wiggly_len}{1.1mm} \fmfset{dot_len}{0.5mm}
			\fmfpen{0.25mm}
			\fmfleft{i1,i2}
			\fmfright{o1,o2}
			\fmf{wiggly,fore=black,tension=1}{i1,n}
			\fmf{wiggly,fore=black,tension=1}{i2,n}
			\fmf{wiggly,fore=black,tension=1}{n,m}
			\fmf{wiggly,fore=black,tension=0.15,left}{m,s,m}
			\fmf{wiggly,fore=black,tension=1}{s,o1}
			\fmf{wiggly,fore=black,tension=1}{s,o2}
			\fmffreeze
			\fmfforce{(-0.15w,0.5h)}{n}
			\fmfforce{(-0.45w,h)}{i1}
			\fmfforce{(-0.45w,0h)}{i2}
			\fmfforce{(1.15w,h)}{o1}
			\fmfforce{(1.15w,0h)}{o2}
		\end{fmfgraph}
	\end{fmffile}
\end{gathered}\,\,\,
+\tfrac{1}{12}
\begin{gathered}
	\begin{fmffile}{wg2L-crystal}
		\begin{fmfgraph}(45,45)
			\fmfset{dash_len}{1.2mm}
			\fmfset{wiggly_len}{1.1mm} \fmfset{dot_len}{0.5mm}
			\fmfpen{0.25mm}
			\fmfsurround{i1,i2,i3,i4,i5,i6}
			\fmf{wiggly,fore=black,tension=1}{i6,v}
			\fmf{wiggly,fore=black,tension=1}{i1,v}
			\fmf{wiggly,fore=black,tension=1}{v,c}
		 	\fmf{wiggly,fore=black,tension=1}{c,u}
			\fmf{phantom,tension=1}{u,i2}
			\fmf{phantom,fore=black,tension=1}{u,i3}
			\fmf{wiggly,fore=black,tension=1}{c,s,i4}
			\fmf{wiggly,fore=black,tension=1}{s,i5}
			\fmffreeze
			\fmfright{x}
			\fmf{wiggly,fore=black}{v,x}
			\fmfforce{(1.1w,0.2h)}{x}
			\fmfforce{(1.05w,0.54h)}{i1}
			\fmfforce{(-.05w,0.54h)}{i4}
		\end{fmfgraph}
	\end{fmffile}
\end{gathered}
+\tfrac{1}{12}\!
\begin{gathered}
	\begin{fmffile}{wg2L-3looptadpole-1PR}
		\begin{fmfgraph}(40,40)
			\fmfset{dash_len}{1.2mm}
			\fmfset{wiggly_len}{1.1mm} \fmfset{dot_len}{0.5mm}
			\fmfpen{0.25mm}
			\fmfleft{i}
			\fmfright{o}
			\fmf{phantom,tension=5}{i,v1}
			\fmf{wiggly,fore=black,tension=2.5}{v2,o}
			\fmf{wiggly,fore=black,left,tension=0.5}{v1,v2,v1}
			\fmf{wiggly,fore=black}{v1,v2}
			\fmffreeze
			\fmfforce{(1.1w,0.5h)}{o}
			\fmf{wiggly,fore=black,tension=0.66,left}{o,o}
		\end{fmfgraph}
	\end{fmffile}
\end{gathered}\,\,\,\,
+\tfrac{1}{12}\!
\begin{gathered}
	\begin{fmffile}{wg2L-3loop2pt-1PR}
		\begin{fmfgraph}(40,40)
			\fmfset{dash_len}{1.2mm}
			\fmfset{wiggly_len}{1.1mm} \fmfset{dot_len}{0.5mm}
			\fmfpen{0.25mm}
			\fmfleft{i}
			\fmfright{o}
			\fmf{phantom,tension=5}{i,v1}
			\fmf{wiggly,fore=black,tension=2.5}{v2,o}
			\fmf{wiggly,fore=black,left,tension=0.5}{v1,v2,v1}
			\fmf{wiggly,fore=black}{v1,v2}
			\fmffreeze
			\fmfforce{(1.1w,0.5h)}{o}
			\fmffreeze
			\fmfright{n,m}
			\fmf{wiggly,fore=black,tension=1}{o,n}
			\fmf{wiggly,fore=black,tension=1}{o,m}
			\fmfforce{(1.35w,0.9h)}{n}
			\fmfforce{(1.35w,0.1h)}{m}
		\end{fmfgraph}
	\end{fmffile}
\end{gathered}\,\,\,\,
+\tfrac{1}{8}\,\,\,\,
\begin{gathered}
	\begin{fmffile}{wg2L-2pt2loop-1PIb}
		\begin{fmfgraph}(50,50)
			\fmfset{dash_len}{1.2mm}
			\fmfset{wiggly_len}{1.1mm} \fmfset{dot_len}{0.5mm}
			\fmfpen{0.25mm}
			\fmftop{t1,t2,t3,t4}
        			\fmfbottom{b1,b2,b3,b4}
        			\fmf{phantom}{t1,v1,b1}
        			\fmf{phantom}{t2,v2,b2}
			\fmf{phantom}{t3,v3,b3}
			\fmf{phantom}{t4,v4,b4}
        			\fmffreeze
			\fmf{wiggly,fore=black,right,tension=0.7}{v1,v2,v1}
        			\fmf{wiggly,fore=black,right,tension=0.7}{v2,v3,v2}
        			\fmf{wiggly,fore=black,tension=3}{v3,v4}
			\fmffreeze
			\fmfforce{(1.1w,0.5h)}{v4}
			\fmfleft{l}
			\fmf{wiggly,fore=black}{v1,l}
			\fmfforce{(-.4w,0.5h)}{l}
		\end{fmfgraph}
	\end{fmffile}
\end{gathered}
+\tfrac{1}{16}
\begin{gathered}
	\begin{fmffile}{wg2L-bubble-doubletadpole-1PR}
		\begin{fmfgraph}(40,40)
			\fmfset{dash_len}{1.2mm}
			\fmfset{wiggly_len}{1.1mm} \fmfset{dot_len}{0.5mm}
			\fmfpen{0.25mm}
			\fmfsurround{a,b,c}
			\fmf{wiggly,fore=black,tension=1}{c,v,a}
			\fmf{wiggly,fore=black,left,tension=1.2}{v,b,v}
			\fmf{wiggly,fore=black,tension=1.3}{c,c}
			\fmf{wiggly,fore=black,tension=1.3}{a,a}
		\end{fmfgraph}
	\end{fmffile}
\end{gathered}\,\,
+\tfrac{1}{4}\,\,\,
\begin{gathered}
	\begin{fmffile}{wg2L-2ptdarttadpole}
		\begin{fmfgraph}(40,40)
			\fmfset{dash_len}{1.2mm}
			\fmfset{wiggly_len}{1.1mm} \fmfset{dot_len}{0.5mm}
			\fmfpen{0.25mm}
			\fmfleft{i,j}
			\fmfright{o}
			\fmf{wiggly,fore=black,tension=5}{i,v1}
			\fmf{wiggly,fore=black,tension=5}{j,v1}
			\fmf{wiggly,fore=black,tension=0.8}{v2,o}
			\fmf{wiggly,fore=black,left,tension=0.08}{v1,v2,v1}
			\fmf{phantom}{v1,v2}
			\fmffreeze
			\fmfforce{(-.2w,0.8h)}{i}
			\fmfforce{(-.2w,0.2h)}{j}
			\fmf{wiggly,fore=black,tension=0.85,left}{i,i}
			\end{fmfgraph}
	\end{fmffile}
\end{gathered}
+\tfrac{1}{4}\,\,\,\,
\begin{gathered}
	\begin{fmffile}{wg2L-4ptdart}
		\begin{fmfgraph}(40,40)
			\fmfset{dash_len}{1.2mm}
			\fmfset{wiggly_len}{1.1mm} \fmfset{dot_len}{0.5mm}
			\fmfpen{0.25mm}
			\fmfleft{i,j}
			\fmfright{o}
			\fmf{wiggly,fore=black,tension=5}{i,v1}
			\fmf{wiggly,fore=black,tension=5}{j,v1}
			\fmf{wiggly,fore=black,tension=0.8}{v2,o}
			\fmf{wiggly,fore=black,left,tension=0.08}{v1,v2,v1}
			\fmf{phantom}{v1,v2}
			\fmffreeze
			\fmfforce{(-.2w,0.8h)}{i}
			\fmfforce{(-.2w,0.2h)}{j}
			\fmfleft{n,m}
			\fmf{wiggly,fore=black,tension=1}{i,n}
			\fmf{wiggly,fore=black,tension=1}{i,m}
			\fmfforce{(-.5w,1.1h)}{n}
			\fmfforce{(-.55w,0.5h)}{m}			
		\end{fmfgraph}
	\end{fmffile}
\end{gathered}
+\tfrac{1}{8}\,
\begin{gathered}
	\begin{fmffile}{wg2L-2pt-doubletadpole-1PR}
		\begin{fmfgraph}(35,35)
			\fmfset{dash_len}{1.2mm}
			\fmfset{wiggly_len}{1.1mm} \fmfset{dot_len}{0.5mm}
			\fmfpen{0.25mm}
			\fmfsurround{a,b,c}
			\fmf{wiggly,fore=black,tension=1}{c,v,a}
			\fmf{wiggly,fore=black,left,tension=0.9}{v,b,v}
			\fmf{wiggly,fore=black,tension=1}{c,c}
			\fmffreeze
			\fmfright{n,m}
			\fmf{wiggly,fore=black,tension=1.3}{a,n}
			\fmf{wiggly,fore=black,tension=1.3}{a,m}
			\fmfforce{(1.3w,0.8h)}{n}
			\fmfforce{(1.1w,0.1h)}{m}	
		\end{fmfgraph}
	\end{fmffile}
\end{gathered}\,\,\\
&+\tfrac{1}{16}\,
\begin{gathered}
	\begin{fmffile}{wg2L-4pt-1PR}
		\begin{fmfgraph}(35,35)
			\fmfset{dash_len}{1.2mm}
			\fmfset{wiggly_len}{1.1mm} \fmfset{dot_len}{0.5mm}
			\fmfpen{0.25mm}
			\fmfsurround{a,b,c}
			\fmf{wiggly,fore=black,tension=1}{c,v,a}
			\fmf{wiggly,fore=black,left,tension=0.8}{v,b,v}
			\fmffreeze
			\fmfright{n,m}
			\fmf{wiggly,fore=black,tension=1.3}{a,n}
			\fmf{wiggly,fore=black,tension=1.3}{a,m}
			\fmfforce{(1.3w,0.8h)}{n}
			\fmfforce{(1.1w,0.1h)}{m}
			\fmffreeze
			\fmfbottom{x,z}
			\fmf{wiggly,fore=black,tension=1.3}{c,x}
			\fmf{wiggly,fore=black,tension=1.3}{c,z}
			\fmfforce{(-.12w,-0.12h)}{x}
			\fmfforce{(.55w,-0.2h)}{z}
		\end{fmfgraph}
	\end{fmffile}
\end{gathered}\,\,
+\tfrac{1}{16}\,
\begin{gathered}
	\begin{fmffile}{wg2L-2pt-doubletadpole-1PRb}
		\begin{fmfgraph}(35,35)
			\fmfset{dash_len}{1.2mm}
			\fmfset{wiggly_len}{1.1mm} \fmfset{dot_len}{0.5mm}
			\fmfpen{0.25mm}
			\fmfsurround{a,b,c}
			\fmf{wiggly,fore=black,tension=1}{c,v,a}
			\fmf{phantom,fore=black,left,tension=1.2}{v,b,v}
			\fmf{wiggly,fore=black,tension=1.1}{c,c}
			\fmf{wiggly,fore=black,tension=1.1}{a,a}
			\fmffreeze
			\fmftop{n,m}
			\fmf{wiggly,fore=black,tension=1}{n,v}
			\fmf{wiggly,fore=black,tension=1}{m,v}
			\fmfforce{(-0.05w,0.77h)}{n}
			\fmfforce{(0.57w,1.1h)}{m}
		\end{fmfgraph}
	\end{fmffile}
\end{gathered}\,\,
+\tfrac{1}{8}\,
\begin{gathered}
	\begin{fmffile}{wg2L-4pt-tadpole-1PRc}
		\begin{fmfgraph}(35,35)
			\fmfset{dash_len}{1.2mm}
			\fmfset{wiggly_len}{1.1mm} \fmfset{dot_len}{0.5mm}
			\fmfpen{0.25mm}
			\fmfsurround{a,b,c}
			\fmf{wiggly,fore=black,tension=1}{c,v,a}
			\fmf{phantom,fore=black,left,tension=1.2}{v,b,v}
			\fmf{wiggly,fore=black,tension=1.1}{a,a}
			\fmffreeze
			\fmftop{n,m}
			\fmf{wiggly,fore=black,tension=1}{n,v}
			\fmf{wiggly,fore=black,tension=1}{m,v}
			\fmfforce{(-0.05w,0.77h)}{n}
			\fmfforce{(0.57w,1.15h)}{m}
			\fmffreeze
			\fmfbottom{x,z}
			\fmf{wiggly,fore=black,tension=1.3}{c,x}
			\fmf{wiggly,fore=black,tension=1.3}{c,z}
			\fmfforce{(-.12w,-0.12h)}{x}
			\fmfforce{(.55w,-0.2h)}{z}
		\end{fmfgraph}
	\end{fmffile}
\end{gathered}\,\,
+\tfrac{1}{16}\,
\begin{gathered}
	\begin{fmffile}{wg2L-6pt-1PR}
		\begin{fmfgraph}(35,35)
			\fmfset{dash_len}{1.2mm}
			\fmfset{wiggly_len}{1.1mm} \fmfset{dot_len}{0.5mm}
			\fmfpen{0.25mm}
			\fmfsurround{a,b,c}
			\fmf{wiggly,fore=black,tension=1}{c,v,a}
			\fmf{phantom,fore=black,left,tension=1.2}{v,b,v}
			\fmffreeze
			\fmftop{n,m}
			\fmf{wiggly,fore=black,tension=1}{n,v}
			\fmf{wiggly,fore=black,tension=1}{m,v}
			\fmfforce{(-0.05w,0.77h)}{n}
			\fmfforce{(0.57w,1.15h)}{m}
			\fmffreeze
			\fmfbottom{x,z}
			\fmf{wiggly,fore=black,tension=1.3}{c,x}
			\fmf{wiggly,fore=black,tension=1.3}{c,z}
			\fmfforce{(-.12w,-0.12h)}{x}
			\fmfforce{(.55w,-0.2h)}{z}
			\fmffreeze
			\fmfright{s,t}
			\fmf{wiggly,fore=black,tension=1.3}{a,s}
			\fmf{wiggly,fore=black,tension=1.3}{a,t}
			\fmfforce{(1.3w,0.8h)}{s}
			\fmfforce{(1.1w,0.1h)}{t}
		\end{fmfgraph}
	\end{fmffile}
\end{gathered}\,\,
\Big)
+\hat{\lambda}^2
\Big(
\tfrac{1}{48}
\begin{gathered}
	\begin{fmffile}{wL2-3loop-bubble}
		\begin{fmfgraph}(50,50)
			\fmfset{dash_len}{1.2mm}
			\fmfset{wiggly_len}{1.1mm} \fmfset{dot_len}{0.5mm}
			\fmfpen{0.25mm}
			\fmfleft{i}
			\fmfright{o}
			\fmf{phantom,tension=10}{i,v1}
			\fmf{phantom,tension=10}{v2,o}
			\fmf{wiggly,left,tension=0.4}{v1,v2,v1}
			\fmf{wiggly,left=0.5}{v1,v2}
			\fmf{wiggly,right=0.5}{v1,v2}
    		\end{fmfgraph}
	\end{fmffile}
\end{gathered}
+\tfrac{1}{16}
\begin{gathered}
	\begin{fmffile}{wL2-bubble}
		\begin{fmfgraph}(35,35)
			\fmfset{dash_len}{1.2mm}
			\fmfset{wiggly_len}{1.1mm} \fmfset{dot_len}{0.5mm}
			\fmfpen{0.25mm}
			\fmftop{t1,t2,t3}
			\fmfbottom{b1,b2,b3}
			\fmf{phantom}{t1,v1,b1}
			\fmf{phantom}{t2,v2,b2}
			\fmf{phantom}{t3,v3,b3}
			\fmffreeze
			\fmf{wiggly,fore=black,right}{v1,v2,v1}
			\fmf{wiggly,fore=black,right}{v2,v3,v2}
			\fmfi{wiggly,fore=black}{fullcircle scaled .5w shifted (1.25w,.5h)}
		\end{fmfgraph}
	\end{fmffile}
\end{gathered}\,\,\,
+\tfrac{1}{12}\,\,
\begin{gathered}
	\begin{fmffile}{wL2-2pt2loop-1PI}
		\begin{fmfgraph}(45,45)
			\fmfset{dash_len}{1.2mm}
			\fmfset{wiggly_len}{1.1mm} \fmfset{dot_len}{0.5mm}
			\fmfpen{0.25mm}
			\fmfleft{i}
			\fmfright{o}
			\fmf{wiggly,fore=black,tension=5}{i,v1}
			\fmf{wiggly,fore=black,tension=5}{v2,o}
			\fmf{wiggly,fore=black,left,tension=0.4}{v1,v2,v1}
			\fmf{wiggly,fore=black}{v1,v2}
			\fmffreeze
			\fmfforce{(-0.2w,0.5h)}{i}
			\fmfforce{(1.2w,0.5h)}{o}
		\end{fmfgraph}
	\end{fmffile}
\end{gathered}\,\,
+\tfrac{1}{8}
\begin{gathered}
	\begin{fmffile}{wL2-2pt2loop-1PIb}
		\begin{fmfgraph}(35,35)
			\fmfset{dash_len}{1.2mm}
			\fmfset{wiggly_len}{1.1mm} \fmfset{dot_len}{0.5mm}
			\fmfpen{0.25mm}
			\fmftop{t1,t2,t3}
			\fmfbottom{b1,b2,b3}
			\fmf{phantom}{t1,v1,b1}
			\fmf{phantom}{t2,v2,b2}
			\fmf{phantom}{t3,v3,b3}
			\fmffreeze
			\fmf{wiggly,fore=black,right}{v1,v2,v1}
			\fmf{wiggly,fore=black,right}{v2,v3,v2}
			\fmffreeze
			\fmfright{x,z}
			\fmf{wiggly,fore=black}{v3,x}
			\fmf{wiggly,fore=black}{v3,z}
			\fmfforce{(1.3w,0.9h)}{x}
			\fmfforce{(1.3w,0.1h)}{z}
		\end{fmfgraph}
	\end{fmffile}
\end{gathered}\,\,\,\\
&+\tfrac{1}{8}\,
\begin{gathered}
	\begin{fmffile}{wL2-2pt2loop-1PR}
		\begin{fmfgraph}(50,50)
			\fmfset{dash_len}{1.2mm}
			\fmfset{wiggly_len}{1.1mm} \fmfset{dot_len}{0.5mm}
			\fmfpen{0.25mm}
			\fmfleft{i}
			\fmfright{o}
			\fmf{wiggly,fore=black}{i,u,v,o}
			\fmf{wiggly,fore=black,tension=0.5,left}{u,u}
			\fmf{wiggly,fore=black,tension=0.5,right}{v,v}
		\end{fmfgraph}
	\end{fmffile}
\end{gathered}\,
+\tfrac{1}{16}\,
\begin{gathered}
	\begin{fmffile}{wL2-4pt1loop-1PI}
		\begin{fmfgraph}(50,50)
			\fmfset{dash_len}{1.2mm}
			\fmfset{wiggly_len}{1.1mm} \fmfset{dot_len}{0.5mm}
			\fmfpen{0.25mm}
			\fmfleft{i1,i2}
			\fmfright{o1,o2}
			\fmf{wiggly,fore=black,tension=1}{i1,u}
			\fmf{wiggly,fore=black,tension=1}{i2,u}
			\fmf{wiggly,fore=black,tension=0.4,left}{u,v,u}
			\fmf{wiggly,fore=black}{v,o1}
			\fmf{wiggly,fore=black}{v,o2}
			\fmffreeze
			\fmfforce{(w,0.8h)}{o1}
			\fmfforce{(w,0.2h)}{o2}
			\fmfforce{(0w,0.8h)}{i1}
			\fmfforce{(0w,0.2h)}{i2}
		\end{fmfgraph}
	\end{fmffile}
\end{gathered}\,
+\tfrac{1}{12}\,\,
\begin{gathered}
	\begin{fmffile}{wL2-4pt1loop31-1PR}
		\begin{fmfgraph}(50,50)
			\fmfset{dash_len}{1.2mm}
			\fmfset{wiggly_len}{1.1mm} \fmfset{dot_len}{0.5mm}
			\fmfpen{0.25mm}
			\fmfleftn{i}{3}
			\fmfright{o}
			\fmf{wiggly,fore=black}{i1,u}
			\fmf{wiggly,fore=black}{i2,u}
			\fmf{wiggly,fore=black}{i3,u}
			\fmf{wiggly,fore=black}{u,v,o}
			\fmf{wiggly,fore=black,tension=0.6,right}{v,v}
			\fmffreeze
			\fmfforce{(0w,0.9h)}{i1}
			\fmfforce{(-.2w,0.5h)}{i2}
			\fmfforce{(0w,0.1h)}{i3}
		\end{fmfgraph}
	\end{fmffile}
\end{gathered}\,
+\tfrac{1}{72}\,\,
\begin{gathered}
	\begin{fmffile}{wL2-6pttree33-1PR}
		\begin{fmfgraph}(45,45)
			\fmfset{dash_len}{1.2mm}
			\fmfset{wiggly_len}{1.1mm} \fmfset{dot_len}{0.5mm}
			\fmfpen{0.25mm}
			\fmfleftn{i}{3}
			\fmfrightn{o}{3}
			\fmf{wiggly,fore=black}{i1,u}
			\fmf{wiggly,fore=black}{i2,u}
			\fmf{wiggly,fore=black}{i3,u}
			\fmf{wiggly,fore=black}{o1,v}
			\fmf{wiggly,fore=black}{o2,v}
			\fmf{wiggly,fore=black}{o3,v}
			\fmf{wiggly,fore=black}{u,v}
			\fmffreeze
			\fmfforce{(0w,0.9h)}{i1}
			\fmfforce{(-.2w,0.5h)}{i2}
			\fmfforce{(0w,0.1h)}{i3}
			\fmfforce{(1w,0.9h)}{o1}
			\fmfforce{(1.2w,0.5h)}{o2}
			\fmfforce{(1w,0.1h)}{o3}
		\end{fmfgraph}
	\end{fmffile}
\end{gathered}\,
\Big)
+\hat{g}\hat{\kappa}\Big(
\tfrac{1}{12}\!
\begin{gathered}
	\begin{fmffile}{wgkappa-bubble-1PIx}
		\begin{fmfgraph}(40,40)
			\fmfset{dash_len}{1.2mm}
			\fmfset{wiggly_len}{1.1mm} \fmfset{dot_len}{0.5mm}
			\fmfpen{0.25mm}
			\fmfleft{i}
			\fmfright{o}
			\fmf{phantom,tension=5}{i,v1}
			\fmf{phantom,tension=5}{v2,o}
			\fmf{wiggly,fore=black,left,tension=0.5}{v1,v2,v1}
			\fmf{wiggly,fore=black}{v1,v2}
			\fmffreeze
			\fmfi{wiggly,fore=black}{fullcircle scaled .55w shifted (1.1w,.5h)}
		\end{fmfgraph}
	\end{fmffile}
\end{gathered}\,\,
+\tfrac{1}{12}\!
\begin{gathered}
	\begin{fmffile}{wgkappa-bubble-2pt-1PI}
		\begin{fmfgraph}(30,30)
			\fmfset{dash_len}{1.2mm}
			\fmfset{wiggly_len}{1.1mm} \fmfset{dot_len}{0.5mm}
			\fmfpen{0.25mm}
			\fmfleft{i}
			\fmfright{o}
			\fmf{phantom,tension=5}{i,v1}
			\fmf{phantom,tension=5}{v2,o}
			\fmf{wiggly,fore=black,left,tension=0.4}{v1,v2,v1}
			\fmf{wiggly,fore=black}{v1,v2}
			\fmffreeze
			\fmfright{o1,o2}
			\fmf{wiggly,fore=black,tension=1}{v2,o1}
			\fmf{wiggly,fore=black,tension=1}{v2,o2}
			\fmfforce{(1.1w,0.9h)}{o1}
			\fmfforce{(1.1w,0.1h)}{o2}
		\end{fmfgraph}
	\end{fmffile}
\end{gathered}\,
+\tfrac{1}{16}\!\!
\begin{gathered}
	\begin{fmffile}{wgkappa-vacuumtadpole-1PR}
		\begin{fmfgraph}(55,55)
			\fmfset{dash_len}{1.2mm}
			\fmfset{wiggly_len}{1.1mm} \fmfset{dot_len}{0.5mm}
			\fmfpen{0.25mm}
			\fmftop{t1,t2,t3}
			\fmfbottom{b1,b2,b3}
			\fmf{phantom}{t1,v1,b1}
			\fmf{phantom}{t2,v2,b2}
			\fmf{phantom}{t3,v3,b3}
			\fmf{wiggly,fore=black,right,tension=1}{v1,v2,v1}
			\fmf{wiggly,fore=black,right,tension=1}{v2,v3,v2}
			\fmf{wiggly,fore=black,tension=1}{v2,b2}
			\fmf{wiggly,fore=black,tension=1}{b2,b2}
		\end{fmfgraph}
	\end{fmffile}
\end{gathered}\!\!
+\tfrac{1}{4}\,
\begin{gathered}
	\begin{fmffile}{wgkappa-2pt2loop-1PIb}
		\begin{fmfgraph}(60,60)
			\fmfset{dash_len}{1.2mm}
			\fmfset{wiggly_len}{1.1mm} \fmfset{dot_len}{0.5mm}
			\fmfpen{0.25mm}
			\fmftop{t1,t2,t3}
			\fmfbottom{b1,b2,b3}
			\fmf{phantom}{t1,v1,b1}
			\fmf{phantom}{t2,v2,b2}
			\fmf{phantom}{t3,v3,b3}
			\fmf{wiggly,fore=black,right,tension=1}{v1,v2,v1}
			\fmf{wiggly,fore=black,right,tension=1}{v2,v3,v2}
			\fmf{wiggly,fore=black,tension=1}{v2,b2}
			\fmffreeze
			\fmfleft{i}
			\fmf{wiggly,fore=black}{i,v1}
			\fmfforce{(-.1w,0.44h)}{i}
		\end{fmfgraph}
	\end{fmffile}
\end{gathered}\!\!
+\tfrac{1}{8}
\begin{gathered}
	\begin{fmffile}{wgkappa-2ptadpole}
		\begin{fmfgraph}(32,32)
			\fmfset{dash_len}{1.2mm}
			\fmfset{wiggly_len}{1.1mm} \fmfset{dot_len}{0.5mm}
			\fmfpen{0.25mm}
			\fmftop{t}
			\fmfbottom{a,b,c}
			\fmf{wiggly,fore=black,tension=1}{a,v}
			\fmf{wiggly,fore=black,tension=1}{b,v}
			\fmf{wiggly,fore=black,tension=1}{c,v}
			\fmf{wiggly,fore=black,tension=1.4,left}{v,t,v}
			\fmf{wiggly,fore=black,tension=0.7}{b,b}
		\end{fmfgraph}
	\end{fmffile}
\end{gathered}\\
&+\tfrac{1}{12}
\begin{gathered}
	\begin{fmffile}{wgkappa-4pt-1PI}
		\begin{fmfgraph}(32,32)
			\fmfset{dash_len}{1.2mm}
			\fmfset{wiggly_len}{1.1mm} \fmfset{dot_len}{0.5mm}
			\fmfpen{0.25mm}
			\fmftop{t}
			\fmfbottom{a,b,c}
			\fmf{wiggly,fore=black,tension=1}{a,v}
			\fmf{wiggly,fore=black,tension=1}{b,v}
			\fmf{wiggly,fore=black,tension=1}{c,v}
			\fmf{wiggly,fore=black,tension=1.4,left}{v,t,v}
			\fmffreeze
			\fmftop{x}
			\fmf{wiggly,fore=black}{t,x}
			\fmfforce{(0.5w,1.45h)}{x}
		\end{fmfgraph}
	\end{fmffile}
\end{gathered}
+\tfrac{1}{48}
\begin{gathered}
	\begin{fmffile}{wgkappa-4pt1looptadpole,1PR}
		\begin{fmfgraph}(22,22)
			\fmfset{dash_len}{1.2mm}
			\fmfset{wiggly_len}{1.1mm} \fmfset{dot_len}{0.5mm}
			\fmfpen{0.25mm}
			\fmfleft{i1,i2}
			\fmfright{o1,o2}
			\fmf{wiggly,fore=black}{i1,v,o2}
			\fmf{wiggly,fore=black}{i2,v,o1}
			\fmffreeze
			\fmfright{x}
			\fmf{wiggly,fore=black}{x,v}
			\fmf{wiggly,fore=black,tension=0.7}{x,x}
			\fmfforce{(1.2w,0.5h)}{x}
		\end{fmfgraph}
	\end{fmffile}
\end{gathered}\,\,\,
+\tfrac{1}{16}\!\!
\begin{gathered}
	\begin{fmffile}{wgkappa-2pt1looptadpole-1PR}
		\begin{fmfgraph}(55,55)
			\fmfset{dash_len}{1.2mm}
			\fmfset{wiggly_len}{1.1mm} \fmfset{dot_len}{0.5mm}
			\fmfpen{0.25mm}
			\fmftop{t1,t2,t3}
			\fmfbottom{b1,b2,b3}
			\fmf{phantom}{t1,v1,b1}
			\fmf{phantom}{t2,v2,b2}
			\fmf{phantom}{t3,v3,b3}
			\fmf{wiggly,fore=black,right,tension=1}{v1,v2,v1}
			\fmf{wiggly,fore=black,right,tension=1}{v2,v3,v2}
			\fmf{wiggly,fore=black,tension=1}{v2,b2}
			\fmffreeze
			\fmfbottom{x,z}
			\fmf{wiggly,fore=black,tension=1}{b2,x}
			\fmf{wiggly,fore=black,tension=1}{b2,z}
			\fmfforce{(0.25w,-.18h)}{x}
			\fmfforce{(0.75w,-.18h)}{z}
		\end{fmfgraph}
	\end{fmffile}
\end{gathered}\!
+\tfrac{1}{8}
\begin{gathered}
	\begin{fmffile}{wgkappa-4ptself1PR}
		\begin{fmfgraph}(32,32)
			\fmfset{dash_len}{1.2mm}
			\fmfset{wiggly_len}{1.1mm} \fmfset{dot_len}{0.5mm}
			\fmfpen{0.25mm}
			\fmftop{t}
			\fmfbottom{a,b,c}
			\fmf{wiggly,fore=black,tension=1}{a,v}
			\fmf{wiggly,fore=black,tension=1}{b,v}
			\fmf{wiggly,fore=black,tension=1}{c,v}
			\fmf{wiggly,fore=black,tension=1.4,left}{v,t,v}
			\fmffreeze
			\fmfbottom{x,z}
			\fmf{wiggly,fore=black}{c,x}
			\fmf{wiggly,fore=black}{c,z}
			\fmfforce{(w,-.4h)}{x}
			\fmfforce{(1.45w,.4h)}{z}
		\end{fmfgraph}
	\end{fmffile}
\end{gathered}\,\,
+\tfrac{1}{48}
\begin{gathered}
	\begin{fmffile}{wgkappa-6pt-1PR}
		\begin{fmfgraph}(40,40)
			\fmfset{dash_len}{1.2mm}
			\fmfset{wiggly_len}{1.1mm} \fmfset{dot_len}{0.5mm}
			\fmfpen{0.25mm}
			\fmfsurroundn{i}{5}
			\fmfright{s,t}
			\fmf{wiggly,fore=black}{i1,c}
			\fmf{wiggly,fore=black}{i2,c}
			\fmf{wiggly,fore=black}{i3,c}
			\fmf{wiggly,fore=black}{i4,c}
			\fmf{wiggly,fore=black}{i5,c}
			\fmf{wiggly,fore=black}{i1,s}
			\fmf{wiggly,fore=black}{i1,t}
			\fmffreeze
			\fmfforce{(1.2w,0.9h)}{s}
			\fmfforce{(1.2w,0.1h)}{t}
		\end{fmfgraph}
	\end{fmffile}
\end{gathered}\,\,
\Big)
-\hat{\gamma}\Big(
\tfrac{1}{48}
\begin{gathered}
	\begin{fmffile}{wgamma-vacuum-1PI}
		\begin{fmfgraph}(50,50)
			\fmfset{dash_len}{1.2mm}
			\fmfset{wiggly_len}{1.1mm} \fmfset{dot_len}{0.5mm}
			\fmfpen{0.25mm}
			\fmfsurroundn{x}{3}
			\fmf{phantom,fore=black}{x1,v}
			\fmf{phantom,fore=black}{x2,v}
			\fmf{phantom,fore=black}{x3,v}
			\fmf{wiggly,fore=black,tension=0.7}{v,v}
			\fmf{wiggly,fore=black,tension=0.7,right}{v,v}
			\fmf{wiggly,fore=black,tension=0.7,left}{v,v}
		\end{fmfgraph}
	\end{fmffile}
\end{gathered}\!
+\tfrac{1}{16}
\begin{gathered}
	\begin{fmffile}{wgamma-2pt1loop-1PI}
		\begin{fmfgraph}(45,45)
			\fmfset{dash_len}{1.2mm}
			\fmfset{wiggly_len}{1.1mm} \fmfset{dot_len}{0.5mm}
			\fmfpen{0.25mm}
			\fmfleft{i}
			\fmfright{o}
			\fmf{wiggly,fore=black,tension=0.7}{i,v,v,o}
			\fmf{wiggly,fore=black,left=90,tension=0.7}{v,v}
		\end{fmfgraph}
	\end{fmffile}
\end{gathered}\!
+\tfrac{1}{48}
\begin{gathered}
	\begin{fmffile}{wgamma-4pt1loop-1PI}
		\begin{fmfgraph}(45,45)
			\fmfset{dash_len}{1.2mm}
			\fmfset{wiggly_len}{1.1mm} \fmfset{dot_len}{0.5mm}
			\fmfpen{0.25mm}
			\fmfsurroundn{x}{8}
			\fmf{phantom,fore=black}{x1,c,x5}
			\fmf{phantom,fore=black}{x2,c,x6}
			\fmf{phantom,fore=black}{x3,c,x7}
			\fmf{phantom,fore=black}{x4,c,x8}
			\fmf{wiggly,fore=black}{x1,c}
			\fmf{wiggly,fore=black}{x8,c}
			\fmf{wiggly,fore=black}{x7,c}
			\fmf{wiggly,fore=black}{x6,c}
			\fmfi{wiggly,fore=black}{fullcircle scaled .38w shifted (0.46w,.58h)}	
		\end{fmfgraph}
	\end{fmffile}
\end{gathered}
+\tfrac{1}{6!}
\begin{gathered}
	\begin{fmffile}{wgamma-6pttree-1PI}
		\begin{fmfgraph}(40,40)
			\fmfset{dash_len}{1.2mm}
			\fmfset{wiggly_len}{1.1mm} \fmfset{dot_len}{0.5mm}
			\fmfpen{0.25mm}
			\fmfsurroundn{x}{6}
			\fmf{wiggly,fore=black}{x1,c,x4}
			\fmf{wiggly,fore=black}{x2,c,x5}
			\fmf{wiggly,fore=black}{x3,c,x6}
		\end{fmfgraph}
	\end{fmffile}
\end{gathered}
\Big)
+\mathcal{O}(\ell^5) .
\end{aligned}
\end{equation}
At every internal vertex there is a $d$-volume integral, $\int d^dz\sqrt{g}$, internal lines are to be interpreted as dressed propagators, $\mathcal{G}(z,w)$, defined in terms of the free curved space-time propagator $G(z,w)$ in (\ref{eq:dressed_prop}), whereas every external line ending at an internal vertex at (say) $z$ represents a dressed propagator with a source, $\int d^dw\sqrt{g}J(w)\mathcal{G}(w,z)$. 

A \emph{very important} note on notation: although we display the couplings, $\hat{g}$, $\hat{\lambda}$, etc., in these Feynman diagrams as though they have been factored out of the loop integrals, it should always be understood that these couplings may be \emph{local} in general~\footnote{Generically, they are local diffeomorphism-invariant combinations of the space-time metric, such as $R_{(d)}$, $R_{\mu\nu}R^{\mu\nu}$, etc., (with bare coefficients).} and in explicit computations they should always be interpreted as being part of the integrands inside the loop integrals. For example,
$$
\hat{g}^2
\begin{gathered}
	\begin{fmffile}{wg2-2loopbubble-1PI}
		\begin{fmfgraph}(40,40)
			\fmfset{dash_len}{1.2mm}
			\fmfset{wiggly_len}{1.1mm} \fmfset{dot_len}{0.5mm}
			\fmfpen{0.25mm}
			\fmfleft{i}
			\fmfright{o}
			\fmf{phantom,tension=5}{i,v1}
			\fmf{phantom,tension=5}{v2,o}
			\fmf{wiggly,fore=black,left,tension=0.4}{v1,v2,v1}
			\fmf{wiggly,fore=black}{v1,v2}
		\end{fmfgraph}
	\end{fmffile}
\end{gathered}\!\dfn \int_{z,w}\hat{g}(z)\hat{g}(w)\mathcal{G}(z,w)^3,
$$
$$
\hat{g}^2\hat{\lambda}\,
\begin{gathered}
	\begin{fmffile}{wg2L-bubble2ptz}
		\begin{fmfgraph}(45,45)
			\fmfset{dash_len}{1.2mm}
			\fmfset{wiggly_len}{1.1mm} \fmfset{dot_len}{0.5mm}
			\fmfpen{0.25mm}
			\fmftop{t1,t2,t3}
			\fmfbottom{b1,b2,b3}
			\fmf{phantom}{t1,v1,b1}
			\fmf{phantom}{t2,v2,b2}
			\fmf{phantom}{t3,v3,b3}
			\fmffreeze
			\fmf{wiggly,fore=black,right}{v1,v2,v1}
			\fmf{wiggly,fore=black}{v2,t3}
			\fmf{wiggly,fore=black}{v2,b3}
			\fmf{wiggly,fore=black,tension=1}{t1,b1}
			\fmfforce{(0.25w,0.7h)}{t1}
			\fmfforce{(0.25w,0.3h)}{b1}
			\fmfforce{(0.9w,0.9h)}{t3}
			\fmfforce{(0.9w,0.1h)}{b3}
		\end{fmfgraph}
	\end{fmffile}
\end{gathered}\,\dfn \int_{z,w,u}\hat{g}(z)\hat{g}(w)\hat{\lambda}(u)\int_vJ(v)\mathcal{G}(v,u)\mathcal{G}(u,z)\mathcal{G}(z,w)^2\mathcal{G}(w,u)\int_{v'}J(v')\mathcal{G}(v',u),
$$
and so on. There will never be an ambiguity in associating a given coupling to a given loop integral when we keep in mind that $\hat{g}$ is associated to a three-point vertex, $\hat{\lambda}$ to a four-point vertex, etc., as discussed above.

Our conventions are such that the $\ell$-dependence in this generating function comes solely from the source and the bare couplings. Since $J\sim \mathcal{O}(\ell^{-1})$, every external leg in a given diagram lowers the order of $\ell$ by one. Recalling the $\ell$-dependences of the couplings, by inspection of $U(J)$ all tree diagrams are $\mathcal{O}(\ell^{-2})$, all one-loop diagrams are 
$\mathcal{O}(\ell^{0})$, all two-loop diagrams are $\mathcal{O}(\ell^2)$, etc., so that a generic $L$-loop diagram in $U(J)$ is $\mathcal{O}(\ell^{2L-2})$.\footnote{If we make the $\ell$-dependence explicit in $J$, we encounter an infinite number of diagrams at any given order in $\ell$, because $J$ lowers the order in $\ell$ by one.} The $\ell$-dependence of the source need not be made explicit, because Green functions are generated by functional derivatives with respect to $J$ (and not with respect to $g_1$), and so the $\ell$-dependence of the source terms does not propagate through to the Green functions. If we consider just the $\ell$-dependences of the couplings then, to see a diagram with $E$ external legs and $L$ loops we would have to include terms up to $\mathcal{O}(\ell^{2L-2+E})$ in the \emph{couplings}.\footnote{On a parallel note, $\ell$ plays the same role as does the string coupling in string theory \cite{Polchinski_v1},$g_s$. In the presence of compact dimensions (and in the fixed-loop momenta representation \cite{D'HokerPhong89}) with compactification radii $R^a$, the role of $g_s$ is replaced by (the T-duality invariant) \cite{SklirosCopelandSaffin13b} $g_{\rm eff}=g_s\prod_{a}(\frac{\ell_s}{R^a})^{\frac{1}{2}}$, with $\ell_s=\sqrt{\alpha'}$ the fundamental string length, so that $L$-loop closed string amplitudes with $E$ vertex operators are $\mathcal{O}(g_{\rm eff}^{2L-2+E})$. } For example, to see a 2-loop diagram ($L=2$) with four external legs ($E=4$) we would need to go up to $\mathcal{O}(\ell^6)$, i.e., we should include terms of order $\lambda^3$, $g^6$, $\kappa^2$, $g^3\kappa$, $\lambda\gamma$, $g^2\gamma$, $\lambda^2g^2$, etc. 

Clearly, $U(J)$ contains a number of tadpole (see below) and more generally cephalopod Feynman diagrams (examples are given on p.~\pageref{cephalopod diagrams}). \emph{All} these can be cancelled by appropriate counterterms, but there is currently no way of knowing what counterterms will do the job. Below we will show that when one starts from a complete normal ordered action, however, the correct counterterms are automatically produced, and so \emph{all} cephalopods are absent, to any finite order in perturbation theory. 

Let us now focus on the tadpoles in $U(J)$, by which we mean diagrams of the form:
$$
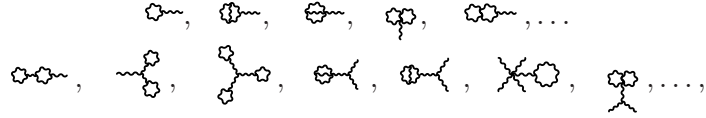
\begin{gathered}
	\begin{fmffile}{wgtadpole-xxx}
		\fmfset{dash_len}{1.2mm}
		\begin{fmfgraph}(43,43)
			\fmfset{dash_len}{1.2mm}
			\fmfset{wiggly_len}{1.1mm} \fmfset{dot_len}{0.5mm}
			\fmfpen{0.25mm}
			\fmfleft{i}
			\fmfright{o}
			\fmf{phantom,tension=5}{i,v1}
			\fmf{wiggly,fore=black,tension=0.8}{v2,o}
			\fmf{wiggly,fore=black,left,tension=0.08}{v1,v2,v1}
			\fmf{phantom}{v1,v2}
		\end{fmfgraph}
	\end{fmffile}
\end{gathered}
,  
\!\!\!\!\!\!
\quad
\begin{gathered}
	\begin{fmffile}{wg3tadpole1-xxx}
		\begin{fmfgraph}(65,65)
			\fmfset{dash_len}{1.2mm}
			\fmfset{wiggly_len}{1.1mm} \fmfset{dot_len}{0.5mm}
			\fmfpen{0.25mm}
			\fmftop{t}
			\fmfbottom{b}
			\fmfleft{l}
			\fmfright{r}
			\fmf{phantom,fore=black,tension=9}{t,u,v,b}
			\fmf{phantom,fore=black,tension=9}{l,s,x,r}
			\fmf{wiggly,fore=black,tension=.01,left}{u,v,u}
			\fmf{phantom,fore=black,tension=0.01}{s,x,s}
			\fmf{wiggly,fore=black,tension=1}{u,v}
			\fmf{wiggly,fore=black,tension=1}{x,r}
		\end{fmfgraph}
	\end{fmffile}
\end{gathered}
,\!\! 
\quad
\begin{gathered}
	\begin{fmffile}{wglambdatadpole-xxx}
		\begin{fmfgraph}(45,45)
			\fmfset{dash_len}{1.2mm}
			\fmfset{wiggly_len}{1.1mm} \fmfset{dot_len}{0.5mm}
			\fmfpen{0.25mm}
			\fmfleft{i}
			\fmfright{o}
			\fmf{phantom,tension=5}{i,v1}
			\fmf{wiggly,fore=black,tension=2.5}{v2,o}
			\fmf{wiggly,fore=black,left,tension=0.5}{v1,v2,v1}
			\fmf{wiggly,fore=black}{v1,v2}
			\fmffreeze
			\fmfforce{(1.1w,0.5h)}{o}
		\end{fmfgraph}
	\end{fmffile}
\end{gathered}
\,, \!\!\!\!
\quad
\begin{gathered}
	\begin{fmffile}{wkappa-tadpole-xxx}
		\begin{fmfgraph}(60,60)
			\fmfset{dash_len}{1.2mm}
			\fmfset{wiggly_len}{1.1mm} \fmfset{dot_len}{0.5mm}
			\fmfpen{0.25mm}
			\fmftop{t1,t2,t3}
			\fmfbottom{b1,b2,b3}
			\fmf{phantom}{t1,v1,b1}
			\fmf{phantom}{t2,v2,b2}
			\fmf{phantom}{t3,v3,b3}
			\fmf{wiggly,fore=black,right,tension=1}{v1,v2,v1}
			\fmf{wiggly,fore=black,right,tension=1}{v2,v3,v2}
			\fmf{wiggly,fore=black,tension=1}{v2,b2}
		\end{fmfgraph}
	\end{fmffile}
\end{gathered}\!\!
, 
\quad
\begin{gathered}
	\begin{fmffile}{wglambda-tadpole-xxx}
		\begin{fmfgraph}(50,50)
			\fmfset{dash_len}{1.2mm}
			\fmfset{wiggly_len}{1.1mm} \fmfset{dot_len}{0.5mm}
			\fmfpen{0.25mm}
			\fmftop{t1,t2,t3,t4}
        			\fmfbottom{b1,b2,b3,b4}
        			\fmf{phantom}{t1,v1,b1}
        			\fmf{phantom}{t2,v2,b2}
			\fmf{phantom}{t3,v3,b3}
			\fmf{phantom}{t4,v4,b4}
        			\fmffreeze
			\fmf{wiggly,fore=black,right,tension=0.7}{v1,v2,v1}
        			\fmf{wiggly,fore=black,right,tension=0.7}{v2,v3,v2}
        			\fmf{wiggly,fore=black,tension=3}{v3,v4}
			\fmffreeze
			\fmfforce{(1.1w,0.5h)}{v4}
		\end{fmfgraph}
	\end{fmffile}
\end{gathered}\,\,,
\dots
$$
\vspace{-0.5cm}
$$
\begin{gathered}
	\begin{fmffile}{wg3-tadpole2-xxx}
		\begin{fmfgraph}(50,50)
			\fmfset{dash_len}{1.2mm}
			\fmfset{wiggly_len}{1.1mm} \fmfset{dot_len}{0.5mm}
			\fmfpen{0.25mm}
			\fmfleft{i}
			\fmfright{o}
			\fmf{phantom,tension=5}{i,v1}
			\fmf{wiggly,fore=black,tension=.04}{v2,o}
			\fmf{wiggly,fore=black,left,tension=0.01}{v1,v3,v1}
			\fmf{wiggly,fore=black,right,tension=0.01}{v2,v4,v2}
			\fmf{wiggly,fore=black,tension=0.03}{v3,v4}
			\fmffreeze
			\fmfforce{(1.2w,0.5h)}{o}
			\fmfforce{(.3w,0.5h)}{v3}
		\end{fmfgraph}
	\end{fmffile}
\end{gathered} \,\,\,,\!\! \!\!
 \,\,\,
\quad
\begin{gathered}
	\begin{fmffile}{wg3-2tadpoles-xxx}
		\begin{fmfgraph}(47,47)
			\fmfset{dash_len}{1.2mm}
			\fmfset{wiggly_len}{1.1mm} \fmfset{dot_len}{0.5mm}
			\fmfpen{0.25mm}
			\fmfleft{i}
			\fmfright{o1,o2}
			\fmf{wiggly,fore=black,tension=1}{i,v1}
			\fmf{phantom,tension=1}{v1,u1,01}
			\fmf{phantom,tension=1}{v1,u2,o2}
			\fmf{wiggly,fore=black,tension=0.8,left}{u1,o1,u1}
			\fmf{wiggly,fore=black,tension=0.4,right}{u2,o2,u2}
			\fmf{wiggly,fore=black,tension=1}{v1,u1}
			\fmf{wiggly,fore=black,tension=1}{v1,u2}
		\end{fmfgraph}
	\end{fmffile}
\end{gathered}
\,\,,\, 
\quad
\begin{gathered}
	\begin{fmffile}{wg4-3tadpole-xxx}
		\begin{fmfgraph}(40,40)
			\fmfset{dash_len}{1.2mm}
			\fmfset{wiggly_len}{1.1mm} \fmfset{dot_len}{0.5mm}
			\fmfpen{0.25mm}
			\fmfsurround{u1,u2,u3}
			\fmf{wiggly,fore=black,tension=1}{u1,v}
			\fmf{wiggly,fore=black,tension=1}{u2,v}
			\fmf{wiggly,fore=black,tension=1}{u3,v}
			\fmf{wiggly,fore=black,tension=1,left}{u1,u1}
			\fmf{wiggly,fore=black,tension=1,left}{u2,u2}
			\fmf{wiggly,fore=black,tension=1,left}{u3,u3}
		\end{fmfgraph}
	\end{fmffile}
\end{gathered}\,\,\,\,
, \!\!\!
\quad
\begin{gathered}
	\begin{fmffile}{wg2L-3loop2pt-1PR-xxx}
		\begin{fmfgraph}(40,40)
			\fmfset{dash_len}{1.2mm}
			\fmfset{wiggly_len}{1.1mm} \fmfset{dot_len}{0.5mm}
			\fmfpen{0.25mm}
			\fmfleft{i}
			\fmfright{o}
			\fmf{phantom,tension=5}{i,v1}
			\fmf{wiggly,fore=black,tension=2.5}{v2,o}
			\fmf{wiggly,fore=black,left,tension=0.5}{v1,v2,v1}
			\fmf{wiggly,fore=black}{v1,v2}
			\fmffreeze
			\fmfforce{(1.1w,0.5h)}{o}
			\fmffreeze
			\fmfright{n,m}
			\fmf{wiggly,fore=black,tension=1}{o,n}
			\fmf{wiggly,fore=black,tension=1}{o,m}
			\fmfforce{(1.35w,0.9h)}{n}
			\fmfforce{(1.35w,0.1h)}{m}
		\end{fmfgraph}
	\end{fmffile}
\end{gathered}
\,\,\,\,\,, \,
\!\! \!\!\!\!\!\!
\quad
\begin{gathered}
	\begin{fmffile}{wg4-tadpole-2pt-xxx}
		\begin{fmfgraph}(60,60)
			\fmfset{dash_len}{1.2mm}
			\fmfset{wiggly_len}{1.1mm} \fmfset{dot_len}{0.5mm}
			\fmfpen{0.25mm}
			\fmftop{t}
			\fmfbottom{b}
			\fmfleft{l}
			\fmfright{r}
			\fmf{phantom,fore=black,tension=9}{t,u,v,b}
			\fmf{phantom,fore=black,tension=9}{l,s,x,r}
			\fmf{wiggly,fore=black,tension=.01,left}{u,v,u}
			\fmf{phantom,fore=black,tension=0.01}{s,x,s}
			\fmf{wiggly,fore=black,tension=1}{u,v}
			\fmf{wiggly,fore=black,tension=1}{x,r}
			\fmffreeze
			\fmfright{a,b}
			\fmf{wiggly,fore=black,tension=1}{r,a}
			\fmf{wiggly,fore=black,tension=1}{r,b}
			\fmffreeze
			\fmfforce{(1.2w,0.8h)}{a}
			\fmfforce{(1.2w,0.2h)}{b}
		\end{fmfgraph}
	\end{fmffile}
\end{gathered}\,\,\,\,\,
, 
\,\,\,
\begin{gathered}
	\begin{fmffile}{wgkappa-4pt1looptadpole,1PR-xxx}
		\begin{fmfgraph}(37,37)
			\fmfset{dash_len}{1.2mm}
			\fmfset{wiggly_len}{1.1mm} \fmfset{dot_len}{0.5mm}
			\fmfpen{0.25mm}
			\fmfleft{i1,i2}
			\fmfright{o1,o2}
			\fmf{wiggly,fore=black}{i1,v,o2}
			\fmf{wiggly,fore=black}{i2,v,o1}
			\fmffreeze
			\fmfright{x}
			\fmf{wiggly,fore=black}{x,v}
			\fmf{wiggly,fore=black,tension=0.7}{x,x}
			\fmfforce{(1.2w,0.5h)}{x}
		\end{fmfgraph}
	\end{fmffile}
\end{gathered}\,\,\,
\,\,\,\,\,, \!\!\!\!
\quad
\begin{gathered}
	\begin{fmffile}{wgkappa-2pt1looptadpole-1PR-xxx}
		\begin{fmfgraph}(60,60)
			\fmfset{dash_len}{1.2mm}
			\fmfset{wiggly_len}{1.1mm} \fmfset{dot_len}{0.5mm}
			\fmfpen{0.25mm}
			\fmftop{t1,t2,t3}
			\fmfbottom{b1,b2,b3}
			\fmf{phantom}{t1,v1,b1}
			\fmf{phantom}{t2,v2,b2}
			\fmf{phantom}{t3,v3,b3}
			\fmf{wiggly,fore=black,right,tension=1}{v1,v2,v1}
			\fmf{wiggly,fore=black,right,tension=1}{v2,v3,v2}
			\fmf{wiggly,fore=black,tension=1}{v2,b2}
			\fmffreeze
			\fmfbottom{x,z}
			\fmf{wiggly,fore=black,tension=1}{b2,x}
			\fmf{wiggly,fore=black,tension=1}{b2,z}
			\fmfforce{(0.25w,-.18h)}{x}
			\fmfforce{(0.75w,-.18h)}{z}
		\end{fmfgraph}
	\end{fmffile}
\end{gathered}\!,
\dots ,
$$
the first line denoting 1-particle irreducible (1PI) tadpole diagrams and the second 1-particle reducible (1PR) tadpole diagrams. We want to construct counterterms to cancel all such diagrams. 
It will be useful to consider first the traditional ``brute force'' method of tadpole cancellation, because later we will show that complete normal ordering leads to a closed-form expression for the relevant counterterm that yields automatically this result. That is, we search for a source counterterm $Y$ in 
$
e^{W(J)}\,\,\propto \,\,e^{\int Y\delta_J}e^{U(J)},
$ 
such that all tadpoles present in $U(J)$, see (\ref{eq:U(Jwiggly)}), are absent in $W(J)$. Omitting the details of this standard procedure, 
making use of the identity,
\begin{equation*}
\begin{aligned}
e^{\int_zY(z)\delta_{J}}e^{U(J)}=e^{\sum_{N=0}^{\infty}\frac{1}{N!}\int_{z_1,\dots,z_N}Y(z_1)\dots Y(z_N)\delta_{J_1}\dots \delta_{J_N}U(J)} \, ,
\end{aligned}
\end{equation*}
one can show that the following is the appropriate source counterterm that incorporates the tadpole cancellation condition up to $\mathcal{O}(\ell^4)$:
\begin{spreadlines}{-.4\baselineskip}
\begin{equation}\label{eq:Ybare}
Y=\hat{g}\Big(\tfrac{1}{2}
\hspace{0.5cm}
\begin{gathered}
	\begin{fmffile}{bubblexas}
		\begin{fmfgraph}(40,40)
			\fmfset{dash_len}{1.2mm}
			\fmfset{wiggly_len}{1.1mm} \fmfset{dot_len}{0.5mm}
			\fmfpen{0.25mm}
			\fmfvn{decor.shape=circle,decor.filled=full, decor.size=3thin}{u}{1}
			\fmfleft{i}
			\fmfright{o}
			\fmf{wiggly,fore=black,tension=5,left,label.dist=1,label=a}{i,u1,i}
			\fmffreeze
			\fmfforce{(-w,0.35h)}{i}
			\fmfforce{(0w,0.35h)}{u1}
			\fmfforce{(1.1w,0.35h)}{o}
			\fmf{phantom,label.dist=0,label=1}{i,u1}
		\end{fmfgraph}\!\!\!\!\!\!
	\end{fmffile}
\end{gathered}\Big)
+\hat{g}^3\Big(
\tfrac{1}{4}\hspace{-.5cm}
\begin{gathered}
	\begin{fmffile}{wg3tav}
		\begin{fmfgraph}(130,130)
			\fmfset{dash_len}{1.2mm}
			\fmfset{wiggly_len}{1.1mm} \fmfset{dot_len}{0.5mm}
			\fmfpen{0.25mm}
			\fmftop{t}
			\fmfbottom{b}
			\fmfleft{l}
			\fmfright{r}
			\fmfv{decor.shape=circle,decor.filled=full, decor.size=3thin}{u}
			\fmf{phantom,fore=black,tension=9}{t,x,v,b}
			\fmf{phantom,fore=black,tension=9}{l,s,u,r}
			\fmf{wiggly,fore=black,tension=.01,left}{x,v,x}
			\fmf{phantom,fore=black,tension=0.01}{s,x,s}
			\fmf{wiggly,fore=black,tension=1}{x,v}
			\fmf{phantom,fore=black,tension=1}{u,r}
			\fmffreeze
			\fmfforce{(0.66w,0.5h)}{u}
		\end{fmfgraph}
	\end{fmffile}
\end{gathered}\hspace{-.5cm}
\Big)
-\hat{g}\hat{\lambda}\Big(
\tfrac{1}{3!}\hspace{-.2cm}
\begin{gathered}
	\begin{fmffile}{wglambdatadpolesdf}
		\begin{fmfgraph}(90,90)
			\fmfset{dash_len}{1.2mm}
			\fmfset{wiggly_len}{1.1mm} \fmfset{dot_len}{0.5mm}
			\fmfpen{0.25mm}
			\fmfleft{i}
			\fmfright{o}
			\fmfv{decor.shape=circle,decor.filled=full, decor.size=3thin}{v2}
			\fmf{phantom,tension=5}{i,v1}
			\fmf{phantom,fore=black,tension=2.5}{v2,o}
			\fmf{wiggly,fore=black,left,tension=0.5}{v1,v2,v1}
			\fmf{wiggly,fore=black}{v1,v2}
			\fmffreeze
			\fmfforce{(1.1w,0.5h)}{o}
		\end{fmfgraph}
	\end{fmffile}
\end{gathered}\!\!\!\!
+\tfrac{1}{4}
\begin{gathered}
	\begin{fmffile}{wlambdabubbled}
		\begin{fmfgraph}(80,80)
			\fmfset{dash_len}{1.2mm}
			\fmfset{wiggly_len}{1.1mm} \fmfset{dot_len}{0.5mm}
			\fmfpen{0.25mm}
			\fmftop{t1,t2,t3}
			\fmfbottom{b1,b2,b3}
			\fmf{phantom}{t1,v1,b1}
			\fmf{phantom}{t2,v2,b2}
			\fmf{phantom}{t3,v3,b3}
			\fmfv{decor.shape=circle,decor.filled=full, decor.size=3thin}{v3}
			\fmffreeze
			\fmf{wiggly,fore=black,right}{v1,v2,v1}
			\fmf{wiggly,fore=black,right}{v2,v3,v2}
		\end{fmfgraph}
	\end{fmffile}
\end{gathered}
\Big)
+\hat{\kappa}\Big(
\tfrac{1}{8}
\begin{gathered}
	\begin{fmffile}{wlambdabubbledk}
		\begin{fmfgraph}(80,80)
			\fmfset{dash_len}{1.2mm}
			\fmfset{wiggly_len}{1.1mm} \fmfset{dot_len}{0.5mm}
			\fmfpen{0.25mm}
			\fmftop{t1,t2,t3}
			\fmfbottom{b1,b2,b3}
			\fmf{phantom}{t1,v1,b1}
			\fmf{phantom}{t2,v2,b2}
			\fmf{phantom}{t3,v3,b3}
			\fmfv{decor.shape=circle,decor.filled=full, decor.size=3thin}{v2}
			\fmffreeze
			\fmf{wiggly,fore=black,right}{v1,v2,v1}
			\fmf{wiggly,fore=black,right}{v2,v3,v2}
		\end{fmfgraph}
	\end{fmffile}
\end{gathered}
\Big)+\mathcal{O}(\ell^5).
\end{equation}
\end{spreadlines}
The dots in these diagrams signify that the associated diagrams are not complete, and the functional derivatives, $\delta_J$, present in $e^{\int Y\delta_J}$ generate the missing lines necessary to complete the vertex. For example, the third diagram represents a dressed tree-level three-point amplitude with external legs merged to a point, say $z$, (that is \emph{not} integrated over here); similar remarks hold for the remaining diagrams. Thus $Y=Y(z)$ is a local counterterm, and because it is a function of the $\hat{g}$, $\hat{\lambda}$, etc., as well as the \emph{dressed} propagator, $\mathcal{G}(z,w)$, it is a function also of the remaining counterterms.

This choice for $Y$ indeed cancels all tadpoles present in $U(J)$ in (\ref{eq:U(Jwiggly)}), as has been verified explicitly. 
We now have an expression for the full tadpole-free generating function (\ref{eq:W(J)inter}) of renormalised connected Green functions:
\begin{equation}\label{eq:W(Jwiggly)}
\begin{aligned}
\tfrac{1}{\hbar}&W(J) = \tfrac{1}{2}
\begin{gathered}
	\begin{fmffile}{wfree}
		\begin{fmfgraph}(50,50)
			\fmfset{dash_len}{1.2mm}
			\fmfset{wiggly_len}{1.1mm} \fmfset{dot_len}{0.5mm}
			\fmfpen{0.25mm}
			\fmfleft{i}
			\fmfright{o}
			\fmf{wiggly,fore=black,tension=5}{i,o}
		\end{fmfgraph}
	\end{fmffile}
\end{gathered}
-\hat{g}
\Big(
\tfrac{1}{3!}
\begin{gathered}
	\begin{fmffile}{wg-3p}
		\begin{fmfgraph}(40,40)
			\fmfset{dash_len}{1.2mm}
			\fmfset{wiggly_len}{1.1mm} \fmfset{dot_len}{0.5mm}
			\fmfpen{0.25mm}
			\fmfleft{i}
			\fmfright{o1,o2}
			\fmf{wiggly,fore=black,tension=5}{i,v1}
			\fmf{wiggly,fore=black,tension=5}{v1,o1}
			\fmf{wiggly,fore=black,tension=5}{v1,o2}
		\end{fmfgraph}
	\end{fmffile}
\end{gathered}
\Big)
+\hat{g}^2
\Big(
\tfrac{1}{12}\!
\begin{gathered}
	\begin{fmffile}{wg2-2loopbubble-1PI}
		\begin{fmfgraph}(40,40)
			\fmfset{dash_len}{1.2mm}
			\fmfset{wiggly_len}{1.1mm} \fmfset{dot_len}{0.5mm}
			\fmfpen{0.25mm}
			\fmfleft{i}
			\fmfright{o}
			\fmf{phantom,tension=5}{i,v1}
			\fmf{phantom,tension=5}{v2,o}
			\fmf{wiggly,fore=black,left,tension=0.4}{v1,v2,v1}
			\fmf{wiggly,fore=black}{v1,v2}
		\end{fmfgraph}
	\end{fmffile}
\end{gathered}\!
+\tfrac{1}{4}\,   
\begin{gathered}
	\begin{fmffile}{wg2-2pt}
		\begin{fmfgraph}(40,40)
			\fmfset{dash_len}{1.2mm}
			\fmfset{wiggly_len}{1.1mm} \fmfset{dot_len}{0.5mm}
			\fmfpen{0.25mm}
			\fmfleft{i}
			\fmfright{o}
			\fmf{wiggly,fore=black,tension=1}{i,v1}
			\fmf{wiggly,fore=black,tension=1}{v2,o}
			\fmf{wiggly,fore=black,left,tension=0.4}{v1,v2,v1}
			\fmffreeze
			\fmfforce{(-.12w,0.5h)}{i}
			\fmfforce{(1.1w,0.5h)}{o}
		\end{fmfgraph}
	\end{fmffile}
\end{gathered}\,
+\tfrac{1}{8}
\begin{gathered}
	\begin{fmffile}{w2-2_g2}
		\begin{fmfgraph}(45,45)
			\fmfset{dash_len}{1.2mm}
			\fmfset{wiggly_len}{1.1mm} \fmfset{dot_len}{0.5mm}
			\fmfpen{0.25mm}
			\fmfsurroundn{i}{4}
			\fmf{wiggly,fore=black}{i1,n,m,i4}
			\fmf{wiggly,fore=black}{i2,n}
			\fmf{wiggly,fore=black}{m,i3}
		\end{fmfgraph}
	\end{fmffile}
\end{gathered}
\Big)
-\hat{\lambda}
\Big(
\tfrac{1}{8}
\begin{gathered}
	\begin{fmffile}{wlambdabubble}
		\begin{fmfgraph}(35,35)
			\fmfset{dash_len}{1.2mm}
			\fmfset{wiggly_len}{1.1mm} \fmfset{dot_len}{0.5mm}
			\fmfpen{0.25mm}
			\fmftop{t1,t2,t3}
			\fmfbottom{b1,b2,b3}
			\fmf{phantom}{t1,v1,b1}
			\fmf{phantom}{t2,v2,b2}
			\fmf{phantom}{t3,v3,b3}
			\fmffreeze
			\fmf{wiggly,fore=black,right}{v1,v2,v1}
			\fmf{wiggly,fore=black,right}{v2,v3,v2}
		\end{fmfgraph}
	\end{fmffile}
\end{gathered}\\
&+\tfrac{1}{4}
\begin{gathered}
	\begin{fmffile}{wlambdaself}
		\begin{fmfgraph}(33,33)
			\fmfset{dash_len}{1.2mm}
			\fmfset{wiggly_len}{1.1mm} \fmfset{dot_len}{0.5mm}
			\fmfpen{0.25mm}
			\fmftop{s}
			\fmfleft{a}
			\fmfright{b}
			\fmf{wiggly,fore=black}{a,v}
			\fmf{wiggly,fore=black}{b,v}
			\fmf{wiggly,fore=black,right,tension=.7}{v,v}
			\fmffreeze
			\fmfforce{(0w,0.2h)}{a}
			\fmfforce{(w,0.2h)}{b}
		\end{fmfgraph}
	\end{fmffile}
\end{gathered}
+\tfrac{1}{4!}
\begin{gathered}
	\begin{fmffile}{wlambdax}
		\begin{fmfgraph}(30,30)
			\fmfset{dash_len}{1.2mm}
			\fmfset{wiggly_len}{1.1mm} \fmfset{dot_len}{0.5mm}
			\fmfpen{0.25mm}
			\fmfleft{i1,i2}
			\fmfright{o1,o2}
			\fmf{wiggly,fore=black}{i1,v,o2}
			\fmf{wiggly,fore=black}{i2,v,o1}
		\end{fmfgraph}
	\end{fmffile}
\end{gathered}
\Big)
-\hat{g}^3
\Big(
\tfrac{1}{4}\,\,
\begin{gathered}
	\begin{fmffile}{wfish}
		\begin{fmfgraph}(40,40)
			\fmfset{dash_len}{1.2mm}
			\fmfset{wiggly_len}{1.1mm} \fmfset{dot_len}{0.5mm}
			\fmfpen{0.25mm}
			\fmfleft{i,j}
			\fmfright{o}
			\fmf{wiggly,fore=black,tension=1}{i,v1}
			\fmf{wiggly,fore=black,tension=1}{j,v1}
			\fmf{wiggly,fore=black}{v1,v2}
			\fmf{wiggly,fore=black,tension=1}{v3,o}
			\fmf{wiggly,fore=black,left,tension=0.3}{v2,v3,v2}
			\fmffreeze
			\fmfforce{(-.2w,0.8h)}{i}
			\fmfforce{(-.2w,0.2h)}{j}
			\fmfforce{(0.05w,0.5h)}{v1}
			\fmfforce{(1.2w,0.5h)}{o}
		\end{fmfgraph}
	\end{fmffile}
\end{gathered}
+\tfrac{1}{8}
\begin{gathered}
	\begin{fmffile}{wg3-crystal}
		\begin{fmfgraph}(45,45)
			\fmfset{dash_len}{1.2mm}
			\fmfset{wiggly_len}{1.1mm} \fmfset{dot_len}{0.5mm}
			\fmfpen{0.25mm}
			\fmfsurround{i1,i2,i3,i4,i5,i6}
			\fmf{wiggly,fore=black,tension=1}{i6,v}
			\fmf{wiggly,fore=black,tension=1}{i1,v}
			\fmf{wiggly,fore=black,tension=1}{v,c}
		 	\fmf{wiggly,fore=black,tension=1}{c,u}
			\fmf{phantom,tension=1}{u,i2}
			\fmf{phantom,fore=black,tension=1}{u,i3}
			\fmf{wiggly,fore=black,tension=1}{c,s,i4}
			\fmf{wiggly,fore=black,tension=1}{s,i5}
		\end{fmfgraph}
	\end{fmffile}
\end{gathered}
+\tfrac{1}{6}
\begin{gathered}
	\begin{fmffile}{wg3-3ptlog}
		\begin{fmfgraph}(35,35)
			\fmfset{dash_len}{1.2mm}
			\fmfset{wiggly_len}{1.1mm} \fmfset{dot_len}{0.5mm}
			\fmfpen{0.25mm}
			\fmfsurroundn{i}{6}
			\fmf{phantom,fore=black}{i1,v,u,i4}
			\fmf{wiggly,fore=black}{v,i1}
			\fmf{phantom,fore=black}{i2,s,t,i5}
			\fmf{wiggly,fore=black}{i5,t}
			\fmf{phantom,fore=black}{i3,w,x,i6}
			\fmf{wiggly,fore=black}{i3,w}
			\fmfi{wiggly,fore=black}{fullcircle scaled .55w shifted (.51w,.5h)}
			\fmffreeze
			\fmfforce{(1.25w,0.5h)}{i1}
			\fmfforce{(0.16w,1.1h)}{i3}
			\fmfforce{(0.16w,-.1h)}{i5}
		\end{fmfgraph}
	\end{fmffile}
\end{gathered}\,\,
\Big)
+\hat{g}\hat{\lambda}\Big(
\tfrac{1}{4}\,
\begin{gathered}
	\begin{fmffile}{wglamba-swim}
		\begin{fmfgraph}(40,40)
			\fmfset{dash_len}{1.2mm}
			\fmfset{wiggly_len}{1.1mm} \fmfset{dot_len}{0.5mm}
			\fmfpen{0.25mm}
			\fmfleft{a}
			\fmfright{f,g}
			\fmf{wiggly,fore=black}{a,v,b}
			\fmf{wiggly,fore=black,tension=.6}{v,v}
			\fmf{wiggly,fore=black,tension=1}{b,f}
			\fmf{wiggly,fore=black,tension=1}{b,g}
			\fmffreeze
			\fmfforce{(1.1w,0.8h)}{f}
			\fmfforce{(1.1w,0.2h)}{g}
			\fmffreeze
			\fmfforce{(-.1w,0.5h)}{a}
		\end{fmfgraph}
	\end{fmffile}
\end{gathered}\,
+\tfrac{1}{4}\,\,
\begin{gathered}
	\begin{fmffile}{wglambda-dart}
		\begin{fmfgraph}(40,40)
			\fmfset{dash_len}{1.2mm}
			\fmfset{wiggly_len}{1.1mm} \fmfset{dot_len}{0.5mm}
			\fmfpen{0.25mm}
			\fmfleft{i,j}
			\fmfright{o}
			\fmf{wiggly,fore=black,tension=5}{i,v1}
			\fmf{wiggly,fore=black,tension=5}{j,v1}
			\fmf{wiggly,fore=black,tension=0.8}{v2,o}
			\fmf{wiggly,fore=black,left,tension=0.08}{v1,v2,v1}
			\fmf{phantom}{v1,v2}
			\fmffreeze
			\fmfforce{(-.2w,0.8h)}{i}
			\fmfforce{(-.2w,0.2h)}{j}
		\end{fmfgraph}
	\end{fmffile}
\end{gathered}\\
&+\tfrac{1}{12}
\begin{gathered}
	\begin{fmffile}{wglambda-5ptree}
		\begin{fmfgraph}(30,30)
			\fmfset{dash_len}{1.2mm}
			\fmfset{wiggly_len}{1.1mm} \fmfset{dot_len}{0.5mm}
			\fmfpen{0.25mm}
			\fmftop{t1,t2}
			\fmfbottom{a,b,c}
			\fmf{wiggly,fore=black,tension=1}{a,v}
			\fmf{wiggly,fore=black,tension=1}{b,v}
			\fmf{wiggly,fore=black,tension=1}{c,v}
			\fmf{wiggly,fore=black,tension=2}{v,t}
			\fmf{wiggly,fore=black,tension=0.7,left,straight}{t,t1}
			\fmf{wiggly,fore=black,tension=0.7,left,straight}{t,t2}
			\fmffreeze
			\fmfforce{(0.5w,-0.2h)}{b}
		\end{fmfgraph}
	\end{fmffile}
\end{gathered}
\Big)
-\hat{\kappa}
\Big(
\tfrac{1}{12}
\begin{gathered}
	\begin{fmffile}{wkappa-3ptself}
		\begin{fmfgraph}(40,40)
			\fmfset{dash_len}{1.2mm}
			\fmfset{wiggly_len}{1.1mm} \fmfset{dot_len}{0.5mm}
			\fmfpen{0.25mm}
			\fmftop{t}
			\fmfbottom{a,b,c}
			\fmf{wiggly,fore=black,tension=1}{a,v}
			\fmf{wiggly,fore=black,tension=1}{b,v}
			\fmf{wiggly,fore=black,tension=1}{c,v}
			\fmf{wiggly,fore=black,tension=1.4,left}{v,t,v}
		\end{fmfgraph}
	\end{fmffile}
\end{gathered}+
\tfrac{1}{5!}
\begin{gathered}
	\begin{fmffile}{wkappa-5pt}
		\begin{fmfgraph}(40,40)
			\fmfset{dash_len}{1.2mm}
			\fmfset{wiggly_len}{1.1mm} \fmfset{dot_len}{0.5mm}
			\fmfpen{0.25mm}
			\fmfsurround{u1,u2,u3,u4,u5}
			\fmf{wiggly,fore=black,tension=1}{u1,v}
			\fmf{wiggly,fore=black,tension=1}{u2,v}
			\fmf{wiggly,fore=black,tension=1}{u3,v}
			\fmf{wiggly,fore=black,tension=1}{u4,v}
			\fmf{wiggly,fore=black,tension=1}{u5,v}
		\end{fmfgraph}
	\end{fmffile}
\end{gathered}
\Big)
+\hat{g}^4
\Big(
\tfrac{1}{48}
\begin{gathered}
	\begin{fmffile}{wg4-6pttree}
		\begin{fmfgraph}(45,45)
			\fmfset{dash_len}{1.2mm}
			\fmfset{wiggly_len}{1.1mm} \fmfset{dot_len}{0.5mm}
			\fmfpen{0.25mm}
			\fmfsurround{u1,u2,u3,u4,u5,u6}
			\fmf{wiggly,fore=black,tension=1}{u1,v}
			\fmf{wiggly,fore=black,tension=1}{u2,v}
			\fmf{wiggly,fore=black,tension=1}{u3,u}
			\fmf{wiggly,fore=black,tension=1}{u4,u}
			\fmf{wiggly,fore=black,tension=1}{u5,s}
			\fmf{wiggly,fore=black,tension=1}{u6,s}
			\fmf{wiggly,fore=black,tension=1,left,straight}{s,c}
			\fmf{wiggly,fore=black,tension=1,left,straight}{u,c}
			\fmf{wiggly,fore=black,tension=1,left,straight}{v,c}
		\end{fmfgraph}
	\end{fmffile}
\end{gathered}
+\tfrac{1}{16}\,
\begin{gathered}
	\begin{fmffile}{wg4-0pt-3loop1PIa}
		\begin{fmfgraph}(27,27)
			\fmfset{dash_len}{1.2mm}
			\fmfset{wiggly_len}{1.1mm} \fmfset{dot_len}{0.5mm}
			\fmfpen{0.25mm}
			\fmfsurround{a,b,c,d}
			\fmf{phantom,fore=black,tension=1,curved}{a,b,c,d,a}
			\fmf{wiggly,fore=black,tension=1,right}{a,b}
			\fmf{wiggly,fore=black,tension=1,right}{c,d}
			\fmf{wiggly,fore=black,tension=1,right}{b,c}
			\fmf{wiggly,fore=black,tension=1,right}{d,a}
			\fmf{wiggly,fore=black,tension=1,straight}{a,b}
			\fmf{wiggly,fore=black,tension=1,straight}{c,d}
		\end{fmfgraph}
	\end{fmffile}
\end{gathered}\,
+\tfrac{1}{16}\,\,\,
\begin{gathered}
	\begin{fmffile}{wg4-4pt-1loop}
		\begin{fmfgraph}(40,40)
			\fmfset{dash_len}{1.2mm}
			\fmfset{wiggly_len}{1.1mm} \fmfset{dot_len}{0.5mm}
			\fmfpen{0.25mm}
			\fmfleft{a,b}
			\fmfright{c,d}
			\fmf{wiggly,fore=black,tension=1}{a,v}
			\fmf{wiggly,fore=black,tension=1}{b,v}
			\fmf{wiggly,fore=black,tension=1}{v,s}
			\fmf{wiggly,fore=black,tension=1}{t,u}
			\fmf{wiggly,fore=black,tension=0.3,left}{s,t,s}
			\fmf{wiggly,fore=black,tension=1}{u,c}
			\fmf{wiggly,fore=black,tension=1}{u,d}	
			\fmffreeze
			\fmfforce{(1.3w,0.8h)}{c}
			\fmfforce{(1.3w,0.2h)}{d}
			\fmfforce{(1w,0.5h)}{u}
			\fmfforce{(-0.06w,0.5h)}{v}
			\fmffreeze
			\fmfforce{(-0.3w,0.8h)}{a}
			\fmfforce{(-0.3w,0.2h)}{b}
		\end{fmfgraph}
	\end{fmffile}
\end{gathered}\,\,\,
+\tfrac{1}{8}\,\,
\begin{gathered}
	\begin{fmffile}{wg4-2pt-1PR-twoloop}
		\begin{fmfgraph}(50,50)
			\fmfset{dash_len}{1.2mm}
			\fmfset{wiggly_len}{1.1mm} \fmfset{dot_len}{0.5mm}
			\fmfpen{0.25mm}
			\fmfleft{i}
			\fmfright{o}
			\fmf{wiggly,fore=black,tension=5}{i,v1}
			\fmf{wiggly,fore=black,tension=5}{v2,o}
			\fmf{wiggly,fore=black,left,tension=0.4}{v1,v3,v1}
			\fmf{wiggly,fore=black,right,tension=0.4}{v2,v4,v2}
			\fmf{wiggly,fore=black}{v3,v4}
			\fmffreeze
			\fmfforce{(-.25w,0.5h)}{i}
			\fmfforce{(1.25w,0.5h)}{o}
			\end{fmfgraph}
	\end{fmffile}
\end{gathered}\,\,\\
&+\tfrac{1}{8}\,\,
\begin{gathered}
	\begin{fmffile}{wg4-4ptlog}
		\begin{fmfgraph}(40,40)
			\fmfset{dash_len}{1.2mm}
			\fmfset{wiggly_len}{1.1mm} \fmfset{dot_len}{0.5mm}
			\fmfpen{0.25mm}
			\fmfsurroundn{i}{4}
			\fmf{phantom,fore=black}{i1,v,u,i3}
			\fmf{wiggly,fore=black}{v,i1}
			\fmf{wiggly,fore=black}{i3,u}
			\fmf{phantom,fore=black}{i2,s,t,i4}
			\fmf{wiggly,fore=black}{i4,t}
			\fmf{wiggly,fore=black}{i2,s}
			\fmfi{wiggly,fore=black}{fullcircle scaled .5w shifted (.51w,.5h)}
			\fmffreeze
			\fmfforce{(1.15w,0.5h)}{i1}
			\fmfforce{(0.5w,1.1h)}{i2}
			\fmfforce{(-.1w,0.5h)}{i3}
			\fmfforce{(0.5w,-.1h)}{i4}
		\end{fmfgraph}
	\end{fmffile}
\end{gathered}\,
+\tfrac{1}{8}\begin{gathered}\,\,
	\begin{fmffile}{wg4-6pttreelong}
		\begin{fmfgraph}(55,55)
			\fmfset{dash_len}{1.2mm}
			\fmfset{wiggly_len}{1.1mm} \fmfset{dot_len}{0.5mm}
			\fmfpen{0.25mm}
			\fmfsurroundn{i}{6}
			\fmf{wiggly,fore=black}{i1,v,u,s,t,i4}
			\fmf{wiggly,fore=black}{i6,v}
			\fmf{wiggly,fore=black}{i4,t}
			\fmf{wiggly,fore=black}{i5,t}
			\fmf{wiggly,fore=black}{i2,u}
			\fmf{wiggly,fore=black}{i3,s}
			\fmffreeze
			\fmfforce{(1.12w,0.4h)}{i1}
			\fmfforce{(-.15w,0.423h)}{i4}
			\fmfforce{(.26w,0.16h)}{i5}
			\fmfforce{(0.67w,0.15h)}{i6}
		\end{fmfgraph}
	\end{fmffile}
\end{gathered}\,\,
+\tfrac{1}{4!}\!\!
\begin{gathered}
	\begin{fmffile}{wpeace}
		\begin{fmfgraph}(75,75)
			\fmfset{dash_len}{1.2mm}
			\fmfset{wiggly_len}{1.1mm} \fmfset{dot_len}{0.5mm}
			\fmfpen{0.25mm}
			\fmfsurroundn{i}{3}
			\fmf{phantom,fore=black}{i1,v,i2}
			\fmf{phantom,fore=black}{i2,u,i3}
			\fmf{phantom,fore=black}{i3,s,i1}
			\fmfi{wiggly,fore=black}{fullcircle scaled .4w shifted (.5w,.5h)}
			\fmf{wiggly,fore=black}{v,c}
			\fmf{wiggly,fore=black}{u,c}
			\fmf{wiggly,fore=black}{s,c}
		\end{fmfgraph}
	\end{fmffile}
\end{gathered}\!\!
+\tfrac{1}{4}
\begin{gathered}
	\begin{fmffile}{wg4-2pt2loop-1PIa}
		\begin{fmfgraph}(60,60)
			\fmfset{dash_len}{1.2mm}
			\fmfset{wiggly_len}{1.1mm} \fmfset{dot_len}{0.5mm}
			\fmfpen{0.25mm}
			\fmftop{t}
			\fmfbottom{b}
			\fmfleft{l}
			\fmfright{r}
			\fmf{phantom,fore=black,tension=9}{t,u,v,b}
			\fmf{phantom,fore=black,tension=9}{l,s,x,r}
			\fmf{wiggly,fore=black,tension=.01,left}{u,v,u}
			\fmf{phantom,fore=black,tension=0.01}{s,x,s}
			\fmf{wiggly,fore=black,tension=1}{u,v}
			\fmf{wiggly,fore=black,tension=1}{x,r}
			\fmf{wiggly,fore=black,tension=1}{l,s}
		\end{fmfgraph}
	\end{fmffile}
\end{gathered}
+\tfrac{1}{4}
\begin{gathered}
	\begin{fmffile}{wg4-2pt2loop-1PIb}
		\begin{fmfgraph}(35,35)
			\fmfset{dash_len}{1.2mm}
			\fmfset{wiggly_len}{1.1mm} \fmfset{dot_len}{0.5mm}
			\fmfpen{0.25mm}
			\fmfleft{i}
			\fmfright{o}
			\fmf{phantom,tension=5}{i,v1}
			\fmf{phantom,tension=5}{v2,o}
			\fmf{wiggly,fore=black,left,tension=0.4}{v1,v2,v1}
			\fmf{wiggly,fore=black}{v1,v2}
			\fmfsurroundn{i}{6}
			\fmf{phantom}{i2,a,c,x1,i5}
			\fmf{phantom}{i3,b,c,x2,i6}
			\fmf{wiggly,fore=black}{x1,i5}
			\fmf{wiggly,fore=black}{x2,i6}
			\fmffreeze
			\fmfforce{(0.1w,-.2h)}{i5}
			\fmfforce{(0.9w,-.2h)}{i6}
			\end{fmfgraph}
	\end{fmffile}
\end{gathered}\,
+\tfrac{1}{4}
\begin{gathered}
	\begin{fmffile}{wg4-4ptlog1PR}
		\begin{fmfgraph}(35,35)
			\fmfset{dash_len}{1.2mm}
			\fmfset{wiggly_len}{1.1mm} \fmfset{dot_len}{0.5mm}
			\fmfpen{0.25mm}
			\fmfsurroundn{i}{6}
			\fmf{phantom,fore=black}{i1,v,u,i4}
			\fmf{wiggly,fore=black}{v,i1}
			\fmf{phantom,fore=black}{i2,s,t,i5}
			\fmf{wiggly,fore=black}{i5,t}
			\fmf{phantom,fore=black}{i3,w,x,i6}
			\fmf{wiggly,fore=black}{i3,w}
			\fmfi{wiggly,fore=black}{fullcircle scaled .55w shifted (.51w,.5h)}
			\fmffreeze
			\fmfforce{(1.25w,0.5h)}{i1}
			\fmfforce{(0.16w,1.1h)}{i3}
			\fmfforce{(0.16w,-.1h)}{i5}
			\fmffreeze
			\fmfright{x,z}
			\fmf{wiggly,fore=black,tension=1}{i1,x}
			\fmf{wiggly,fore=black,tension=1}{i1,z}
			\fmfforce{(1.5w,0.9h)}{z}
			\fmfforce{(1.5w,0.1h)}{x}
		\end{fmfgraph}
	\end{fmffile}
\end{gathered}\,\,\,\,
+\tfrac{1}{4}\,\,\,\,\,
\begin{gathered}
	\begin{fmffile}{wg4-4pt-1loopb}
		\begin{fmfgraph}(40,40)
			\fmfset{dash_len}{1.2mm}
			\fmfset{wiggly_len}{1.1mm} \fmfset{dot_len}{0.5mm}
			\fmfpen{0.25mm}
			\fmfsurroundn{i}{4}
			\fmf{wiggly,fore=black}{i1,n,m,i4}
			\fmf{wiggly,fore=black}{i2,n}
			\fmf{wiggly,fore=black}{m,i3}
			\fmffreeze
			\fmfleft{x}
			\fmf{wiggly,fore=black,tension=10,left}{i3,u,i3}
			\fmf{wiggly,fore=black}{u,x}
			\fmfforce{(-.4w,0.6h)}{u}
			\fmfforce{(-.75w,0.75h)}{x}
		\end{fmfgraph}
	\end{fmffile}
\end{gathered}
\Big)
-\hat{g}^2\hat{\lambda}\\
&\times\Big(
\tfrac{1}{8}
\begin{gathered}
	\begin{fmffile}{wg2L-bubble}
		\begin{fmfgraph}(45,45)
			\fmfset{dash_len}{1.2mm}
			\fmfset{wiggly_len}{1.1mm} \fmfset{dot_len}{0.5mm}
			\fmfpen{0.25mm}
			\fmftop{t1,t2,t3}
			\fmfbottom{b1,b2,b3}
			\fmf{phantom}{t1,v1,b1}
			\fmf{phantom}{t2,v2,b2}
			\fmf{phantom}{t3,v3,b3}
			\fmffreeze
			\fmf{wiggly,fore=black,right}{v1,v2,v1}
			\fmf{wiggly,fore=black,right}{v2,v3,v2}
			\fmf{wiggly,fore=black,tension=1}{t1,b1}
			\fmfforce{(0.25w,0.7h)}{t1}
			\fmfforce{(0.25w,0.3h)}{b1}
		\end{fmfgraph}
	\end{fmffile}
\end{gathered}
+\tfrac{1}{8}
\begin{gathered}
	\begin{fmffile}{wg2L-bubble2pt}
		\begin{fmfgraph}(45,45)
			\fmfset{dash_len}{1.2mm}
			\fmfset{wiggly_len}{1.1mm} \fmfset{dot_len}{0.5mm}
			\fmfpen{0.25mm}
			\fmftop{t1,t2,t3}
			\fmfbottom{b1,b2,b3}
			\fmf{phantom}{t1,v1,b1}
			\fmf{phantom}{t2,v2,b2}
			\fmf{phantom}{t3,v3,b3}
			\fmffreeze
			\fmf{wiggly,fore=black,right}{v1,v2,v1}
			\fmf{wiggly,fore=black}{v2,t3}
			\fmf{wiggly,fore=black}{v2,b3}
			\fmf{wiggly,fore=black,tension=1}{t1,b1}
			\fmfforce{(0.25w,0.7h)}{t1}
			\fmfforce{(0.25w,0.3h)}{b1}
			\fmfforce{(0.9w,0.9h)}{t3}
			\fmfforce{(0.9w,0.1h)}{b3}
		\end{fmfgraph}
	\end{fmffile}
\end{gathered}
+\tfrac{1}{4}\,\,\,
\begin{gathered}
	\begin{fmffile}{wg2L-fishself}
		\begin{fmfgraph}(45,45)
			\fmfset{dash_len}{1.2mm}
			\fmfset{wiggly_len}{1.1mm} \fmfset{dot_len}{0.5mm}
			\fmfpen{0.25mm}
			\fmfleft{j}
			\fmfright{o}
			\fmf{wiggly,fore=black,tension=1}{j,v1}
			\fmf{wiggly,fore=black}{v1,v2}
			\fmf{wiggly,fore=black,tension=1}{v3,o}
			\fmf{wiggly,fore=black,left,tension=0.3}{v2,v3,v2}
			\fmf{wiggly,fore=black,tension=0.6}{v1,v1}
			\fmffreeze
			\fmfforce{(-.2w,0.2h)}{j}
			\fmfforce{(0.05w,0.5h)}{v1}
			\fmfforce{(1.2w,0.5h)}{o}
			\end{fmfgraph}
	\end{fmffile}
\end{gathered}\,
+\tfrac{1}{12}\,\,\,
\begin{gathered}
	\begin{fmffile}{wg2L-4ptfish}
		\begin{fmfgraph}(40,40)
			\fmfset{dash_len}{1.2mm}
			\fmfset{wiggly_len}{1.1mm} \fmfset{dot_len}{0.5mm}
			\fmfpen{0.25mm}
			\fmfleftn{i}{3}
			\fmfright{o}
			\fmf{wiggly,fore=black,tension=1}{i1,v}
			\fmf{wiggly,fore=black,tension=1}{i2,v}
			\fmf{wiggly,fore=black,tension=1}{i3,v}
			\fmf{wiggly,fore=black,tension=0.5}{v,u}
			\fmf{wiggly,fore=black,tension=0.2,left}{u,x,u}
			\fmf{wiggly,fore=black,tension=1}{x,o}
			\fmffreeze
			\fmfforce{(-.1w,0.9h)}{i1}
			\fmfforce{(-.1w,0.1h)}{i3}
			\fmfforce{(-.35w,0.5h)}{i2}
			\fmfforce{(1.3w,0.5h)}{o}
		\end{fmfgraph}\,\,
	\end{fmffile}
\end{gathered}
+\tfrac{1}{8}\,\,
\begin{gathered}
	\begin{fmffile}{wg2L-3loopbubble-1PI}
		\begin{fmfgraph}(27,27)
			\fmfset{dash_len}{1.2mm}
			\fmfset{wiggly_len}{1.1mm} \fmfset{dot_len}{0.5mm}
			\fmfpen{0.25mm}
			\fmfsurroundn{i}{3}
			\fmf{wiggly,fore=black,tension=1,right=1}{i1,i2}
			\fmf{wiggly,fore=black,tension=1,right=1}{i2,i3}
			\fmf{wiggly,fore=black,tension=1,right=0.8}{i3,i1}
			\fmf{wiggly,fore=black,tension=1}{i1,i2}
			\fmf{wiggly,fore=black,tension=1}{i2,i3}
		\end{fmfgraph}
	\end{fmffile}
\end{gathered}\,\,
+\tfrac{1}{4}\,\,\,\,\,
\begin{gathered}
	\begin{fmffile}{wg2L-2ptdoublebubble-1PI}
		\begin{fmfgraph}(17,17)
			\fmfset{dash_len}{1.2mm}
			\fmfset{wiggly_len}{1.1mm} \fmfset{dot_len}{0.5mm}
			\fmfpen{0.25mm}
			\fmfright{o1,o2}
			\fmfleft{i}
			\fmf{wiggly,fore=black,tension=1,right=0.63}{i,o1}
			\fmf{wiggly,fore=black,tension=1,right=0.6}{o1,o2}
			\fmf{wiggly,fore=black,tension=1,right=0.63}{o2,i}
			\fmf{wiggly,fore=black,tension=5,left}{i,u,i}
			\fmffreeze
			\fmfforce{(-1.2w,0.5h)}{u}
			\fmffreeze
			\fmfright{x1,x2}
			\fmf{wiggly,fore=black,tension=1}{o1,x2}
			\fmf{wiggly,fore=black,tension=1}{o2,x1}
			\fmfforce{(1.5w,1.8h)}{x1}
			\fmfforce{(1.5w,-0.8h)}{x2}
		\end{fmfgraph}
	\end{fmffile}
\end{gathered}\,\,\,
+\tfrac{1}{4}\,\,\,\,\,
\begin{gathered}
	\begin{fmffile}{wg2L-4pt1loop-1PI}
		\begin{fmfgraph}(17,17)
			\fmfset{dash_len}{1.2mm}
			\fmfset{wiggly_len}{1.1mm} \fmfset{dot_len}{0.5mm}
			\fmfpen{0.25mm}
			\fmfright{o1,o2}
			\fmfleft{i}
			\fmf{wiggly,fore=black,tension=1,right=0.63}{i,o1}
			\fmf{wiggly,fore=black,tension=1,right=0.6}{o1,o2}
			\fmf{wiggly,fore=black,tension=1,right=0.63}{o2,i}
			\fmffreeze
			\fmfleft{s1,s2}
			\fmf{wiggly,fore=black,tension=1}{i,s2}
			\fmf{wiggly,fore=black,tension=1}{i,s1}
			\fmfforce{(-0.8w,1.6h)}{s1}
			\fmfforce{(-0.8w,-0.6h)}{s2}
			\fmffreeze
			\fmfright{x1,x2}
			\fmf{wiggly,fore=black,tension=1}{o1,x2}
			\fmf{wiggly,fore=black,tension=1}{o2,x1}
			\fmfforce{(1.4w,2h)}{x1}
			\fmfforce{(1.4w,-h)}{x2}
		\end{fmfgraph}
	\end{fmffile}
\end{gathered}\,\,\,
+\tfrac{1}{2}
\begin{gathered}
	\begin{fmffile}{wg2L-2pt2loop-1PI}
		\begin{fmfgraph}(40,40)
			\fmfset{dash_len}{1.2mm}
			\fmfset{wiggly_len}{1.1mm} \fmfset{dot_len}{0.5mm}
			\fmfpen{0.25mm}
			\fmfsurroundn{i}{4}
			\fmfi{wiggly,fore=black}{fullcircle scaled .63w shifted (0.5w,.5h)}
			\fmf{phantom,fore=black}{i1,u,v,i3}
			\fmf{wiggly,fore=black}{i1,v}
			\fmf{phantom,fore=black}{i2,s,t,i4}
			\fmf{wiggly,fore=black}{i2,s}
			\fmffreeze
			\fmfforce{(0.5w,1.2h)}{i2}
			\fmfforce{(0.2w,0.5h)}{v}
			\fmfforce{(1.3w,0.5h)}{i1}
		\end{fmfgraph}
	\end{fmffile}
\end{gathered}\,\,\\
&+\tfrac{1}{4}\,\,\,\,\,
\begin{gathered}
	\begin{fmffile}{wg2L-4pt-1loopc}
		\begin{fmfgraph}(45,45)
			\fmfset{dash_len}{1.2mm}
			\fmfset{wiggly_len}{1.1mm} \fmfset{dot_len}{0.5mm}
			\fmfpen{0.25mm}
			\fmfsurroundn{i}{4}
			\fmf{wiggly,fore=black}{i1,n,m,i4}
			\fmf{wiggly,fore=black}{i2,n}
			\fmf{wiggly,fore=black}{m,i3}
			\fmffreeze
			\fmfleft{x}
			\fmf{wiggly,fore=black,tension=0.6,right}{i3,i3}
			\fmf{wiggly,fore=black}{i3,x}
			\fmfforce{(-.4w,0.63h)}{x}
		\end{fmfgraph}
	\end{fmffile}
\end{gathered}
+\tfrac{1}{8}\,\,\,
\begin{gathered}
	\begin{fmffile}{wg2L-4pt1loopwiggly}
		\begin{fmfgraph}(35,35)
			\fmfset{dash_len}{1.2mm}
			\fmfset{wiggly_len}{1.1mm} \fmfset{dot_len}{0.5mm}
			\fmfpen{0.25mm}
			\fmfleft{i1,i2}
			\fmfright{o1,o2}
			\fmf{wiggly,fore=black,tension=1}{i1,n}
			\fmf{wiggly,fore=black,tension=1}{i2,n}
			\fmf{wiggly,fore=black,tension=1}{n,m}
			\fmf{wiggly,fore=black,tension=0.15,left}{m,s,m}
			\fmf{wiggly,fore=black,tension=1}{s,o1}
			\fmf{wiggly,fore=black,tension=1}{s,o2}
			\fmffreeze
			\fmfforce{(-0.15w,0.5h)}{n}
			\fmfforce{(-0.45w,h)}{i1}
			\fmfforce{(-0.45w,0h)}{i2}
			\fmfforce{(1.15w,h)}{o1}
			\fmfforce{(1.15w,0h)}{o2}
		\end{fmfgraph}
	\end{fmffile}
\end{gathered}\,\,\,
+\tfrac{1}{12}
\begin{gathered}
	\begin{fmffile}{wg2L-crystal}
		\begin{fmfgraph}(45,45)
			\fmfset{dash_len}{1.2mm}
			\fmfset{wiggly_len}{1.1mm} \fmfset{dot_len}{0.5mm}
			\fmfpen{0.25mm}
			\fmfsurround{i1,i2,i3,i4,i5,i6}
			\fmf{wiggly,fore=black,tension=1}{i6,v}
			\fmf{wiggly,fore=black,tension=1}{i1,v}
			\fmf{wiggly,fore=black,tension=1}{v,c}
		 	\fmf{wiggly,fore=black,tension=1}{c,u}
			\fmf{phantom,tension=1}{u,i2}
			\fmf{phantom,fore=black,tension=1}{u,i3}
			\fmf{wiggly,fore=black,tension=1}{c,s,i4}
			\fmf{wiggly,fore=black,tension=1}{s,i5}
			\fmffreeze
			\fmfright{x}
			\fmf{wiggly,fore=black}{v,x}
			\fmfforce{(1.1w,0.2h)}{x}
			\fmfforce{(1.05w,0.54h)}{i1}
			\fmfforce{(-.05w,0.54h)}{i4}
		\end{fmfgraph}
	\end{fmffile}
\end{gathered}
+\tfrac{1}{8}\,\,\,\,
\begin{gathered}
	\begin{fmffile}{wg2L-2pt2loop-1PIb}
		\begin{fmfgraph}(50,50)
			\fmfset{dash_len}{1.2mm}
			\fmfset{wiggly_len}{1.1mm} \fmfset{dot_len}{0.5mm}
			\fmfpen{0.25mm}
			\fmftop{t1,t2,t3,t4}
        			\fmfbottom{b1,b2,b3,b4}
        			\fmf{phantom}{t1,v1,b1}
        			\fmf{phantom}{t2,v2,b2}
			\fmf{phantom}{t3,v3,b3}
			\fmf{phantom}{t4,v4,b4}
        			\fmffreeze
			\fmf{wiggly,fore=black,right,tension=0.7}{v1,v2,v1}
        			\fmf{wiggly,fore=black,right,tension=0.7}{v2,v3,v2}
        			\fmf{wiggly,fore=black,tension=3}{v3,v4}
			\fmffreeze
			\fmfforce{(1.1w,0.5h)}{v4}
			\fmfleft{l}
			\fmf{wiggly,fore=black}{v1,l}
			\fmfforce{(-.4w,0.5h)}{l}
		\end{fmfgraph}
	\end{fmffile}
\end{gathered}
+\tfrac{1}{4}\,\,\,\,
\begin{gathered}
	\begin{fmffile}{wg2L-4ptdart}
		\begin{fmfgraph}(40,40)
			\fmfset{dash_len}{1.2mm}
			\fmfset{wiggly_len}{1.1mm} \fmfset{dot_len}{0.5mm}
			\fmfpen{0.25mm}
			\fmfleft{i,j}
			\fmfright{o}
			\fmf{wiggly,fore=black,tension=5}{i,v1}
			\fmf{wiggly,fore=black,tension=5}{j,v1}
			\fmf{wiggly,fore=black,tension=0.8}{v2,o}
			\fmf{wiggly,fore=black,left,tension=0.08}{v1,v2,v1}
			\fmf{phantom}{v1,v2}
			\fmffreeze
			\fmfforce{(-.2w,0.8h)}{i}
			\fmfforce{(-.2w,0.2h)}{j}
			\fmfleft{n,m}
			\fmf{wiggly,fore=black,tension=1}{i,n}
			\fmf{wiggly,fore=black,tension=1}{i,m}
			\fmfforce{(-.5w,1.1h)}{n}
			\fmfforce{(-.55w,0.5h)}{m}			
		\end{fmfgraph}
	\end{fmffile}
\end{gathered}
+\tfrac{1}{16}\,
\begin{gathered}
	\begin{fmffile}{wg2L-4pt-1PR}
		\begin{fmfgraph}(35,35)
			\fmfset{dash_len}{1.2mm}
			\fmfset{wiggly_len}{1.1mm} \fmfset{dot_len}{0.5mm}
			\fmfpen{0.25mm}
			\fmfsurround{a,b,c}
			\fmf{wiggly,fore=black,tension=1}{c,v,a}
			\fmf{wiggly,fore=black,left,tension=0.8}{v,b,v}
			\fmffreeze
			\fmfright{n,m}
			\fmf{wiggly,fore=black,tension=1.3}{a,n}
			\fmf{wiggly,fore=black,tension=1.3}{a,m}
			\fmfforce{(1.3w,0.8h)}{n}
			\fmfforce{(1.1w,0.1h)}{m}
			\fmffreeze
			\fmfbottom{x,z}
			\fmf{wiggly,fore=black,tension=1.3}{c,x}
			\fmf{wiggly,fore=black,tension=1.3}{c,z}
			\fmfforce{(-.12w,-0.12h)}{x}
			\fmfforce{(.55w,-0.2h)}{z}
		\end{fmfgraph}
	\end{fmffile}
\end{gathered}\,\,
+\tfrac{1}{16}\,
\begin{gathered}
	\begin{fmffile}{wg2L-6pt-1PR}
		\begin{fmfgraph}(35,35)
			\fmfset{dash_len}{1.2mm}
			\fmfset{wiggly_len}{1.1mm} \fmfset{dot_len}{0.5mm}
			\fmfpen{0.25mm}
			\fmfsurround{a,b,c}
			\fmf{wiggly,fore=black,tension=1}{c,v,a}
			\fmf{phantom,fore=black,left,tension=1.2}{v,b,v}
			\fmffreeze
			\fmftop{n,m}
			\fmf{wiggly,fore=black,tension=1}{n,v}
			\fmf{wiggly,fore=black,tension=1}{m,v}
			\fmfforce{(-0.05w,0.77h)}{n}
			\fmfforce{(0.57w,1.15h)}{m}
			\fmffreeze
			\fmfbottom{x,z}
			\fmf{wiggly,fore=black,tension=1.3}{c,x}
			\fmf{wiggly,fore=black,tension=1.3}{c,z}
			\fmfforce{(-.12w,-0.12h)}{x}
			\fmfforce{(.55w,-0.2h)}{z}
			\fmffreeze
			\fmfright{s,t}
			\fmf{wiggly,fore=black,tension=1.3}{a,s}
			\fmf{wiggly,fore=black,tension=1.3}{a,t}
			\fmfforce{(1.3w,0.8h)}{s}
			\fmfforce{(1.1w,0.1h)}{t}
		\end{fmfgraph}
	\end{fmffile}
\end{gathered}\,\,
\Big)\\
&+\hat{\lambda}^2
\Big(
\tfrac{1}{48}
\begin{gathered}
	\begin{fmffile}{wL2-3loop-bubble}
		\begin{fmfgraph}(50,50)
			\fmfset{dash_len}{1.2mm}
			\fmfset{wiggly_len}{1.1mm} \fmfset{dot_len}{0.5mm}
			\fmfpen{0.25mm}
			\fmfleft{i}
			\fmfright{o}
			\fmf{phantom,tension=10}{i,v1}
			\fmf{phantom,tension=10}{v2,o}
			\fmf{wiggly,left,tension=0.4}{v1,v2,v1}
			\fmf{wiggly,left=0.5}{v1,v2}
			\fmf{wiggly,right=0.5}{v1,v2}
    		\end{fmfgraph}
	\end{fmffile}
\end{gathered}
+\tfrac{1}{16}
\begin{gathered}
	\begin{fmffile}{wL2-bubble}
		\begin{fmfgraph}(35,35)
			\fmfset{dash_len}{1.2mm}
			\fmfset{wiggly_len}{1.1mm} \fmfset{dot_len}{0.5mm}
			\fmfpen{0.25mm}
			\fmftop{t1,t2,t3}
			\fmfbottom{b1,b2,b3}
			\fmf{phantom}{t1,v1,b1}
			\fmf{phantom}{t2,v2,b2}
			\fmf{phantom}{t3,v3,b3}
			\fmffreeze
			\fmf{wiggly,fore=black,right}{v1,v2,v1}
			\fmf{wiggly,fore=black,right}{v2,v3,v2}
			\fmfi{wiggly,fore=black}{fullcircle scaled .5w shifted (1.25w,.5h)}
		\end{fmfgraph}
	\end{fmffile}
\end{gathered}\,\,\,
+\tfrac{1}{12}\,\,
\begin{gathered}
	\begin{fmffile}{wL2-2pt2loop-1PI}
		\begin{fmfgraph}(45,45)
			\fmfset{dash_len}{1.2mm}
			\fmfset{wiggly_len}{1.1mm} \fmfset{dot_len}{0.5mm}
			\fmfpen{0.25mm}
			\fmfleft{i}
			\fmfright{o}
			\fmf{wiggly,fore=black,tension=5}{i,v1}
			\fmf{wiggly,fore=black,tension=5}{v2,o}
			\fmf{wiggly,fore=black,left,tension=0.4}{v1,v2,v1}
			\fmf{wiggly,fore=black}{v1,v2}
			\fmffreeze
			\fmfforce{(-0.2w,0.5h)}{i}
			\fmfforce{(1.2w,0.5h)}{o}
		\end{fmfgraph}
	\end{fmffile}
\end{gathered}\,\,
+\tfrac{1}{8}
\begin{gathered}
	\begin{fmffile}{wL2-2pt2loop-1PIb}
		\begin{fmfgraph}(35,35)
			\fmfset{dash_len}{1.2mm}
			\fmfset{wiggly_len}{1.1mm} \fmfset{dot_len}{0.5mm}
			\fmfpen{0.25mm}
			\fmftop{t1,t2,t3}
			\fmfbottom{b1,b2,b3}
			\fmf{phantom}{t1,v1,b1}
			\fmf{phantom}{t2,v2,b2}
			\fmf{phantom}{t3,v3,b3}
			\fmffreeze
			\fmf{wiggly,fore=black,right}{v1,v2,v1}
			\fmf{wiggly,fore=black,right}{v2,v3,v2}
			\fmffreeze
			\fmfright{x,z}
			\fmf{wiggly,fore=black}{v3,x}
			\fmf{wiggly,fore=black}{v3,z}
			\fmfforce{(1.3w,0.9h)}{x}
			\fmfforce{(1.3w,0.1h)}{z}
		\end{fmfgraph}
	\end{fmffile}
\end{gathered}\,\,\, 
+\tfrac{1}{8}\,
\begin{gathered}
	\begin{fmffile}{wL2-2pt2loop-1PR}
		\begin{fmfgraph}(50,50)
			\fmfset{dash_len}{1.2mm}
			\fmfset{wiggly_len}{1.1mm} \fmfset{dot_len}{0.5mm}
			\fmfpen{0.25mm}
			\fmfleft{i}
			\fmfright{o}
			\fmf{wiggly,fore=black}{i,u,v,o}
			\fmf{wiggly,fore=black,tension=0.5,left}{u,u}
			\fmf{wiggly,fore=black,tension=0.5,right}{v,v}
		\end{fmfgraph}
	\end{fmffile}
\end{gathered}\,
+\tfrac{1}{16}\,
\begin{gathered}
	\begin{fmffile}{wL2-4pt1loop-1PI}
		\begin{fmfgraph}(50,50)
			\fmfset{dash_len}{1.2mm}
			\fmfset{wiggly_len}{1.1mm} \fmfset{dot_len}{0.5mm}
			\fmfpen{0.25mm}
			\fmfleft{i1,i2}
			\fmfright{o1,o2}
			\fmf{wiggly,fore=black,tension=1}{i1,u}
			\fmf{wiggly,fore=black,tension=1}{i2,u}
			\fmf{wiggly,fore=black,tension=0.4,left}{u,v,u}
			\fmf{wiggly,fore=black}{v,o1}
			\fmf{wiggly,fore=black}{v,o2}
			\fmffreeze
			\fmfforce{(w,0.8h)}{o1}
			\fmfforce{(w,0.2h)}{o2}
			\fmfforce{(0w,0.8h)}{i1}
			\fmfforce{(0w,0.2h)}{i2}
		\end{fmfgraph}
	\end{fmffile}
\end{gathered}\,
+\tfrac{1}{12}\,\,
\begin{gathered}
	\begin{fmffile}{wL2-4pt1loop31-1PR}
		\begin{fmfgraph}(50,50)
			\fmfset{dash_len}{1.2mm}
			\fmfset{wiggly_len}{1.1mm} \fmfset{dot_len}{0.5mm}
			\fmfpen{0.25mm}
			\fmfleftn{i}{3}
			\fmfright{o}
			\fmf{wiggly,fore=black}{i1,u}
			\fmf{wiggly,fore=black}{i2,u}
			\fmf{wiggly,fore=black}{i3,u}
			\fmf{wiggly,fore=black}{u,v,o}
			\fmf{wiggly,fore=black,tension=0.6,right}{v,v}
			\fmffreeze
			\fmfforce{(0w,0.9h)}{i1}
			\fmfforce{(-.2w,0.5h)}{i2}
			\fmfforce{(0w,0.1h)}{i3}
		\end{fmfgraph}
	\end{fmffile}
\end{gathered}\,\\
&+\tfrac{1}{72}\,\,
\begin{gathered}
	\begin{fmffile}{wL2-6pttree33-1PR}
		\begin{fmfgraph}(45,45)
			\fmfset{dash_len}{1.2mm}
			\fmfset{wiggly_len}{1.1mm} \fmfset{dot_len}{0.5mm}
			\fmfpen{0.25mm}
			\fmfleftn{i}{3}
			\fmfrightn{o}{3}
			\fmf{wiggly,fore=black}{i1,u}
			\fmf{wiggly,fore=black}{i2,u}
			\fmf{wiggly,fore=black}{i3,u}
			\fmf{wiggly,fore=black}{o1,v}
			\fmf{wiggly,fore=black}{o2,v}
			\fmf{wiggly,fore=black}{o3,v}
			\fmf{wiggly,fore=black}{u,v}
			\fmffreeze
			\fmfforce{(0w,0.9h)}{i1}
			\fmfforce{(-.2w,0.5h)}{i2}
			\fmfforce{(0w,0.1h)}{i3}
			\fmfforce{(1w,0.9h)}{o1}
			\fmfforce{(1.2w,0.5h)}{o2}
			\fmfforce{(1w,0.1h)}{o3}
		\end{fmfgraph}
	\end{fmffile}
\end{gathered}\,
\Big)
+\hat{g}\hat{\kappa}\Big(
\tfrac{1}{12}\!
\begin{gathered}
	\begin{fmffile}{wgkappa-bubble-1PI}
		\begin{fmfgraph}(30,30)
			\fmfset{dash_len}{1.2mm}
			\fmfset{wiggly_len}{1.1mm} \fmfset{dot_len}{0.5mm}
			\fmfpen{0.25mm}
			\fmfleft{i}
			\fmfright{o}
			\fmf{phantom,tension=5}{i,v1}
			\fmf{phantom,tension=5}{v2,o}
			\fmf{wiggly,fore=black,left,tension=0.5}{v1,v2,v1}
			\fmf{wiggly,fore=black}{v1,v2}
			\fmffreeze
			\fmfi{wiggly,fore=black}{fullcircle scaled .55w shifted (1.1w,.5h)}
		\end{fmfgraph}
	\end{fmffile}
\end{gathered}\,\,
+\tfrac{1}{12}\!
\begin{gathered}
	\begin{fmffile}{wgkappa-bubble-2pt-1PI}
		\begin{fmfgraph}(35,35)
			\fmfset{dash_len}{1.2mm}
			\fmfset{wiggly_len}{1.1mm} \fmfset{dot_len}{0.5mm}
			\fmfpen{0.25mm}
			\fmfleft{i}
			\fmfright{o}
			\fmf{phantom,tension=5}{i,v1}
			\fmf{phantom,tension=5}{v2,o}
			\fmf{wiggly,fore=black,left,tension=0.4}{v1,v2,v1}
			\fmf{wiggly,fore=black}{v1,v2}
			\fmffreeze
			\fmfright{o1,o2}
			\fmf{wiggly,fore=black,tension=1}{v2,o1}
			\fmf{wiggly,fore=black,tension=1}{v2,o2}
			\fmfforce{(1.1w,0.9h)}{o1}
			\fmfforce{(1.1w,0.1h)}{o2}
		\end{fmfgraph}
	\end{fmffile}
\end{gathered}\,
+\tfrac{1}{4}\,
\begin{gathered}
	\begin{fmffile}{wgkappa-2pt2loop-1PIb}
		\begin{fmfgraph}(60,60)
			\fmfset{dash_len}{1.2mm}
			\fmfset{wiggly_len}{1.1mm} \fmfset{dot_len}{0.5mm}
			\fmfpen{0.25mm}
			\fmftop{t1,t2,t3}
			\fmfbottom{b1,b2,b3}
			\fmf{phantom}{t1,v1,b1}
			\fmf{phantom}{t2,v2,b2}
			\fmf{phantom}{t3,v3,b3}
			\fmf{wiggly,fore=black,right,tension=1}{v1,v2,v1}
			\fmf{wiggly,fore=black,right,tension=1}{v2,v3,v2}
			\fmf{wiggly,fore=black,tension=1}{v2,b2}
			\fmffreeze
			\fmfleft{i}
			\fmf{wiggly,fore=black}{i,v1}
			\fmfforce{(-.1w,0.44h)}{i}
		\end{fmfgraph}
	\end{fmffile}
\end{gathered}\!\!
+\tfrac{1}{12}
\begin{gathered}
	\begin{fmffile}{wgkappa-4pt-1PI}
		\begin{fmfgraph}(32,32)
			\fmfset{dash_len}{1.2mm}
			\fmfset{wiggly_len}{1.1mm} \fmfset{dot_len}{0.5mm}
			\fmfpen{0.25mm}
			\fmftop{t}
			\fmfbottom{a,b,c}
			\fmf{wiggly,fore=black,tension=1}{a,v}
			\fmf{wiggly,fore=black,tension=1}{b,v}
			\fmf{wiggly,fore=black,tension=1}{c,v}
			\fmf{wiggly,fore=black,tension=1.4,left}{v,t,v}
			\fmffreeze
			\fmftop{x}
			\fmf{wiggly,fore=black}{t,x}
			\fmfforce{(0.5w,1.45h)}{x}
		\end{fmfgraph}
	\end{fmffile}
\end{gathered}
+\tfrac{1}{8}
\begin{gathered}
	\begin{fmffile}{wgkappa-4ptself1PR}
		\begin{fmfgraph}(32,32)
			\fmfset{dash_len}{1.2mm}
			\fmfset{wiggly_len}{1.1mm} \fmfset{dot_len}{0.5mm}
			\fmfpen{0.25mm}
			\fmftop{t}
			\fmfbottom{a,b,c}
			\fmf{wiggly,fore=black,tension=1}{a,v}
			\fmf{wiggly,fore=black,tension=1}{b,v}
			\fmf{wiggly,fore=black,tension=1}{c,v}
			\fmf{wiggly,fore=black,tension=1.4,left}{v,t,v}
			\fmffreeze
			\fmfbottom{x,z}
			\fmf{wiggly,fore=black}{c,x}
			\fmf{wiggly,fore=black}{c,z}
			\fmfforce{(w,-.4h)}{x}
			\fmfforce{(1.45w,.4h)}{z}
		\end{fmfgraph}
	\end{fmffile}
\end{gathered}\,\,
+\tfrac{1}{48}
\begin{gathered}
	\begin{fmffile}{wgkappa-6pt-1PR}
		\begin{fmfgraph}(40,40)
			\fmfset{dash_len}{1.2mm}
			\fmfset{wiggly_len}{1.1mm} \fmfset{dot_len}{0.5mm}
			\fmfpen{0.25mm}
			\fmfsurroundn{i}{5}
			\fmfright{s,t}
			\fmf{wiggly,fore=black}{i1,c}
			\fmf{wiggly,fore=black}{i2,c}
			\fmf{wiggly,fore=black}{i3,c}
			\fmf{wiggly,fore=black}{i4,c}
			\fmf{wiggly,fore=black}{i5,c}
			\fmf{wiggly,fore=black}{i1,s}
			\fmf{wiggly,fore=black}{i1,t}
			\fmffreeze
			\fmfforce{(1.2w,0.9h)}{s}
			\fmfforce{(1.2w,0.1h)}{t}
		\end{fmfgraph}
	\end{fmffile}
\end{gathered}\,\,
\Big)\\
&-\hat{\gamma}\Big(
\tfrac{1}{48}
\begin{gathered}
	\begin{fmffile}{wgamma-vacuum-1PI}
		\begin{fmfgraph}(50,50)
			\fmfset{dash_len}{1.2mm}
			\fmfset{wiggly_len}{1.1mm} \fmfset{dot_len}{0.5mm}
			\fmfpen{0.25mm}
			\fmfsurroundn{x}{3}
			\fmf{phantom,fore=black}{x1,v}
			\fmf{phantom,fore=black}{x2,v}
			\fmf{phantom,fore=black}{x3,v}
			\fmf{wiggly,fore=black,tension=0.7}{v,v}
			\fmf{wiggly,fore=black,tension=0.7,right}{v,v}
			\fmf{wiggly,fore=black,tension=0.7,left}{v,v}
		\end{fmfgraph}
	\end{fmffile}
\end{gathered}\!
+\tfrac{1}{16}
\begin{gathered}
	\begin{fmffile}{wgamma-2pt1loop-1PI}
		\begin{fmfgraph}(45,45)
			\fmfset{dash_len}{1.2mm}
			\fmfset{wiggly_len}{1.1mm} \fmfset{dot_len}{0.5mm}
			\fmfpen{0.25mm}
			\fmfleft{i}
			\fmfright{o}
			\fmf{wiggly,fore=black,tension=0.7}{i,v,v,o}
			\fmf{wiggly,fore=black,left=90,tension=0.7}{v,v}
		\end{fmfgraph}
	\end{fmffile}
\end{gathered}\!
+\tfrac{1}{48}
\begin{gathered}
	\begin{fmffile}{wgamma-4pt1loop-1PI}
		\begin{fmfgraph}(45,45)
			\fmfset{dash_len}{1.2mm}
			\fmfset{wiggly_len}{1.1mm} \fmfset{dot_len}{0.5mm}
			\fmfpen{0.25mm}
			\fmfsurroundn{x}{8}
			\fmf{phantom,fore=black}{x1,c,x5}
			\fmf{phantom,fore=black}{x2,c,x6}
			\fmf{phantom,fore=black}{x3,c,x7}
			\fmf{phantom,fore=black}{x4,c,x8}
			\fmf{wiggly,fore=black}{x1,c}
			\fmf{wiggly,fore=black}{x8,c}
			\fmf{wiggly,fore=black}{x7,c}
			\fmf{wiggly,fore=black}{x6,c}
			\fmfi{wiggly,fore=black}{fullcircle scaled .38w shifted (0.46w,.58h)}	
		\end{fmfgraph}
	\end{fmffile}
\end{gathered}
+\tfrac{1}{6!}
\begin{gathered}
	\begin{fmffile}{wgamma-6pttree-1PI}
		\begin{fmfgraph}(40,40)
			\fmfset{dash_len}{1.2mm}
			\fmfset{wiggly_len}{1.1mm} \fmfset{dot_len}{0.5mm}
			\fmfpen{0.25mm}
			\fmfsurroundn{x}{6}
			\fmf{wiggly,fore=black}{x1,c,x4}
			\fmf{wiggly,fore=black}{x2,c,x5}
			\fmf{wiggly,fore=black}{x3,c,x6}
		\end{fmfgraph}
	\end{fmffile}
\end{gathered}
\Big)
+\mathcal{O}(\ell^5)+\ln N
-\smallint \hat{\Lambda}\\
&+\tfrac{1}{2}\Big(\hspace{0.4cm}
\begin{gathered}
	\begin{fmffile}{bubble1wer}
		\begin{fmfgraph}(30,30)
			\fmfset{dash_len}{1.2mm}
			\fmfset{wiggly_len}{1.1mm} \fmfset{dot_len}{0.5mm}
			\fmfpen{0.25mm}
			\fmfvn{decor.shape=circle,decor.filled=shaded, decor.size=3.5thin}{u}{1}
			\fmfleft{i}
			\fmfright{o}
			\fmf{dashes,fore=black,tension=5,left}{i,u1,i}
			\fmffreeze
			\fmfforce{(-w,0.35h)}{i}
			\fmfforce{(0w,0.35h)}{u1}
			\fmfforce{(1.1w,0.35h)}{o}
		\end{fmfgraph}\!\!\!\!
	\end{fmffile}
\end{gathered}\hspace{0cm}
+
\tfrac{1}{2}\hspace{0.45cm}
\begin{gathered}
	\begin{fmffile}{bubble2wer}
		\begin{fmfgraph}(30,30)
			\fmfset{dash_len}{1.2mm}
			\fmfset{wiggly_len}{1.1mm} \fmfset{dot_len}{0.5mm}
			\fmfpen{0.25mm}
			\fmfvn{decor.shape=circle,decor.filled=shaded, decor.size=3.5thin}{u}{2}
			\fmfleft{i}
			\fmfright{o}
			\fmf{dashes,fore=black,tension=5,left}{i,u1,u2,i}
			\fmffreeze
			\fmfforce{(-w,0.35h)}{i}
			\fmfforce{(0w,0.35h)}{u1}
			\fmfforce{(-1w,0.35h)}{u2}
			\fmfforce{(1.1w,0.35h)}{o}
		\end{fmfgraph}\!\!\!\!
	\end{fmffile}
\end{gathered}\hspace{0cm}
+
\tfrac{1}{3}\!\!
\begin{gathered}
	\begin{fmffile}{bubble3ert}
		\begin{fmfgraph}(60,60)
			\fmfset{dash_len}{1.2mm}
			\fmfset{wiggly_len}{1.1mm} \fmfset{dot_len}{0.5mm}
			\fmfpen{0.25mm}
			\fmfvn{decor.shape=circle,decor.filled=shaded, decor.size=3.5thin}{x}{3}
			\fmfsurroundn{u}{6}
			\fmf{phantom,fore=black,tension=1}{u1,x1,c,v,u4}
			\fmf{phantom,fore=black,tension=1}{u2,u,c,x3,u5}
			\fmf{phantom,fore=black,tension=1}{u3,x2,c,t,u6}
			\fmffreeze
			\fmf{dashes,fore=black,tension=1,right=.7}{x1,x2}
			\fmf{dashes,fore=black,tension=1,right=.7}{x2,x3}
			\fmf{dashes,fore=black,tension=1,right=.7}{x3,x1}
		\end{fmfgraph}\!\!
	\end{fmffile}
\end{gathered}
+\dots+
\hspace{0.35cm}
\begin{gathered}
	\begin{fmffile}{bubble1byu}
		\begin{fmfgraph}(30,30)
			\fmfset{dash_len}{1.2mm}
			\fmfset{wiggly_len}{1.1mm} \fmfset{dot_len}{0.5mm}
			\fmfpen{0.25mm}
			\fmfvn{decor.shape=square,decor.filled=shaded, decor.size=3.5thin}{u}{1}
			\fmfleft{i}
			\fmfright{o}
			\fmf{plain,fore=black,tension=5,left}{i,u1,i}
			\fmffreeze
			\fmfforce{(-w,0.35h)}{i}
			\fmfforce{(0w,0.35h)}{u1}
			\fmfforce{(1.1w,0.35h)}{o}
		\end{fmfgraph}\!\!\!\!
	\end{fmffile}
\end{gathered}
+
\tfrac{1}{2}\hspace{0.45cm}
\begin{gathered}
	\begin{fmffile}{bubble2bhj}
		\begin{fmfgraph}(30,30)
			\fmfset{dash_len}{1.2mm}
			\fmfset{wiggly_len}{1.1mm} \fmfset{dot_len}{0.5mm}
			\fmfpen{0.25mm}
			\fmfvn{decor.shape=square,decor.filled=shaded, decor.size=3.5thin}{u}{2}
			\fmfleft{i}
			\fmfright{o}
			\fmf{plain,fore=black,tension=5,left}{i,u1,u2,i}
			\fmffreeze
			\fmfforce{(-w,0.35h)}{i}
			\fmfforce{(0w,0.35h)}{u1}
			\fmfforce{(-1w,0.35h)}{u2}
			\fmfforce{(1.1w,0.35h)}{o}
		\end{fmfgraph}\!\!\!\!
	\end{fmffile}
\end{gathered}
+
\tfrac{1}{3}\!\!
\begin{gathered}
	\begin{fmffile}{bubble3bhj}
		\begin{fmfgraph}(60,60)
			\fmfset{dash_len}{1.2mm}
			\fmfset{wiggly_len}{1.1mm} \fmfset{dot_len}{0.5mm}
			\fmfpen{0.25mm}
			\fmfvn{decor.shape=square,decor.filled=shaded, decor.size=3.5thin}{x}{3}
			\fmfsurroundn{u}{6}
			\fmf{phantom,fore=black,tension=1}{u1,x1,c,v,u4}
			\fmf{phantom,fore=black,tension=1}{u2,u,c,x3,u5}
			\fmf{phantom,fore=black,tension=1}{u3,x2,c,t,u6}
			\fmffreeze
			\fmf{plain,fore=black,tension=1,right=.7}{x1,x2}
			\fmf{plain,fore=black,tension=1,right=.7}{x2,x3}
			\fmf{plain,fore=black,tension=1,right=.7}{x3,x1}
		\end{fmfgraph}\!\!
	\end{fmffile}
\end{gathered}
+\dots\Big) .
\end{aligned}
\end{equation}
Notice that the source counterterm has left every remaining diagram (including combinatorial factors) completely intact.


One point we want to emphasise is that mass and wave-function renormalisation counterterms must be taken into account before cancelling tadpoles. Had we cancelled all tadpoles before incorporating the mass and wave-function renormalisation contributions, these latter counterterms would have reintroduced tadpoles into $W(J)$, and consequently $W(J)$ would not have been tadpole-free.

Secondly, the explicit expression for the source counterterm (\ref{eq:Ybare}) will clearly need to be modified when considering higher-order contributions in $\ell$, and the conventional procedure we have outlined does not offer much insight into the general solution for $Y$ that ensures tadpole cancellation to all orders in perturbation theory. This Section should be viewed as a precursor to Sec.~\!\ref{sec:CNO} where we derive the (conjecturally) exact to any finite order in perturbation theory definition of the source counterterm $Y$ without relying on a truncation at any particular order in $\ell$. 

Needless to say, with this choice of source counterterm, we can rest assured that the vacuum about which we are doing perturbation theory is indeed a true minimum (up to $\mathcal{O}(\ell^4)$) of the full quantum effective action, so that $\frac{\delta \Gamma}{\delta \varphi}|_{\varphi=0}=0$.

\section{Complete Normal Ordering}\label{sec:CNO}
We now introduce the notion of `complete normal ordering', that will be a generalisation of conventional normal ordering. It is well known \cite{Coleman75} that normal ordering the bare action of a given interacting field theory results in Green functions that are free from internal vertices that contain propagators that begin and end on the same vertex. In this manner certain but \emph{not} all tadpole diagrams are cancelled, as are various Feynman diagrams involving self-interactions. `Complete normal ordering' the bare action, on the contrary, as we will show, subtracts \emph{all} tadpole diagrams to any finite loop order in perturbation theory and, more generally, subtracts all cephalopod Feynman diagrams\footnote{We recall that cephalopod diagrams were defined in subsection 2.1.}. 

We begin in the following subsection with the definition of `complete normal ordering', then go on to provide a combinatorial interpretation.

\subsection{Definition}\label{sec:CNO-dfn}
Denote by $\mathcal{O}(\phi)$ some (local or non-local) combination of renormalised elementary fields, $\phi$. Complete normal ordering, denoted by $\mathcal{O}(\phi)\rightarrow \,\n\,\mathcal{O}(\phi)\n$, is defined by:
\begin{equation}\label{eq:nmathcalFn2}
\boxed{\n \,\mathcal{O}(\phi)\,\n=\mathcal{O}(\delta_X)\,e^{-\tfrac{1}{\hbar}\left[W(X)-W(0)\right]+\int_zX(z)\phi(z)}\Big|_{X=0}} \, ,
\end{equation}
where $\int_z$ and  $\delta_X$ were defined below (\ref{eq:W(J)01}) and above (\ref{eq:W(J)inter}) respectively, and $W(X)$ is the \emph{full} generating function of renormalised connected Green functions, $G_N(z_1,\dots,z_N)$, see (\ref{eq:W(J) GN}). This expression is the promised appropriate generalisation of (\ref{eq:NO1}). It can be massaged into a somewhat more transparent form by a straightforward application of the infinite-dimensional generalisation of the finite-dimensional identity (see p.~\!152 in \cite{Coleman}), $G\big(\partial_{\bf X}\big)F\big({\bf X}\big)=F\big(\partial_{\bf Y}\big)G\big({\bf Y}\big)\,e^{({\bf X},{\bf Y})}|_{{\bf Y}=0}$, 
for generic vectors ${\bf X}$, ${\bf Y}$, with $({\bf X},{\bf Y})$ the natural inner product of the space. This leads to the equivalent expression:
\begin{equation}\label{eq:nmathcalFn}
\n \mathcal{O}(\phi)\,\n=\exp\Big(-\sum_{N=2}^{\infty}\frac{1}{N!}\int_{z_1}\dots\int_{z_N}G_N(z_1,\dots,z_N)\frac{\delta}{\delta \phi(z_1)}\dots\frac{\delta}{\delta \phi(z_N)}\Big)\,\mathcal{O}(\phi),
\end{equation}
making it clear that complete normal ordering a given operator is equivalent to subtracting from it all possible contractions using the full Green function. In the following subsection we will be completely explicit and provide a pictorial representation of complete normal-ordered monomials.

If we denote expectation values in the \emph{full} theory by, 
$\big\langle\dots\big\rangle\dfn \int \mathcal{D}\phi \,e^{-\frac{1}{\hbar}I_B(Z^{\frac{1}{2}}\phi)|_{J=0}}(\dots)$, 
it becomes almost immediate that the expectation value of a complete normal ordered product vanishes when $\mathcal{O}(\delta_X)\cdot1=0$:
\begin{equation}\label{eq:<nFn>=0}
\big\langle\!\n \mathcal{O}(\phi)\!\n\! \big\rangle=e^{\tfrac{1}{\hbar}W(0)}\mathcal{O}(\delta_X)\cdot1.
\end{equation}
To prove (\ref{eq:<nFn>=0}), insert the operator (\ref{eq:nmathcalFn2}) directly into the path integral, keeping (for now) the source term, $-\int J\phi$, in the action generic, and denote the resulting expectation value by $\langle \dots\rangle_J$ to distinguish it from $\langle\dots\rangle$ above:
\begin{equation}\label{eq:<nOn>a}
\begin{aligned}
\big\langle\!\n \mathcal{O}(\phi)\!\n \!\big\rangle_J&=\int \mathcal{D}\phi \,e^{-\frac{1}{\hbar}I_B(Z^{\frac{1}{2}}\phi)}\n \,\!\mathcal{O}(\phi)\,\n\\
&=\int \mathcal{D}\phi \,e^{-\frac{1}{\hbar}I_B(Z^{\frac{1}{2}}\phi)}\mathcal{O}(\delta_X)e^{-\tfrac{1}{\hbar}\left[W(X)-W(0)\right]+\int_zX(z)\phi(z)}\Big|_{X=0}\\
&=\mathcal{O}(\delta_X)e^{\tfrac{1}{\hbar}W(J+X)-\tfrac{1}{\hbar}W(X)}\Big|_{X=0}e^{\tfrac{1}{\hbar}W(0)}.
\end{aligned}
\end{equation}
Hence, for vanishing source, $J=0$, (\ref{eq:<nFn>=0}) follows directly. That this relation does not vanish in the presence of external sources is an important consistency check: functional derivatives with respect to $J$ in (\ref{eq:<nOn>a}) generate correlation functions of the form, 
$
\big\langle\!\n \mathcal{O}(\phi)\!\n\! \phi(z_1)\phi(z_2)\dots\big\rangle,
$ 
and there should clearly be non-vanishing contributions coming from contractions between $\n \,\mathcal{O}(\phi)\,\n$ and the remaining $\phi$ insertions, as well as contractions between the $\phi$ insertions. Conversely, this is \emph{not} the case for functional derivatives with respect to the local couplings, because these conspire to cancel precisely when $J=0$, so everything is self-consistent. So completely normal ordered products' vacuum expectation values, $\big\langle\!\n \mathcal{O}(\phi)\!\n\! \big\rangle$, vanish in the fully interacting theory when $\mathcal{O}(\delta_X)\cdot1=0$.

Complete normal ordering also has a unique inverse, $\n\,\mathcal{O}(\phi)\n\,\rightarrow \mathcal{O}(\phi)$, 
\begin{equation}\label{eq:mathcalF}
\mathcal{O}(\phi)=\exp\Big(\sum_{N=2}^{\infty}\frac{1}{N!}\int_{z_1}\dots\int_{z_N}G_N(z_1,\dots,z_N)\frac{\delta}{\delta \phi(z_1)}\dots\frac{\delta}{\delta \phi(z_N)}\Big)\,\n \mathcal{O}(\phi)\,\n \, .
\end{equation}
In terms of $W(X)$ this reads: $\mathcal{O}(\phi)=\mathcal{O}(\delta_X)\, e^{\tfrac{1}{\hbar}\left[W(X)-W(0)\right]}\n e^{\int_zX(z)\phi(z)}\n|_{X=0}$. 

A few comments are in order. Just like normal ordering, complete normal ordering is also not unique. One can replace the $N$-point Green functions, $G_N(z_1,\dots,z_N)$, that appear in the defining relation $\n\,\mathcal{O}(\phi)\,\n$, by shifted Green functions: 
$$
G_N(z_1,\dots,z_N)\rightarrow G'_N(z_1,\dots,z_N)=G_N(z_1,\dots,z_N)+\Delta_N(z_1,\dots,z_N),
$$
and different choices of $\Delta_N$ give different complete normal ordering prescriptions. As we shall see, a specific prescription is required to cancel cephalopods completely (i.e., both the infinite and finite parts of these diagrams), namely $\Delta_N=0$. However, an alternative subtraction scheme is to choose $\Delta_N=-G_N$ for all $N\neq 2$, and $\Delta_2=-G_2+\mathcal{G}$, with $\mathcal{G}$ the free propagator, as above. With this choice of scheme complete normal ordering reduces to the usual normal ordering, $\n\,\mathcal{O}(\phi)\,\n\,\rightarrow \,\,:\!\mathcal{O}(\phi)\!\!:$, hence making it clear that normal ordering is a particular case of the more general definition of complete normal ordering in a particular scheme. We emphasise that complete normal ordering does not change the quantum field theory in any observable way, and it can always be undone by a particular choice of counterterms, as we will see below.

\subsection{Combinatorial Interpretation}\label{sec:CI-CNO}

According to (\ref{eq:nmathcalFn}), the completely normal ordered expression for a given \emph{local} monomial, $\n\,\mathcal{O}(\phi)\,\n$, is obtained by adding to $\mathcal{O}(\phi)$, all possible contractions of fields in $\mathcal{O}(\phi)$, using the `negative' of the full $N$-point Green functions at coincident points, $-G_N$, for the contractions. The remaining non-contracted terms are implicitly in time order.

It is useful to give a pictorial representation of this procedure. Consider a collection of dots (each of which will be depicted by
`$\!\!\!\!\!
\begin{gathered}
	\begin{fmffile}{1pt-self-0}
		\begin{fmfgraph}(45,45)
			\fmfset{dash_len}{1.2mm}
			\fmfset{wiggly_len}{1.1mm} \fmfset{dot_len}{0.5mm}
			\fmfpen{0.25mm}
			\fmfsurroundn{i}{1}
			\fmfvn{decor.shape=circle,decor.filled=full, decor.size=2.5thin}{i}{1}
		\end{fmfgraph}
	\end{fmffile}
\end{gathered}\,\,\,$'). If $N$ of these dots are disconnected they will 
denote the (implicitly) time-ordered monomial at coincident space-time points, $\phi^N$. If  $M$ of the remaining dots are connected by a line they will denote the negative of the full renormalised connected Green function at coincident points, $-G_M$, with\footnote{These quantities are divergent generically and will require some regularisation procedure to make sense of them, but we wish to proceed in a scheme-independent manner for now. An explicit regularisation procedure will be required when we discuss beta functions.} $G_M\dfn \lim_{\{z_1,z_2,\dots\}\rightarrow z}G_M(z_1,\dots,z_M),$ and $G_M(z_1,\dots,z_M)$ defined in (\ref{eq:W(J) GN}). There may in general be a number of disconnected collections of continuous lines. At most two lines can end on a dot, while a continuous line cannot begin and end on the same dot if it does not also connect to other dots. With these rules and identifications, the `completely normal-ordered' monomial, $\n\,\phi^N\n,$ with the fields at coincident points  is obtained by summing all partitions\footnote{By partition we mean a collection of disjoint subsets of a given set, with the union of subsets equal to the entire original set.} of the set of $N$ dots, in which the different ways of connecting (all or a subset of) dots by lines distinguish one partition from another. 

We proceed by example, considering the cases $N=2,\dots,6$ (the cases $N=0,1$ being trivial: $\n\,\phi^0\,\n=1$, $\n\,\phi\,\n=\phi$). 
The 2 partitions of the set of $N=2$ identical elements are:
\begin{spreadlines}{0.6\baselineskip}\label{comb}
\begin{equation*}
\begin{aligned}
&
\bigg\{
\begin{array}{l}
\begin{gathered}
	\begin{fmffile}{2pt-self-0}
		\begin{fmfgraph}(45,45)
			\fmfset{dash_len}{1.2mm}
			\fmfset{wiggly_len}{1.1mm} \fmfset{dot_len}{0.5mm}
			\fmfpen{0.25mm}
			\fmfsurroundn{i}{2}
			\fmfvn{decor.shape=circle,decor.filled=full, decor.size=2.5thin}{i}{2}
		\end{fmfgraph}
	\end{fmffile}
\end{gathered}
\phantom{\Bigg\{}\end{array}\!\!\!\bigg\}
\quad \Leftrightarrow\quad\phi^2 \, ,
\\
&
\bigg\{
\begin{array}{l}
\begin{gathered}
	\begin{fmffile}{2pt-self-1}
		\begin{fmfgraph}(45,45)
			\fmfset{dash_len}{1.2mm}
			\fmfset{wiggly_len}{1.1mm} \fmfset{dot_len}{0.5mm}
			\fmfpen{0.25mm}
			\fmfsurroundn{i}{2}
			\fmfvn{decor.shape=circle,decor.filled=full, decor.size=2.5thin}{i}{2}
			\fmf{plain,fore=black}{i1,i2}
		\end{fmfgraph}
	\end{fmffile}
\end{gathered}
\phantom{\Bigg\{}\end{array}\!\!\!\bigg\}
\quad \Leftrightarrow\quad -G_2 \, ,
\end{aligned}
\end{equation*}
\end{spreadlines}
and according to the above we should add these to obtain the complete normal-ordered monomial: $\n\,\phi^2\n=\phi^2-G_2$. 
We would like to emphasise that we are using the `full' interacting and renormalised (connected) 2-point Green function on the right-hand side of this relation. 
Similarly, the 5 partitions of the set of $N=3$ identical elements are:
\begin{spreadlines}{0.6\baselineskip}
\begin{equation*}
\begin{aligned}
&
\bigg\{
\begin{array}{l}
\begin{gathered}
	\begin{fmffile}{3pt-self-0}
		\begin{fmfgraph}(45,45)
			\fmfset{dash_len}{1.2mm}
			\fmfset{wiggly_len}{1.1mm} \fmfset{dot_len}{0.5mm}
			\fmfpen{0.25mm}
			\fmfsurroundn{i}{3}
			\fmfvn{decor.shape=circle,decor.filled=full, decor.size=2.5thin}{i}{3}
		\end{fmfgraph}
	\end{fmffile}
\end{gathered}
\phantom{\Bigg\{}\end{array}\!\!\!\bigg\}
\quad \Leftrightarrow\quad \phi^3 \, ,
\\
&
\bigg\{
\begin{array}{l}
\begin{gathered}
	\begin{fmffile}{3pt-self-1}
		\begin{fmfgraph}(45,45)
			\fmfset{dash_len}{1.2mm}
			\fmfset{wiggly_len}{1.1mm} \fmfset{dot_len}{0.5mm}
			\fmfpen{0.25mm}
			\fmfsurroundn{i}{3}
			\fmfvn{decor.shape=circle,decor.filled=full, decor.size=2.5thin}{i}{3}
			\fmf{plain,fore=black}{i1,i2}
		\end{fmfgraph}
	\end{fmffile}
\end{gathered}\hspace{0.7cm}
\begin{gathered}
	\begin{fmffile}{3pt-self-2}
		\begin{fmfgraph}(45,45)
			\fmfset{dash_len}{1.2mm}
			\fmfset{wiggly_len}{1.1mm} \fmfset{dot_len}{0.5mm}
			\fmfpen{0.25mm}
			\fmfsurroundn{i}{3}
			\fmfvn{decor.shape=circle,decor.filled=full, decor.size=2.5thin}{i}{3}
			\fmf{plain,fore=black}{i2,i3}
		\end{fmfgraph}
	\end{fmffile}
\end{gathered}\hspace{0.7cm}
\begin{gathered}
	\begin{fmffile}{3pt-self-3}
		\begin{fmfgraph}(45,45)
			\fmfset{dash_len}{1.2mm}
			\fmfset{wiggly_len}{1.1mm} \fmfset{dot_len}{0.5mm}
			\fmfpen{0.25mm}
			\fmfsurroundn{i}{3}
			\fmfvn{decor.shape=circle,decor.filled=full, decor.size=2.5thin}{i}{3}
			\fmf{plain,fore=black}{i1,i3}
		\end{fmfgraph}
	\end{fmffile}
\end{gathered}
\phantom{\Bigg\{}\end{array}\!\!\!\bigg\}
\quad \Leftrightarrow\quad -3\,G_2\,\phi \, ,
\\
&
\bigg\{
\begin{array}{l}
\begin{gathered}
	\begin{fmffile}{3pt-self-f}
		\begin{fmfgraph}(45,45)
			\fmfset{dash_len}{1.2mm}
			\fmfset{wiggly_len}{1.1mm} \fmfset{dot_len}{0.5mm}
			\fmfpen{0.25mm}
			\fmfsurroundn{i}{3}
			\fmfvn{decor.shape=circle,decor.filled=full, decor.size=2.5thin}{i}{3}
			\fmf{plain,fore=black}{i1,i3}
			\fmf{plain,fore=black}{i2,i3}
			\fmf{plain,fore=black}{i1,i2}
		\end{fmfgraph}
	\end{fmffile}
\end{gathered}
\phantom{\Bigg\{}\end{array}\!\!\!\bigg\}
\quad \Leftrightarrow\quad -G_3 \, ,
\end{aligned}
\end{equation*}
\end{spreadlines}
and summing these gives the expression $\n\,\phi^3\n=\phi^3-3G_2\phi-G_3$ \label{eq:*phi3*}
for the complete normal-ordered product. The 15 partitions of a set of $N=4$ identical elements are:
\begin{spreadlines}{0.6\baselineskip}
\begin{equation*}
\begin{aligned}
&
\bigg\{
\begin{array}{l}
\begin{gathered}
	\begin{fmffile}{4pt-self-0}
		\begin{fmfgraph}(45,45)
			\fmfset{dash_len}{1.2mm}
			\fmfset{wiggly_len}{1.1mm} \fmfset{dot_len}{0.5mm}
			\fmfpen{0.25mm}
			\fmfsurroundn{i}{4}
			\fmfvn{decor.shape=circle,decor.filled=full, decor.size=2.5thin}{i}{4}
		\end{fmfgraph}
	\end{fmffile}
\end{gathered}
\phantom{\Bigg\{}\end{array}\!\!\!\bigg\}
\quad \Leftrightarrow\quad \phi^4 \, ,
\\
&
\left\{
\begin{array}{l}
\begin{gathered}
	\begin{fmffile}{4pt-self-1}
		\begin{fmfgraph}(45,45)
			\fmfset{dash_len}{1.2mm}
			\fmfset{wiggly_len}{1.1mm} \fmfset{dot_len}{0.5mm}
			\fmfpen{0.25mm}
			\fmfsurroundn{i}{4}
			\fmfvn{decor.shape=circle,decor.filled=full, decor.size=2.5thin}{i}{4}
			\fmf{plain,fore=black}{i1,i2}
		\end{fmfgraph}
	\end{fmffile}
\end{gathered}\hspace{0.7cm}
\begin{gathered}
	\begin{fmffile}{4pt-self-2}
		\begin{fmfgraph}(45,45)
			\fmfset{dash_len}{1.2mm}
			\fmfset{wiggly_len}{1.1mm} \fmfset{dot_len}{0.5mm}
			\fmfpen{0.25mm}
			\fmfsurroundn{i}{4}
			\fmfvn{decor.shape=circle,decor.filled=full, decor.size=2.5thin}{i}{4}
			\fmf{plain,fore=black}{i2,i3}
		\end{fmfgraph}
	\end{fmffile}
\end{gathered}\hspace{0.7cm}
\begin{gathered}
	\begin{fmffile}{4pt-self-3}
		\begin{fmfgraph}(45,45)
			\fmfset{dash_len}{1.2mm}
			\fmfset{wiggly_len}{1.1mm} \fmfset{dot_len}{0.5mm}
			\fmfpen{0.25mm}
			\fmfsurroundn{i}{4}
			\fmfvn{decor.shape=circle,decor.filled=full, decor.size=2.5thin}{i}{4}
			\fmf{plain,fore=black}{i3,i4}
		\end{fmfgraph}
	\end{fmffile}
\end{gathered}\hspace{0.7cm}
\begin{gathered}
	\begin{fmffile}{4pt-self-4}
		\begin{fmfgraph}(45,45)
			\fmfset{dash_len}{1.2mm}
			\fmfset{wiggly_len}{1.1mm} \fmfset{dot_len}{0.5mm}
			\fmfpen{0.25mm}
			\fmfsurroundn{i}{4}
			\fmfvn{decor.shape=circle,decor.filled=full, decor.size=2.5thin}{i}{4}
			\fmf{plain,fore=black}{i4,i1}
		\end{fmfgraph}
	\end{fmffile}
\end{gathered}\phantom{\Bigg\{}\\
\end{array}\!\!\!\right\}
\quad \Leftrightarrow\quad -6\,G_2\phi^2 \, ,
\\
&
\left\{
\begin{array}{l}
\begin{gathered}
	\begin{fmffile}{4pt-self-3p-1}
		\begin{fmfgraph}(45,45)
			\fmfset{dash_len}{1.2mm}
			\fmfset{wiggly_len}{1.1mm} \fmfset{dot_len}{0.5mm}
			\fmfpen{0.25mm}
			\fmfsurroundn{i}{4}
			\fmfvn{decor.shape=circle,decor.filled=full, decor.size=2.5thin}{i}{4}
			\fmf{plain,fore=black}{i1,i2,i3,i1}
		\end{fmfgraph}
	\end{fmffile}
\end{gathered}\hspace{0.7cm}
\begin{gathered}
	\begin{fmffile}{4pt-self-3p-2}
		\begin{fmfgraph}(45,45)
			\fmfset{dash_len}{1.2mm}
			\fmfset{wiggly_len}{1.1mm} \fmfset{dot_len}{0.5mm}
			\fmfpen{0.25mm}
			\fmfsurroundn{i}{4}
			\fmfvn{decor.shape=circle,decor.filled=full, decor.size=2.5thin}{i}{4}
			\fmf{plain,fore=black}{i2,i3,i4,i2}
		\end{fmfgraph}
	\end{fmffile}
\end{gathered}\hspace{0.7cm}
\begin{gathered}
	\begin{fmffile}{4pt-self-3p-3}
		\begin{fmfgraph}(45,45)
			\fmfset{dash_len}{1.2mm}
			\fmfset{wiggly_len}{1.1mm} \fmfset{dot_len}{0.5mm}
			\fmfpen{0.25mm}
			\fmfsurroundn{i}{4}
			\fmfvn{decor.shape=circle,decor.filled=full, decor.size=2.5thin}{i}{4}
			\fmf{plain,fore=black}{i3,i4,i1,i3}
		\end{fmfgraph}
	\end{fmffile}
\end{gathered}\hspace{0.7cm}
\begin{gathered}
	\begin{fmffile}{4pt-self-3p-4}
		\begin{fmfgraph}(45,45)
			\fmfset{dash_len}{1.2mm}
			\fmfset{wiggly_len}{1.1mm} \fmfset{dot_len}{0.5mm}
			\fmfpen{0.25mm}
			\fmfsurroundn{i}{4}
			\fmfvn{decor.shape=circle,decor.filled=full, decor.size=2.5thin}{i}{4}
			\fmf{plain,fore=black}{i4,i1,i2,i4}
		\end{fmfgraph}
	\end{fmffile}
\end{gathered}
\phantom{\Bigg\{}\end{array}\!\!\!\right\}
\quad \Leftrightarrow\quad -4\,G_3\,\phi \, ,
\\
&
\left\{
\begin{array}{l}
\begin{gathered}
	\begin{fmffile}{4pt-self-22p-1}
		\begin{fmfgraph}(45,45)
			\fmfset{dash_len}{1.2mm}
			\fmfset{wiggly_len}{1.1mm} \fmfset{dot_len}{0.5mm}
			\fmfpen{0.25mm}
			\fmfsurroundn{i}{4}
			\fmfvn{decor.shape=circle,decor.filled=full, decor.size=2.5thin}{i}{4}
			\fmf{plain,fore=black}{i1,i2}
			\fmf{plain,fore=black}{i3,i4}
		\end{fmfgraph}
	\end{fmffile}
\end{gathered}\hspace{0.7cm}
\begin{gathered}
	\begin{fmffile}{4pt-self-22p-2}
		\begin{fmfgraph}(45,45)
			\fmfset{dash_len}{1.2mm}
			\fmfset{wiggly_len}{1.1mm} \fmfset{dot_len}{0.5mm}
			\fmfpen{0.25mm}
			\fmfsurroundn{i}{4}
			\fmfvn{decor.shape=circle,decor.filled=full, decor.size=2.5thin}{i}{4}
			\fmf{plain,fore=black}{i2,i3}
			\fmf{plain,fore=black}{i1,i4}
		\end{fmfgraph}
	\end{fmffile}
\end{gathered}\phantom{\Bigg\{}\\
\begin{gathered}
	\begin{fmffile}{4pt-self-22p-3}
		\begin{fmfgraph}(45,45)
			\fmfset{dash_len}{1.2mm}
			\fmfset{wiggly_len}{1.1mm} \fmfset{dot_len}{0.5mm}
			\fmfpen{0.25mm}
			\fmfsurroundn{i}{4}
			\fmfvn{decor.shape=circle,decor.filled=full, decor.size=2.5thin}{i}{4}
			\fmf{plain,fore=black}{i1,i3}
			\fmf{plain,fore=black}{i2,i4}
		\end{fmfgraph}
	\end{fmffile}
\end{gathered}
\phantom{\Bigg\{}\end{array}\!\!\!\right\}
\quad \Leftrightarrow\quad 3\,G_2^2
\\
&
\bigg\{
\begin{array}{l}
\begin{gathered}
	\begin{fmffile}{4pt-self-4p-1}
		\begin{fmfgraph}(45,45)
			\fmfset{dash_len}{1.2mm}
			\fmfset{wiggly_len}{1.1mm} \fmfset{dot_len}{0.5mm}
			\fmfpen{0.25mm}
			\fmfsurroundn{i}{4}
			\fmfvn{decor.shape=circle,decor.filled=full, decor.size=2.5thin}{i}{4}
			\fmf{plain,fore=black}{i1,i2,i3,i4,i1}
		\end{fmfgraph}
	\end{fmffile}
\end{gathered}
\phantom{\Bigg\{}\end{array}\!\!\!\bigg\}
\quad \Leftrightarrow\quad -G_4 \, ,
\end{aligned}
\end{equation*}
\end{spreadlines}
leading to the following expression
\begin{equation}\label{eq:*phi4*}
\begin{aligned}
\n\,\phi^4&\n=\phi^4-6G_2\phi^2-4G_3\phi+3G_2^2-G_4
\end{aligned}
\end{equation}
The 52 partitions of the set of $N=5$ identical elements with the above identifications are:
\begin{spreadlines}{0.6\baselineskip}
\begin{equation*}
\begin{aligned}
&
\bigg\{
\begin{array}{l}
\begin{gathered}
	\begin{fmffile}{5pt-self-0}
		\begin{fmfgraph}(45,45)
			\fmfset{dash_len}{1.2mm}
			\fmfset{wiggly_len}{1.1mm} \fmfset{dot_len}{0.5mm}
			\fmfpen{0.25mm}
			\fmfsurroundn{i}{5}
			\fmfvn{decor.shape=circle,decor.filled=full, decor.size=2.5thin}{i}{5}
		\end{fmfgraph}
	\end{fmffile}
\end{gathered}
\phantom{\Bigg\{}\end{array}\!\!\!\bigg\}
\quad \Leftrightarrow\quad \phi^5 \, ,
\\
&
\left\{
\begin{array}{l}
\begin{gathered}
	\begin{fmffile}{5pt-self-2p-1}
		\begin{fmfgraph}(45,45)
			\fmfset{dash_len}{1.2mm}
			\fmfset{wiggly_len}{1.1mm} \fmfset{dot_len}{0.5mm}
			\fmfpen{0.25mm}
			\fmfsurroundn{i}{5}
			\fmfvn{decor.shape=circle,decor.filled=full, decor.size=2.5thin}{i}{5}
			\fmf{plain,fore=black}{i1,i2}
		\end{fmfgraph}
	\end{fmffile}
\end{gathered}\hspace{0.7cm}
\begin{gathered}
	\begin{fmffile}{5pt-self-2p-2}
		\begin{fmfgraph}(45,45)
			\fmfset{dash_len}{1.2mm}
			\fmfset{wiggly_len}{1.1mm} \fmfset{dot_len}{0.5mm}
			\fmfpen{0.25mm}
			\fmfsurroundn{i}{5}
			\fmfvn{decor.shape=circle,decor.filled=full, decor.size=2.5thin}{i}{5}
			\fmf{plain,fore=black}{i2,i3}
		\end{fmfgraph}
	\end{fmffile}
\end{gathered}\hspace{0.7cm}
\begin{gathered}
	\begin{fmffile}{5pt-self-2p-3}
		\begin{fmfgraph}(45,45)
			\fmfset{dash_len}{1.2mm}
			\fmfset{wiggly_len}{1.1mm} \fmfset{dot_len}{0.5mm}
			\fmfpen{0.25mm}
			\fmfsurroundn{i}{5}
			\fmfvn{decor.shape=circle,decor.filled=full, decor.size=2.5thin}{i}{5}
			\fmf{plain,fore=black}{i3,i4}
		\end{fmfgraph}
	\end{fmffile}
\end{gathered}\hspace{0.7cm}
\begin{gathered}
	\begin{fmffile}{5pt-self-2p-4}
		\begin{fmfgraph}(45,45)
			\fmfset{dash_len}{1.2mm}
			\fmfset{wiggly_len}{1.1mm} \fmfset{dot_len}{0.5mm}
			\fmfpen{0.25mm}
			\fmfsurroundn{i}{5}
			\fmfvn{decor.shape=circle,decor.filled=full, decor.size=2.5thin}{i}{5}
			\fmf{plain,fore=black}{i4,i5}
		\end{fmfgraph}
	\end{fmffile}
\end{gathered}\hspace{0.7cm}
\begin{gathered}
	\begin{fmffile}{5pt-self-2p-5}
		\begin{fmfgraph}(45,45)
			\fmfset{dash_len}{1.2mm}
			\fmfset{wiggly_len}{1.1mm} \fmfset{dot_len}{0.5mm}
			\fmfpen{0.25mm}
			\fmfsurroundn{i}{5}
			\fmfvn{decor.shape=circle,decor.filled=full, decor.size=2.5thin}{i}{5}
			\fmf{plain,fore=black}{i5,i1}
		\end{fmfgraph}
	\end{fmffile}
\end{gathered}\phantom{\Bigg\{}\\
\end{array}\right\}
\quad \Leftrightarrow\quad-10\,G_2\phi^3 \, ,
\\
&
\left\{
\begin{array}{l}
\begin{gathered}
	\begin{fmffile}{5pt-self-3p-1}
		\begin{fmfgraph}(45,45)
			\fmfset{dash_len}{1.2mm}
			\fmfset{wiggly_len}{1.1mm} \fmfset{dot_len}{0.5mm}
			\fmfpen{0.25mm}
			\fmfsurroundn{i}{5}
			\fmfvn{decor.shape=circle,decor.filled=full, decor.size=2.5thin}{i}{5}
			\fmf{plain,fore=black}{i1,i2,i3,i1}
		\end{fmfgraph}
	\end{fmffile}
\end{gathered}\hspace{0.7cm}
\begin{gathered}
	\begin{fmffile}{5pt-self-3p-2}
		\begin{fmfgraph}(45,45)
			\fmfset{dash_len}{1.2mm}
			\fmfset{wiggly_len}{1.1mm} \fmfset{dot_len}{0.5mm}
			\fmfpen{0.25mm}
			\fmfsurroundn{i}{5}
			\fmfvn{decor.shape=circle,decor.filled=full, decor.size=2.5thin}{i}{5}
			\fmf{plain,fore=black}{i2,i3,i4,i2}
		\end{fmfgraph}
	\end{fmffile}
\end{gathered}\hspace{0.7cm}
\begin{gathered}
	\begin{fmffile}{5pt-self-3p-3}
		\begin{fmfgraph}(45,45)
			\fmfset{dash_len}{1.2mm}
			\fmfset{wiggly_len}{1.1mm} \fmfset{dot_len}{0.5mm}
			\fmfpen{0.25mm}
			\fmfsurroundn{i}{5}
			\fmfvn{decor.shape=circle,decor.filled=full, decor.size=2.5thin}{i}{5}
			\fmf{plain,fore=black}{i3,i4,i5,i3}
		\end{fmfgraph}
	\end{fmffile}
\end{gathered}\hspace{0.7cm}
\begin{gathered}
	\begin{fmffile}{5pt-self-3p-4}
		\begin{fmfgraph}(45,45)
			\fmfset{dash_len}{1.2mm}
			\fmfset{wiggly_len}{1.1mm} \fmfset{dot_len}{0.5mm}
			\fmfpen{0.25mm}
			\fmfsurroundn{i}{5}
			\fmfvn{decor.shape=circle,decor.filled=full, decor.size=2.5thin}{i}{5}
			\fmf{plain,fore=black}{i4,i5,i1,i4}
		\end{fmfgraph}
	\end{fmffile}
\end{gathered}\hspace{0.7cm}
\begin{gathered}
	\begin{fmffile}{5pt-self-3p-5}
		\begin{fmfgraph}(45,45)
			\fmfset{dash_len}{1.2mm}
			\fmfset{wiggly_len}{1.1mm} \fmfset{dot_len}{0.5mm}
			\fmfpen{0.25mm}
			\fmfsurroundn{i}{5}
			\fmfvn{decor.shape=circle,decor.filled=full, decor.size=2.5thin}{i}{5}
			\fmf{plain,fore=black}{i5,i1,i2,i5}
		\end{fmfgraph}
	\end{fmffile}
\end{gathered}\phantom{\Bigg\{}\\
\end{array}\right\}
\quad \Leftrightarrow\quad-10\,G_3\phi^2 \, ,
\\
&
\left\{
\begin{array}{l}
\begin{gathered}
	\begin{fmffile}{5pt-self-22p-1}
		\begin{fmfgraph}(45,45)
			\fmfset{dash_len}{1.2mm}
			\fmfset{wiggly_len}{1.1mm} \fmfset{dot_len}{0.5mm}
			\fmfpen{0.25mm}
			\fmfsurroundn{i}{5}
			\fmfvn{decor.shape=circle,decor.filled=full, decor.size=2.5thin}{i}{5}
			\fmf{plain,fore=black}{i1,i2}
			\fmf{plain,fore=black}{i3,i4}
		\end{fmfgraph}
	\end{fmffile}
\end{gathered}\hspace{0.7cm}
\begin{gathered}
	\begin{fmffile}{5pt-self-22p-2}
		\begin{fmfgraph}(45,45)
			\fmfset{dash_len}{1.2mm}
			\fmfset{wiggly_len}{1.1mm} \fmfset{dot_len}{0.5mm}
			\fmfpen{0.25mm}
			\fmfsurroundn{i}{5}
			\fmfvn{decor.shape=circle,decor.filled=full, decor.size=2.5thin}{i}{5}
			\fmf{plain,fore=black}{i2,i3}
			\fmf{plain,fore=black}{i4,i5}
		\end{fmfgraph}
	\end{fmffile}
\end{gathered}\hspace{0.7cm}
\begin{gathered}
	\begin{fmffile}{5pt-self-22p-3}
		\begin{fmfgraph}(45,45)
			\fmfset{dash_len}{1.2mm}
			\fmfset{wiggly_len}{1.1mm} \fmfset{dot_len}{0.5mm}
			\fmfpen{0.25mm}
			\fmfsurroundn{i}{5}
			\fmfvn{decor.shape=circle,decor.filled=full, decor.size=2.5thin}{i}{5}
			\fmf{plain,fore=black}{i3,i4}
			\fmf{plain,fore=black}{i5,i1}
		\end{fmfgraph}
	\end{fmffile}
\end{gathered}\hspace{0.7cm}
\begin{gathered}
	\begin{fmffile}{5pt-self-22p-4}
		\begin{fmfgraph}(45,45)
			\fmfset{dash_len}{1.2mm}
			\fmfset{wiggly_len}{1.1mm} \fmfset{dot_len}{0.5mm}
			\fmfpen{0.25mm}
			\fmfsurroundn{i}{5}
			\fmfvn{decor.shape=circle,decor.filled=full, decor.size=2.5thin}{i}{5}
			\fmf{plain,fore=black}{i4,i5}
			\fmf{plain,fore=black}{i1,i2}
		\end{fmfgraph}
	\end{fmffile}
\end{gathered}\hspace{0.7cm}
\begin{gathered}
	\begin{fmffile}{5pt-self-22p-5}
		\begin{fmfgraph}(45,45)
			\fmfset{dash_len}{1.2mm}
			\fmfset{wiggly_len}{1.1mm} \fmfset{dot_len}{0.5mm}
			\fmfpen{0.25mm}
			\fmfsurroundn{i}{5}
			\fmfvn{decor.shape=circle,decor.filled=full, decor.size=2.5thin}{i}{5}
			\fmf{plain,fore=black}{i5,i1}
			\fmf{plain,fore=black}{i2,i3}
		\end{fmfgraph}
	\end{fmffile}
\end{gathered}\phantom{\Bigg\{}\\
\end{array}\right\}
\quad \Leftrightarrow\quad15\,G_2^2\,\phi \, ,
\\
&
\left\{
\begin{array}{l}
\begin{gathered}
	\begin{fmffile}{5pt-self-4p-1}
		\begin{fmfgraph}(45,45)
			\fmfset{dash_len}{1.2mm}
			\fmfset{wiggly_len}{1.1mm} \fmfset{dot_len}{0.5mm}
			\fmfpen{0.25mm}
			\fmfsurroundn{i}{5}
			\fmfvn{decor.shape=circle,decor.filled=full, decor.size=2.5thin}{i}{5}
			\fmf{plain,fore=black}{i1,i2,i3,i4,i1}
		\end{fmfgraph}
	\end{fmffile}
\end{gathered}\hspace{0.7cm}
\begin{gathered}
	\begin{fmffile}{5pt-self-4p-2}
		\begin{fmfgraph}(45,45)
			\fmfset{dash_len}{1.2mm}
			\fmfset{wiggly_len}{1.1mm} \fmfset{dot_len}{0.5mm}
			\fmfpen{0.25mm}
			\fmfsurroundn{i}{5}
			\fmfvn{decor.shape=circle,decor.filled=full, decor.size=2.5thin}{i}{5}
			\fmf{plain,fore=black}{i2,i3,i4,i5,i2}
		\end{fmfgraph}
	\end{fmffile}
\end{gathered}\hspace{0.7cm}
\begin{gathered}
	\begin{fmffile}{5pt-self-4p-3}
		\begin{fmfgraph}(45,45)
			\fmfset{dash_len}{1.2mm}
			\fmfset{wiggly_len}{1.1mm} \fmfset{dot_len}{0.5mm}
			\fmfpen{0.25mm}
			\fmfsurroundn{i}{5}
			\fmfvn{decor.shape=circle,decor.filled=full, decor.size=2.5thin}{i}{5}
			\fmf{plain,fore=black}{i3,i4,i5,i1,i3}
		\end{fmfgraph}
	\end{fmffile}
\end{gathered}\hspace{0.7cm}
\begin{gathered}
	\begin{fmffile}{5pt-self-4p-4}
		\begin{fmfgraph}(45,45)
			\fmfset{dash_len}{1.2mm}
			\fmfset{wiggly_len}{1.1mm} \fmfset{dot_len}{0.5mm}
			\fmfpen{0.25mm}
			\fmfsurroundn{i}{5}
			\fmfvn{decor.shape=circle,decor.filled=full, decor.size=2.5thin}{i}{5}
			\fmf{plain,fore=black}{i4,i5,i1,i2,i4}
		\end{fmfgraph}
	\end{fmffile}
\end{gathered}\hspace{0.7cm}
\begin{gathered}
	\begin{fmffile}{5pt-self-4p-5}
		\begin{fmfgraph}(45,45)
			\fmfset{dash_len}{1.2mm}
			\fmfset{wiggly_len}{1.1mm} \fmfset{dot_len}{0.5mm}
			\fmfpen{0.25mm}
			\fmfsurroundn{i}{5}
			\fmfvn{decor.shape=circle,decor.filled=full, decor.size=2.5thin}{i}{5}
			\fmf{plain,fore=black}{i5,i1,i2,i3,i5}
		\end{fmfgraph}
	\end{fmffile}
\end{gathered}
\phantom{\Bigg\{}\end{array}\right\}
\quad \Leftrightarrow\quad-5\,G_4\,\phi \, ,\\
&
\left\{
\begin{array}{l}
\begin{gathered}
	\begin{fmffile}{5pt-self-23p-1}
		\begin{fmfgraph}(45,45)
			\fmfset{dash_len}{1.2mm}
			\fmfset{wiggly_len}{1.1mm} \fmfset{dot_len}{0.5mm}
			\fmfpen{0.25mm}
			\fmfsurroundn{i}{5}
			\fmfvn{decor.shape=circle,decor.filled=full, decor.size=2.5thin}{i}{5}
			\fmf{plain,fore=black}{i1,i2,i3,i1}
			\fmf{plain,fore=black}{i4,i5}
		\end{fmfgraph}
	\end{fmffile}
\end{gathered}\hspace{0.7cm}
\begin{gathered}
	\begin{fmffile}{5pt-self-23p-2}
		\begin{fmfgraph}(45,45)
			\fmfset{dash_len}{1.2mm}
			\fmfset{wiggly_len}{1.1mm} \fmfset{dot_len}{0.5mm}
			\fmfpen{0.25mm}
			\fmfsurroundn{i}{5}
			\fmfvn{decor.shape=circle,decor.filled=full, decor.size=2.5thin}{i}{5}
			\fmf{plain,fore=black}{i2,i3,i4,i2}
			\fmf{plain,fore=black}{i5,i1}
		\end{fmfgraph}
	\end{fmffile}
\end{gathered}\hspace{0.7cm}
\begin{gathered}
	\begin{fmffile}{5pt-self-23p-3}
		\begin{fmfgraph}(45,45)
			\fmfset{dash_len}{1.2mm}
			\fmfset{wiggly_len}{1.1mm} \fmfset{dot_len}{0.5mm}
			\fmfpen{0.25mm}
			\fmfsurroundn{i}{5}
			\fmfvn{decor.shape=circle,decor.filled=full, decor.size=2.5thin}{i}{5}
			\fmf{plain,fore=black}{i3,i4,i5,i3}
			\fmf{plain,fore=black}{i1,i2}
		\end{fmfgraph}
	\end{fmffile}
\end{gathered}\hspace{0.7cm}
\begin{gathered}
	\begin{fmffile}{5pt-self-23p-4}
		\begin{fmfgraph}(45,45)
			\fmfset{dash_len}{1.2mm}
			\fmfset{wiggly_len}{1.1mm} \fmfset{dot_len}{0.5mm}
			\fmfpen{0.25mm}
			\fmfsurroundn{i}{5}
			\fmfvn{decor.shape=circle,decor.filled=full, decor.size=2.5thin}{i}{5}
			\fmf{plain,fore=black}{i4,i5,i1,i4}
			\fmf{plain,fore=black}{i2,i3}
		\end{fmfgraph}
	\end{fmffile}
\end{gathered}\hspace{0.7cm}
\begin{gathered}
	\begin{fmffile}{5pt-self-23p-5}
		\begin{fmfgraph}(45,45)
			\fmfset{dash_len}{1.2mm}
			\fmfset{wiggly_len}{1.1mm} \fmfset{dot_len}{0.5mm}
			\fmfpen{0.25mm}
			\fmfsurroundn{i}{5}
			\fmfvn{decor.shape=circle,decor.filled=full, decor.size=2.5thin}{i}{5}
			\fmf{plain,fore=black}{i5,i1,i2,i5}
			\fmf{plain,fore=black}{i3,i4}
		\end{fmfgraph}
	\end{fmffile}
\end{gathered}\phantom{\Bigg\{}\\
\end{array}\right\}
\quad \Leftrightarrow\quad10\,G_2G_3
\\
&
\bigg\{
\begin{array}{l}
\begin{gathered}
	\begin{fmffile}{5pt-self-5p}
		\begin{fmfgraph}(45,45)
			\fmfset{dash_len}{1.2mm}
			\fmfset{wiggly_len}{1.1mm} \fmfset{dot_len}{0.5mm}
			\fmfpen{0.25mm}
			\fmfsurroundn{i}{5}
			\fmfvn{decor.shape=circle,decor.filled=full, decor.size=2.5thin}{i}{5}
			\fmf{plain,fore=black}{i1,i2,i3,i4,i5,i1}
		\end{fmfgraph}
	\end{fmffile}
\end{gathered}
\phantom{\Bigg\{}\end{array}\!\!\!\bigg\}
\quad \Leftrightarrow\quad -G_5 \, ,
\end{aligned}
\end{equation*}
\end{spreadlines}
so that the complete normal-ordered product, $\n\phi^5\n$, is given by:
\begin{equation}\label{eq:*phi5*}
\begin{aligned}
\n\phi^5&\n=\phi^5-10G_2\phi^3-10G_3\phi^2+(15G_2^2-5G_4)\phi+(10G_2G_3-G_5),
\end{aligned}
\end{equation} 
and finally the 203 partitions of $N=6$ identical elements read:
\begin{spreadlines}{0.6\baselineskip}
\begin{equation*}
\begin{aligned}
&
\bigg\{
\begin{array}{l}
\begin{gathered}
	\begin{fmffile}{6pt-self-0}
		\begin{fmfgraph}(45,45)
			\fmfset{dash_len}{1.2mm}
			\fmfset{wiggly_len}{1.1mm} \fmfset{dot_len}{0.5mm}
			\fmfpen{0.25mm}
			\fmfsurroundn{i}{6}
			\fmfvn{decor.shape=circle,decor.filled=full, decor.size=2.5thin}{i}{6}
		\end{fmfgraph}
	\end{fmffile}
\end{gathered}
\phantom{\Bigg\{}\end{array}\!\!\!\bigg\}
\quad \Leftrightarrow\quad \phi^6 \, ,
\\
&
\left\{
\begin{array}{l}
\begin{gathered}
	\begin{fmffile}{6pt-self-2p-1}
		\begin{fmfgraph}(45,45)
			\fmfset{dash_len}{1.2mm}
			\fmfset{wiggly_len}{1.1mm} \fmfset{dot_len}{0.5mm}
			\fmfpen{0.25mm}
			\fmfsurroundn{i}{6}
			\fmfvn{decor.shape=circle,decor.filled=full, decor.size=2.5thin}{i}{6}
			\fmf{plain,fore=black}{i1,i2}
		\end{fmfgraph}
	\end{fmffile}
\end{gathered}\hspace{0.7cm}
\begin{gathered}
	\begin{fmffile}{6pt-self-2p-2}
		\begin{fmfgraph}(45,45)
			\fmfset{dash_len}{1.2mm}
			\fmfset{wiggly_len}{1.1mm} \fmfset{dot_len}{0.5mm}
			\fmfpen{0.25mm}
			\fmfsurroundn{i}{6}
			\fmfvn{decor.shape=circle,decor.filled=full, decor.size=2.5thin}{i}{6}
			\fmf{plain,fore=black}{i2,i3}
		\end{fmfgraph}
	\end{fmffile}
\end{gathered}\hspace{0.7cm}
\begin{gathered}
	\begin{fmffile}{6pt-self-2p-3}
		\begin{fmfgraph}(45,45)
			\fmfset{dash_len}{1.2mm}
			\fmfset{wiggly_len}{1.1mm} \fmfset{dot_len}{0.5mm}
			\fmfpen{0.25mm}
			\fmfsurroundn{i}{6}
			\fmfvn{decor.shape=circle,decor.filled=full, decor.size=2.5thin}{i}{6}
			\fmf{plain,fore=black}{i3,i4}
		\end{fmfgraph}
	\end{fmffile}
\end{gathered}\hspace{0.7cm}
\begin{gathered}
	\begin{fmffile}{6pt-self-2p-4}
		\begin{fmfgraph}(45,45)
			\fmfset{dash_len}{1.2mm}
			\fmfset{wiggly_len}{1.1mm} \fmfset{dot_len}{0.5mm}
			\fmfpen{0.25mm}
			\fmfsurroundn{i}{6}
			\fmfvn{decor.shape=circle,decor.filled=full, decor.size=2.5thin}{i}{6}
			\fmf{plain,fore=black}{i4,i5}
		\end{fmfgraph}
	\end{fmffile}
\end{gathered}\hspace{0.7cm}
\begin{gathered}
	\begin{fmffile}{6pt-self-2p-5}
		\begin{fmfgraph}(45,45)
			\fmfset{dash_len}{1.2mm}
			\fmfset{wiggly_len}{1.1mm} \fmfset{dot_len}{0.5mm}
			\fmfpen{0.25mm}
			\fmfsurroundn{i}{6}
			\fmfvn{decor.shape=circle,decor.filled=full, decor.size=2.5thin}{i}{6}
			\fmf{plain,fore=black}{i5,i6}
		\end{fmfgraph}
	\end{fmffile}
\end{gathered}\hspace{0.7cm}
\begin{gathered}
	\begin{fmffile}{6pt-self-2p-6}
		\begin{fmfgraph}(45,45)
			\fmfset{dash_len}{1.2mm}
			\fmfset{wiggly_len}{1.1mm} \fmfset{dot_len}{0.5mm}
			\fmfpen{0.25mm}
			\fmfsurroundn{i}{6}
			\fmfvn{decor.shape=circle,decor.filled=full, decor.size=2.5thin}{i}{6}
			\fmf{plain,fore=black}{i6,i1}
		\end{fmfgraph}
	\end{fmffile}
\end{gathered}
\phantom{\Bigg\{}\\
\end{array}\right\}
\quad \Leftrightarrow\quad-15\,G_2\phi^4 \, 
\\
&
\left\{
\begin{array}{l}
\begin{gathered}
	\begin{fmffile}{6pt-self-3p-1}
		\begin{fmfgraph}(45,45)
			\fmfset{dash_len}{1.2mm}
			\fmfset{wiggly_len}{1.1mm} \fmfset{dot_len}{0.5mm}
			\fmfpen{0.25mm}
			\fmfsurroundn{i}{6}
			\fmfvn{decor.shape=circle,decor.filled=full, decor.size=2.5thin}{i}{6}
			\fmf{plain,fore=black}{i1,i2,i3,i1}
		\end{fmfgraph}
	\end{fmffile}
\end{gathered}\hspace{0.7cm}
\begin{gathered}
	\begin{fmffile}{6pt-self-3p-2}
		\begin{fmfgraph}(45,45)
			\fmfset{dash_len}{1.2mm}
			\fmfset{wiggly_len}{1.1mm} \fmfset{dot_len}{0.5mm}
			\fmfpen{0.25mm}
			\fmfsurroundn{i}{6}
			\fmfvn{decor.shape=circle,decor.filled=full, decor.size=2.5thin}{i}{6}
			\fmf{plain,fore=black}{i2,i3,i4,i2}
		\end{fmfgraph}
	\end{fmffile}
\end{gathered}\hspace{0.7cm}
\begin{gathered}
	\begin{fmffile}{6pt-self-3p-3}
		\begin{fmfgraph}(45,45)
			\fmfset{dash_len}{1.2mm}
			\fmfset{wiggly_len}{1.1mm} \fmfset{dot_len}{0.5mm}
			\fmfpen{0.25mm}
			\fmfsurroundn{i}{6}
			\fmfvn{decor.shape=circle,decor.filled=full, decor.size=2.5thin}{i}{6}
			\fmf{plain,fore=black}{i3,i4,i5,i3}
		\end{fmfgraph}
	\end{fmffile}
\end{gathered}\hspace{0.7cm}
\begin{gathered}
	\begin{fmffile}{6pt-self-3p-4}
		\begin{fmfgraph}(45,45)
			\fmfset{dash_len}{1.2mm}
			\fmfset{wiggly_len}{1.1mm} \fmfset{dot_len}{0.5mm}
			\fmfpen{0.25mm}
			\fmfsurroundn{i}{6}
			\fmfvn{decor.shape=circle,decor.filled=full, decor.size=2.5thin}{i}{6}
			\fmf{plain,fore=black}{i4,i5,i6,i4}
		\end{fmfgraph}
	\end{fmffile}
\end{gathered}\hspace{0.7cm}
\begin{gathered}
	\begin{fmffile}{6pt-self-3p-5}
		\begin{fmfgraph}(45,45)
			\fmfset{dash_len}{1.2mm}
			\fmfset{wiggly_len}{1.1mm} \fmfset{dot_len}{0.5mm}
			\fmfpen{0.25mm}
			\fmfsurroundn{i}{6}
			\fmfvn{decor.shape=circle,decor.filled=full, decor.size=2.5thin}{i}{6}
			\fmf{plain,fore=black}{i5,i6,i1,i5}
		\end{fmfgraph}
	\end{fmffile}
\end{gathered}\hspace{0.7cm}
\begin{gathered}
	\begin{fmffile}{6pt-self-3p-6}
		\begin{fmfgraph}(45,45)
			\fmfset{dash_len}{1.2mm}
			\fmfset{wiggly_len}{1.1mm} \fmfset{dot_len}{0.5mm}
			\fmfpen{0.25mm}
			\fmfsurroundn{i}{6}
			\fmfvn{decor.shape=circle,decor.filled=full, decor.size=2.5thin}{i}{6}
			\fmf{plain,fore=black}{i6,i1,i2,i6}
		\end{fmfgraph}
	\end{fmffile}
\end{gathered}
\phantom{\Bigg\{}\\
\end{array}\right\}
\quad \Leftrightarrow\quad-20\,G_3\phi^3 \, ,
\\
&
\left\{
\begin{array}{l}
\begin{gathered}
	\begin{fmffile}{6pt-self-22p-1}
		\begin{fmfgraph}(45,45)
			\fmfset{dash_len}{1.2mm}
			\fmfset{wiggly_len}{1.1mm} \fmfset{dot_len}{0.5mm}
			\fmfpen{0.25mm}
			\fmfsurroundn{i}{6}
			\fmfvn{decor.shape=circle,decor.filled=full, decor.size=2.5thin}{i}{6}
			\fmf{plain,fore=black}{i1,i2}
			\fmf{plain,fore=black}{i6,i3}
		\end{fmfgraph}
	\end{fmffile}
\end{gathered}\hspace{0.7cm}
\begin{gathered}
	\begin{fmffile}{6pt-self-22p-2}
		\begin{fmfgraph}(45,45)
			\fmfset{dash_len}{1.2mm}
			\fmfset{wiggly_len}{1.1mm} \fmfset{dot_len}{0.5mm}
			\fmfpen{0.25mm}
			\fmfsurroundn{i}{6}
			\fmfvn{decor.shape=circle,decor.filled=full, decor.size=2.5thin}{i}{6}
			\fmf{plain,fore=black}{i2,i3}
			\fmf{plain,fore=black}{i1,i4}
		\end{fmfgraph}
	\end{fmffile}
\end{gathered}\hspace{0.7cm}
\begin{gathered}
	\begin{fmffile}{6pt-self-22p-3}
		\begin{fmfgraph}(45,45)
			\fmfset{dash_len}{1.2mm}
			\fmfset{wiggly_len}{1.1mm} \fmfset{dot_len}{0.5mm}
			\fmfpen{0.25mm}
			\fmfsurroundn{i}{6}
			\fmfvn{decor.shape=circle,decor.filled=full, decor.size=2.5thin}{i}{6}
			\fmf{plain,fore=black}{i3,i4}
			\fmf{plain,fore=black}{i2,i5}
		\end{fmfgraph}
	\end{fmffile}
\end{gathered}\hspace{0.7cm}
\begin{gathered}
	\begin{fmffile}{6pt-self-22p-4}
		\begin{fmfgraph}(45,45)
			\fmfset{dash_len}{1.2mm}
			\fmfset{wiggly_len}{1.1mm} \fmfset{dot_len}{0.5mm}
			\fmfpen{0.25mm}
			\fmfsurroundn{i}{6}
			\fmfvn{decor.shape=circle,decor.filled=full, decor.size=2.5thin}{i}{6}
			\fmf{plain,fore=black}{i4,i5}
			\fmf{plain,fore=black}{i3,i6}
		\end{fmfgraph}
	\end{fmffile}
\end{gathered}\hspace{0.7cm}
\begin{gathered}
	\begin{fmffile}{6pt-self-22p-5}
		\begin{fmfgraph}(45,45)
			\fmfset{dash_len}{1.2mm}
			\fmfset{wiggly_len}{1.1mm} \fmfset{dot_len}{0.5mm}
			\fmfpen{0.25mm}
			\fmfsurroundn{i}{6}
			\fmfvn{decor.shape=circle,decor.filled=full, decor.size=2.5thin}{i}{6}
			\fmf{plain,fore=black}{i5,i6}
			\fmf{plain,fore=black}{i4,i1}
		\end{fmfgraph}
	\end{fmffile}
\end{gathered}\hspace{0.7cm}
\begin{gathered}
	\begin{fmffile}{6pt-self-22p-6}
		\begin{fmfgraph}(45,45)
			\fmfset{dash_len}{1.2mm}
			\fmfset{wiggly_len}{1.1mm} \fmfset{dot_len}{0.5mm}
			\fmfpen{0.25mm}
			\fmfsurroundn{i}{6}
			\fmfvn{decor.shape=circle,decor.filled=full, decor.size=2.5thin}{i}{6}
			\fmf{plain,fore=black}{i6,i1}
			\fmf{plain,fore=black}{i5,i2}
		\end{fmfgraph}
	\end{fmffile}
\end{gathered}
\phantom{\Bigg\{}\\
\end{array}\right\}
\quad \Leftrightarrow\quad45\,G_2^2\phi^2 \, ,
\\
\end{aligned}
\end{equation*}
\end{spreadlines}

\begin{spreadlines}{0.6\baselineskip}
\begin{equation*}
\begin{aligned}
&
\left\{
\begin{array}{l}
\begin{gathered}
	\begin{fmffile}{6pt-self-4p-1}
		\begin{fmfgraph}(45,45)
			\fmfset{dash_len}{1.2mm}
			\fmfset{wiggly_len}{1.1mm} \fmfset{dot_len}{0.5mm}
			\fmfpen{0.25mm}
			\fmfsurroundn{i}{6}
			\fmfvn{decor.shape=circle,decor.filled=full, decor.size=2.5thin}{i}{6}
			\fmf{plain,fore=black}{i1,i2,i3,i4,i1}
		\end{fmfgraph}
	\end{fmffile}
\end{gathered}\hspace{0.7cm}
\begin{gathered}
	\begin{fmffile}{6pt-self-4p-2}
		\begin{fmfgraph}(45,45)
			\fmfset{dash_len}{1.2mm}
			\fmfset{wiggly_len}{1.1mm} \fmfset{dot_len}{0.5mm}
			\fmfpen{0.25mm}
			\fmfsurroundn{i}{6}
			\fmfvn{decor.shape=circle,decor.filled=full, decor.size=2.5thin}{i}{6}
			\fmf{plain,fore=black}{i2,i3,i4,i5,i2}
		\end{fmfgraph}
	\end{fmffile}
\end{gathered}\hspace{0.7cm}
\begin{gathered}
	\begin{fmffile}{6pt-self-4p-3}
		\begin{fmfgraph}(45,45)
			\fmfset{dash_len}{1.2mm}
			\fmfset{wiggly_len}{1.1mm} \fmfset{dot_len}{0.5mm}
			\fmfpen{0.25mm}
			\fmfsurroundn{i}{6}
			\fmfvn{decor.shape=circle,decor.filled=full, decor.size=2.5thin}{i}{6}
			\fmf{plain,fore=black}{i3,i4,i5,i6,i3}
		\end{fmfgraph}
	\end{fmffile}
\end{gathered}\hspace{0.7cm}
\begin{gathered}
	\begin{fmffile}{6pt-self-4p-4}
		\begin{fmfgraph}(45,45)
			\fmfset{dash_len}{1.2mm}
			\fmfset{wiggly_len}{1.1mm} \fmfset{dot_len}{0.5mm}
			\fmfpen{0.25mm}
			\fmfsurroundn{i}{6}
			\fmfvn{decor.shape=circle,decor.filled=full, decor.size=2.5thin}{i}{6}
			\fmf{plain,fore=black}{i4,i5,i6,i1,i4}
		\end{fmfgraph}
	\end{fmffile}
\end{gathered}\hspace{0.7cm}
\begin{gathered}
	\begin{fmffile}{6pt-self-4p-5}
		\begin{fmfgraph}(45,45)
			\fmfset{dash_len}{1.2mm}
			\fmfset{wiggly_len}{1.1mm} \fmfset{dot_len}{0.5mm}
			\fmfpen{0.25mm}
			\fmfsurroundn{i}{6}
			\fmfvn{decor.shape=circle,decor.filled=full, decor.size=2.5thin}{i}{6}
			\fmf{plain,fore=black}{i5,i6,i1,i2,i5}
		\end{fmfgraph}
	\end{fmffile}
\end{gathered}\hspace{0.7cm}
\begin{gathered}
	\begin{fmffile}{6pt-self-4p-6}
		\begin{fmfgraph}(45,45)
			\fmfset{dash_len}{1.2mm}
			\fmfset{wiggly_len}{1.1mm} \fmfset{dot_len}{0.5mm}
			\fmfpen{0.25mm}
			\fmfsurroundn{i}{6}
			\fmfvn{decor.shape=circle,decor.filled=full, decor.size=2.5thin}{i}{6}
			\fmf{plain,fore=black}{i6,i1,i2,i3,i6}
		\end{fmfgraph}
	\end{fmffile}
\end{gathered}
\phantom{\Bigg\{}\\
\end{array}\right\}
\quad \Leftrightarrow\quad -6\,G_5\,\phi \, ,
\\
&
\left\{
\begin{array}{l}
\begin{gathered}
	\begin{fmffile}{6pt-self-32p-1}
		\begin{fmfgraph}(45,45)
			\fmfset{dash_len}{1.2mm}
			\fmfset{wiggly_len}{1.1mm} \fmfset{dot_len}{0.5mm}
			\fmfpen{0.25mm}
			\fmfsurroundn{i}{6}
			\fmfvn{decor.shape=circle,decor.filled=full, decor.size=2.5thin}{i}{6}
			\fmf{plain,fore=black}{i1,i2,i3,i1}
			\fmf{plain,fore=black}{i6,i4}
		\end{fmfgraph}
	\end{fmffile}
\end{gathered}\hspace{0.7cm}
\begin{gathered}
	\begin{fmffile}{6pt-self-32p-2}
		\begin{fmfgraph}(45,45)
			\fmfset{dash_len}{1.2mm}
			\fmfset{wiggly_len}{1.1mm} \fmfset{dot_len}{0.5mm}
			\fmfpen{0.25mm}
			\fmfsurroundn{i}{6}
			\fmfvn{decor.shape=circle,decor.filled=full, decor.size=2.5thin}{i}{6}
			\fmf{plain,fore=black}{i2,i3,i4,i2}
			\fmf{plain,fore=black}{i1,i5}
		\end{fmfgraph}
	\end{fmffile}
\end{gathered}\hspace{0.7cm}
\begin{gathered}
	\begin{fmffile}{6pt-self-32p-3}
		\begin{fmfgraph}(45,45)
			\fmfset{dash_len}{1.2mm}
			\fmfset{wiggly_len}{1.1mm} \fmfset{dot_len}{0.5mm}
			\fmfpen{0.25mm}
			\fmfsurroundn{i}{6}
			\fmfvn{decor.shape=circle,decor.filled=full, decor.size=2.5thin}{i}{6}
			\fmf{plain,fore=black}{i3,i4,i5,i3}
			\fmf{plain,fore=black}{i2,i6}
		\end{fmfgraph}
	\end{fmffile}
\end{gathered}\hspace{0.7cm}
\begin{gathered}
	\begin{fmffile}{6pt-self-32p-4}
		\begin{fmfgraph}(45,45)
			\fmfset{dash_len}{1.2mm}
			\fmfset{wiggly_len}{1.1mm} \fmfset{dot_len}{0.5mm}
			\fmfpen{0.25mm}
			\fmfsurroundn{i}{6}
			\fmfvn{decor.shape=circle,decor.filled=full, decor.size=2.5thin}{i}{6}
			\fmf{plain,fore=black}{i4,i5,i6,i4}
			\fmf{plain,fore=black}{i3,i1}
		\end{fmfgraph}
	\end{fmffile}
\end{gathered}\hspace{0.7cm}
\begin{gathered}
	\begin{fmffile}{6pt-self-32p-5}
		\begin{fmfgraph}(45,45)
			\fmfset{dash_len}{1.2mm}
			\fmfset{wiggly_len}{1.1mm} \fmfset{dot_len}{0.5mm}
			\fmfpen{0.25mm}
			\fmfsurroundn{i}{6}
			\fmfvn{decor.shape=circle,decor.filled=full, decor.size=2.5thin}{i}{6}
			\fmf{plain,fore=black}{i5,i6,i1,i5}
			\fmf{plain,fore=black}{i4,i2}
		\end{fmfgraph}
	\end{fmffile}
\end{gathered}\hspace{0.7cm}
\begin{gathered}
	\begin{fmffile}{6pt-self-32p-6}
		\begin{fmfgraph}(45,45)
			\fmfset{dash_len}{1.2mm}
			\fmfset{wiggly_len}{1.1mm} \fmfset{dot_len}{0.5mm}
			\fmfpen{0.25mm}
			\fmfsurroundn{i}{6}
			\fmfvn{decor.shape=circle,decor.filled=full, decor.size=2.5thin}{i}{6}
			\fmf{plain,fore=black}{i6,i1,i2,i6}
			\fmf{plain,fore=black}{i5,i3}
		\end{fmfgraph}
	\end{fmffile}
\end{gathered}
\phantom{\Bigg\{}\\
\end{array}\right\}
\quad \Leftrightarrow\quad60\,G_2G_3\,\phi \, ,
\\
&
\left\{
\begin{array}{l}
\begin{gathered}
	\begin{fmffile}{6pt-self-33p-1b}
		\begin{fmfgraph}(45,45)
			\fmfset{dash_len}{1.2mm}
			\fmfset{wiggly_len}{1.1mm} \fmfset{dot_len}{0.5mm}
			\fmfpen{0.25mm}
			\fmfsurroundn{i}{6}
			\fmfvn{decor.shape=circle,decor.filled=full, decor.size=2.5thin}{i}{6}
			\fmf{plain,fore=black}{i1,i3,i4,i1}
			\fmf{plain,fore=black}{i2,i5,i6,i2}
		\end{fmfgraph}
	\end{fmffile}
\end{gathered}\hspace{0.7cm}
\begin{gathered}
	\begin{fmffile}{6pt-self-33p-2b}
		\begin{fmfgraph}(45,45)
			\fmfset{dash_len}{1.2mm}
			\fmfset{wiggly_len}{1.1mm} \fmfset{dot_len}{0.5mm}
			\fmfpen{0.25mm}
			\fmfsurroundn{i}{6}
			\fmfvn{decor.shape=circle,decor.filled=full, decor.size=2.5thin}{i}{6}
			\fmf{plain,fore=black}{i2,i4,i5,i2}
			\fmf{plain,fore=black}{i3,i6,i1,i3}
		\end{fmfgraph}
	\end{fmffile}
\end{gathered}\hspace{0.7cm}
\begin{gathered}
	\begin{fmffile}{6pt-self-33p-3b}
		\begin{fmfgraph}(45,45)
			\fmfset{dash_len}{1.2mm}
			\fmfset{wiggly_len}{1.1mm} \fmfset{dot_len}{0.5mm}
			\fmfpen{0.25mm}
			\fmfsurroundn{i}{6}
			\fmfvn{decor.shape=circle,decor.filled=full, decor.size=2.5thin}{i}{6}
			\fmf{plain,fore=black}{i3,i5,i6,i3}
			\fmf{plain,fore=black}{i4,i1,i2,i4}
		\end{fmfgraph}
	\end{fmffile}
\end{gathered}\hspace{0.7cm}
\begin{gathered}
	\begin{fmffile}{6pt-self-33p-4b}
		\begin{fmfgraph}(45,45)
			\fmfset{dash_len}{1.2mm}
			\fmfset{wiggly_len}{1.1mm} \fmfset{dot_len}{0.5mm}
			\fmfpen{0.25mm}
			\fmfsurroundn{i}{6}
			\fmfvn{decor.shape=circle,decor.filled=full, decor.size=2.5thin}{i}{6}
			\fmf{plain,fore=black}{i4,i6,i1,i4}
			\fmf{plain,fore=black}{i5,i2,i3,i5}
		\end{fmfgraph}
	\end{fmffile}
\end{gathered}\hspace{0.7cm}
\begin{gathered}
	\begin{fmffile}{6pt-self-33p-5b}
		\begin{fmfgraph}(45,45)
			\fmfset{dash_len}{1.2mm}
			\fmfset{wiggly_len}{1.1mm} \fmfset{dot_len}{0.5mm}
			\fmfpen{0.25mm}
			\fmfsurroundn{i}{6}
			\fmfvn{decor.shape=circle,decor.filled=full, decor.size=2.5thin}{i}{6}
			\fmf{plain,fore=black}{i5,i1,i2,i5}
			\fmf{plain,fore=black}{i6,i3,i4,i6}
		\end{fmfgraph}
	\end{fmffile}
\end{gathered}\hspace{0.7cm}
\begin{gathered}
	\begin{fmffile}{6pt-self-33p-6b}
		\begin{fmfgraph}(45,45)
			\fmfset{dash_len}{1.2mm}
			\fmfset{wiggly_len}{1.1mm} \fmfset{dot_len}{0.5mm}
			\fmfpen{0.25mm}
			\fmfsurroundn{i}{6}
			\fmfvn{decor.shape=circle,decor.filled=full, decor.size=2.5thin}{i}{6}
			\fmf{plain,fore=black}{i6,i2,i3,i6}
			\fmf{plain,fore=black}{i1,i4,i5,i1}
		\end{fmfgraph}
	\end{fmffile}
\end{gathered}
\phantom{\Bigg\{}\\
\end{array}\right\}
\quad \Leftrightarrow\quad10\,G_3^2 \, ,
\\
\end{aligned}
\end{equation*}
\end{spreadlines}

\begin{spreadlines}{0.6\baselineskip}
\begin{equation*}
\begin{aligned}
&
\left\{
\begin{array}{l}
\begin{gathered}
	\begin{fmffile}{6pt-self-222p-1}
		\begin{fmfgraph}(45,45)
			\fmfset{dash_len}{1.2mm}
			\fmfset{wiggly_len}{1.1mm} \fmfset{dot_len}{0.5mm}
			\fmfpen{0.25mm}
			\fmfsurroundn{i}{6}
			\fmfvn{decor.shape=circle,decor.filled=full, decor.size=2.5thin}{i}{6}
			\fmf{plain,fore=black}{i2,i4}
			\fmf{plain,fore=black}{i3,i5}
			\fmf{plain,fore=black}{i1,i6}
		\end{fmfgraph}
	\end{fmffile}
\end{gathered}\hspace{0.7cm}
\begin{gathered}
	\begin{fmffile}{6pt-self-222p-2}
		\begin{fmfgraph}(45,45)
			\fmfset{dash_len}{1.2mm}
			\fmfset{wiggly_len}{1.1mm} \fmfset{dot_len}{0.5mm}
			\fmfpen{0.25mm}
			\fmfsurroundn{i}{6}
			\fmfvn{decor.shape=circle,decor.filled=full, decor.size=2.5thin}{i}{6}
			\fmf{plain,fore=black}{i3,i5}
			\fmf{plain,fore=black}{i4,i6}
			\fmf{plain,fore=black}{i2,i1}
		\end{fmfgraph}
	\end{fmffile}
\end{gathered}\hspace{0.7cm}
\begin{gathered}
	\begin{fmffile}{6pt-self-222p-3}
		\begin{fmfgraph}(45,45)
			\fmfset{dash_len}{1.2mm}
			\fmfset{wiggly_len}{1.1mm} \fmfset{dot_len}{0.5mm}
			\fmfpen{0.25mm}
			\fmfsurroundn{i}{6}
			\fmfvn{decor.shape=circle,decor.filled=full, decor.size=2.5thin}{i}{6}
			\fmf{plain,fore=black}{i4,i6}
			\fmf{plain,fore=black}{i5,i1}
			\fmf{plain,fore=black}{i3,i2}
		\end{fmfgraph}
	\end{fmffile}
\end{gathered}\hspace{0.7cm}
\begin{gathered}
	\begin{fmffile}{6pt-self-222p-4}
		\begin{fmfgraph}(45,45)
			\fmfset{dash_len}{1.2mm}
			\fmfset{wiggly_len}{1.1mm} \fmfset{dot_len}{0.5mm}
			\fmfpen{0.25mm}
			\fmfsurroundn{i}{6}
			\fmfvn{decor.shape=circle,decor.filled=full, decor.size=2.5thin}{i}{6}
			\fmf{plain,fore=black}{i5,i1}
			\fmf{plain,fore=black}{i6,i2}
			\fmf{plain,fore=black}{i4,i3}
		\end{fmfgraph}
	\end{fmffile}
\end{gathered}\hspace{0.7cm}
\begin{gathered}
	\begin{fmffile}{6pt-self-222p-5}
		\begin{fmfgraph}(45,45)
			\fmfset{dash_len}{1.2mm}
			\fmfset{wiggly_len}{1.1mm} \fmfset{dot_len}{0.5mm}
			\fmfpen{0.25mm}
			\fmfsurroundn{i}{6}
			\fmfvn{decor.shape=circle,decor.filled=full, decor.size=2.5thin}{i}{6}
			\fmf{plain,fore=black}{i6,i2}
			\fmf{plain,fore=black}{i1,i3}
			\fmf{plain,fore=black}{i5,i4}
		\end{fmfgraph}
	\end{fmffile}
\end{gathered}\hspace{0.7cm}
\begin{gathered}
	\begin{fmffile}{6pt-self-222p-6}
		\begin{fmfgraph}(45,45)
			\fmfset{dash_len}{1.2mm}
			\fmfset{wiggly_len}{1.1mm} \fmfset{dot_len}{0.5mm}
			\fmfpen{0.25mm}
			\fmfsurroundn{i}{6}
			\fmfvn{decor.shape=circle,decor.filled=full, decor.size=2.5thin}{i}{6}
			\fmf{plain,fore=black}{i1,i3}
			\fmf{plain,fore=black}{i2,i4}
			\fmf{plain,fore=black}{i6,i5}
		\end{fmfgraph}
	\end{fmffile}
\end{gathered}
\phantom{\Bigg\{}\\
\end{array}\right\}
\quad \Leftrightarrow\quad-15\,G_2^3 \, ,
\\
&
\left\{
\begin{array}{l}
\begin{gathered}
	\begin{fmffile}{6pt-self-42p-1}
		\begin{fmfgraph}(45,45)
			\fmfset{dash_len}{1.2mm}
			\fmfset{wiggly_len}{1.1mm} \fmfset{dot_len}{0.5mm}
			\fmfpen{0.25mm}
			\fmfsurroundn{i}{6}
			\fmfvn{decor.shape=circle,decor.filled=full, decor.size=2.5thin}{i}{6}
			\fmf{plain,fore=black}{i1,i2,i3,i4,i1}
			\fmf{plain,fore=black}{i5,i6}
		\end{fmfgraph}
	\end{fmffile}
\end{gathered}\hspace{0.7cm}
\begin{gathered}
	\begin{fmffile}{6pt-self-42p-2}
		\begin{fmfgraph}(45,45)
			\fmfset{dash_len}{1.2mm}
			\fmfset{wiggly_len}{1.1mm} \fmfset{dot_len}{0.5mm}
			\fmfpen{0.25mm}
			\fmfsurroundn{i}{6}
			\fmfvn{decor.shape=circle,decor.filled=full, decor.size=2.5thin}{i}{6}
			\fmf{plain,fore=black}{i2,i3,i4,i5,i2}
			\fmf{plain,fore=black}{i6,i1}
		\end{fmfgraph}
	\end{fmffile}
\end{gathered}\hspace{0.7cm}
\begin{gathered}
	\begin{fmffile}{6pt-self-42p-3}
		\begin{fmfgraph}(45,45)
			\fmfset{dash_len}{1.2mm}
			\fmfset{wiggly_len}{1.1mm} \fmfset{dot_len}{0.5mm}
			\fmfpen{0.25mm}
			\fmfsurroundn{i}{6}
			\fmfvn{decor.shape=circle,decor.filled=full, decor.size=2.5thin}{i}{6}
			\fmf{plain,fore=black}{i3,i4,i5,i6,i3}
			\fmf{plain,fore=black}{i1,i2}
		\end{fmfgraph}
	\end{fmffile}
\end{gathered}\hspace{0.7cm}
\begin{gathered}
	\begin{fmffile}{6pt-self-42p-4}
		\begin{fmfgraph}(45,45)
			\fmfset{dash_len}{1.2mm}
			\fmfset{wiggly_len}{1.1mm} \fmfset{dot_len}{0.5mm}
			\fmfpen{0.25mm}
			\fmfsurroundn{i}{6}
			\fmfvn{decor.shape=circle,decor.filled=full, decor.size=2.5thin}{i}{6}
			\fmf{plain,fore=black}{i4,i5,i6,i1,i4}
			\fmf{plain,fore=black}{i2,i3}
		\end{fmfgraph}
	\end{fmffile}
\end{gathered}\hspace{0.7cm}
\begin{gathered}
	\begin{fmffile}{6pt-self-42p-5}
		\begin{fmfgraph}(45,45)
			\fmfset{dash_len}{1.2mm}
			\fmfset{wiggly_len}{1.1mm} \fmfset{dot_len}{0.5mm}
			\fmfpen{0.25mm}
			\fmfsurroundn{i}{6}
			\fmfvn{decor.shape=circle,decor.filled=full, decor.size=2.5thin}{i}{6}
			\fmf{plain,fore=black}{i5,i6,i1,i2,i5}
			\fmf{plain,fore=black}{i3,i4}
		\end{fmfgraph}
	\end{fmffile}
\end{gathered}\hspace{0.7cm}
\begin{gathered}
	\begin{fmffile}{6pt-self-42p-6}
		\begin{fmfgraph}(45,45)
			\fmfset{dash_len}{1.2mm}
			\fmfset{wiggly_len}{1.1mm} \fmfset{dot_len}{0.5mm}
			\fmfpen{0.25mm}
			\fmfsurroundn{i}{6}
			\fmfvn{decor.shape=circle,decor.filled=full, decor.size=2.5thin}{i}{6}
			\fmf{plain,fore=black}{i6,i1,i2,i3,i6}
			\fmf{plain,fore=black}{i4,i5}
		\end{fmfgraph}
	\end{fmffile}
\end{gathered}
\phantom{\Bigg\{}\\
\end{array}\right\}
\quad \Leftrightarrow\quad 15\,G_2G_4 \, ,
\\
&
\bigg\{
\begin{array}{l}
\begin{gathered}
	\begin{fmffile}{6pt-self-6p}
		\begin{fmfgraph}(45,45)
			\fmfset{dash_len}{1.2mm}
			\fmfset{wiggly_len}{1.1mm} \fmfset{dot_len}{0.5mm}
			\fmfpen{0.25mm}
			\fmfsurroundn{i}{6}
			\fmfvn{decor.shape=circle,decor.filled=full, decor.size=2.5thin}{i}{6}
			\fmf{plain,fore=black}{i1,i2,i3,i4,i5,i6,i1}
		\end{fmfgraph}
	\end{fmffile}
\end{gathered}
\phantom{\Bigg\{}\end{array}\!\!\!\bigg\}
\quad \Leftrightarrow\quad -G_6 \, ,
\end{aligned}
\end{equation*}
\end{spreadlines}
so that the complete normal ordered monomial is given explicitly by:
\begin{equation}\label{eq:*phi6*}
\begin{aligned}
\n\,\phi^6&\n=\phi^6-15\,G_2\phi^4-20\,G_3\phi^3+(45\,G_2^2-15\,G_4)\phi^2\\
&\qquad+(60\,G_2G_3-6\,G_5)\phi+(10\,G_3^2-15\,G_2^3+15\,G_2G_4-G_6).
\end{aligned}
\end{equation}

These results can easily be generalised to the case of derivative interactions. For example, consider the local term $\n\,\phi^2(\nabla\phi)^2\n$. There are still 15 partitions as in the case of $\n\,\phi^4\n$ discussed above, but the presence of derivative terms breaks some of the degeneracy. If we denote the derivative insertion, $\nabla \phi$, by a circle, `$\!\!\!\!\!
\begin{gathered}
	\begin{fmffile}{1pt-self-0d}
		\begin{fmfgraph}(45,45)
			\fmfset{dash_len}{1.2mm}
			\fmfset{wiggly_len}{1.1mm} \fmfset{dot_len}{0.5mm}
			\fmfpen{0.25mm}
			\fmfsurroundn{i}{1}
			\fmfvn{decor.shape=circle,decor.filled=empty, decor.size=2.5thin}{i}{1}
		\end{fmfgraph}
	\end{fmffile}
\end{gathered}\,\,\,$', and a $\phi$ insertion by a dot as previously, `$\!\!\!\!\!
\begin{gathered}
	\begin{fmffile}{1pt-self-0}
		\begin{fmfgraph}(45,45)
			\fmfset{dash_len}{1.2mm}
			\fmfset{wiggly_len}{1.1mm} \fmfset{dot_len}{0.5mm}
			\fmfpen{0.25mm}
			\fmfsurroundn{i}{1}
			\fmfvn{decor.shape=circle,decor.filled=full, decor.size=2.5thin}{i}{1}
		\end{fmfgraph}
	\end{fmffile}
\end{gathered}\,\,\,$', we are led to the following results:
\begin{spreadlines}{0.6\baselineskip}
\begin{equation*}
\begin{aligned}
&
\bigg\{
\begin{array}{l}
\begin{gathered}
	\begin{fmffile}{4pt-self-der-0}
		\begin{fmfgraph}(45,45)
			\fmfset{dash_len}{1.2mm}
			\fmfset{wiggly_len}{1.1mm} \fmfset{dot_len}{0.5mm}
			\fmfpen{0.25mm}
			\fmfsurroundn{i}{4}
			\fmfvn{decor.shape=circle,decor.filled=full, decor.size=2.5thin}{i}{4}
			\fmfv{decor.shape=circle,decor.filled=empty, decor.size=2.5thin}{i1}
			\fmfv{decor.shape=circle,decor.filled=empty, decor.size=2.5thin}{i2}
		\end{fmfgraph}
	\end{fmffile}
\end{gathered}
\phantom{\Bigg\{}\end{array}\!\!\!\bigg\}
\quad \Leftrightarrow\quad \phi^2(\nabla\phi)^2 \, ,
\\
&
\left\{
\begin{array}{l}
\begin{gathered}
	\begin{fmffile}{4pt-self-der-1}
		\begin{fmfgraph}(45,45)
			\fmfset{dash_len}{1.2mm}
			\fmfset{wiggly_len}{1.1mm} \fmfset{dot_len}{0.5mm}
			\fmfpen{0.25mm}
			\fmfsurroundn{i}{4}
			\fmfvn{decor.shape=circle,decor.filled=full, decor.size=2.5thin}{i}{4}
			\fmfv{decor.shape=circle,decor.filled=empty, decor.size=2.5thin}{i1}
			\fmfv{decor.shape=circle,decor.filled=empty, decor.size=2.5thin}{i2}
			\fmf{plain,fore=black}{i1,i2}
		\end{fmfgraph}
	\end{fmffile}
\end{gathered}\hspace{0.7cm}
\begin{gathered}
	\begin{fmffile}{4pt-self-der-2}
		\begin{fmfgraph}(45,45)
			\fmfset{dash_len}{1.2mm}
			\fmfset{wiggly_len}{1.1mm} \fmfset{dot_len}{0.5mm}
			\fmfpen{0.25mm}
			\fmfsurroundn{i}{4}
			\fmfvn{decor.shape=circle,decor.filled=full, decor.size=2.5thin}{i}{4}
			\fmfv{decor.shape=circle,decor.filled=empty, decor.size=2.5thin}{i1}
			\fmfv{decor.shape=circle,decor.filled=empty, decor.size=2.5thin}{i2}
			\fmf{plain,fore=black}{i2,i3}
		\end{fmfgraph}
	\end{fmffile}
\end{gathered}\hspace{0.7cm}
\begin{gathered}
	\begin{fmffile}{4pt-self-der-3}
		\begin{fmfgraph}(45,45)
			\fmfset{dash_len}{1.2mm}
			\fmfset{wiggly_len}{1.1mm} \fmfset{dot_len}{0.5mm}
			\fmfpen{0.25mm}
			\fmfsurroundn{i}{4}
			\fmfvn{decor.shape=circle,decor.filled=full, decor.size=2.5thin}{i}{4}
			\fmfv{decor.shape=circle,decor.filled=empty, decor.size=2.5thin}{i1}
			\fmfv{decor.shape=circle,decor.filled=empty, decor.size=2.5thin}{i2}
			\fmf{plain,fore=black}{i3,i4}
		\end{fmfgraph}
	\end{fmffile}
\end{gathered}\hspace{0.7cm}
\begin{gathered}
	\begin{fmffile}{4pt-self-der-4}
		\begin{fmfgraph}(45,45)
			\fmfset{dash_len}{1.2mm}
			\fmfset{wiggly_len}{1.1mm} \fmfset{dot_len}{0.5mm}
			\fmfpen{0.25mm}
			\fmfsurroundn{i}{4}
			\fmfvn{decor.shape=circle,decor.filled=full, decor.size=2.5thin}{i}{4}
			\fmfv{decor.shape=circle,decor.filled=empty, decor.size=2.5thin}{i1}
			\fmfv{decor.shape=circle,decor.filled=empty, decor.size=2.5thin}{i2}
			\fmf{plain,fore=black}{i4,i1}
		\end{fmfgraph}
	\end{fmffile}
\end{gathered}\phantom{\Bigg\{}\\
\begin{gathered}
	\begin{fmffile}{4pt-self-der-6c}
		\begin{fmfgraph}(45,45)
			\fmfset{dash_len}{1.2mm}
			\fmfset{wiggly_len}{1.1mm} \fmfset{dot_len}{0.5mm}
			\fmfpen{0.25mm}
			\fmfsurroundn{i}{4}
			\fmfvn{decor.shape=circle,decor.filled=full, decor.size=2.5thin}{i}{4}
			\fmfv{decor.shape=circle,decor.filled=empty, decor.size=2.5thin}{i1}
			\fmfv{decor.shape=circle,decor.filled=empty, decor.size=2.5thin}{i2}
			\fmf{plain,fore=black}{i1,i3}
		\end{fmfgraph}
	\end{fmffile}
\end{gathered}\hspace{0.7cm}
\begin{gathered}
	\begin{fmffile}{4pt-self-der-7c}
		\begin{fmfgraph}(45,45)
			\fmfset{dash_len}{1.2mm}
			\fmfset{wiggly_len}{1.1mm} \fmfset{dot_len}{0.5mm}
			\fmfpen{0.25mm}
			\fmfsurroundn{i}{4}
			\fmfvn{decor.shape=circle,decor.filled=full, decor.size=2.5thin}{i}{4}
			\fmfv{decor.shape=circle,decor.filled=empty, decor.size=2.5thin}{i1}
			\fmfv{decor.shape=circle,decor.filled=empty, decor.size=2.5thin}{i2}
			\fmf{plain,fore=black}{i2,i4}
		\end{fmfgraph}
	\end{fmffile}
\end{gathered}
\phantom{\Bigg\{}\end{array}\!\!\!\right\}
\quad \Leftrightarrow\quad \,-(\nabla^2G_2)\phi^2-4(\nabla G_2)\phi\nabla\phi-G_2(\nabla\phi)^2 \, ,
\\
&
\left\{
\begin{array}{l}
\begin{gathered}
	\begin{fmffile}{4pt-self-der-3p-1}
		\begin{fmfgraph}(45,45)
			\fmfset{dash_len}{1.2mm}
			\fmfset{wiggly_len}{1.1mm} \fmfset{dot_len}{0.5mm}
			\fmfpen{0.25mm}
			\fmfsurroundn{i}{4}
			\fmfvn{decor.shape=circle,decor.filled=full, decor.size=2.5thin}{i}{4}
			\fmfv{decor.shape=circle,decor.filled=empty, decor.size=2.5thin}{i1}
			\fmfv{decor.shape=circle,decor.filled=empty, decor.size=2.5thin}{i2}
			\fmf{plain,fore=black}{i1,i2,i3,i1}
		\end{fmfgraph}
	\end{fmffile}
\end{gathered}\hspace{0.7cm}
\begin{gathered}
	\begin{fmffile}{4pt-self-der-3p-2}
		\begin{fmfgraph}(45,45)
			\fmfset{dash_len}{1.2mm}
			\fmfset{wiggly_len}{1.1mm} \fmfset{dot_len}{0.5mm}
			\fmfpen{0.25mm}
			\fmfsurroundn{i}{4}
			\fmfvn{decor.shape=circle,decor.filled=full, decor.size=2.5thin}{i}{4}
			\fmfv{decor.shape=circle,decor.filled=empty, decor.size=2.5thin}{i1}
			\fmfv{decor.shape=circle,decor.filled=empty, decor.size=2.5thin}{i2}
			\fmf{plain,fore=black}{i2,i3,i4,i2}
		\end{fmfgraph}
	\end{fmffile}
\end{gathered}\hspace{0.7cm}
\begin{gathered}
	\begin{fmffile}{4pt-self-der-3p-3}
		\begin{fmfgraph}(45,45)
			\fmfset{dash_len}{1.2mm}
			\fmfset{wiggly_len}{1.1mm} \fmfset{dot_len}{0.5mm}
			\fmfpen{0.25mm}
			\fmfsurroundn{i}{4}
			\fmfvn{decor.shape=circle,decor.filled=full, decor.size=2.5thin}{i}{4}
			\fmfv{decor.shape=circle,decor.filled=empty, decor.size=2.5thin}{i1}
			\fmfv{decor.shape=circle,decor.filled=empty, decor.size=2.5thin}{i2}
			\fmf{plain,fore=black}{i3,i4,i1,i3}
		\end{fmfgraph}
	\end{fmffile}
\end{gathered}\hspace{0.7cm}
\begin{gathered}
	\begin{fmffile}{4pt-self-der-3p-4}
		\begin{fmfgraph}(45,45)
			\fmfset{dash_len}{1.2mm}
			\fmfset{wiggly_len}{1.1mm} \fmfset{dot_len}{0.5mm}
			\fmfpen{0.25mm}
			\fmfsurroundn{i}{4}
			\fmfvn{decor.shape=circle,decor.filled=full, decor.size=2.5thin}{i}{4}
			\fmfv{decor.shape=circle,decor.filled=empty, decor.size=2.5thin}{i1}
			\fmfv{decor.shape=circle,decor.filled=empty, decor.size=2.5thin}{i2}
			\fmf{plain,fore=black}{i4,i1,i2,i4}
		\end{fmfgraph}
	\end{fmffile}
\end{gathered}
\phantom{\Bigg\{}\end{array}\!\!\!\right\}
\quad \Leftrightarrow\quad -2\,\nabla^2G_3\,\phi-2\,\nabla G_3\nabla \phi \, ,
\\
&
\left\{
\begin{array}{l}
\begin{gathered}
	\begin{fmffile}{4pt-self-der-22p-1}
		\begin{fmfgraph}(45,45)
			\fmfset{dash_len}{1.2mm}
			\fmfset{wiggly_len}{1.1mm} \fmfset{dot_len}{0.5mm}
			\fmfpen{0.25mm}
			\fmfsurroundn{i}{4}
			\fmfvn{decor.shape=circle,decor.filled=full, decor.size=2.5thin}{i}{4}
			\fmfv{decor.shape=circle,decor.filled=empty, decor.size=2.5thin}{i1}
			\fmfv{decor.shape=circle,decor.filled=empty, decor.size=2.5thin}{i2}
			\fmf{plain,fore=black}{i1,i2}
			\fmf{plain,fore=black}{i3,i4}
		\end{fmfgraph}
	\end{fmffile}
\end{gathered}\hspace{0.7cm}
\begin{gathered}
	\begin{fmffile}{4pt-self-der-22p-2}
		\begin{fmfgraph}(45,45)
			\fmfset{dash_len}{1.2mm}
			\fmfset{wiggly_len}{1.1mm} \fmfset{dot_len}{0.5mm}
			\fmfpen{0.25mm}
			\fmfsurroundn{i}{4}
			\fmfvn{decor.shape=circle,decor.filled=full, decor.size=2.5thin}{i}{4}
			\fmfv{decor.shape=circle,decor.filled=empty, decor.size=2.5thin}{i1}
			\fmfv{decor.shape=circle,decor.filled=empty, decor.size=2.5thin}{i2}
			\fmf{plain,fore=black}{i2,i3}
			\fmf{plain,fore=black}{i1,i4}
		\end{fmfgraph}
	\end{fmffile}
\end{gathered}\phantom{\Bigg\{}\\
\begin{gathered}
	\begin{fmffile}{4pt-self-der-22p-3}
		\begin{fmfgraph}(45,45)
			\fmfset{dash_len}{1.2mm}
			\fmfset{wiggly_len}{1.1mm} \fmfset{dot_len}{0.5mm}
			\fmfpen{0.25mm}
			\fmfsurroundn{i}{4}
			\fmfvn{decor.shape=circle,decor.filled=full, decor.size=2.5thin}{i}{4}
			\fmfv{decor.shape=circle,decor.filled=empty, decor.size=2.5thin}{i1}
			\fmfv{decor.shape=circle,decor.filled=empty, decor.size=2.5thin}{i2}
			\fmf{plain,fore=black}{i1,i3}
			\fmf{plain,fore=black}{i2,i4}
		\end{fmfgraph}
	\end{fmffile}
\end{gathered}
\phantom{\Bigg\{}\end{array}\!\!\!\right\}
\quad \Leftrightarrow\quad G_2\nabla^2 G_2+2(\nabla G_2)^2 \, ,
\\
&
\bigg\{
\begin{array}{l}
\begin{gathered}
	\begin{fmffile}{4pt-self-der-4p-1}
		\begin{fmfgraph}(45,45)
			\fmfset{dash_len}{1.2mm}
			\fmfset{wiggly_len}{1.1mm} \fmfset{dot_len}{0.5mm}
			\fmfpen{0.25mm}
			\fmfsurroundn{i}{4}
			\fmfvn{decor.shape=circle,decor.filled=full, decor.size=2.5thin}{i}{4}
			\fmfv{decor.shape=circle,decor.filled=empty, decor.size=2.5thin}{i1}
			\fmfv{decor.shape=circle,decor.filled=empty, decor.size=2.5thin}{i2}
			\fmf{plain,fore=black}{i1,i2,i3,i4,i1}
		\end{fmfgraph}
	\end{fmffile}
\end{gathered}
\phantom{\Bigg\{}\end{array}\!\!\!\bigg\}
\quad \Leftrightarrow\quad -\nabla^2G_4 \, ,
\end{aligned}
\end{equation*}
\end{spreadlines}
leading to the following expression for the complete normal ordered quantity:
\begin{equation}\label{eq:*phi2dphi2*}
\begin{aligned}
\n\phi^2(\nabla\phi)^2&\n=\phi^2(\nabla\phi)^2-(\nabla^2G_2)\phi^2-4(\nabla G_2)\phi\nabla\phi-G_2(\nabla\phi)^2\\
&\quad-2\,(\nabla^2G_3)\,\phi-2\,(\nabla G_3)\nabla \phi+G_2\nabla^2 G_2+2(\nabla G_2)^2 -\nabla^2G_4.
\end{aligned}
\end{equation}
The notation here is such that $\nabla^2G_N\dfn \lim_{\{z_j\}\rightarrow z}\nabla_1\nabla_{2}G_N(z_1,\dots,z_N)$, 
for all $N=2,3,\dots$, we recall that the Green function is symmetric with respect to its arguments, and we note that space-time index contractions are implicit. 

Returning now to the case of non-derivative interactions, we notice that the general result for any positive integer $N$ can be expressed compactly in terms of complete Bell polynomials\footnote{Complete Bell polynomials are defined in (\ref{eq:Bn genfunc_foot}).
} \cite{Riordan58,RomanRota78}, $B_N(a_1,\dots,a_N)$, when the identifications $a_1=\phi$ and $a_{j>1}=-G_j$ are made:
\begin{equation}\label{eq:nphiNn}
 \boxed{\n\,\phi^N\n=B_N(\phi,-G_2,\dots,-G_N)}
\end{equation}
for all $N=0,1,2,3,\dots$. This expression is the promised generalisation of (\ref{eq:Bell::phiN}), the latter being obtained from the former by setting the couplings in the Green functions to zero so that only the free 2-point propagator, $\mathcal{G}\equiv \mathcal{G}(z,z)$, remains.

{\it The simple formula (\ref{eq:nphiNn}) is one of the main results of the paper.}

One may invert (\ref{eq:nphiNn}), to express time-ordered monomials in terms of complete normal ordered monomials and Green functions. A simple way to do this is by flipping the sign of all Green functions while swapping all $\phi^K\leftrightarrow \n\phi^K\n$, as seen by comparing (\ref{eq:nmathcalFn}) to (\ref{eq:mathcalF}), $\phi^N=\n B_N(\phi,G_2,\dots,G_N)\n$, 
where note that for a $\phi$-independent quantity $c$: $\n \,c\,\phi^k\n\,=c\,\n \!\phi^k\n\,$ and $\n1\n=1$. 

A useful corollary that follows directly from the complete Bell polynomial generating function is:
\begin{equation}\label{eq:nexpgphin}
\n\exp\big(\mathfrak{g}\,\phi\big)\n\,=\exp\bigg(\mathfrak{g}\,\phi-\sum_{N=2}^{\infty}\frac{1}{N!}\,\mathfrak{g}^NG_N\bigg),
\end{equation}
generalising the classic free-field normal ordering expression, $:\!e^{\mathfrak{g}\,\phi}\!:\,=\exp\big(\mathfrak{g}\,\phi-\frac{1}{2}\,\mathfrak{g}^2\mathcal{G}\big)$. 
Clearly, the latter again follows from the full result (\ref{eq:nexpgphin}) when we set the couplings in the $N$-point Green functions, $G_N$, to zero.

\subsection{Counterterm Interpretation}\label{sec:counterterms}
We next explain explicitly how to derive the counterterms that are induced from complete normal ordering the bare action, thus generalising (\ref{eq:deltagn}). In turn, this procedure proves that complete normal ordering does not change the quantum field theory in any observable manner, given that it can always be undone by an appropriate choice of local counterterms. 

So suppose we have an interaction term $\frac{1}{N!}g_N\phi^N$ in the action of interest. According to (\ref{eq:nphiNn}), using the defining relation (\ref{eq:Bn genfunc_foot}) for complete Bell polynomials we learn that:
$$
\frac{1}{N!}\n\,\phi^N\n=\sum_{n=0}^{N}\frac{1}{n!(N-n)!}B_{N-n}(0,-G_2,\dots,-G_{N-n})\phi^n,
$$
according to which the complete normal ordered version of the interaction, $\frac{1}{N!}g_N\n \phi^N\n\,\,$, equals $\frac{1}{N!}g_N\phi^N$ plus a counterterm polynomial, $\sum_{n=0}^{N-1}\frac{1}{n!}\delta_{\n} g_n\phi^n$, with:
\begin{equation}\label{eq:deltagn cno}
\boxed{\delta_{\n}  g_n=\frac{g_N}{(N-n)!}B_{N-n}(0,-G_2,\dots,-G_{N-n})},
\end{equation}
for $0\leq n<N$. These are the counterterms induced by complete normal ordering, the appropriate generalisation of (\ref{eq:deltagn}). Generically there will be various interaction terms, corresponding to a sum over $N$, and so various terms will contribute to the \emph{same} $\delta_{\n}  g_n$. After some elementary rearrangement we can put the result into the form:
\begin{equation}\label{eq:counterterms gen}
\begin{aligned}
\sum_{N=0}^{\infty}\frac{1}{N!}g_N\n\phi^N\n\,
&= \sum_{N=0}^{\infty}\frac{1}{N!}\big(g_N+\delta_{\n} g_N\big)\phi^N,
\end{aligned}
\end{equation}
where this more complete counterterm, $\delta_{\n} g_N$, reads:
\begin{equation}\label{eq:delta g N}
\delta_{\n} g_N=\sum_{r=1}^{\infty}\frac{1}{r!}g_{N+r}B_r(0,-G_2,\dots,-G_r).
\end{equation}
If we do not wish to complete normal order the mass (i.e.~the $N=2$) term on the left hand side of (\ref{eq:counterterms gen}) we should add the term $\frac{1}{2}g_2G_2\delta_{N,0}$ to the right hand side of (\ref{eq:delta g N}). 

It will be useful to display explicitly the results for all counterterms, see also comments on notation on p.~\pageref{coupling notation}. From (\ref{eq:delta g N}), and if we do not complete normal order the mass term, we find:
\begin{equation}\label{eq:counterterms GNs}
\begin{aligned}
\delta_{\n}  \Lambda&=g\big(\!-\tfrac{1}{6}G_3\big)+\lambda\big(\tfrac{1}{8}G_2^2-\tfrac{1}{24}G_4\big)+\kappa\big(\tfrac{1}{12}G_2G_3-\tfrac{1}{5!}G_5\big)\\
&\qquad +\gamma\big(\tfrac{1}{48}G_2G_4+\tfrac{1}{72}G_3^2-\tfrac{1}{48}G_2^3-\tfrac{1}{6!}G_6\big)+\mathcal{O}(\ell^5)\phantom{\Big[}\\
Y\,&=g\big(\tfrac{1}{2}G_2\big)+\lambda\big(\tfrac{1}{3!}G_3\big)+\kappa\big(\tfrac{1}{24}G_4-\tfrac{1}{8}G_2^2\big)\\
&\qquad+\gamma\big(\tfrac{1}{5!}G_5-\tfrac{1}{12}G_2G_3\big)+\mathcal{O}(\ell^5)\phantom{\Big[},\\
\delta_{\n}  m^2\,&=\lambda\big(\!-\tfrac{1}{2}G_2\big)+\kappa\big(\!-\tfrac{1}{6}G_3\big)+\gamma\big(\!-\tfrac{1}{24}G_4+\tfrac{1}{8}G_2^2\big)+\mathcal{O}(\ell^5)\phantom{\Big[},\\
\delta_{\n}  g\,&=\kappa\big(\!-\tfrac{1}{2}G_2\big)+\gamma\big(\!-\tfrac{1}{6}G_3\big)+\mathcal{O}(\ell^5)\phantom{\Big[},\\
\delta_{\n}  \lambda\,&=\gamma\big(\!-\tfrac{1}{2}G_2\big)+\mathcal{O}(\ell^5)\phantom{\Big[},\\
\delta_{\n}  \kappa\,&=\mathcal{O}(\ell^5)\phantom{\Big[},\\
\delta_{\n}  \gamma\,&=\mathcal{O}(\ell^5)\phantom{\Big[},\\
&\,\,\,\vdots
\end{aligned}
\end{equation}
Substituting the counterterms (\ref{eq:counterterms GNs}) back into (\ref{eq:W(Jwiggly)}) is equivalent to having started from a path integral with a complete normal ordered bare action. 

Before we move on to substituting these counterterms into the generating function of interest, $W(J)$, let us make one final important remark: we have discussed complete normal ordering for \emph{interaction} terms in a given action. Indeed, when we do not want to complete normal order the kinetic term in the theory of interest, the above counterterms are the complete set of complete normal ordering counterterms. We will refer to this as \emph{weak} complete normal ordering. When in addition we also complete normal order the kinetic term (i.e.~when we complete normal order the \emph{full} action), we will refer to this as \emph{strong} complete normal ordering. The generating functions of the two types are related by a shift of the vacuum diagrams,
\begin{equation}\label{eq:Ws=Ww+Q4}
W(J)_{\rm strong} = W(J)_{\rm weak}+Q_4,
\end{equation}
where $Q_4=\int \frac{1}{2}(\Delta+m^2)G_2-\int\frac{1}{2}\!\wf \!G_2$. We initially discuss weak complete normal ordering, and in Sec.~\ref{sec:VC} (where we discuss the vacuum diagrams in detail) we will see that strong complete normal ordering is more natural. But of course when vacuum diagrams are not physical (e.g.~for flat spacetime and in the absence of other fields) there is no distinction between the two types of complete normal ordering.

\section{Cephalopod Cancellation}\label{sec:CC}

Recall that we want to determine the resulting expression for $W(J)$ in perturbation theory up to and including $\mathcal{O}(\ell^4)$, see (\ref{eq:W(Jwiggly)}). By inspection, (in weak complete normal ordering) this implies we first require $G_2$ up to $\mathcal{O}(\ell^2)$, $G_3$ up to $\mathcal{O}(\ell^3)$, $G_4$ up to $\mathcal{O}(\ell^2)$, $G_5$ up to $\mathcal{O}(\ell)$, and $G_6$ up to $\mathcal{O}(\ell^0)$, and we determine these next. These expressions will then be substituted into (\ref{eq:counterterms GNs}), and subsequently into (\ref{eq:W(Jwiggly)}). If we were instead to adopt the strong form of complete normal ordering, the above is sufficient with one modification, that we would require $G_2$ up to an including $\mathcal{O}(\ell^4)$. 

From the definition of full renormalised connected Green functions in (\ref{eq:W(J) GN}) we see that we need to take functional derivatives of $\frac{1}{\hbar}W(J)$, see (\ref{eq:W(Jwiggly)}), and then point merge the external legs,
$$
G_N=\lim_{\{z_1,\dots,z_N\}\rightarrow z}G_N(z_1,\dots,z_N).
$$
This point-merging procedure will typically give rise to divergences and some regularisation prescription is necessary, such as dimensional regularisation. The usual procedure in quantum field theory is then to  adopt a scheme, such as minimal subtraction, whereby (up to overlapping divergences that need particular care) one only keeps the divergent contribution in the above point merging limit. Here, however, we are subtracting the \emph{full} $N$-point Green function at coincident points (independently of whether it is even infinite or not), and so this is an oversubtraction scheme whereby we shift the \emph{renormalised} couplings by finite terms, in addition to subtracting the infinities. 
 
The procedure of extracting the $G_N$ from $W(J)$ in (\ref{eq:W(Jwiggly)}) is now completely straightforward given the above conventions, and we give the results directly:
\vspace{-1cm}

\begin{spreadlines}{-.5\baselineskip}
\begin{equation*}\label{eq:G_N wiggly}
\begin{aligned}
G_2=&\,\,
\hspace{0.5cm}
\begin{gathered}
	\begin{fmffile}{wbubblex}
		\begin{fmfgraph}(40,40)
			\fmfset{dash_len}{1.2mm}
			\fmfset{wiggly_len}{1.1mm} \fmfset{dot_len}{0.5mm}
			\fmfpen{0.25mm}
			\fmfvn{decor.shape=circle,decor.filled=full, decor.size=3thin}{u}{1}
			\fmfleft{i}
			\fmfright{o}
			\fmf{wiggly,fore=black,tension=5,left}{i,u1,i}
			\fmffreeze
			\fmfforce{(-w,0.35h)}{i}
			\fmfforce{(0w,0.35h)}{u1}
			\fmfforce{(1.1w,0.35h)}{o}
		\end{fmfgraph}\!\!\!\!\!\!
	\end{fmffile}
\end{gathered}
+\hat{g}^2
\Big(\tfrac{1}{2}
\hspace{-.5cm}
\begin{gathered}
	\begin{fmffile}{wg3tav}
		\begin{fmfgraph}(130,130)
			\fmfset{dash_len}{1.2mm}
			\fmfset{wiggly_len}{1.1mm} \fmfset{dot_len}{0.5mm}
			\fmfpen{0.25mm}
			\fmftop{t}
			\fmfbottom{b}
			\fmfleft{l}
			\fmfright{r}
			\fmfv{decor.shape=circle,decor.filled=full, decor.size=3thin}{u}
			\fmf{phantom,fore=black,tension=9}{t,x,v,b}
			\fmf{phantom,fore=black,tension=9}{l,s,u,r}
			\fmf{wiggly,fore=black,tension=.01,left}{x,v,x}
			\fmf{phantom,fore=black,tension=0.01}{s,x,s}
			\fmf{wiggly,fore=black,tension=1}{x,v}
			\fmf{phantom,fore=black,tension=1}{u,r}
			\fmffreeze
			\fmfforce{(0.66w,0.5h)}{u}
		\end{fmfgraph}
	\end{fmffile}
\end{gathered}\hspace{-.5cm}
\Big)
-\hat{\lambda}
\Big(\tfrac{1}{2}
\begin{gathered}
	\begin{fmffile}{wlambdabubbled}
		\begin{fmfgraph}(80,80)
			\fmfset{dash_len}{1.2mm}
			\fmfset{wiggly_len}{1.1mm} \fmfset{dot_len}{0.5mm}
			\fmfpen{0.25mm}
			\fmftop{t1,t2,t3}
			\fmfbottom{b1,b2,b3}
			\fmf{phantom}{t1,v1,b1}
			\fmf{phantom}{t2,v2,b2}
			\fmf{phantom}{t3,v3,b3}
			\fmfv{decor.shape=circle,decor.filled=full, decor.size=3thin}{v3}
			\fmffreeze
			\fmf{wiggly,fore=black,right}{v1,v2,v1}
			\fmf{wiggly,fore=black,right}{v2,v3,v2}
		\end{fmfgraph}
	\end{fmffile}
\end{gathered}
\Big)
+\hat{g}^4\Big(
\tfrac{1}{4}\,\,
\begin{gathered}
	\begin{fmffile}{wg4-0pt-3loop1PIaG32a}
		\begin{fmfgraph}(37,37)
			\fmfset{dash_len}{1.2mm}
			\fmfset{wiggly_len}{1.1mm} \fmfset{dot_len}{0.5mm}
			\fmfpen{0.25mm}
			\fmfsurround{a,b,c,d}
			\fmf{phantom,fore=black,tension=1,curved}{a,b,c,d,a}
			\fmf{wiggly,fore=black,tension=1,right}{a,b}
			\fmf{wiggly,fore=black,tension=1,right}{c,d}
			\fmf{wiggly,fore=black,tension=1,right}{b,c}
			\fmf{wiggly,fore=black,tension=1,right}{d,a}
			\fmf{wiggly,fore=black,tension=1,straight}{a,b}
			\fmf{wiggly,fore=black,tension=1,straight}{c,d}
			\fmffreeze
			\fmfv{decor.shape=circle,decor.filled=full, decor.size=3thin}{u}
			\fmfforce{(0.97w,0.02h)}{u}
		\end{fmfgraph}
	\end{fmffile}
\end{gathered}\,\,
+\tfrac{1}{2}\!\!\!\!
\begin{gathered}
	\begin{fmffile}{wpeacehc}
		\begin{fmfgraph}(100,100)
			\fmfset{dash_len}{1.2mm}
			\fmfset{wiggly_len}{1.1mm} \fmfset{dot_len}{0.5mm}
			\fmfpen{0.25mm}
			\fmfsurroundn{i}{3}
			\fmf{phantom,fore=black}{i1,v,i2}
			\fmf{phantom,fore=black}{i2,u,i3}
			\fmf{phantom,fore=black}{i3,s,i1}
			\fmfi{wiggly,fore=black}{fullcircle scaled .4w shifted (.5w,.5h)}
			\fmf{wiggly,fore=black}{v,c}
			\fmf{wiggly,fore=black}{u,c}
			\fmf{wiggly,fore=black}{s,c}
			\fmffreeze
			\fmfv{decor.shape=circle,decor.filled=full, decor.size=3thin}{h}
			\fmfforce{(0.7w,0.5h)}{h}
		\end{fmfgraph}
	\end{fmffile}
\end{gathered}\!\!\!\!
+\tfrac{1}{2}\,\,
\begin{gathered}
	\begin{fmffile}{wg4-0pt-3loop1PIaG32b}
		\begin{fmfgraph}(37,37)
			\fmfset{dash_len}{1.2mm}
			\fmfset{wiggly_len}{1.1mm} \fmfset{dot_len}{0.5mm}
			\fmfpen{0.25mm}
			\fmfsurround{a,b,c,d}
			\fmf{phantom,fore=black,tension=1,curved}{a,b,c,d,a}
			\fmf{wiggly,fore=black,tension=1,right}{a,b}
			\fmf{wiggly,fore=black,tension=1,right}{c,d}
			\fmf{wiggly,fore=black,tension=1,right}{b,c}
			\fmf{wiggly,fore=black,tension=1,right}{d,a}
			\fmf{wiggly,fore=black,tension=1,straight}{a,b}
			\fmf{wiggly,fore=black,tension=1,straight}{c,d}
			\fmffreeze
			\fmfv{decor.shape=circle,decor.filled=full, decor.size=3thin}{u}
			\fmfforce{(0.99w,0.98h)}{u}
		\end{fmfgraph}
	\end{fmffile}
\end{gathered}\,\,
\Big)\\[0.1cm]
&-\hat{g}^2\hat{\lambda}\Big(
\tfrac{1}{4}
\begin{gathered}
	\begin{fmffile}{wg2L-bubblezb}
		\begin{fmfgraph}(80,80)
			\fmfset{dash_len}{1.2mm}
			\fmfset{wiggly_len}{1.1mm} \fmfset{dot_len}{0.5mm}
			\fmfpen{0.25mm}
			\fmftop{t1,t2,t3}
			\fmfbottom{b1,b2,b3}
			\fmf{phantom}{t1,v1,b1}
			\fmf{phantom}{t2,v2,b2}
			\fmf{phantom}{t3,v3,b3}
			\fmffreeze
			\fmf{wiggly,fore=black,right}{v1,v2,v1}
			\fmf{wiggly,fore=black,right}{v2,v3,v2}
			\fmf{wiggly,fore=black,tension=1}{t1,b1}
			\fmfforce{(0.25w,0.7h)}{t1}
			\fmfforce{(0.25w,0.3h)}{b1}
			\fmffreeze
			\fmfv{decor.shape=circle,decor.filled=full, decor.size=3thin}{u}
			\fmfforce{(1w,0.5h)}{u}
		\end{fmfgraph}
	\end{fmffile}
\end{gathered}
+
\tfrac{1}{2}
\begin{gathered}
	\begin{fmffile}{wg2L-bubblezbb}
		\begin{fmfgraph}(80,80)
			\fmfset{dash_len}{1.2mm}
			\fmfset{wiggly_len}{1.1mm} \fmfset{dot_len}{0.5mm}
			\fmfpen{0.25mm}
			\fmftop{t1,t2,t3}
			\fmfbottom{b1,b2,b3}
			\fmf{phantom}{t1,v1,b1}
			\fmf{phantom}{t2,v2,b2}
			\fmf{phantom}{t3,v3,b3}
			\fmffreeze
			\fmf{wiggly,fore=black,right}{v1,v2,v1}
			\fmf{wiggly,fore=black,right}{v2,v3,v2}
			\fmf{wiggly,fore=black,tension=1}{t1,b1}
			\fmfforce{(0.25w,0.7h)}{t1}
			\fmfforce{(0.25w,0.3h)}{b1}
			\fmffreeze
			\fmfv{decor.shape=circle,decor.filled=full, decor.size=3thin}{u}
			\fmfforce{(0.4w,0.31h)}{u}
		\end{fmfgraph}
	\end{fmffile}
\end{gathered}
+
\tfrac{1}{2}
\begin{gathered}
	\begin{fmffile}{wg2L-bubblezbbb}
		\begin{fmfgraph}(80,80)
			\fmfset{dash_len}{1.2mm}
			\fmfset{wiggly_len}{1.1mm} \fmfset{dot_len}{0.5mm}
			\fmfpen{0.25mm}
			\fmftop{t1,t2,t3}
			\fmfbottom{b1,b2,b3}
			\fmf{phantom}{t1,v1,b1}
			\fmf{phantom}{t2,v2,b2}
			\fmf{phantom}{t3,v3,b3}
			\fmffreeze
			\fmf{wiggly,fore=black,right}{v1,v2,v1}
			\fmf{wiggly,fore=black,right}{v2,v3,v2}
			\fmf{wiggly,fore=black,tension=1}{t1,b1}
			\fmfforce{(0.25w,0.7h)}{t1}
			\fmfforce{(0.25w,0.3h)}{b1}
			\fmffreeze
			\fmfv{decor.shape=circle,decor.filled=full, decor.size=3thin}{u}
			\fmfforce{(0w,0.5h)}{u}
		\end{fmfgraph}
	\end{fmffile}
\end{gathered}
+
\,\,
\begin{gathered}
	\begin{fmffile}{wg2L-3loopbubble-1PIhc}
		\begin{fmfgraph}(40,40)
			\fmfset{dash_len}{1.2mm}
			\fmfset{wiggly_len}{1.1mm} \fmfset{dot_len}{0.5mm}
			\fmfpen{0.25mm}
			\fmfsurroundn{i}{3}
			\fmf{wiggly,fore=black,tension=1,right=1}{i1,i2}
			\fmf{wiggly,fore=black,tension=1,right=1}{i2,i3}
			\fmf{wiggly,fore=black,tension=1,right=0.8}{i3,i1}
			\fmf{wiggly,fore=black,tension=1}{i1,i2}
			\fmf{wiggly,fore=black,tension=1}{i2,i3}
			\fmffreeze
			\fmfv{decor.shape=circle,decor.filled=full, decor.size=3thin}{u}
			\fmfforce{(0.9w,1.05h)}{u}
		\end{fmfgraph}
	\end{fmffile}
\end{gathered}\,\,
+\tfrac{1}{4}\,\,
\begin{gathered}
	\begin{fmffile}{wg2L-3loopbubble-1PIhd}
		\begin{fmfgraph}(40,40)
			\fmfset{dash_len}{1.2mm}
			\fmfset{wiggly_len}{1.1mm} \fmfset{dot_len}{0.5mm}
			\fmfpen{0.25mm}
			\fmfsurroundn{i}{3}
			\fmf{wiggly,fore=black,tension=1,right=1}{i1,i2}
			\fmf{wiggly,fore=black,tension=1,right=1}{i2,i3}
			\fmf{wiggly,fore=black,tension=1,right=0.8}{i3,i1}
			\fmf{wiggly,fore=black,tension=1}{i1,i2}
			\fmf{wiggly,fore=black,tension=1}{i2,i3}
			\fmffreeze
			\fmfv{decor.shape=circle,decor.filled=full, decor.size=3thin}{u}
			\fmfforce{(0.8w,0.0h)}{u}
		\end{fmfgraph}
	\end{fmffile}
\end{gathered}\,\,\Big)
+\hat{\lambda}^2\Big(
\tfrac{1}{6}\!
\begin{gathered}
	\begin{fmffile}{wL2-3loop-bubbleG42b}
		\begin{fmfgraph}(75,75)
			\fmfset{dash_len}{1.2mm}
			\fmfset{wiggly_len}{1.1mm} \fmfset{dot_len}{0.5mm}
			\fmfpen{0.25mm}
			\fmfleft{i}
			\fmfright{o}
			\fmf{phantom,tension=10}{i,v1}
			\fmf{phantom,tension=10}{v2,o}
			\fmf{wiggly,left,tension=0.4}{v1,v2,v1}
			\fmf{wiggly,left=0.5}{v1,v2}
			\fmf{wiggly,right=0.5}{v1,v2}
			\fmffreeze
			\fmfv{decor.shape=circle,decor.filled=full, decor.size=3thin}{u}
			\fmfforce{(0.52w,0.17h)}{u}
    		\end{fmfgraph}
	\end{fmffile}
\end{gathered}\!
\Big)\\[0.4cm]
&+\hat{g}\hat{\kappa}\Big(
\tfrac{1}{6}\!\!
\begin{gathered}
	\begin{fmffile}{wgkappa-bubble-1PIxb}
		\begin{fmfgraph}(70,70)
			\fmfset{dash_len}{1.2mm}
			\fmfset{wiggly_len}{1.1mm} \fmfset{dot_len}{0.5mm}
			\fmfpen{0.25mm}
			\fmfleft{i}
			\fmfright{o}
			\fmf{phantom,tension=5}{i,v1}
			\fmf{phantom,tension=5}{v2,o}
			\fmf{wiggly,fore=black,left,tension=0.5}{v1,v2,v1}
			\fmf{wiggly,fore=black}{v1,v2}
			\fmffreeze
			\fmfi{wiggly,fore=black}{fullcircle scaled .55w shifted (1.1w,.5h)}
			\fmffreeze
			\fmfv{decor.shape=circle,decor.filled=full, decor.size=3thin}{u}
			\fmfforce{(1.37w,0.5h)}{u}
		\end{fmfgraph}
	\end{fmffile}
\end{gathered}\,\,\,\,\,
+\tfrac{1}{2}\!\!
\begin{gathered}
	\begin{fmffile}{wgkappa-bubble-1PIxbb}
		\begin{fmfgraph}(70,70)
			\fmfset{dash_len}{1.2mm}
			\fmfset{wiggly_len}{1.1mm} \fmfset{dot_len}{0.5mm}
			\fmfpen{0.25mm}
			\fmfleft{i}
			\fmfright{o}
			\fmf{phantom,tension=5}{i,v1}
			\fmf{phantom,tension=5}{v2,o}
			\fmf{wiggly,fore=black,left,tension=0.5}{v1,v2,v1}
			\fmf{wiggly,fore=black}{v1,v2}
			\fmffreeze
			\fmfi{wiggly,fore=black}{fullcircle scaled .55w shifted (1.1w,.5h)}
			\fmffreeze
			\fmfv{decor.shape=circle,decor.filled=full, decor.size=3thin}{u}
			\fmfforce{(.5w,0.23h)}{u}
		\end{fmfgraph}
	\end{fmffile}
\end{gathered}\,\,\,\,\,
\Big)
-\hat{\gamma}\Big(
\tfrac{1}{8}
\begin{gathered}
	\begin{fmffile}{wgamma-vacuum-1PIb}
		\begin{fmfgraph}(70,70)
			\fmfset{dash_len}{1.2mm}
			\fmfset{wiggly_len}{1.1mm} \fmfset{dot_len}{0.5mm}
			\fmfpen{0.25mm}
			\fmfsurroundn{x}{3}
			\fmf{phantom,fore=black}{x1,v}
			\fmf{phantom,fore=black}{x2,v}
			\fmf{phantom,fore=black}{x3,v}
			\fmf{wiggly,fore=black,tension=0.7}{v,v}
			\fmf{wiggly,fore=black,tension=0.7,right}{v,v}
			\fmf{wiggly,fore=black,tension=0.7,left}{v,v}
			\fmffreeze
			\fmfv{decor.shape=circle,decor.filled=full, decor.size=3thin}{u}
			\fmfforce{(0w,0.5h)}{u}
		\end{fmfgraph}
	\end{fmffile}
\end{gathered}\!
\Big)+\mathcal{O}(\ell^6) \, ,\\[0.45cm]
G_3=&\,-\hat{g}\Big(\!\!
\begin{gathered}
	\begin{fmffile}{wg2-2loopbubble-1PIG3}
		\begin{fmfgraph}(70,70)
			\fmfset{dash_len}{1.2mm}
			\fmfset{wiggly_len}{1.1mm} \fmfset{dot_len}{0.5mm}
			\fmfpen{0.25mm}
			\fmfleft{i}
			\fmfright{o}
			\fmf{phantom,tension=5}{i,v1}
			\fmf{phantom,tension=5}{v2,o}
			\fmf{wiggly,fore=black,left,tension=0.4}{v1,v2,v1}
			\fmf{wiggly,fore=black}{v1,v2}
			\fmfv{decor.shape=circle,decor.filled=full, decor.size=3thin}{v2}
		\end{fmfgraph}
	\end{fmffile}
\end{gathered}\!\Big)
-\hat{g}^3\Big(
\tfrac{3}{2}\,\,
\begin{gathered}
	\begin{fmffile}{wg4-0pt-3loop1PIaG3}
		\begin{fmfgraph}(37,37)
			\fmfset{dash_len}{1.2mm}
			\fmfset{wiggly_len}{1.1mm} \fmfset{dot_len}{0.5mm}
			\fmfpen{0.25mm}
			\fmfsurround{a,b,c,d}
			\fmf{phantom,fore=black,tension=1,curved}{a,b,c,d,a}
			\fmf{wiggly,fore=black,tension=1,right}{a,b}
			\fmf{wiggly,fore=black,tension=1,right}{c,d}
			\fmf{wiggly,fore=black,tension=1,right}{b,c}
			\fmf{wiggly,fore=black,tension=1,right}{d,a}
			\fmf{wiggly,fore=black,tension=1,straight}{a,b}
			\fmf{wiggly,fore=black,tension=1,straight}{c,d}
			\fmfv{decor.shape=circle,decor.filled=full, decor.size=3thin}{a}
		\end{fmfgraph}
	\end{fmffile}
\end{gathered}\,\,
+\!\!\!\!\!
\begin{gathered}
	\begin{fmffile}{wpeaceG3}
		\begin{fmfgraph}(105,105)
			\fmfset{dash_len}{1.2mm}
			\fmfset{wiggly_len}{1.1mm} \fmfset{dot_len}{0.5mm}
			\fmfpen{0.25mm}
			\fmfsurroundn{i}{3}
			\fmf{phantom,fore=black}{i1,v,i2}
			\fmf{phantom,fore=black}{i2,u,i3}
			\fmf{phantom,fore=black}{i3,s,i1}
			\fmfi{wiggly,fore=black}{fullcircle scaled .4w shifted (.5w,.5h)}
			\fmf{wiggly,fore=black}{v,c}
			\fmf{wiggly,fore=black}{u,c}
			\fmf{wiggly,fore=black}{s,c}
			\fmfv{decor.shape=circle,decor.filled=full, decor.size=3thin}{v}
		\end{fmfgraph}
	\end{fmffile}
\end{gathered}\!\!\!\!\!
\Big)
+\hat{g}\hat{\lambda}
\Big(\tfrac{3}{2}\,\,
\begin{gathered}
	\begin{fmffile}{wg2L-bubbleG3}
		\begin{fmfgraph}(70,70)
			\fmfset{dash_len}{1.2mm}
			\fmfset{wiggly_len}{1.1mm} \fmfset{dot_len}{0.5mm}
			\fmfpen{0.25mm}
			\fmftop{t1,t2,t3}
			\fmfbottom{b1,b2,b3}
			\fmf{phantom}{t1,v1,b1}
			\fmf{phantom}{t2,v2,b2}
			\fmf{phantom}{t3,v3,b3}
			\fmffreeze
			\fmf{wiggly,fore=black,right}{v1,v2,v1}
			\fmf{wiggly,fore=black,right}{v2,v3,v2}
			\fmf{wiggly,fore=black,tension=1}{t1,b1}
			\fmfforce{(0.25w,0.7h)}{t1}
			\fmfforce{(0.25w,0.3h)}{b1}
			\fmfv{decor.shape=circle,decor.filled=full, decor.size=3thin}{b1}
		\end{fmfgraph}
	\end{fmffile}
\end{gathered}
+\tfrac{3}{2}
\,\,
\begin{gathered}
	\begin{fmffile}{wg2L-3loopbubble-1PIG3}
		\begin{fmfgraph}(35,35)
			\fmfset{dash_len}{1.2mm}
			\fmfset{wiggly_len}{1.1mm} \fmfset{dot_len}{0.5mm}
			\fmfpen{0.25mm}
			\fmfsurroundn{i}{3}
			\fmf{wiggly,fore=black,tension=1,right=1}{i1,i2}
			\fmf{wiggly,fore=black,tension=1,right=1}{i2,i3}
			\fmf{wiggly,fore=black,tension=1,right=0.8}{i3,i1}
			\fmf{wiggly,fore=black,tension=1}{i1,i2}
			\fmf{wiggly,fore=black,tension=1}{i2,i3}
			\fmfv{decor.shape=circle,decor.filled=full, decor.size=3thin}{i1}
		\end{fmfgraph}
	\end{fmffile}
\end{gathered}\,\,
\Big)
-\hat{\kappa}\Big(\tfrac{1}{2}\!
\begin{gathered}
	\begin{fmffile}{wgkappa-bubble-1PIG3}
		\begin{fmfgraph}(60,60)
			\fmfset{dash_len}{1.2mm}
			\fmfset{wiggly_len}{1.1mm} \fmfset{dot_len}{0.5mm}
			\fmfpen{0.25mm}
			\fmfleft{i}
			\fmfright{o}
			\fmf{phantom,tension=5}{i,v1}
			\fmf{phantom,tension=5}{v2,o}
			\fmf{wiggly,fore=black,left,tension=0.5}{v1,v2,v1}
			\fmf{wiggly,fore=black}{v1,v2}
			\fmffreeze
			\fmfi{wiggly,fore=black}{fullcircle scaled .55w shifted (1.1w,.5h)}
			\fmfv{decor.shape=circle,decor.filled=full, decor.size=3thin}{v1}
		\end{fmfgraph}
	\end{fmffile}
\end{gathered}\,\,\,\,\Big)+\mathcal{O}(\ell^5) \, ,\\
G_4=&\,\,\hat{g}^2\Big(3
\,\,
\begin{gathered}
	\begin{fmffile}{wg2L-3loopbubble-1PIG4}
		\begin{fmfgraph}(35,35)
			\fmfset{dash_len}{1.2mm}
			\fmfset{wiggly_len}{1.1mm} \fmfset{dot_len}{0.5mm}
			\fmfpen{0.25mm}
			\fmfsurroundn{i}{3}
			\fmf{wiggly,fore=black,tension=1,right=1}{i1,i2}
			\fmf{wiggly,fore=black,tension=1,right=1}{i2,i3}
			\fmf{wiggly,fore=black,tension=1,right=0.8}{i3,i1}
			\fmf{wiggly,fore=black,tension=1}{i1,i2}
			\fmf{wiggly,fore=black,tension=1}{i2,i3}
			\fmfv{decor.shape=circle,decor.filled=full, decor.size=3thin}{i2}
		\end{fmfgraph}
	\end{fmffile}
\end{gathered}\,\,
\Big)-
\hat{\lambda}\Big(\!
\begin{gathered}
	\begin{fmffile}{wL2-3loop-bubbleG4}
		\begin{fmfgraph}(75,75)
			\fmfset{dash_len}{1.2mm}
			\fmfset{wiggly_len}{1.1mm} \fmfset{dot_len}{0.5mm}
			\fmfpen{0.25mm}
			\fmfleft{i}
			\fmfright{o}
			\fmf{phantom,tension=10}{i,v1}
			\fmf{phantom,tension=10}{v2,o}
			\fmf{wiggly,left,tension=0.4}{v1,v2,v1}
			\fmf{wiggly,left=0.5}{v1,v2}
			\fmf{wiggly,right=0.5}{v1,v2}
			\fmfv{decor.shape=circle,decor.filled=full, decor.size=3thin}{v2}
    		\end{fmfgraph}
	\end{fmffile}
\end{gathered}\!
\Big)+\mathcal{O}(\ell^4) \, , \phantom{\Big(\tfrac{1}{2}
\hspace{-.5cm}
\begin{gathered}
	\begin{fmffile}{wg3tav}
		\begin{fmfgraph}(130,130)
			\fmfset{dash_len}{1.2mm}
			\fmfset{wiggly_len}{1.1mm} \fmfset{dot_len}{0.5mm}
			\fmfpen{0.25mm}
			\fmftop{t}
			\fmfbottom{b}
			\fmfleft{l}
			\fmfright{r}
			\fmfv{decor.shape=circle,decor.filled=full, decor.size=3thin}{u}
			\fmf{phantom,fore=black,tension=9}{t,x,v,b}
			\fmf{phantom,fore=black,tension=9}{l,s,u,r}
			\fmf{wiggly,fore=black,tension=.01,left}{x,v,x}
			\fmf{phantom,fore=black,tension=0.01}{s,x,s}
			\fmf{wiggly,fore=black,tension=1}{x,v}
			\fmf{phantom,fore=black,tension=1}{u,r}
			\fmffreeze
			\fmfforce{(0.66w,0.5h)}{u}
		\end{fmfgraph}
	\end{fmffile}
\end{gathered}\hspace{-.5cm}
\Big)}\\
G_5=&\,\,\mathcal{O}(\ell^3) \, , \phantom{\Big(\tfrac{1}{2}
\hspace{-.5cm}
\begin{gathered}
	\begin{fmffile}{wg3tav}
		\begin{fmfgraph}(130,130)
			\fmfset{dash_len}{1.2mm}
			\fmfset{wiggly_len}{1.1mm} \fmfset{dot_len}{0.5mm}
			\fmfpen{0.25mm}
			\fmftop{t}
			\fmfbottom{b}
			\fmfleft{l}
			\fmfright{r}
			\fmfv{decor.shape=circle,decor.filled=full, decor.size=3thin}{u}
			\fmf{phantom,fore=black,tension=9}{t,x,v,b}
			\fmf{phantom,fore=black,tension=9}{l,s,u,r}
			\fmf{wiggly,fore=black,tension=.01,left}{x,v,x}
			\fmf{phantom,fore=black,tension=0.01}{s,x,s}
			\fmf{wiggly,fore=black,tension=1}{x,v}
			\fmf{phantom,fore=black,tension=1}{u,r}
			\fmffreeze
			\fmfforce{(0.66w,0.5h)}{u}
		\end{fmfgraph}
	\end{fmffile}
\end{gathered}\hspace{-.5cm}
\Big)} \\
G_6=&\,\,\mathcal{O}(\ell^4) \, ,\phantom{\Big(\tfrac{1}{2}
\hspace{-.5cm}
\begin{gathered}
	\begin{fmffile}{wg3tav}
		\begin{fmfgraph}(130,130)
			\fmfset{dash_len}{1.2mm}
			\fmfset{wiggly_len}{1.1mm} \fmfset{dot_len}{0.5mm}
			\fmfpen{0.25mm}
			\fmftop{t}
			\fmfbottom{b}
			\fmfleft{l}
			\fmfright{r}
			\fmfv{decor.shape=circle,decor.filled=full, decor.size=3thin}{u}
			\fmf{phantom,fore=black,tension=9}{t,x,v,b}
			\fmf{phantom,fore=black,tension=9}{l,s,u,r}
			\fmf{wiggly,fore=black,tension=.01,left}{x,v,x}
			\fmf{phantom,fore=black,tension=0.01}{s,x,s}
			\fmf{wiggly,fore=black,tension=1}{x,v}
			\fmf{phantom,fore=black,tension=1}{u,r}
			\fmffreeze
			\fmfforce{(0.66w,0.5h)}{u}
		\end{fmfgraph}
	\end{fmffile}
\end{gathered}\hspace{-.5cm}
\Big)}
\end{aligned}
\end{equation*}
\end{spreadlines}

\vspace{-0.5cm}
\setlength{\parindent}{0cm}
where the dots here indicate the point at which the external legs have merged to a point, and we only keep terms that contribute up to and including $\mathcal{O}(\ell^4)$ in $W(J)$. 

Substituting these into the counterterm expressions (\ref{eq:counterterms GNs}) leads to
\begin{spreadlines}{-0.3\baselineskip}
\begin{equation}\label{eq:delta g etc }
\begin{aligned}
\delta_{\n} \Lambda&=g\bigg[\hat{g}\Big(\tfrac{1}{6}\!\!
\begin{gathered}
	\begin{fmffile}{wg2-2loopbubble-1PIG32}
		\begin{fmfgraph}(70,70)
			\fmfset{dash_len}{1.2mm}
			\fmfset{wiggly_len}{1.1mm} \fmfset{dot_len}{0.5mm}
			\fmfpen{0.25mm}
			\fmfleft{i}
			\fmfright{o}
			\fmf{phantom,tension=5}{i,v1}
			\fmf{phantom,tension=5}{v2,o}
			\fmf{wiggly,fore=black,left,tension=0.4}{v1,v2,v1}
			\fmf{wiggly,fore=black}{v1,v2}
		\end{fmfgraph}
	\end{fmffile}
\end{gathered}\!\Big)
+\hat{g}^3\Big(
\tfrac{1}{4}\,\,
\begin{gathered}
	\begin{fmffile}{wg4-0pt-3loop1PIaG32}
		\begin{fmfgraph}(37,37)
			\fmfset{dash_len}{1.2mm}
			\fmfset{wiggly_len}{1.1mm} \fmfset{dot_len}{0.5mm}
			\fmfpen{0.25mm}
			\fmfsurround{a,b,c,d}
			\fmf{phantom,fore=black,tension=1,curved}{a,b,c,d,a}
			\fmf{wiggly,fore=black,tension=1,right}{a,b}
			\fmf{wiggly,fore=black,tension=1,right}{c,d}
			\fmf{wiggly,fore=black,tension=1,right}{b,c}
			\fmf{wiggly,fore=black,tension=1,right}{d,a}
			\fmf{wiggly,fore=black,tension=1,straight}{a,b}
			\fmf{wiggly,fore=black,tension=1,straight}{c,d}
		\end{fmfgraph}
	\end{fmffile}
\end{gathered}\,\,
+\tfrac{1}{6}\!\!\!\!\!
\begin{gathered}
	\begin{fmffile}{wpeaceG32}
		\begin{fmfgraph}(105,105)
			\fmfset{dash_len}{1.2mm}
			\fmfset{wiggly_len}{1.1mm} \fmfset{dot_len}{0.5mm}
			\fmfpen{0.25mm}
			\fmfsurroundn{i}{3}
			\fmf{phantom,fore=black}{i1,v,i2}
			\fmf{phantom,fore=black}{i2,u,i3}
			\fmf{phantom,fore=black}{i3,s,i1}
			\fmfi{wiggly,fore=black}{fullcircle scaled .4w shifted (.5w,.5h)}
			\fmf{wiggly,fore=black}{v,c}
			\fmf{wiggly,fore=black}{u,c}
			\fmf{wiggly,fore=black}{s,c}
		\end{fmfgraph}
	\end{fmffile}
\end{gathered}\!\!\!\!\!
\Big)
-\hat{g}\hat{\lambda}
\Big(\tfrac{1}{4}\,\,
\begin{gathered}
	\begin{fmffile}{wg2L-bubbleG32}
		\begin{fmfgraph}(70,70)
			\fmfset{dash_len}{1.2mm}
			\fmfset{wiggly_len}{1.1mm} \fmfset{dot_len}{0.5mm}
			\fmfpen{0.25mm}
			\fmftop{t1,t2,t3}
			\fmfbottom{b1,b2,b3}
			\fmf{phantom}{t1,v1,b1}
			\fmf{phantom}{t2,v2,b2}
			\fmf{phantom}{t3,v3,b3}
			\fmffreeze
			\fmf{wiggly,fore=black,right}{v1,v2,v1}
			\fmf{wiggly,fore=black,right}{v2,v3,v2}
			\fmf{wiggly,fore=black,tension=1}{t1,b1}
			\fmfforce{(0.25w,0.7h)}{t1}
			\fmfforce{(0.25w,0.3h)}{b1}
		\end{fmfgraph}
	\end{fmffile}
\end{gathered}
+\tfrac{1}{4}
\,\,
\begin{gathered}
	\begin{fmffile}{wg2L-3loopbubble-1PIG32}
		\begin{fmfgraph}(35,35)
			\fmfset{dash_len}{1.2mm}
			\fmfset{wiggly_len}{1.1mm} \fmfset{dot_len}{0.5mm}
			\fmfpen{0.25mm}
			\fmfsurroundn{i}{3}
			\fmf{wiggly,fore=black,tension=1,right=1}{i1,i2}
			\fmf{wiggly,fore=black,tension=1,right=1}{i2,i3}
			\fmf{wiggly,fore=black,tension=1,right=0.8}{i3,i1}
			\fmf{wiggly,fore=black,tension=1}{i1,i2}
			\fmf{wiggly,fore=black,tension=1}{i2,i3}
		\end{fmfgraph}
	\end{fmffile}
\end{gathered}\,\,
\Big)
+\hat{\kappa}\Big(\tfrac{1}{12}\!
\begin{gathered}
	\begin{fmffile}{wgkappa-bubble-1PIG32}
		\begin{fmfgraph}(60,60)
			\fmfset{dash_len}{1.2mm}
			\fmfset{wiggly_len}{1.1mm} \fmfset{dot_len}{0.5mm}
			\fmfpen{0.25mm}
			\fmfleft{i}
			\fmfright{o}
			\fmf{phantom,tension=5}{i,v1}
			\fmf{phantom,tension=5}{v2,o}
			\fmf{wiggly,fore=black,left,tension=0.5}{v1,v2,v1}
			\fmf{wiggly,fore=black}{v1,v2}
			\fmffreeze
			\fmfi{wiggly,fore=black}{fullcircle scaled .55w shifted (1.1w,.5h)}
		\end{fmfgraph}
	\end{fmffile}
\end{gathered}\,\,\,\,\Big)\bigg]\\
&+\lambda\bigg\{\tfrac{1}{8}\Big[
\hspace{0.5cm}
\begin{gathered}
	\begin{fmffile}{wbubblex2}
		\begin{fmfgraph}(40,40)
			\fmfset{dash_len}{1.2mm}
			\fmfset{wiggly_len}{1.1mm} \fmfset{dot_len}{0.5mm}
			\fmfpen{0.25mm}
			\fmfleft{i}
			\fmfright{o}
			\fmf{wiggly,fore=black,tension=5,left}{i,u1,i}
			\fmffreeze
			\fmfforce{(-w,0.35h)}{i}
			\fmfforce{(0w,0.35h)}{u1}
			\fmfforce{(1.1w,0.35h)}{o}
		\end{fmfgraph}\!\!\!\!\!\!
	\end{fmffile}
\end{gathered}
+\hat{g}^2
\Big(\tfrac{1}{2}
\hspace{-.5cm}
\begin{gathered}
	\begin{fmffile}{wg3tav2}
		\begin{fmfgraph}(130,130)
			\fmfset{dash_len}{1.2mm}
			\fmfset{wiggly_len}{1.1mm} \fmfset{dot_len}{0.5mm}
			\fmfpen{0.25mm}
			\fmftop{t}
			\fmfbottom{b}
			\fmfleft{l}
			\fmfright{r}
			\fmf{phantom,fore=black,tension=9}{t,x,v,b}
			\fmf{phantom,fore=black,tension=9}{l,s,u,r}
			\fmf{wiggly,fore=black,tension=.01,left}{x,v,x}
			\fmf{phantom,fore=black,tension=0.01}{s,x,s}
			\fmf{wiggly,fore=black,tension=1}{x,v}
			\fmf{phantom,fore=black,tension=1}{u,r}
			\fmffreeze
			\fmfforce{(0.66w,0.5h)}{u}
		\end{fmfgraph}
	\end{fmffile}
\end{gathered}\hspace{-.5cm}
\Big)
-\hat{\lambda}
\Big(\tfrac{1}{2}
\begin{gathered}
	\begin{fmffile}{wlambdabubbled2}
		\begin{fmfgraph}(80,80)
			\fmfset{dash_len}{1.2mm}
			\fmfset{wiggly_len}{1.1mm} \fmfset{dot_len}{0.5mm}
			\fmfpen{0.25mm}
			\fmftop{t1,t2,t3}
			\fmfbottom{b1,b2,b3}
			\fmf{phantom}{t1,v1,b1}
			\fmf{phantom}{t2,v2,b2}
			\fmf{phantom}{t3,v3,b3}
			\fmffreeze
			\fmf{wiggly,fore=black,right}{v1,v2,v1}
			\fmf{wiggly,fore=black,right}{v2,v3,v2}
		\end{fmfgraph}
	\end{fmffile}
\end{gathered}
\Big)\Big]^2-\hat{g}^2\Big(\tfrac{1}{8}
\,\,
\begin{gathered}
	\begin{fmffile}{wg2L-3loopbubble-1PIG42}
		\begin{fmfgraph}(35,35)
			\fmfset{dash_len}{1.2mm}
			\fmfset{wiggly_len}{1.1mm} \fmfset{dot_len}{0.5mm}
			\fmfpen{0.25mm}
			\fmfsurroundn{i}{3}
			\fmf{wiggly,fore=black,tension=1,right=1}{i1,i2}
			\fmf{wiggly,fore=black,tension=1,right=1}{i2,i3}
			\fmf{wiggly,fore=black,tension=1,right=0.8}{i3,i1}
			\fmf{wiggly,fore=black,tension=1}{i1,i2}
			\fmf{wiggly,fore=black,tension=1}{i2,i3}
		\end{fmfgraph}
	\end{fmffile}
\end{gathered}\,\,
\Big)+
\hat{\lambda}\Big(\tfrac{1}{24}\!
\begin{gathered}
	\begin{fmffile}{wL2-3loop-bubbleG42}
		\begin{fmfgraph}(75,75)
			\fmfset{dash_len}{1.2mm}
			\fmfset{wiggly_len}{1.1mm} \fmfset{dot_len}{0.5mm}
			\fmfpen{0.25mm}
			\fmfleft{i}
			\fmfright{o}
			\fmf{phantom,tension=10}{i,v1}
			\fmf{phantom,tension=10}{v2,o}
			\fmf{wiggly,left,tension=0.4}{v1,v2,v1}
			\fmf{wiggly,left=0.5}{v1,v2}
			\fmf{wiggly,right=0.5}{v1,v2}
    		\end{fmfgraph}
	\end{fmffile}
\end{gathered}\!
\Big)\bigg\}\\
&+\kappa\bigg\{\tfrac{1}{12}\Big[\hspace{0.5cm}
\begin{gathered}
	\begin{fmffile}{wbubblex2}
		\begin{fmfgraph}(40,40)
			\fmfset{dash_len}{1.2mm}
			\fmfset{wiggly_len}{1.1mm} \fmfset{dot_len}{0.5mm}
			\fmfpen{0.25mm}
			\fmfleft{i}
			\fmfright{o}
			\fmf{wiggly,fore=black,tension=5,left}{i,u1,i}
			\fmffreeze
			\fmfforce{(-w,0.35h)}{i}
			\fmfforce{(0w,0.35h)}{u1}
			\fmfforce{(1.1w,0.35h)}{o}
		\end{fmfgraph}\!\!\!\!\!\!
	\end{fmffile}
\end{gathered}
+\mathcal{O}(\ell^2)
\Big]\Big[-\hat{g}\Big(\!\!
\begin{gathered}
	\begin{fmffile}{wg2-2loopbubble-1PIG32}
		\begin{fmfgraph}(70,70)
			\fmfset{dash_len}{1.2mm}
			\fmfset{wiggly_len}{1.1mm} \fmfset{dot_len}{0.5mm}
			\fmfpen{0.25mm}
			\fmfleft{i}
			\fmfright{o}
			\fmf{phantom,tension=5}{i,v1}
			\fmf{phantom,tension=5}{v2,o}
			\fmf{wiggly,fore=black,left,tension=0.4}{v1,v2,v1}
			\fmf{wiggly,fore=black}{v1,v2}
		\end{fmfgraph}
	\end{fmffile}
\end{gathered}\!\Big)+\mathcal{O}(\ell^3)\Big]+\mathcal{O}(\ell^3)\bigg\}\phantom{\Bigg\}}\\
&
+\gamma\Big[\!-\tfrac{1}{48}\Big(\hspace{0.5cm}
\begin{gathered}
	\begin{fmffile}{wbubblex2}
		\begin{fmfgraph}(40,40)
			\fmfset{dash_len}{1.2mm}
			\fmfset{wiggly_len}{1.1mm} \fmfset{dot_len}{0.5mm}
			\fmfpen{0.25mm}
			\fmfleft{i}
			\fmfright{o}
			\fmf{wiggly,fore=black,tension=5,left}{i,u1,i}
			\fmffreeze
			\fmfforce{(-w,0.35h)}{i}
			\fmfforce{(0w,0.35h)}{u1}
			\fmfforce{(1.1w,0.35h)}{o}
		\end{fmfgraph}\!\!\!\!\!\!
	\end{fmffile}
\end{gathered}\Big)^3+\mathcal{O}(\ell^2)\Big]+\mathcal{O}(\ell^5) \, , \phantom{\Bigg\}}\\
Y\,&=g\bigg\{\tfrac{1}{2}\Big[\hspace{0.5cm}
\begin{gathered}
	\begin{fmffile}{wbubblex}
		\begin{fmfgraph}(40,40)
			\fmfset{dash_len}{1.2mm}
			\fmfset{wiggly_len}{1.1mm} \fmfset{dot_len}{0.5mm}
			\fmfpen{0.25mm}
			\fmfvn{decor.shape=circle,decor.filled=full, decor.size=3thin}{u}{1}
			\fmfleft{i}
			\fmfright{o}
			\fmf{wiggly,fore=black,tension=5,left}{i,u1,i}
			\fmffreeze
			\fmfforce{(-w,0.35h)}{i}
			\fmfforce{(0w,0.35h)}{u1}
			\fmfforce{(1.1w,0.35h)}{o}
		\end{fmfgraph}\!\!\!\!\!\!
	\end{fmffile}
\end{gathered}
+\hat{g}^2
\Big(\tfrac{1}{2}
\hspace{-.5cm}
\begin{gathered}
	\begin{fmffile}{wg3tav}
		\begin{fmfgraph}(130,130)
			\fmfset{dash_len}{1.2mm}
			\fmfset{wiggly_len}{1.1mm} \fmfset{dot_len}{0.5mm}
			\fmfpen{0.25mm}
			\fmftop{t}
			\fmfbottom{b}
			\fmfleft{l}
			\fmfright{r}
			\fmfv{decor.shape=circle,decor.filled=full, decor.size=3thin}{u}
			\fmf{phantom,fore=black,tension=9}{t,x,v,b}
			\fmf{phantom,fore=black,tension=9}{l,s,u,r}
			\fmf{wiggly,fore=black,tension=.01,left}{x,v,x}
			\fmf{phantom,fore=black,tension=0.01}{s,x,s}
			\fmf{wiggly,fore=black,tension=1}{x,v}
			\fmf{phantom,fore=black,tension=1}{u,r}
			\fmffreeze
			\fmfforce{(0.66w,0.5h)}{u}
		\end{fmfgraph}
	\end{fmffile}
\end{gathered}\hspace{-.5cm}
\Big)
-\hat{\lambda}
\Big(\tfrac{1}{2}
\begin{gathered}
	\begin{fmffile}{wlambdabubbled}
		\begin{fmfgraph}(80,80)
			\fmfset{dash_len}{1.2mm}
			\fmfset{wiggly_len}{1.1mm} \fmfset{dot_len}{0.5mm}
			\fmfpen{0.25mm}
			\fmftop{t1,t2,t3}
			\fmfbottom{b1,b2,b3}
			\fmf{phantom}{t1,v1,b1}
			\fmf{phantom}{t2,v2,b2}
			\fmf{phantom}{t3,v3,b3}
			\fmfv{decor.shape=circle,decor.filled=full, decor.size=3thin}{v3}
			\fmffreeze
			\fmf{wiggly,fore=black,right}{v1,v2,v1}
			\fmf{wiggly,fore=black,right}{v2,v3,v2}
		\end{fmfgraph}
	\end{fmffile}
\end{gathered}
\Big)
+\mathcal{O}(\ell^4) \Big]\bigg\}\phantom{\Bigg[}\\
&+\lambda\bigg\{\tfrac{1}{3!}\Big[-\hat{g}\Big(\!\!
\begin{gathered}
	\begin{fmffile}{wg2-2loopbubble-1PIG3}
		\begin{fmfgraph}(70,70)
			\fmfset{dash_len}{1.2mm}
			\fmfset{wiggly_len}{1.1mm} \fmfset{dot_len}{0.5mm}
			\fmfpen{0.25mm}
			\fmfleft{i}
			\fmfright{o}
			\fmf{phantom,tension=5}{i,v1}
			\fmf{phantom,tension=5}{v2,o}
			\fmf{wiggly,fore=black,left,tension=0.4}{v1,v2,v1}
			\fmf{wiggly,fore=black}{v1,v2}
			\fmfv{decor.shape=circle,decor.filled=full, decor.size=3thin}{v2}
		\end{fmfgraph}
	\end{fmffile}
\end{gathered}\!\Big)+\mathcal{O}(\ell^3)\Big]\bigg\}
+\kappa\bigg\{-\tfrac{1}{8}\Big[\hspace{0.5cm}
\begin{gathered}
	\begin{fmffile}{wbubblex}
		\begin{fmfgraph}(40,40)
			\fmfset{dash_len}{1.2mm}
			\fmfset{wiggly_len}{1.1mm} \fmfset{dot_len}{0.5mm}
			\fmfpen{0.25mm}
			\fmfvn{decor.shape=circle,decor.filled=full, decor.size=3thin}{u}{1}
			\fmfleft{i}
			\fmfright{o}
			\fmf{wiggly,fore=black,tension=5,left}{i,u1,i}
			\fmffreeze
			\fmfforce{(-w,0.35h)}{i}
			\fmfforce{(0w,0.35h)}{u1}
			\fmfforce{(1.1w,0.35h)}{o}
		\end{fmfgraph}\!\!\!\!\!\!
	\end{fmffile}
\end{gathered}\Big]^2+\mathcal{O}(\ell^2)\bigg\}+\mathcal{O}(\ell^5) \, , \phantom{\Bigg[}\\
\delta_{\n}  m^2\,&=\lambda\bigg\{\!-\tfrac{1}{2}\Big[\hspace{0.5cm}
\begin{gathered}
	\begin{fmffile}{wbubblex}
		\begin{fmfgraph}(40,40)
			\fmfset{dash_len}{1.2mm}
			\fmfset{wiggly_len}{1.1mm} \fmfset{dot_len}{0.5mm}
			\fmfpen{0.25mm}
			\fmfvn{decor.shape=circle,decor.filled=full, decor.size=3thin}{u}{1}
			\fmfleft{i}
			\fmfright{o}
			\fmf{wiggly,fore=black,tension=5,left}{i,u1,i}
			\fmffreeze
			\fmfforce{(-w,0.35h)}{i}
			\fmfforce{(0w,0.35h)}{u1}
			\fmfforce{(1.1w,0.35h)}{o}
		\end{fmfgraph}\!\!\!\!\!\!
	\end{fmffile}
\end{gathered}
+\hat{g}^2
\Big(\tfrac{1}{2}
\hspace{-.5cm}
\begin{gathered}
	\begin{fmffile}{wg3tav}
		\begin{fmfgraph}(130,130)
			\fmfset{dash_len}{1.2mm}
			\fmfset{wiggly_len}{1.1mm} \fmfset{dot_len}{0.5mm}
			\fmfpen{0.25mm}
			\fmftop{t}
			\fmfbottom{b}
			\fmfleft{l}
			\fmfright{r}
			\fmfv{decor.shape=circle,decor.filled=full, decor.size=3thin}{u}
			\fmf{phantom,fore=black,tension=9}{t,x,v,b}
			\fmf{phantom,fore=black,tension=9}{l,s,u,r}
			\fmf{wiggly,fore=black,tension=.01,left}{x,v,x}
			\fmf{phantom,fore=black,tension=0.01}{s,x,s}
			\fmf{wiggly,fore=black,tension=1}{x,v}
			\fmf{phantom,fore=black,tension=1}{u,r}
			\fmffreeze
			\fmfforce{(0.66w,0.5h)}{u}
		\end{fmfgraph}
	\end{fmffile}
\end{gathered}\hspace{-.5cm}
\Big)
-\hat{\lambda}
\Big(\tfrac{1}{2}
\begin{gathered}
	\begin{fmffile}{wlambdabubbled}
		\begin{fmfgraph}(80,80)
			\fmfset{dash_len}{1.2mm}
			\fmfset{wiggly_len}{1.1mm} \fmfset{dot_len}{0.5mm}
			\fmfpen{0.25mm}
			\fmftop{t1,t2,t3}
			\fmfbottom{b1,b2,b3}
			\fmf{phantom}{t1,v1,b1}
			\fmf{phantom}{t2,v2,b2}
			\fmf{phantom}{t3,v3,b3}
			\fmfv{decor.shape=circle,decor.filled=full, decor.size=3thin}{v3}
			\fmffreeze
			\fmf{wiggly,fore=black,right}{v1,v2,v1}
			\fmf{wiggly,fore=black,right}{v2,v3,v2}
		\end{fmfgraph}
	\end{fmffile}
\end{gathered}
\Big)
+\mathcal{O}(\ell^4)\Big]\bigg\}\phantom{\Bigg[}\\
&+\kappa\bigg\{\hat{g}\Big(\tfrac{1}{6}\!\!
\begin{gathered}
	\begin{fmffile}{wg2-2loopbubble-1PIG3}
		\begin{fmfgraph}(70,70)
			\fmfset{dash_len}{1.2mm}
			\fmfset{wiggly_len}{1.1mm} \fmfset{dot_len}{0.5mm}
			\fmfpen{0.25mm}
			\fmfleft{i}
			\fmfright{o}
			\fmf{phantom,tension=5}{i,v1}
			\fmf{phantom,tension=5}{v2,o}
			\fmf{wiggly,fore=black,left,tension=0.4}{v1,v2,v1}
			\fmf{wiggly,fore=black}{v1,v2}
			\fmfv{decor.shape=circle,decor.filled=full, decor.size=3thin}{v2}
		\end{fmfgraph}
	\end{fmffile}
\end{gathered}\!\Big)+\mathcal{O}(\ell^3)\bigg\}
+\gamma\bigg\{\tfrac{1}{8}\Big[\hspace{0.5cm}
\begin{gathered}
	\begin{fmffile}{wbubblex}
		\begin{fmfgraph}(40,40)
			\fmfset{dash_len}{1.2mm}
			\fmfset{wiggly_len}{1.1mm} \fmfset{dot_len}{0.5mm}
			\fmfpen{0.25mm}
			\fmfvn{decor.shape=circle,decor.filled=full, decor.size=3thin}{u}{1}
			\fmfleft{i}
			\fmfright{o}
			\fmf{wiggly,fore=black,tension=5,left}{i,u1,i}
			\fmffreeze
			\fmfforce{(-w,0.35h)}{i}
			\fmfforce{(0w,0.35h)}{u1}
			\fmfforce{(1.1w,0.35h)}{o}
		\end{fmfgraph}\!\!\!\!\!\!
	\end{fmffile}
\end{gathered}
+\mathcal{O}(\ell^2)\Big]^2\bigg\}+\mathcal{O}(\ell^5) \, , \phantom{\Bigg[}\\
\delta_{\n}  g\,&=-\kappa\Big(\tfrac{1}{2}\hspace{0.5cm}
\begin{gathered}
	\begin{fmffile}{wbubblex}
		\begin{fmfgraph}(40,40)
			\fmfset{dash_len}{1.2mm}
			\fmfset{wiggly_len}{1.1mm} \fmfset{dot_len}{0.5mm}
			\fmfpen{0.25mm}
			\fmfvn{decor.shape=circle,decor.filled=full, decor.size=3thin}{u}{1}
			\fmfleft{i}
			\fmfright{o}
			\fmf{wiggly,fore=black,tension=5,left}{i,u1,i}
			\fmffreeze
			\fmfforce{(-w,0.35h)}{i}
			\fmfforce{(0w,0.35h)}{u1}
			\fmfforce{(1.1w,0.35h)}{o}
		\end{fmfgraph}\!\!\!\!\!\!
	\end{fmffile}
\end{gathered}\Big)+\mathcal{O}(\ell^5) \, , \phantom{\Bigg[}\\
\delta_{\n}  \lambda\,&=-\gamma\Big(\tfrac{1}{2}\hspace{0.5cm}
\begin{gathered}
	\begin{fmffile}{wbubblex}
		\begin{fmfgraph}(40,40)
			\fmfset{dash_len}{1.2mm}
			\fmfset{wiggly_len}{1.1mm} \fmfset{dot_len}{0.5mm}
			\fmfpen{0.25mm}
			\fmfvn{decor.shape=circle,decor.filled=full, decor.size=3thin}{u}{1}
			\fmfleft{i}
			\fmfright{o}
			\fmf{wiggly,fore=black,tension=5,left}{i,u1,i}
			\fmffreeze
			\fmfforce{(-w,0.35h)}{i}
			\fmfforce{(0w,0.35h)}{u1}
			\fmfforce{(1.1w,0.35h)}{o}
		\end{fmfgraph}\!\!\!\!\!\!
	\end{fmffile}
\end{gathered}\Big)+\mathcal{O}(\ell^5) \, , \phantom{\Bigg[}\\
\delta_{\n}  \kappa\,&=\mathcal{O}(\ell^5) \, , \phantom{\Bigg[}\\
\delta_{\n}  \gamma\,&=\mathcal{O}(\ell^5) \, , \phantom{\Bigg[}\\
&\,\,\,\vdots\phantom{\Bigg[}
\end{aligned}
\end{equation}
\end{spreadlines}
We note that the vacuum counterterm diagrams are already complete, and hence there are no `dots' present in these (although of course some of these will be modified by the explicit counterterms in $\hat{g}$, $\hat{\lambda}$, etc.). 
We would  like ultimately to express these counterterms in terms of the renormalised couplings, $g_N$, and these are related to the couplings appearing explicitly via the relations (\ref{eq:ghatN}) (recall also the footnote on p.~\!\pageref{foot:couplings} and also the couplings notation explained on p.~\!\pageref{coupling notation}). With this objective in mind, we first write the above counterterms in terms of the quantities $g$, $\lambda$, $\kappa$, $\gamma$, (related to the aforementioned couplings via $\hat{g}=g+\delta_{\n}  g$, $\hat{\lambda}=\lambda+\delta_{\n}  \lambda$, etc.) to the order of interest, $\mathcal{O}(\ell^4)$, starting from the source counterterm. 

\subsection{Tadpole Counterterms}\label{sec:TCT}
\setlength{\parindent}{1cm}
Let us begin with the source counterterm, $Y$. In (\ref{eq:Ybare}) we determined what the explicit expression for this counterterm must be in order to cancel all tadpoles up to $\mathcal{O}(\ell^4)$, and we now show that complete normal ordering which leads to the explicit expression for $Y$ in (\ref{eq:delta g etc }) is equal to (\ref{eq:Ybare}), allowing us to conclude that complete normal ordering cancels all tadpoles, a result which is expected to hold to any finite order in perturbation theory. Indeed, (\ref{eq:Ybare}) and (\ref{eq:delta g etc }) are equivalent on account of the counterterms (\ref{eq:delta g etc }): given that $\hat{g}=g+\delta_{\n}  g$, $\hat{\lambda}=\lambda+\delta_{\n}  \lambda$, etc., (\ref{eq:Ybare}) can equivalently be written as:
\begin{spreadlines}{-.4\baselineskip}
\begin{equation}\label{eq:Ybareb}
\begin{aligned}
Y&=\hat{g}\Big(\tfrac{1}{2}
\hspace{0.5cm}
\begin{gathered}
	\begin{fmffile}{bubblexas}
		\begin{fmfgraph}(40,40)
			\fmfset{dash_len}{1.2mm}
			\fmfset{wiggly_len}{1.1mm} \fmfset{dot_len}{0.5mm}
			\fmfpen{0.25mm}
			\fmfvn{decor.shape=circle,decor.filled=full, decor.size=3thin}{u}{1}
			\fmfleft{i}
			\fmfright{o}
			\fmf{wiggly,fore=black,tension=5,left,label.dist=1,label=a}{i,u1,i}
			\fmffreeze
			\fmfforce{(-w,0.35h)}{i}
			\fmfforce{(0w,0.35h)}{u1}
			\fmfforce{(1.1w,0.35h)}{o}
			\fmf{phantom,label.dist=0,label=1}{i,u1}
		\end{fmfgraph}\!\!\!\!\!\!
	\end{fmffile}
\end{gathered}\Big)
+\hat{g}^3\Big(
\tfrac{1}{4}\hspace{-.5cm}
\begin{gathered}
	\begin{fmffile}{wg3tav}
		\begin{fmfgraph}(130,130)
			\fmfset{dash_len}{1.2mm}
			\fmfset{wiggly_len}{1.1mm} \fmfset{dot_len}{0.5mm}
			\fmfpen{0.25mm}
			\fmftop{t}
			\fmfbottom{b}
			\fmfleft{l}
			\fmfright{r}
			\fmfv{decor.shape=circle,decor.filled=full, decor.size=3thin}{u}
			\fmf{phantom,fore=black,tension=9}{t,x,v,b}
			\fmf{phantom,fore=black,tension=9}{l,s,u,r}
			\fmf{wiggly,fore=black,tension=.01,left}{x,v,x}
			\fmf{phantom,fore=black,tension=0.01}{s,x,s}
			\fmf{wiggly,fore=black,tension=1}{x,v}
			\fmf{phantom,fore=black,tension=1}{u,r}
			\fmffreeze
			\fmfforce{(0.66w,0.5h)}{u}
		\end{fmfgraph}
	\end{fmffile}
\end{gathered}\hspace{-.5cm}
\Big)
-\hat{g}\hat{\lambda}\Big(
\tfrac{1}{3!}\hspace{-.2cm}
\begin{gathered}
	\begin{fmffile}{wglambdatadpolesdf}
		\begin{fmfgraph}(90,90)
			\fmfset{dash_len}{1.2mm}
			\fmfset{wiggly_len}{1.1mm} \fmfset{dot_len}{0.5mm}
			\fmfpen{0.25mm}
			\fmfleft{i}
			\fmfright{o}
			\fmfv{decor.shape=circle,decor.filled=full, decor.size=3thin}{v2}
			\fmf{phantom,tension=5}{i,v1}
			\fmf{phantom,fore=black,tension=2.5}{v2,o}
			\fmf{wiggly,fore=black,left,tension=0.5}{v1,v2,v1}
			\fmf{wiggly,fore=black}{v1,v2}
			\fmffreeze
			\fmfforce{(1.1w,0.5h)}{o}
		\end{fmfgraph}
	\end{fmffile}
\end{gathered}\!\!\!\!
+\tfrac{1}{4}
\begin{gathered}
	\begin{fmffile}{wlambdabubbled}
		\begin{fmfgraph}(80,80)
			\fmfset{dash_len}{1.2mm}
			\fmfset{wiggly_len}{1.1mm} \fmfset{dot_len}{0.5mm}
			\fmfpen{0.25mm}
			\fmftop{t1,t2,t3}
			\fmfbottom{b1,b2,b3}
			\fmf{phantom}{t1,v1,b1}
			\fmf{phantom}{t2,v2,b2}
			\fmf{phantom}{t3,v3,b3}
			\fmfv{decor.shape=circle,decor.filled=full, decor.size=3thin}{v3}
			\fmffreeze
			\fmf{wiggly,fore=black,right}{v1,v2,v1}
			\fmf{wiggly,fore=black,right}{v2,v3,v2}
		\end{fmfgraph}
	\end{fmffile}
\end{gathered}
\Big)
+\hat{\kappa}\Big(
\tfrac{1}{8}
\begin{gathered}
	\begin{fmffile}{wlambdabubbledk}
		\begin{fmfgraph}(80,80)
			\fmfset{dash_len}{1.2mm}
			\fmfset{wiggly_len}{1.1mm} \fmfset{dot_len}{0.5mm}
			\fmfpen{0.25mm}
			\fmftop{t1,t2,t3}
			\fmfbottom{b1,b2,b3}
			\fmf{phantom}{t1,v1,b1}
			\fmf{phantom}{t2,v2,b2}
			\fmf{phantom}{t3,v3,b3}
			\fmfv{decor.shape=circle,decor.filled=full, decor.size=3thin}{v2}
			\fmffreeze
			\fmf{wiggly,fore=black,right}{v1,v2,v1}
			\fmf{wiggly,fore=black,right}{v2,v3,v2}
		\end{fmfgraph}
	\end{fmffile}
\end{gathered}
\Big)+\mathcal{O}(\ell^5)\\
&=g\Big(\tfrac{1}{2}
\hspace{0.5cm}
\begin{gathered}
	\begin{fmffile}{bubblexas}
		\begin{fmfgraph}(40,40)
			\fmfset{dash_len}{1.2mm}
			\fmfset{wiggly_len}{1.1mm} \fmfset{dot_len}{0.5mm}
			\fmfpen{0.25mm}
			\fmfvn{decor.shape=circle,decor.filled=full, decor.size=3thin}{u}{1}
			\fmfleft{i}
			\fmfright{o}
			\fmf{wiggly,fore=black,tension=5,left,label.dist=1,label=a}{i,u1,i}
			\fmffreeze
			\fmfforce{(-w,0.35h)}{i}
			\fmfforce{(0w,0.35h)}{u1}
			\fmfforce{(1.1w,0.35h)}{o}
			\fmf{phantom,label.dist=0,label=1}{i,u1}
		\end{fmfgraph}\!\!\!\!\!\!
	\end{fmffile}
\end{gathered}\Big)
+g^3\Big(
\tfrac{1}{4}\hspace{-.5cm}
\begin{gathered}
	\begin{fmffile}{wg3tav}
		\begin{fmfgraph}(130,130)
			\fmfset{dash_len}{1.2mm}
			\fmfset{wiggly_len}{1.1mm} \fmfset{dot_len}{0.5mm}
			\fmfpen{0.25mm}
			\fmftop{t}
			\fmfbottom{b}
			\fmfleft{l}
			\fmfright{r}
			\fmfv{decor.shape=circle,decor.filled=full, decor.size=3thin}{u}
			\fmf{phantom,fore=black,tension=9}{t,x,v,b}
			\fmf{phantom,fore=black,tension=9}{l,s,u,r}
			\fmf{wiggly,fore=black,tension=.01,left}{x,v,x}
			\fmf{phantom,fore=black,tension=0.01}{s,x,s}
			\fmf{wiggly,fore=black,tension=1}{x,v}
			\fmf{phantom,fore=black,tension=1}{u,r}
			\fmffreeze
			\fmfforce{(0.66w,0.5h)}{u}
		\end{fmfgraph}
	\end{fmffile}
\end{gathered}\hspace{-.5cm}
\Big)
-g\lambda\Big(
\tfrac{1}{3!}\hspace{-.2cm}
\begin{gathered}
	\begin{fmffile}{wglambdatadpolesdf}
		\begin{fmfgraph}(90,90)
			\fmfset{dash_len}{1.2mm}
			\fmfset{wiggly_len}{1.1mm} \fmfset{dot_len}{0.5mm}
			\fmfpen{0.25mm}
			\fmfleft{i}
			\fmfright{o}
			\fmfv{decor.shape=circle,decor.filled=full, decor.size=3thin}{v2}
			\fmf{phantom,tension=5}{i,v1}
			\fmf{phantom,fore=black,tension=2.5}{v2,o}
			\fmf{wiggly,fore=black,left,tension=0.5}{v1,v2,v1}
			\fmf{wiggly,fore=black}{v1,v2}
			\fmffreeze
			\fmfforce{(1.1w,0.5h)}{o}
		\end{fmfgraph}
	\end{fmffile}
\end{gathered}\!\!\!\!
+\tfrac{1}{4}
\begin{gathered}
	\begin{fmffile}{wlambdabubbled}
		\begin{fmfgraph}(80,80)
			\fmfset{dash_len}{1.2mm}
			\fmfset{wiggly_len}{1.1mm} \fmfset{dot_len}{0.5mm}
			\fmfpen{0.25mm}
			\fmftop{t1,t2,t3}
			\fmfbottom{b1,b2,b3}
			\fmf{phantom}{t1,v1,b1}
			\fmf{phantom}{t2,v2,b2}
			\fmf{phantom}{t3,v3,b3}
			\fmfv{decor.shape=circle,decor.filled=full, decor.size=3thin}{v3}
			\fmffreeze
			\fmf{wiggly,fore=black,right}{v1,v2,v1}
			\fmf{wiggly,fore=black,right}{v2,v3,v2}
		\end{fmfgraph}
	\end{fmffile}
\end{gathered}
\Big)
-\kappa\Big(
\tfrac{1}{8}
\begin{gathered}
	\begin{fmffile}{wlambdabubbledk}
		\begin{fmfgraph}(80,80)
			\fmfset{dash_len}{1.2mm}
			\fmfset{wiggly_len}{1.1mm} \fmfset{dot_len}{0.5mm}
			\fmfpen{0.25mm}
			\fmftop{t1,t2,t3}
			\fmfbottom{b1,b2,b3}
			\fmf{phantom}{t1,v1,b1}
			\fmf{phantom}{t2,v2,b2}
			\fmf{phantom}{t3,v3,b3}
			\fmfv{decor.shape=circle,decor.filled=full, decor.size=3thin}{v2}
			\fmffreeze
			\fmf{wiggly,fore=black,right}{v1,v2,v1}
			\fmf{wiggly,fore=black,right}{v2,v3,v2}
		\end{fmfgraph}
	\end{fmffile}
\end{gathered}
\Big)+\mathcal{O}(\ell^5),
\end{aligned}
\end{equation}
\end{spreadlines}
where we note that to this order in $\ell$ the only effect of the counterterms $\delta_{\n}  g$, $\delta_{\n}  \lambda$, and $\delta_{\n}  \kappa$ (appearing in $\hat{g}$, $\hat{\lambda}$ and $\hat{\kappa}$ respectively in the first equality in (\ref{eq:Ybareb})) is to flip the sign of the term proportional to $\kappa$. Similarly, up to the order we are working in the $Y$ counterterm in (\ref{eq:delta g etc }), all counterterms $\delta_{\n}  g$, $\delta_{\n}  \lambda$ and $\delta_{\n}  \kappa$ lead to higher-order contributions, and so the $Y$ counterterm determined by the complete normal ordering prescription is \emph{precisely} that required to cancel all tadpoles up to $\mathcal{O}(\ell^4)$. 

To see how the relevant manipulations are carried out explicitly, consider the simple example
\begin{spreadlines}{-1\baselineskip}
\begin{equation}\label{eq:g2sewingexample}
\begin{aligned}
\hat{g}^2
\hspace{-.5cm}
\begin{gathered}
	\begin{fmffile}{wg3tav}
		\begin{fmfgraph}(130,130)
			\fmfset{dash_len}{1.2mm}
			\fmfset{wiggly_len}{1.1mm} \fmfset{dot_len}{0.5mm}
			\fmfpen{0.25mm}
			\fmftop{t}
			\fmfbottom{b}
			\fmfleft{l}
			\fmfright{r}
			\fmfv{decor.shape=circle,decor.filled=full, decor.size=3thin}{u}
			\fmf{phantom,fore=black,tension=9}{t,x,v,b}
			\fmf{phantom,fore=black,tension=9}{l,s,u,r}
			\fmf{wiggly,fore=black,tension=.01,left}{x,v,x}
			\fmf{phantom,fore=black,tension=0.01}{s,x,s}
			\fmf{wiggly,fore=black,tension=1}{x,v}
			\fmf{phantom,fore=black,tension=1}{u,r}
			\fmffreeze
			\fmfforce{(0.66w,0.5h)}{u}
		\end{fmfgraph}
	\end{fmffile}
\end{gathered}\hspace{-.5cm}
=&\,
\Big(g+\delta_{\n}  g\Big)^2
\hspace{-.5cm}
\begin{gathered}
	\begin{fmffile}{wg3tav}
		\begin{fmfgraph}(130,130)
			\fmfset{dash_len}{1.2mm}
			\fmfset{wiggly_len}{1.1mm} \fmfset{dot_len}{0.5mm}
			\fmfpen{0.25mm}
			\fmftop{t}
			\fmfbottom{b}
			\fmfleft{l}
			\fmfright{r}
			\fmfv{decor.shape=circle,decor.filled=full, decor.size=3thin}{u}
			\fmf{phantom,fore=black,tension=9}{t,x,v,b}
			\fmf{phantom,fore=black,tension=9}{l,s,u,r}
			\fmf{wiggly,fore=black,tension=.01,left}{x,v,x}
			\fmf{phantom,fore=black,tension=0.01}{s,x,s}
			\fmf{wiggly,fore=black,tension=1}{x,v}
			\fmf{phantom,fore=black,tension=1}{u,r}
			\fmffreeze
			\fmfforce{(0.66w,0.5h)}{u}
		\end{fmfgraph}
	\end{fmffile}
\end{gathered}\hspace{-.5cm}\\
=&\Big(g^2+2g\delta_{\n}  g+\delta_{\n}  g^2\Big)
\hspace{-.5cm}
\begin{gathered}
	\begin{fmffile}{wg3tav}
		\begin{fmfgraph}(130,130)
			\fmfset{dash_len}{1.2mm}
			\fmfset{wiggly_len}{1.1mm} \fmfset{dot_len}{0.5mm}
			\fmfpen{0.25mm}
			\fmftop{t}
			\fmfbottom{b}
			\fmfleft{l}
			\fmfright{r}
			\fmfv{decor.shape=circle,decor.filled=full, decor.size=3thin}{u}
			\fmf{phantom,fore=black,tension=9}{t,x,v,b}
			\fmf{phantom,fore=black,tension=9}{l,s,u,r}
			\fmf{wiggly,fore=black,tension=.01,left}{x,v,x}
			\fmf{phantom,fore=black,tension=0.01}{s,x,s}
			\fmf{wiggly,fore=black,tension=1}{x,v}
			\fmf{phantom,fore=black,tension=1}{u,r}
			\fmffreeze
			\fmfforce{(0.66w,0.5h)}{u}
		\end{fmfgraph}
	\end{fmffile}
\end{gathered}\hspace{-.5cm}\\
=&\,\Big(g^2
-g\kappa
\hspace{0.5cm}
\begin{gathered}
	\begin{fmffile}{wbubblexk}
		\begin{fmfgraph}(40,40)
			\fmfset{dash_len}{1.2mm}
			\fmfset{wiggly_len}{1.1mm} \fmfset{dot_len}{0.5mm}
			\fmfpen{0.25mm}
			\fmfvn{decor.shape=hexagram,decor.filled=full, decor.size=4thin}{u}{1}
			\fmfleft{i}
			\fmfright{o}
			\fmf{wiggly,fore=black,tension=5,left}{i,u1,i}
			\fmffreeze
			\fmfforce{(-w,0.35h)}{i}
			\fmfforce{(0w,0.35h)}{u1}
			\fmfforce{(1.1w,0.35h)}{o}
		\end{fmfgraph}\!\!\!\!\!\!
	\end{fmffile}
\end{gathered}
+\mathcal{O}(\ell^6)\Big)
\hspace{-.5cm}
\begin{gathered}
	\begin{fmffile}{wg3tavi}
		\begin{fmfgraph}(130,130)
			\fmfset{dash_len}{1.2mm}
			\fmfset{wiggly_len}{1.1mm} \fmfset{dot_len}{0.5mm}
			\fmfpen{0.25mm}
			\fmftop{t}
			\fmfbottom{b}
			\fmfleft{l}
			\fmfright{r}
			\fmfv{decor.shape=circle,decor.filled=full, decor.size=3thin}{u}
			\fmfv{decor.shape=hexagram,decor.filled=full, decor.size=4thin}{v}
			\fmf{phantom,fore=black,tension=9}{t,x,v,b}
			\fmf{phantom,fore=black,tension=9}{l,s,u,r}
			\fmf{wiggly,fore=black,tension=.01,left}{x,v,x}
			\fmf{phantom,fore=black,tension=0.01}{s,x,s}
			\fmf{wiggly,fore=black,tension=1}{x,v}
			\fmf{phantom,fore=black,tension=1}{u,r}
			\fmffreeze
			\fmfforce{(0.66w,0.5h)}{u}
		\end{fmfgraph}
	\end{fmffile}
\end{gathered}\hspace{-.5cm}\\
=&\,
g^2
\hspace{-.5cm}
\begin{gathered}
	\begin{fmffile}{wg3tav}
		\begin{fmfgraph}(130,130)
			\fmfset{dash_len}{1.2mm}
			\fmfset{wiggly_len}{1.1mm} \fmfset{dot_len}{0.5mm}
			\fmfpen{0.25mm}
			\fmftop{t}
			\fmfbottom{b}
			\fmfleft{l}
			\fmfright{r}
			\fmfv{decor.shape=circle,decor.filled=full, decor.size=3thin}{u}
			\fmf{phantom,fore=black,tension=9}{t,x,v,b}
			\fmf{phantom,fore=black,tension=9}{l,s,u,r}
			\fmf{wiggly,fore=black,tension=.01,left}{x,v,x}
			\fmf{phantom,fore=black,tension=0.01}{s,x,s}
			\fmf{wiggly,fore=black,tension=1}{x,v}
			\fmf{phantom,fore=black,tension=1}{u,r}
			\fmffreeze
			\fmfforce{(0.66w,0.5h)}{u}
		\end{fmfgraph}
	\end{fmffile}
\end{gathered}\hspace{-.5cm}
-g\kappa\!\!\!
\begin{gathered}
	\begin{fmffile}{wgkappa-bubble-1PIp}
		\begin{fmfgraph}(80,80)
			\fmfset{dash_len}{1.2mm}
			\fmfset{wiggly_len}{1.1mm} \fmfset{dot_len}{0.5mm}
			\fmfpen{0.25mm}
			\fmfleft{i}
			\fmfright{o}
			\fmf{phantom,tension=5}{i,v1}
			\fmf{phantom,tension=5}{v2,o}
			\fmf{wiggly,fore=black,left,tension=0.5}{v1,v2,v1}
			\fmf{wiggly,fore=black}{v1,v2}
			\fmffreeze
			\fmfi{wiggly,fore=black}{fullcircle scaled .55w shifted (1.1w,.5h)}
			\fmfv{decor.shape=circle,decor.filled=full, decor.size=3thin}{i}
			\fmfforce{(0.48w,0.23h)}{i}
			\end{fmfgraph}
	\end{fmffile}
\end{gathered}\,\,\,\,\,
+\mathcal{O}(\ell^6),
\end{aligned}
\end{equation}
\end{spreadlines}
where in the third equality we made use of the explicit expression for $\delta_{\n}  g$ in (\ref{eq:delta g etc }) and in the fourth equality we sewed the associated incomplete Feynman diagrams at the vertices denoted by 
`
$\!\!\!\!\!
\begin{fmffile}{star}
\begin{fmfgraph}(30,30)
\fmfsurroundn{v}{1}
\fmfv{decor.shape=hexagram,decor.filled=full, decor.size=4thin}{v1}
\end{fmfgraph}
\end{fmffile}
$
'. 
The remaining incomplete vertex in the third equality is denoted by 
`
$\!\!\!\!\!
\begin{fmffile}{circle}
\begin{fmfgraph}(30,30)
\fmfsurroundn{v}{1}
\fmfv{decor.shape=circle,decor.filled=full, decor.size=4thin}{v1}
\end{fmfgraph}
\end{fmffile}
$\,\,'. 
The latter completes the Feynman diagrams associated to the counterterms in (\ref{eq:W(Jwiggly)}). A second example is the following,
\begin{spreadlines}{-.1\baselineskip}
\begin{equation}\label{eq:lambdasewingexample}
\begin{aligned}
\hat{\lambda}
\begin{gathered}
	\begin{fmffile}{wlambdabubbled}
		\begin{fmfgraph}(80,80)
			\fmfset{dash_len}{1.2mm}
			\fmfset{wiggly_len}{1.1mm} \fmfset{dot_len}{0.5mm}
			\fmfpen{0.25mm}
			\fmftop{t1,t2,t3}
			\fmfbottom{b1,b2,b3}
			\fmf{phantom}{t1,v1,b1}
			\fmf{phantom}{t2,v2,b2}
			\fmf{phantom}{t3,v3,b3}
			\fmfv{decor.shape=circle,decor.filled=full, decor.size=3thin}{v3}
			\fmffreeze
			\fmf{wiggly,fore=black,right}{v1,v2,v1}
			\fmf{wiggly,fore=black,right}{v2,v3,v2}
		\end{fmfgraph}
	\end{fmffile}
\end{gathered}
=&\,
\Big(\lambda
+\delta_{\n}  \lambda
\Big)
\begin{gathered}
	\begin{fmffile}{wlambdabubbledkka}
		\begin{fmfgraph}(80,80)
			\fmfset{dash_len}{1.2mm}
			\fmfset{wiggly_len}{1.1mm} \fmfset{dot_len}{0.5mm}
			\fmfpen{0.25mm}
			\fmftop{t1,t2,t3}
			\fmfbottom{b1,b2,b3}
			\fmf{phantom}{t1,v1,b1}
			\fmf{phantom}{t2,v2,b2}
			\fmf{phantom}{t3,v3,b3}
			\fmfv{decor.shape=circle,decor.filled=full, decor.size=3thin}{v3}
			\fmffreeze
			\fmf{wiggly,fore=black,right}{v1,v2,v1}
			\fmf{wiggly,fore=black,right}{v2,v3,v2}
		\end{fmfgraph}
	\end{fmffile}
\end{gathered}\\
=&\,
\Big(\lambda
-\gamma
\tfrac{1}{2}
\hspace{0.5cm}
\begin{gathered}
	\begin{fmffile}{wbubblexg}
		\begin{fmfgraph}(40,40)
			\fmfset{dash_len}{1.2mm}
			\fmfset{wiggly_len}{1.1mm} \fmfset{dot_len}{0.5mm}
			\fmfpen{0.25mm}
			\fmfvn{decor.shape=hexagram,decor.filled=full, decor.size=4thin}{u}{1}
			\fmfleft{i}
			\fmfright{o}
			\fmf{wiggly,fore=black,tension=5,left}{i,u1,i}
			\fmffreeze
			\fmfforce{(-w,0.35h)}{i}
			\fmfforce{(0w,0.35h)}{u1}
			\fmfforce{(1.1w,0.35h)}{o}
		\end{fmfgraph}\!\!\!\!\!\!
	\end{fmffile}
\end{gathered}
+\mathcal{O}(\ell^6)
\Big)
\begin{gathered}
	\begin{fmffile}{wlambdabubbledkk}
		\begin{fmfgraph}(80,80)
			\fmfset{dash_len}{1.2mm}
			\fmfset{wiggly_len}{1.1mm} \fmfset{dot_len}{0.5mm}
			\fmfpen{0.25mm}
			\fmftop{t1,t2,t3}
			\fmfbottom{b1,b2,b3}
			\fmf{phantom}{t1,v1,b1}
			\fmf{phantom}{t2,v2,b2}
			\fmf{phantom}{t3,v3,b3}
			\fmfv{decor.shape=circle,decor.filled=full, decor.size=3thin}{v3}
			\fmfv{decor.shape=hexagram,decor.filled=full, decor.size=4thin}{v2}
			\fmffreeze
			\fmf{wiggly,fore=black,right}{v1,v2,v1}
			\fmf{wiggly,fore=black,right}{v2,v3,v2}
		\end{fmfgraph}
	\end{fmffile}
\end{gathered}\\
=&\,
\lambda
\begin{gathered}
	\begin{fmffile}{wlambdabubbled}
		\begin{fmfgraph}(80,80)
			\fmfset{dash_len}{1.2mm}
			\fmfset{wiggly_len}{1.1mm} \fmfset{dot_len}{0.5mm}
			\fmfpen{0.25mm}
			\fmftop{t1,t2,t3}
			\fmfbottom{b1,b2,b3}
			\fmf{phantom}{t1,v1,b1}
			\fmf{phantom}{t2,v2,b2}
			\fmf{phantom}{t3,v3,b3}
			\fmfv{decor.shape=circle,decor.filled=full, decor.size=3thin}{v3}
			\fmffreeze
			\fmf{wiggly,fore=black,right}{v1,v2,v1}
			\fmf{wiggly,fore=black,right}{v2,v3,v2}
		\end{fmfgraph}
	\end{fmffile}
\end{gathered}
-\gamma
\Big(\tfrac{1}{2}\,
\begin{gathered}
	\begin{fmffile}{wgamma-vacuum-1PIkk}
		\begin{fmfgraph}(90,90)
			\fmfset{dash_len}{1.2mm}
			\fmfset{wiggly_len}{1.1mm} \fmfset{dot_len}{0.5mm}
			\fmfpen{0.25mm}
			\fmfsurroundn{x}{3}
			\fmf{phantom,fore=black}{x1,v}
			\fmf{phantom,fore=black}{x2,v}
			\fmf{phantom,fore=black}{x3,v}
			\fmf{wiggly,fore=black,tension=0.7}{v,v}
			\fmf{wiggly,fore=black,tension=0.7,right}{v,v}
			\fmf{wiggly,fore=black,tension=0.7,left}{v,v}
			\fmffreeze
			\fmfbottom{b}
			\fmfforce{(0w,0.5h)}{b}
			\fmfv{decor.shape=circle,decor.filled=full, decor.size=3thin}{b}
			\end{fmfgraph}
	\end{fmffile}
\end{gathered}
\Big)+\mathcal{O}(\ell^6),
\end{aligned}
\end{equation}
\end{spreadlines}
where the manipulations are precisely analogous to the above example. Both of these examples appear in the mass renormalisation counterterm, $\delta_{\n}  m^2$, in (\ref{eq:delta g etc }), and we focus on this next. 

\subsection{Remaining Counterterms}\label{sec:RCT}
We note that the wiggly propagator as defined in (\ref{eq:dressed_prop}) and (\ref{eq:deltam2 deltaZ}) also contains mass renormalisation counterterms, and therefore we have a recursive structure in the $\delta_{\n}  m^2$ equation in (\ref{eq:delta g etc }),
\begin{equation}\label{eq:deltam2}
\begin{aligned}
\delta_{\n}  m^2\,&=\lambda\bigg\{\!-\tfrac{1}{2}\Big[\hspace{0.5cm}
\begin{gathered}
	\begin{fmffile}{wbubblex}
		\begin{fmfgraph}(40,40)
			\fmfset{dash_len}{1.2mm}
			\fmfset{wiggly_len}{1.1mm} \fmfset{dot_len}{0.5mm}
			\fmfpen{0.25mm}
			\fmfvn{decor.shape=circle,decor.filled=full, decor.size=3thin}{u}{1}
			\fmfleft{i}
			\fmfright{o}
			\fmf{wiggly,fore=black,tension=5,left}{i,u1,i}
			\fmffreeze
			\fmfforce{(-w,0.35h)}{i}
			\fmfforce{(0w,0.35h)}{u1}
			\fmfforce{(1.1w,0.35h)}{o}
		\end{fmfgraph}\!\!\!\!\!\!
	\end{fmffile}
\end{gathered}
+\hat{g}^2
\Big(\tfrac{1}{2}
\hspace{-.5cm}
\begin{gathered}
	\begin{fmffile}{wg3tav}
		\begin{fmfgraph}(130,130)
			\fmfset{dash_len}{1.2mm}
			\fmfset{wiggly_len}{1.1mm} \fmfset{dot_len}{0.5mm}
			\fmfpen{0.25mm}
			\fmftop{t}
			\fmfbottom{b}
			\fmfleft{l}
			\fmfright{r}
			\fmfv{decor.shape=circle,decor.filled=full, decor.size=3thin}{u}
			\fmf{phantom,fore=black,tension=9}{t,x,v,b}
			\fmf{phantom,fore=black,tension=9}{l,s,u,r}
			\fmf{wiggly,fore=black,tension=.01,left}{x,v,x}
			\fmf{phantom,fore=black,tension=0.01}{s,x,s}
			\fmf{wiggly,fore=black,tension=1}{x,v}
			\fmf{phantom,fore=black,tension=1}{u,r}
			\fmffreeze
			\fmfforce{(0.66w,0.5h)}{u}
		\end{fmfgraph}
	\end{fmffile}
\end{gathered}\hspace{-.5cm}
\Big)
-\hat{\lambda}
\Big(\tfrac{1}{2}
\begin{gathered}
	\begin{fmffile}{wlambdabubbled}
		\begin{fmfgraph}(80,80)
			\fmfset{dash_len}{1.2mm}
			\fmfset{wiggly_len}{1.1mm} \fmfset{dot_len}{0.5mm}
			\fmfpen{0.25mm}
			\fmftop{t1,t2,t3}
			\fmfbottom{b1,b2,b3}
			\fmf{phantom}{t1,v1,b1}
			\fmf{phantom}{t2,v2,b2}
			\fmf{phantom}{t3,v3,b3}
			\fmfv{decor.shape=circle,decor.filled=full, decor.size=3thin}{v3}
			\fmffreeze
			\fmf{wiggly,fore=black,right}{v1,v2,v1}
			\fmf{wiggly,fore=black,right}{v2,v3,v2}
		\end{fmfgraph}
	\end{fmffile}
\end{gathered}
\Big)
+\mathcal{O}(\ell^4)\Big]\bigg\}\phantom{\Bigg[}\\
&+\kappa\bigg\{\hat{g}\Big(\tfrac{1}{6}\!\!
\begin{gathered}
	\begin{fmffile}{wg2-2loopbubble-1PIG3}
		\begin{fmfgraph}(70,70)
			\fmfset{dash_len}{1.2mm}
			\fmfset{wiggly_len}{1.1mm} \fmfset{dot_len}{0.5mm}
			\fmfpen{0.25mm}
			\fmfleft{i}
			\fmfright{o}
			\fmf{phantom,tension=5}{i,v1}
			\fmf{phantom,tension=5}{v2,o}
			\fmf{wiggly,fore=black,left,tension=0.4}{v1,v2,v1}
			\fmf{wiggly,fore=black}{v1,v2}
			\fmfv{decor.shape=circle,decor.filled=full, decor.size=3thin}{v2}
		\end{fmfgraph}
	\end{fmffile}
\end{gathered}\!\Big)+\mathcal{O}(\ell^3)\bigg\}
+\gamma\bigg\{\tfrac{1}{8}\Big[\hspace{0.5cm}
\begin{gathered}
	\begin{fmffile}{wbubblex}
		\begin{fmfgraph}(40,40)
			\fmfset{dash_len}{1.2mm}
			\fmfset{wiggly_len}{1.1mm} \fmfset{dot_len}{0.5mm}
			\fmfpen{0.25mm}
			\fmfvn{decor.shape=circle,decor.filled=full, decor.size=3thin}{u}{1}
			\fmfleft{i}
			\fmfright{o}
			\fmf{wiggly,fore=black,tension=5,left}{i,u1,i}
			\fmffreeze
			\fmfforce{(-w,0.35h)}{i}
			\fmfforce{(0w,0.35h)}{u1}
			\fmfforce{(1.1w,0.35h)}{o}
		\end{fmfgraph}\!\!\!\!\!\!
	\end{fmffile}
\end{gathered}
+\mathcal{O}(\ell^2)\Big]^2\bigg\}+\mathcal{O}(\ell^5).\phantom{\Bigg[}
\end{aligned}
\end{equation}
This recursive relation can easily be solved within perturbation theory. In particular, we require $\delta_{\n}  m^2$ up to $\mathcal{O}(\ell^4)$, while $\delta_{\n}  m^2\sim \mathcal{O}(\ell^2)$, implying that the mass renormalisation contributions inside the wiggly propagators in (\ref{eq:deltam2}) can all be replaced by dashed propagators, see (\ref{eq:dressed_prop}), because most of the explicit couplings appearing are already $\mathcal{O}(\ell^4)$, 
e.g.~$\lambda \hat{g}^2
\hspace{-.5cm}
\begin{gathered}
	\begin{fmffile}{wg3tavx}
		\begin{fmfgraph}(130,130)
			\fmfset{dash_len}{1.2mm}
			\fmfset{wiggly_len}{1.1mm} \fmfset{dot_len}{0.5mm}
			\fmfpen{0.25mm}
			\fmftop{t}
			\fmfbottom{b}
			\fmfleft{l}
			\fmfright{r}
			\fmfv{decor.shape=circle,decor.filled=full, decor.size=3thin}{u}
			\fmf{phantom,fore=black,tension=9}{t,x,v,b}
			\fmf{phantom,fore=black,tension=9}{l,s,u,r}
			\fmf{wiggly,fore=black,tension=.01,left}{x,v,x}
			\fmf{phantom,fore=black,tension=0.01}{s,x,s}
			\fmf{wiggly,fore=black,tension=1}{x,v}
			\fmf{phantom,fore=black,tension=1}{u,r}
			\fmffreeze
			\fmfforce{(0.66w,0.5h)}{u}
			\fmfforce{(0.5w,0.5h)}{t}
			\fmfforce{(0.5w,0.5h)}{b}
		\end{fmfgraph}
	\end{fmffile}
\end{gathered}\hspace{-.5cm}
=\lambda g^2
\hspace{-.5cm}
\begin{gathered}
	\begin{fmffile}{dg3tavx}
		\begin{fmfgraph}(130,130)
			\fmfset{dash_len}{1.2mm}
			\fmfset{wiggly_len}{1.1mm} \fmfset{dot_len}{0.5mm}
			\fmfpen{0.25mm}
			\fmftop{t}
			\fmfbottom{b}
			\fmfleft{l}
			\fmfright{r}
			\fmfv{decor.shape=circle,decor.filled=full, decor.size=3thin}{u}
			\fmf{phantom,fore=black,tension=9}{t,x,v,b}
			\fmf{phantom,fore=black,tension=9}{l,s,u,r}
			\fmf{dashes,fore=black,tension=.01,left}{x,v,x}
			\fmf{phantom,fore=black,tension=0.01}{s,x,s}
			\fmf{dashes,fore=black,tension=1}{x,v}
			\fmf{phantom,fore=black,tension=1}{u,r}
			\fmffreeze
			\fmfforce{(0.66w,0.5h)}{u}
		\end{fmfgraph}
	\end{fmffile}
\end{gathered}\hspace{-.5cm}+\mathcal{O}(\ell^6)$, \emph{except} for the first diagram that is proportional to $\lambda\sim \mathcal{O}(\ell^2)$ where there is a contribution of the form:
\begin{equation}\label{eq:wigglyexample}
\begin{aligned}
\hspace{0.5cm}
\begin{gathered}
	\begin{fmffile}{wbubblex}
		\begin{fmfgraph}(40,40)
			\fmfset{dash_len}{1.2mm}
			\fmfset{wiggly_len}{1.1mm} \fmfset{dot_len}{0.5mm}
			\fmfpen{0.25mm}
			\fmfvn{decor.shape=circle,decor.filled=full, decor.size=3thin}{u}{1}
			\fmfleft{i}
			\fmfright{o}
			\fmf{wiggly,fore=black,tension=5,left}{i,u1,i}
			\fmffreeze
			\fmfforce{(-w,0.35h)}{i}
			\fmfforce{(0w,0.35h)}{u1}
			\fmfforce{(1.1w,0.35h)}{o}
		\end{fmfgraph}\!\!\!\!\!\!
	\end{fmffile}
\end{gathered}
&=
\hspace{0.5cm}
\begin{gathered}
	\begin{fmffile}{dbubblexdd}
		\begin{fmfgraph}(40,40)
			\fmfset{dash_len}{1.2mm}
			\fmfset{wiggly_len}{1.1mm} \fmfset{dot_len}{0.5mm}
			\fmfpen{0.25mm}
			\fmfvn{decor.shape=circle,decor.filled=full, decor.size=3thin}{u}{1}
			\fmfleft{i}
			\fmfright{o}
			\fmf{dashes,fore=black,tension=5,left}{i,u1,i}
			\fmffreeze
			\fmfforce{(-w,0.35h)}{i}
			\fmfforce{(0w,0.35h)}{u1}
			\fmfforce{(1.1w,0.35h)}{o}
		\end{fmfgraph}\!\!\!\!\!\!
	\end{fmffile}
\end{gathered}
+
\hspace{0.6cm}
\begin{gathered}
	\begin{fmffile}{dbubble2b}
		\begin{fmfgraph}(40,40)
			\fmfset{dash_len}{1.2mm}
			\fmfset{wiggly_len}{1.1mm} \fmfset{dot_len}{0.5mm}
			\fmfpen{0.25mm}
			\fmfvn{decor.shape=circle,decor.filled=shaded, decor.size=5thin}{u}{2}
			\fmfvn{decor.shape=circle,decor.filled=full, decor.size=3thin}{u}{1}
			\fmfleft{i}
			\fmfright{o}
			\fmf{dashes,fore=black,tension=5,left}{i,u1,u2,i}
			\fmffreeze
			\fmfforce{(-w,0.35h)}{i}
			\fmfforce{(0w,0.35h)}{u1}
			\fmfforce{(-1w,0.35h)}{u2}
			\fmfforce{(1.1w,0.35h)}{o}
		\end{fmfgraph}\!\!\!\!
	\end{fmffile}
\end{gathered}
+
\!\!
\begin{gathered}
	\begin{fmffile}{ddbubble3}
		\begin{fmfgraph}(80,80)
			\fmfset{dash_len}{1.2mm}
			\fmfset{wiggly_len}{1.1mm} \fmfset{dot_len}{0.5mm}
			\fmfpen{0.25mm}
			\fmfvn{decor.shape=circle,decor.filled=shaded, decor.size=5thin}{x}{3}
			\fmfvn{decor.shape=circle,decor.filled=full, decor.size=3thin}{x}{1}
			\fmfsurroundn{u}{6}
			\fmf{phantom,fore=black,tension=1}{u1,x1,c,v,u4}
			\fmf{phantom,fore=black,tension=1}{u2,u,c,x3,u5}
			\fmf{phantom,fore=black,tension=1}{u3,x2,c,t,u6}
			\fmffreeze
			\fmf{dashes,fore=black,tension=1,right=.7}{x1,x2}
			\fmf{dashes,fore=black,tension=1,right=.7}{x2,x3}
			\fmf{dashes,fore=black,tension=1,right=.7}{x3,x1}
		\end{fmfgraph}\!\!
	\end{fmffile}
\end{gathered}
+\dots
\\
&=
\hspace{0.5cm}
\begin{gathered}
	\begin{fmffile}{dbubblexdd}
		\begin{fmfgraph}(40,40)
			\fmfset{dash_len}{1.2mm}
			\fmfset{wiggly_len}{1.1mm} \fmfset{dot_len}{0.5mm}
			\fmfpen{0.25mm}
			\fmfvn{decor.shape=circle,decor.filled=full, decor.size=3thin}{u}{1}
			\fmfleft{i}
			\fmfright{o}
			\fmf{dashes,fore=black,tension=5,left}{i,u1,i}
			\fmffreeze
			\fmfforce{(-w,0.35h)}{i}
			\fmfforce{(0w,0.35h)}{u1}
			\fmfforce{(1.1w,0.35h)}{o}
		\end{fmfgraph}\!\!\!\!\!\!
	\end{fmffile}
\end{gathered}
+\lambda
\Big(\tfrac{1}{2}
\begin{gathered}
	\begin{fmffile}{dlambdabubbled}
		\begin{fmfgraph}(80,80)
			\fmfset{dash_len}{1.2mm}
			\fmfset{wiggly_len}{1.1mm} \fmfset{dot_len}{0.5mm}
			\fmfpen{0.25mm}
			\fmftop{t1,t2,t3}
			\fmfbottom{b1,b2,b3}
			\fmf{phantom}{t1,v1,b1}
			\fmf{phantom}{t2,v2,b2}
			\fmf{phantom}{t3,v3,b3}
			\fmfv{decor.shape=circle,decor.filled=full, decor.size=3thin}{v3}
			\fmffreeze
			\fmf{dashes,fore=black,right}{v1,v2,v1}
			\fmf{dashes,fore=black,right}{v2,v3,v2}
		\end{fmfgraph}
	\end{fmffile}
\end{gathered}\Big)+\mathcal{O}(\ell^4).
\end{aligned}
\end{equation}
In the first equality we made use of the definition (\ref{eq:dressed_prop}) and in the latter we made use of (\ref{eq:deltam2 deltaZ}) and the $\mathcal{O}(\ell^2)$ term in (\ref{eq:deltam2}). Taking these considerations into account we can now solve for $\delta_{\n}  m^2$ to the order of interest in (\ref{eq:deltam2}) leading to,
\begin{equation}\label{eq:deltam2b}
\begin{aligned}
\delta_{\n}  m^2\,&=-\lambda\Big(\tfrac{1}{2}\hspace{0.5cm}
\begin{gathered}
	\begin{fmffile}{ddbubblex}
		\begin{fmfgraph}(40,40)
			\fmfset{dash_len}{1.2mm}
			\fmfset{wiggly_len}{1.1mm} \fmfset{dot_len}{0.5mm}
			\fmfpen{0.25mm}
			\fmfvn{decor.shape=circle,decor.filled=full, decor.size=3thin}{u}{1}
			\fmfleft{i}
			\fmfright{o}
			\fmf{dashes,fore=black,tension=5,left}{i,u1,i}
			\fmffreeze
			\fmfforce{(-w,0.35h)}{i}
			\fmfforce{(0w,0.35h)}{u1}
			\fmfforce{(1.1w,0.35h)}{o}
		\end{fmfgraph}\!\!\!\!\!\!
	\end{fmffile}
\end{gathered}\Big)
-g^2\lambda
\Big(\tfrac{1}{4}
\hspace{-.5cm}
\begin{gathered}
	\begin{fmffile}{ddg3tav}
		\begin{fmfgraph}(130,130)
			\fmfset{dash_len}{1.2mm}
			\fmfset{wiggly_len}{1.1mm} \fmfset{dot_len}{0.5mm}
			\fmfpen{0.25mm}
			\fmftop{t}
			\fmfbottom{b}
			\fmfleft{l}
			\fmfright{r}
			\fmfv{decor.shape=circle,decor.filled=full, decor.size=3thin}{u}
			\fmf{phantom,fore=black,tension=9}{t,x,v,b}
			\fmf{phantom,fore=black,tension=9}{l,s,u,r}
			\fmf{dashes,fore=black,tension=.01,left}{x,v,x}
			\fmf{phantom,fore=black,tension=0.01}{s,x,s}
			\fmf{dashes,fore=black,tension=1}{x,v}
			\fmf{phantom,fore=black,tension=1}{u,r}
			\fmffreeze
			\fmfforce{(0.66w,0.5h)}{u}
		\end{fmfgraph}
	\end{fmffile}
\end{gathered}\hspace{-.5cm}
\Big)
+\kappa g\Big(\tfrac{1}{6}\!\!
\begin{gathered}
	\begin{fmffile}{dg2-2loopbubble-1PIG3}
		\begin{fmfgraph}(70,70)
			\fmfset{dash_len}{1.2mm}
			\fmfset{wiggly_len}{1.1mm} \fmfset{dot_len}{0.5mm}
			\fmfpen{0.25mm}
			\fmfleft{i}
			\fmfright{o}
			\fmf{phantom,tension=5}{i,v1}
			\fmf{phantom,tension=5}{v2,o}
			\fmf{dashes,fore=black,left,tension=0.4}{v1,v2,v1}
			\fmf{dashes,fore=black}{v1,v2}
			\fmfv{decor.shape=circle,decor.filled=full, decor.size=3thin}{v2}
		\end{fmfgraph}
	\end{fmffile}
\end{gathered}\!\Big)
+\gamma\Big(\tfrac{1}{8}
\begin{gathered}
	\begin{fmffile}{dlambdabubbledk}
		\begin{fmfgraph}(80,80)
			\fmfset{dash_len}{1.2mm}
			\fmfset{wiggly_len}{1.1mm} \fmfset{dot_len}{0.5mm}
			\fmfpen{0.25mm}
			\fmftop{t1,t2,t3}
			\fmfbottom{b1,b2,b3}
			\fmf{phantom}{t1,v1,b1}
			\fmf{phantom}{t2,v2,b2}
			\fmf{phantom}{t3,v3,b3}
			\fmfv{decor.shape=circle,decor.filled=full, decor.size=3thin}{v2}
			\fmffreeze
			\fmf{dashes,fore=black,right}{v1,v2,v1}
			\fmf{dashes,fore=black,right}{v2,v3,v2}
		\end{fmfgraph}
	\end{fmffile}
\end{gathered}
\Big)
+\mathcal{O}(\ell^6).\phantom{\Bigg[}
\end{aligned}
\end{equation}
This we can now substitute into the dressed propagators of the various counterterms in (\ref{eq:delta g etc }), in order to obtain explicit expressions for the remaining counterterms, $\delta_{\n} \Lambda$, $Y$, $\delta_{\n}  g$, $\delta_{\n}  \lambda$, etc. 

\setlength{\parindent}{1cm}
Carrying this out explicitly, we collect the above results and find that the full set of counterterms \emph{induced by (weak) complete normal ordering} up to $\mathcal{O}(\ell^4)$ reads:
\begin{spreadlines}{-0.3\baselineskip}
\begin{equation}\label{eq:delta g etc b}
\begin{aligned}
\delta_{\n}  \Lambda&=g^2\Big(\tfrac{1}{6}\!\!
\begin{gathered}
	\begin{fmffile}{dg2-2loopbubble-1PIG32}
		\begin{fmfgraph}(70,70)
			\fmfset{dash_len}{1.2mm}
			\fmfset{wiggly_len}{1.1mm} \fmfset{dot_len}{0.5mm}
			\fmfpen{0.25mm}
			\fmfleft{i}
			\fmfright{o}
			\fmf{phantom,tension=5}{i,v1}
			\fmf{phantom,tension=5}{v2,o}
			\fmf{dashes,fore=black,left,tension=0.4}{v1,v2,v1}
			\fmf{dashes,fore=black}{v1,v2}
		\end{fmfgraph}
	\end{fmffile}
\end{gathered}\!\Big)
+g^4\Big(
\tfrac{1}{4}\,\,
\begin{gathered}
	\begin{fmffile}{dg4-0pt-3loop1PIaG32}
		\begin{fmfgraph}(37,37)
			\fmfset{dash_len}{1.2mm}
			\fmfset{wiggly_len}{1.1mm} \fmfset{dot_len}{0.5mm}
			\fmfpen{0.25mm}
			\fmfsurround{a,b,c,d}
			\fmf{phantom,fore=black,tension=1,curved}{a,b,c,d,a}
			\fmf{dashes,fore=black,tension=1,right}{a,b}
			\fmf{dashes,fore=black,tension=1,right}{c,d}
			\fmf{dashes,fore=black,tension=1,right}{b,c}
			\fmf{dashes,fore=black,tension=1,right}{d,a}
			\fmf{dashes,fore=black,tension=1,straight}{a,b}
			\fmf{dashes,fore=black,tension=1,straight}{c,d}
		\end{fmfgraph}
	\end{fmffile}
\end{gathered}\,\,
+\tfrac{1}{6}\!\!\!\!\!
\begin{gathered}
	\begin{fmffile}{dpeaceG32}
		\begin{fmfgraph}(105,105)
			\fmfset{dash_len}{1.2mm}
			\fmfset{wiggly_len}{1.1mm} \fmfset{dot_len}{0.5mm}
			\fmfpen{0.25mm}
			\fmfsurroundn{i}{3}
			\fmf{phantom,fore=black}{i1,v,i2}
			\fmf{phantom,fore=black}{i2,u,i3}
			\fmf{phantom,fore=black}{i3,s,i1}
			\fmfi{dashes,fore=black}{fullcircle scaled .4w shifted (.5w,.5h)}
			\fmf{dashes,fore=black}{v,c}
			\fmf{dashes,fore=black}{u,c}
			\fmf{dashes,fore=black}{s,c}
		\end{fmfgraph}
	\end{fmffile}
\end{gathered}\!\!\!\!\!
\Big)
+g^2\lambda
\Big(\tfrac{1}{8}\,\,
\begin{gathered}
	\begin{fmffile}{dg2L-bubbleG32}
		\begin{fmfgraph}(70,70)
			\fmfset{dash_len}{1.2mm}
			\fmfset{wiggly_len}{1.1mm} \fmfset{dot_len}{0.5mm}
			\fmfpen{0.25mm}
			\fmftop{t1,t2,t3}
			\fmfbottom{b1,b2,b3}
			\fmf{phantom}{t1,v1,b1}
			\fmf{phantom}{t2,v2,b2}
			\fmf{phantom}{t3,v3,b3}
			\fmffreeze
			\fmf{dashes,fore=black,right}{v1,v2,v1}
			\fmf{dashes,fore=black,right}{v2,v3,v2}
			\fmf{dashes,fore=black,tension=1}{t1,b1}
			\fmfforce{(0.25w,0.7h)}{t1}
			\fmfforce{(0.25w,0.3h)}{b1}
		\end{fmfgraph}
	\end{fmffile}
\end{gathered}
-\tfrac{3}{8}
\,\,
\begin{gathered}
	\begin{fmffile}{dg2L-3loopbubble-1PIG32}
		\begin{fmfgraph}(35,35)
			\fmfset{dash_len}{1.2mm}
			\fmfset{wiggly_len}{1.1mm} \fmfset{dot_len}{0.5mm}
			\fmfpen{0.25mm}
			\fmfsurroundn{i}{3}
			\fmf{dashes,fore=black,tension=1,right=1}{i1,i2}
			\fmf{dashes,fore=black,tension=1,right=1}{i2,i3}
			\fmf{dashes,fore=black,tension=1,right=0.8}{i3,i1}
			\fmf{dashes,fore=black,tension=1}{i1,i2}
			\fmf{dashes,fore=black,tension=1}{i2,i3}
		\end{fmfgraph}
	\end{fmffile}
\end{gathered}\,\,
\Big)
-g\kappa\Big(\tfrac{1}{12}\!
\begin{gathered}
	\begin{fmffile}{dgkappa-bubble-1PIG32}
		\begin{fmfgraph}(60,60)
			\fmfset{dash_len}{1.2mm}
			\fmfset{wiggly_len}{1.1mm} \fmfset{dot_len}{0.5mm}
			\fmfpen{0.25mm}
			\fmfleft{i}
			\fmfright{o}
			\fmf{phantom,tension=5}{i,v1}
			\fmf{phantom,tension=5}{v2,o}
			\fmf{dashes,fore=black,left,tension=0.5}{v1,v2,v1}
			\fmf{dashes,fore=black}{v1,v2}
			\fmffreeze
			\fmfi{dashes,fore=black}{fullcircle scaled .55w shifted (1.1w,.5h)}
		\end{fmfgraph}
	\end{fmffile}
\end{gathered}\,\,\,\,\Big)\\
&
+\lambda
\Big(\tfrac{1}{8}
\begin{gathered}
	\begin{fmffile}{dlambdabubbled2}
		\begin{fmfgraph}(80,80)
			\fmfset{dash_len}{1.2mm}
			\fmfset{wiggly_len}{1.1mm} \fmfset{dot_len}{0.5mm}
			\fmfpen{0.25mm}
			\fmftop{t1,t2,t3}
			\fmfbottom{b1,b2,b3}
			\fmf{phantom}{t1,v1,b1}
			\fmf{phantom}{t2,v2,b2}
			\fmf{phantom}{t3,v3,b3}
			\fmffreeze
			\fmf{dashes,fore=black,right}{v1,v2,v1}
			\fmf{dashes,fore=black,right}{v2,v3,v2}
		\end{fmfgraph}
	\end{fmffile}
\end{gathered}
\Big)+
\lambda^2\Big(\tfrac{1}{24}\!
\begin{gathered}
	\begin{fmffile}{dL2-3loop-bubbleG42}
		\begin{fmfgraph}(75,75)
			\fmfset{dash_len}{1.2mm}
			\fmfset{wiggly_len}{1.1mm} \fmfset{dot_len}{0.5mm}
			\fmfpen{0.25mm}
			\fmfleft{i}
			\fmfright{o}
			\fmf{phantom,tension=10}{i,v1}
			\fmf{phantom,tension=10}{v2,o}
			\fmf{dashes,left,tension=0.4}{v1,v2,v1}
			\fmf{dashes,left=0.5}{v1,v2}
			\fmf{dashes,right=0.5}{v1,v2}
    		\end{fmfgraph}
	\end{fmffile}
\end{gathered}\!
\Big)
-\gamma
\Big(\tfrac{1}{48}\,
\begin{gathered}
	\begin{fmffile}{dgamma-vacuum-1PIkk}
		\begin{fmfgraph}(80,80)
			\fmfset{dash_len}{1.2mm}
			\fmfset{wiggly_len}{1.1mm} \fmfset{dot_len}{0.5mm}
			\fmfpen{0.25mm}
			\fmfsurroundn{x}{3}
			\fmf{phantom,fore=black}{x1,v}
			\fmf{phantom,fore=black}{x2,v}
			\fmf{phantom,fore=black}{x3,v}
			\fmf{dashes,fore=black,tension=0.7}{v,v}
			\fmf{dashes,fore=black,tension=0.7,right}{v,v}
			\fmf{dashes,fore=black,tension=0.7,left}{v,v}
			\fmffreeze
			\fmfbottom{b}
			\fmfforce{(0w,0.5h)}{b}
			\end{fmfgraph}
	\end{fmffile}
\end{gathered}\!\Big)
+\mathcal{O}(\ell^6) \, , \phantom{\Bigg\}}\\
Y\,&=g\Big(\tfrac{1}{2}
\hspace{0.5cm}
\begin{gathered}
	\begin{fmffile}{dbubblexas}
		\begin{fmfgraph}(40,40)
			\fmfset{dash_len}{1.2mm}
			\fmfset{wiggly_len}{1.1mm} \fmfset{dot_len}{0.5mm}
			\fmfpen{0.25mm}
			\fmfvn{decor.shape=circle,decor.filled=full, decor.size=3thin}{u}{1}
			\fmfleft{i}
			\fmfright{o}
			\fmf{dashes,fore=black,tension=5,left,label.dist=1,label=a}{i,u1,i}
			\fmffreeze
			\fmfforce{(-w,0.35h)}{i}
			\fmfforce{(0w,0.35h)}{u1}
			\fmfforce{(1.1w,0.35h)}{o}
			\fmf{phantom,label.dist=0,label=1}{i,u1}
		\end{fmfgraph}\!\!\!\!\!\!
	\end{fmffile}
\end{gathered}\Big)
+g^3\Big(
\tfrac{1}{4}\hspace{-.5cm}
\begin{gathered}
	\begin{fmffile}{dg3tav}
		\begin{fmfgraph}(130,130)
			\fmfset{dash_len}{1.2mm}
			\fmfset{wiggly_len}{1.1mm} \fmfset{dot_len}{0.5mm}
			\fmfpen{0.25mm}
			\fmftop{t}
			\fmfbottom{b}
			\fmfleft{l}
			\fmfright{r}
			\fmfv{decor.shape=circle,decor.filled=full, decor.size=3thin}{u}
			\fmf{phantom,fore=black,tension=9}{t,x,v,b}
			\fmf{phantom,fore=black,tension=9}{l,s,u,r}
			\fmf{dashes,fore=black,tension=.01,left}{x,v,x}
			\fmf{phantom,fore=black,tension=0.01}{s,x,s}
			\fmf{dashes,fore=black,tension=1}{x,v}
			\fmf{phantom,fore=black,tension=1}{u,r}
			\fmffreeze
			\fmfforce{(0.66w,0.5h)}{u}
		\end{fmfgraph}
	\end{fmffile}
\end{gathered}\hspace{-.5cm}
\Big)
-g\lambda\Big(
\tfrac{1}{3!}\hspace{-.2cm}
\begin{gathered}
	\begin{fmffile}{dglambdatadpolesdf}
		\begin{fmfgraph}(90,90)
			\fmfset{dash_len}{1.2mm}
			\fmfset{wiggly_len}{1.1mm} \fmfset{dot_len}{0.5mm}
			\fmfpen{0.25mm}
			\fmfleft{i}
			\fmfright{o}
			\fmfv{decor.shape=circle,decor.filled=full, decor.size=3thin}{v2}
			\fmf{phantom,tension=5}{i,v1}
			\fmf{phantom,fore=black,tension=2.5}{v2,o}
			\fmf{dashes,fore=black,left,tension=0.5}{v1,v2,v1}
			\fmf{dashes,fore=black}{v1,v2}
			\fmffreeze
			\fmfforce{(1.1w,0.5h)}{o}
		\end{fmfgraph}
	\end{fmffile}
\end{gathered}\!\!\!\!
\Big)
-\kappa\Big(
\tfrac{1}{8}
\begin{gathered}
	\begin{fmffile}{dlambdabubbledk}
		\begin{fmfgraph}(80,80)
			\fmfset{dash_len}{1.2mm}
			\fmfset{wiggly_len}{1.1mm} \fmfset{dot_len}{0.5mm}
			\fmfpen{0.25mm}
			\fmftop{t1,t2,t3}
			\fmfbottom{b1,b2,b3}
			\fmf{phantom}{t1,v1,b1}
			\fmf{phantom}{t2,v2,b2}
			\fmf{phantom}{t3,v3,b3}
			\fmfv{decor.shape=circle,decor.filled=full, decor.size=3thin}{v2}
			\fmffreeze
			\fmf{dashes,fore=black,right}{v1,v2,v1}
			\fmf{dashes,fore=black,right}{v2,v3,v2}
		\end{fmfgraph}
	\end{fmffile}
\end{gathered}
\Big)+\mathcal{O}(\ell^5) \, , \phantom{\Bigg[}\\
\delta_{\n}  m^2\,&=-\lambda\Big(\tfrac{1}{2}\hspace{0.5cm}
\begin{gathered}
	\begin{fmffile}{ddbubblex}
		\begin{fmfgraph}(40,40)
			\fmfset{dash_len}{1.2mm}
			\fmfset{wiggly_len}{1.1mm} \fmfset{dot_len}{0.5mm}
			\fmfpen{0.25mm}
			\fmfvn{decor.shape=circle,decor.filled=full, decor.size=3thin}{u}{1}
			\fmfleft{i}
			\fmfright{o}
			\fmf{dashes,fore=black,tension=5,left}{i,u1,i}
			\fmffreeze
			\fmfforce{(-w,0.35h)}{i}
			\fmfforce{(0w,0.35h)}{u1}
			\fmfforce{(1.1w,0.35h)}{o}
		\end{fmfgraph}\!\!\!\!\!\!
	\end{fmffile}
\end{gathered}\Big)
-g^2\lambda
\Big(\tfrac{1}{4}
\hspace{-.5cm}
\begin{gathered}
	\begin{fmffile}{ddg3tav}
		\begin{fmfgraph}(130,130)
			\fmfset{dash_len}{1.2mm}
			\fmfset{wiggly_len}{1.1mm} \fmfset{dot_len}{0.5mm}
			\fmfpen{0.25mm}
			\fmftop{t}
			\fmfbottom{b}
			\fmfleft{l}
			\fmfright{r}
			\fmfv{decor.shape=circle,decor.filled=full, decor.size=3thin}{u}
			\fmf{phantom,fore=black,tension=9}{t,x,v,b}
			\fmf{phantom,fore=black,tension=9}{l,s,u,r}
			\fmf{dashes,fore=black,tension=.01,left}{x,v,x}
			\fmf{phantom,fore=black,tension=0.01}{s,x,s}
			\fmf{dashes,fore=black,tension=1}{x,v}
			\fmf{phantom,fore=black,tension=1}{u,r}
			\fmffreeze
			\fmfforce{(0.66w,0.5h)}{u}
		\end{fmfgraph}
	\end{fmffile}
\end{gathered}\hspace{-.5cm}
\Big)
+\kappa g\Big(\tfrac{1}{6}\!\!
\begin{gathered}
	\begin{fmffile}{dg2-2loopbubble-1PIG3}
		\begin{fmfgraph}(70,70)
			\fmfset{dash_len}{1.2mm}
			\fmfset{wiggly_len}{1.1mm} \fmfset{dot_len}{0.5mm}
			\fmfpen{0.25mm}
			\fmfleft{i}
			\fmfright{o}
			\fmf{phantom,tension=5}{i,v1}
			\fmf{phantom,tension=5}{v2,o}
			\fmf{dashes,fore=black,left,tension=0.4}{v1,v2,v1}
			\fmf{dashes,fore=black}{v1,v2}
			\fmfv{decor.shape=circle,decor.filled=full, decor.size=3thin}{v2}
		\end{fmfgraph}
	\end{fmffile}
\end{gathered}\!\Big)
+\gamma\Big(\tfrac{1}{8}
\begin{gathered}
	\begin{fmffile}{dlambdabubbledk}
		\begin{fmfgraph}(80,80)
			\fmfset{dash_len}{1.2mm}
			\fmfset{wiggly_len}{1.1mm} \fmfset{dot_len}{0.5mm}
			\fmfpen{0.25mm}
			\fmftop{t1,t2,t3}
			\fmfbottom{b1,b2,b3}
			\fmf{phantom}{t1,v1,b1}
			\fmf{phantom}{t2,v2,b2}
			\fmf{phantom}{t3,v3,b3}
			\fmfv{decor.shape=circle,decor.filled=full, decor.size=3thin}{v2}
			\fmffreeze
			\fmf{dashes,fore=black,right}{v1,v2,v1}
			\fmf{dashes,fore=black,right}{v2,v3,v2}
		\end{fmfgraph}
	\end{fmffile}
\end{gathered}
\Big)
+\mathcal{O}(\ell^6) \, , \phantom{\Bigg[}\\
\delta_{\n}  g\,&=-\kappa\Big(\tfrac{1}{2}\hspace{0.5cm}
\begin{gathered}
	\begin{fmffile}{dbubblexz}
		\begin{fmfgraph}(40,40)
			\fmfset{dash_len}{1.2mm}
			\fmfset{wiggly_len}{1.1mm} \fmfset{dot_len}{0.5mm}
			\fmfpen{0.25mm}
			\fmfvn{decor.shape=circle,decor.filled=full, decor.size=3thin}{u}{1}
			\fmfleft{i}
			\fmfright{o}
			\fmf{dashes,fore=black,tension=5,left}{i,u1,i}
			\fmffreeze
			\fmfforce{(-w,0.35h)}{i}
			\fmfforce{(0w,0.35h)}{u1}
			\fmfforce{(1.1w,0.35h)}{o}
		\end{fmfgraph}\!\!\!\!\!\!
	\end{fmffile}
\end{gathered}\Big)+\mathcal{O}(\ell^5) \, , \phantom{\Bigg[}\\
\delta_{\n}  \lambda\,&=-\gamma\Big(\tfrac{1}{2}\hspace{0.5cm}
\begin{gathered}
	\begin{fmffile}{dbubblexz}
		\begin{fmfgraph}(40,40)
			\fmfset{dash_len}{1.2mm}
			\fmfset{wiggly_len}{1.1mm} \fmfset{dot_len}{0.5mm}
			\fmfpen{0.25mm}
			\fmfvn{decor.shape=circle,decor.filled=full, decor.size=3thin}{u}{1}
			\fmfleft{i}
			\fmfright{o}
			\fmf{dashes,fore=black,tension=5,left}{i,u1,i}
			\fmffreeze
			\fmfforce{(-w,0.35h)}{i}
			\fmfforce{(0w,0.35h)}{u1}
			\fmfforce{(1.1w,0.35h)}{o}
		\end{fmfgraph}\!\!\!\!\!\!
	\end{fmffile}
\end{gathered}\Big)+\mathcal{O}(\ell^5) \, , \phantom{\Bigg[}\\
\delta_{\n}  \kappa\,&=\mathcal{O}(\ell^5) \, , \phantom{\Bigg[}\\
\delta_{\n}  \gamma\,&=\mathcal{O}(\ell^5) \, ,\phantom{\Bigg[}\\
&\,\,\,\vdots\phantom{\Bigg[}
\end{aligned}
\end{equation}
\end{spreadlines}
Once again here, the incomplete vertices are denoted by black dots. A useful consistency check is to note that, say, the $\delta_{\n}  g$ term in the above (which in the bare action is associated to a $\phi^3$ vertex) is now getting a contribution from a $\phi^5$ vertex, via the $\kappa$ coupling. These two vertices differ by the number of lines that meet at their respective vertices, and the difference (namely 2) is precisely the number of lines that meet at the incomplete vertex of the Feynman diagram multiplying $\kappa$ in the above. Similar remarks hold for all counterterms and couplings, and incomplete Feynman diagrams on the left- and right-hand sides of the above relations, showing that every incomplete Feynman diagram has precisely the correct number of legs meeting at every incomplete vertex.

\subsection{Vacuum Contribution}\label{sec:VC}
\setlength{\parindent}{1cm}
The vacuum contribution requires particular care. Recall that below (\ref{eq:counterterms GNs}) we mentioned two inequivalent forms of complete normal ordering: the `strong' and `weak' forms. In the former we complete normal order the full action and in the latter we only complete normal order the interaction terms. The two differ by a $J$-independent shift in $W(J)$, see (\ref{eq:Ws=Ww+Q4}). 

Let us now determine the full vacuum term in $W(J)$, see (\ref{eq:W(Jwiggly)}), in the context of complete normal ordering (of the weak form, see the discussion following (\ref{eq:counterterms GNs})). This is a sum of three contributions:
\begin{itemize}
\item[(a)] the wave-function and mass-renormalisation vacuum contributions in $W(J)$, denoted by $Q_1$ in (\ref{eq:Q1});
\item[(b)] the terms in $\hat{\Lambda}$ in $W(J)$ (the counterterm associated to which is given in the first equality in (\ref{eq:delta g etc b})), which we shall call $Q_2\dfn -\delta_{\n}\Lambda$ (see below); 
\item[(c)] the bubble diagrams that appear explicitly in $W(J)$, which we shall call $Q_3$ (see below); 
\end{itemize}
If we were to adopt instead the strong form of complete normal ordering, the only thing that would change in the entire calculation is that we would also have a vacuum contribution coming from the complete normal ordering of the kinetic term, and so we would have one more contribution:
\begin{itemize}
\item[(d)] the vacuum contributions associated to adopting the `strong' form of complete normal ordering, which we shall denote by $Q_4$. (To adopt the `weak' form of complete normal ordering is to set $Q_4=0$.)
\end{itemize}
The total vacuum contribution will be denoted by:
\begin{equation}\label{eq:QWQS}
\begin{aligned}
&Q_{\rm W}\dfn Q_1+Q_2+Q_3\qquad{\rm (weak \,\,form)} \, , \\[0.2cm]
&Q_{\rm S}\dfn Q_1+Q_2+Q_3+Q_4\qquad{\rm (strong \,\,form)} \, .
\end{aligned}
\end{equation}

We now determine $Q_{\rm W}$ and $Q_{\rm S}$ explicitly, starting from $Q_1$. Substituting the mass counterterm, $\begin{gathered}
	\begin{fmffile}{circle-shaded}
		\begin{fmfgraph}(40,40)
			\fmfset{dash_len}{1.2mm}
			\fmfset{wiggly_len}{1.1mm} \fmfset{dot_len}{0.5mm}
			\fmfpen{0.25mm}
			\fmfvn{decor.shape=circle,decor.filled=shaded, decor.size=5thin}{u}{1}
			\fmfleft{i}
			\fmfright{o}
			\fmf{phantom,fore=black,tension=5}{i,u1,o}
			\fmffreeze
			\fmfforce{(0w,0.35h)}{i}
			\fmfforce{(0w,0.35h)}{u1}
			\fmfforce{(0w,0.35h)}{o}
		\end{fmfgraph}
	\end{fmffile}
\end{gathered}\!\!\!\!\!
=-\delta_{\n} m^2$, of (\ref{eq:delta g etc b}) into (\ref{eq:Q1}) leads to:
\begin{equation*}\label{eq:Q1b}
\begin{aligned}
Q_1\,\,&\!\!\dfn 
\tfrac{1}{2}\Big(\hspace{0.5cm}\begin{gathered}
	\begin{fmffile}{bubble1}
		\begin{fmfgraph}(40,40)
			\fmfset{dash_len}{1.2mm}
			\fmfset{wiggly_len}{1.1mm} \fmfset{dot_len}{0.5mm}
			\fmfpen{0.25mm}
			\fmfvn{decor.shape=circle,decor.filled=shaded, decor.size=5thin}{u}{1}
			\fmfleft{i}
			\fmfright{o}
			\fmf{dashes,fore=black,tension=5,left}{i,u1,i}
			\fmffreeze
			\fmfforce{(-w,0.35h)}{i}
			\fmfforce{(0w,0.35h)}{u1}
			\fmfforce{(1.1w,0.35h)}{o}
		\end{fmfgraph}\!\!\!\!
	\end{fmffile}
\end{gathered}\hspace{0cm}
+
\tfrac{1}{2}\hspace{0.6cm}
\begin{gathered}
	\begin{fmffile}{bubble2}
		\begin{fmfgraph}(40,40)
			\fmfset{dash_len}{1.2mm}
			\fmfset{wiggly_len}{1.1mm} \fmfset{dot_len}{0.5mm}
			\fmfpen{0.25mm}
			\fmfvn{decor.shape=circle,decor.filled=shaded, decor.size=5thin}{u}{2}
			\fmfleft{i}
			\fmfright{o}
			\fmf{dashes,fore=black,tension=5,left}{i,u1,u2,i}
			\fmffreeze
			\fmfforce{(-w,0.35h)}{i}
			\fmfforce{(0w,0.35h)}{u1}
			\fmfforce{(-1w,0.35h)}{u2}
			\fmfforce{(1.1w,0.35h)}{o}
		\end{fmfgraph}\!\!\!\!
	\end{fmffile}
\end{gathered}\hspace{0cm}
+
\tfrac{1}{3}\!\!
\begin{gathered}
	\begin{fmffile}{bubble3}
		\begin{fmfgraph}(80,80)
			\fmfset{dash_len}{1.2mm}
			\fmfset{wiggly_len}{1.1mm} \fmfset{dot_len}{0.5mm}
			\fmfpen{0.25mm}
			\fmfvn{decor.shape=circle,decor.filled=shaded, decor.size=5thin}{x}{3}
			\fmfsurroundn{u}{6}
			\fmf{phantom,fore=black,tension=1}{u1,x1,c,v,u4}
			\fmf{phantom,fore=black,tension=1}{u2,u,c,x3,u5}
			\fmf{phantom,fore=black,tension=1}{u3,x2,c,t,u6}
			\fmffreeze
			\fmf{dashes,fore=black,tension=1,right=.7}{x1,x2}
			\fmf{dashes,fore=black,tension=1,right=.7}{x2,x3}
			\fmf{dashes,fore=black,tension=1,right=.7}{x3,x1}
		\end{fmfgraph}\!\!
	\end{fmffile}
\end{gathered}
+\dots+
\hspace{0.5cm}
\begin{gathered}
	\begin{fmffile}{bubble1b}
		\begin{fmfgraph}(40,40)
			\fmfset{dash_len}{1.2mm}
			\fmfset{wiggly_len}{1.1mm} \fmfset{dot_len}{0.5mm}
			\fmfpen{0.25mm}
			\fmfvn{decor.shape=square,decor.filled=shaded, decor.size=5thin}{u}{1}
			\fmfleft{i}
			\fmfright{o}
			\fmf{plain,fore=black,tension=5,left}{i,u1,i}
			\fmffreeze
			\fmfforce{(-w,0.35h)}{i}
			\fmfforce{(0w,0.35h)}{u1}
			\fmfforce{(1.1w,0.35h)}{o}
		\end{fmfgraph}\!\!\!\!
	\end{fmffile}
\end{gathered}
+
\tfrac{1}{2}\hspace{0.6cm}
\begin{gathered}
	\begin{fmffile}{bubble2b}
		\begin{fmfgraph}(40,40)
			\fmfset{dash_len}{1.2mm}
			\fmfset{wiggly_len}{1.1mm} \fmfset{dot_len}{0.5mm}
			\fmfpen{0.25mm}
			\fmfvn{decor.shape=square,decor.filled=shaded, decor.size=5thin}{u}{2}
			\fmfleft{i}
			\fmfright{o}
			\fmf{plain,fore=black,tension=5,left}{i,u1,u2,i}
			\fmffreeze
			\fmfforce{(-w,0.35h)}{i}
			\fmfforce{(0w,0.35h)}{u1}
			\fmfforce{(-1w,0.35h)}{u2}
			\fmfforce{(1.1w,0.35h)}{o}
		\end{fmfgraph}\!\!\!\!
	\end{fmffile}
\end{gathered}
+
\tfrac{1}{3}\!\!
\begin{gathered}
	\begin{fmffile}{bubble3b}
		\begin{fmfgraph}(80,80)
			\fmfset{dash_len}{1.2mm}
			\fmfset{wiggly_len}{1.1mm} \fmfset{dot_len}{0.5mm}
			\fmfpen{0.25mm}
			\fmfvn{decor.shape=square,decor.filled=shaded, decor.size=5thin}{x}{3}
			\fmfsurroundn{u}{6}
			\fmf{phantom,fore=black,tension=1}{u1,x1,c,v,u4}
			\fmf{phantom,fore=black,tension=1}{u2,u,c,x3,u5}
			\fmf{phantom,fore=black,tension=1}{u3,x2,c,t,u6}
			\fmffreeze
			\fmf{plain,fore=black,tension=1,right=.7}{x1,x2}
			\fmf{plain,fore=black,tension=1,right=.7}{x2,x3}
			\fmf{plain,fore=black,tension=1,right=.7}{x3,x1}
		\end{fmfgraph}\!\!
	\end{fmffile}
\end{gathered}
+\dots\Big)\\
&=\lambda
\Big(\tfrac{1}{4}
\begin{gathered}
	\begin{fmffile}{dlambdabubbled2}
		\begin{fmfgraph}(80,80)
			\fmfset{dash_len}{1.2mm}
			\fmfset{wiggly_len}{1.1mm} \fmfset{dot_len}{0.5mm}
			\fmfpen{0.25mm}
			\fmftop{t1,t2,t3}
			\fmfbottom{b1,b2,b3}
			\fmf{phantom}{t1,v1,b1}
			\fmf{phantom}{t2,v2,b2}
			\fmf{phantom}{t3,v3,b3}
			\fmffreeze
			\fmf{dashes,fore=black,right}{v1,v2,v1}
			\fmf{dashes,fore=black,right}{v2,v3,v2}
		\end{fmfgraph}
	\end{fmffile}
\end{gathered}
\Big)+\lambda^2\Big(\tfrac{1}{16}
\begin{gathered}
	\begin{fmffile}{dL2-bubble}
		\begin{fmfgraph}(70,70)
			\fmfset{dash_len}{1.2mm}
			\fmfset{wiggly_len}{1.1mm} \fmfset{dot_len}{0.5mm}
			\fmfpen{0.25mm}
			\fmftop{t1,t2,t3}
			\fmfbottom{b1,b2,b3}
			\fmf{phantom}{t1,v1,b1}
			\fmf{phantom}{t2,v2,b2}
			\fmf{phantom}{t3,v3,b3}
			\fmffreeze
			\fmf{dashes,fore=black,right}{v1,v2,v1}
			\fmf{dashes,fore=black,right}{v2,v3,v2}
			\fmfi{dashes,fore=black}{fullcircle scaled .5w shifted (1.25w,.5h)}
		\end{fmfgraph}
	\end{fmffile}
\end{gathered}\,\,\,\,\,\,\Big)
-g\kappa\Big(\tfrac{1}{12}\!
\begin{gathered}
	\begin{fmffile}{dgkappa-bubble-1PIG32}
		\begin{fmfgraph}(60,60)
			\fmfset{dash_len}{1.2mm}
			\fmfset{wiggly_len}{1.1mm} \fmfset{dot_len}{0.5mm}
			\fmfpen{0.25mm}
			\fmfleft{i}
			\fmfright{o}
			\fmf{phantom,tension=5}{i,v1}
			\fmf{phantom,tension=5}{v2,o}
			\fmf{dashes,fore=black,left,tension=0.5}{v1,v2,v1}
			\fmf{dashes,fore=black}{v1,v2}
			\fmffreeze
			\fmfi{dashes,fore=black}{fullcircle scaled .55w shifted (1.1w,.5h)}
		\end{fmfgraph}
	\end{fmffile}
\end{gathered}\,\,\,\,\Big)
+g^2\lambda
\Big(\tfrac{1}{8}\,
\begin{gathered}
	\begin{fmffile}{dg2L-bubbleG32}
		\begin{fmfgraph}(70,70)
			\fmfset{dash_len}{1.2mm}
			\fmfset{wiggly_len}{1.1mm} \fmfset{dot_len}{0.5mm}
			\fmfpen{0.25mm}
			\fmftop{t1,t2,t3}
			\fmfbottom{b1,b2,b3}
			\fmf{phantom}{t1,v1,b1}
			\fmf{phantom}{t2,v2,b2}
			\fmf{phantom}{t3,v3,b3}
			\fmffreeze
			\fmf{dashes,fore=black,right}{v1,v2,v1}
			\fmf{dashes,fore=black,right}{v2,v3,v2}
			\fmf{dashes,fore=black,tension=1}{t1,b1}
			\fmfforce{(0.25w,0.7h)}{t1}
			\fmfforce{(0.25w,0.3h)}{b1}
		\end{fmfgraph}
	\end{fmffile}
\end{gathered}\Big)\\
&\hspace{0.4cm}-\gamma
\Big(\tfrac{1}{16}\,
\begin{gathered}
	\begin{fmffile}{dgamma-vacuum-1PIkk}
		\begin{fmfgraph}(80,80)
			\fmfset{dash_len}{1.2mm}
			\fmfset{wiggly_len}{1.1mm} \fmfset{dot_len}{0.5mm}
			\fmfpen{0.25mm}
			\fmfsurroundn{x}{3}
			\fmf{phantom,fore=black}{x1,v}
			\fmf{phantom,fore=black}{x2,v}
			\fmf{phantom,fore=black}{x3,v}
			\fmf{dashes,fore=black,tension=0.7}{v,v}
			\fmf{dashes,fore=black,tension=0.7,right}{v,v}
			\fmf{dashes,fore=black,tension=0.7,left}{v,v}
			\fmffreeze
			\fmfbottom{b}
			\fmfforce{(0w,0.5h)}{b}
			\end{fmfgraph}
	\end{fmffile}
\end{gathered}\!\Big)+\tfrac{1}{2}\Big(
\hspace{0.5cm}
\begin{gathered}
	\begin{fmffile}{bubble1b}
		\begin{fmfgraph}(40,40)
			\fmfset{dash_len}{1.2mm}
			\fmfset{wiggly_len}{1.1mm} \fmfset{dot_len}{0.5mm}
			\fmfpen{0.25mm}
			\fmfvn{decor.shape=square,decor.filled=shaded, decor.size=5thin}{u}{1}
			\fmfleft{i}
			\fmfright{o}
			\fmf{plain,fore=black,tension=5,left}{i,u1,i}
			\fmffreeze
			\fmfforce{(-w,0.35h)}{i}
			\fmfforce{(0w,0.35h)}{u1}
			\fmfforce{(1.1w,0.35h)}{o}
		\end{fmfgraph}\!\!\!\!
	\end{fmffile}
\end{gathered}
+
\tfrac{1}{2}\hspace{0.6cm}
\begin{gathered}
	\begin{fmffile}{bubble2b}
		\begin{fmfgraph}(40,40)
			\fmfset{dash_len}{1.2mm}
			\fmfset{wiggly_len}{1.1mm} \fmfset{dot_len}{0.5mm}
			\fmfpen{0.25mm}
			\fmfvn{decor.shape=square,decor.filled=shaded, decor.size=5thin}{u}{2}
			\fmfleft{i}
			\fmfright{o}
			\fmf{plain,fore=black,tension=5,left}{i,u1,u2,i}
			\fmffreeze
			\fmfforce{(-w,0.35h)}{i}
			\fmfforce{(0w,0.35h)}{u1}
			\fmfforce{(-1w,0.35h)}{u2}
			\fmfforce{(1.1w,0.35h)}{o}
		\end{fmfgraph}\!\!\!\!
	\end{fmffile}
\end{gathered}
+
\tfrac{1}{3}\!\!
\begin{gathered}
	\begin{fmffile}{bubble3b}
		\begin{fmfgraph}(80,80)
			\fmfset{dash_len}{1.2mm}
			\fmfset{wiggly_len}{1.1mm} \fmfset{dot_len}{0.5mm}
			\fmfpen{0.25mm}
			\fmfvn{decor.shape=square,decor.filled=shaded, decor.size=5thin}{x}{3}
			\fmfsurroundn{u}{6}
			\fmf{phantom,fore=black,tension=1}{u1,x1,c,v,u4}
			\fmf{phantom,fore=black,tension=1}{u2,u,c,x3,u5}
			\fmf{phantom,fore=black,tension=1}{u3,x2,c,t,u6}
			\fmffreeze
			\fmf{plain,fore=black,tension=1,right=.7}{x1,x2}
			\fmf{plain,fore=black,tension=1,right=.7}{x2,x3}
			\fmf{plain,fore=black,tension=1,right=.7}{x3,x1}
		\end{fmfgraph}\!\!
	\end{fmffile}
\end{gathered}
+\dots\Big)+\mathcal{O}(\ell^6).
\end{aligned}
\end{equation*}
Next consider $Q_2\dfn -\delta_{\n}\Lambda$; this was derived above, see (\ref{eq:delta g etc b}), 
\begin{spreadlines}{-0.3\baselineskip}
\begin{equation*}\label{eq:delta g etc bb}
\begin{aligned}
Q_2&=-g^2\Big(\tfrac{1}{6}\!\!
\begin{gathered}
	\begin{fmffile}{dg2-2loopbubble-1PIG32}
		\begin{fmfgraph}(70,70)
			\fmfset{dash_len}{1.2mm}
			\fmfset{wiggly_len}{1.1mm} \fmfset{dot_len}{0.5mm}
			\fmfpen{0.25mm}
			\fmfleft{i}
			\fmfright{o}
			\fmf{phantom,tension=5}{i,v1}
			\fmf{phantom,tension=5}{v2,o}
			\fmf{dashes,fore=black,left,tension=0.4}{v1,v2,v1}
			\fmf{dashes,fore=black}{v1,v2}
		\end{fmfgraph}
	\end{fmffile}
\end{gathered}\!\Big)
-g^4\Big(
\tfrac{1}{4}\,\,
\begin{gathered}
	\begin{fmffile}{dg4-0pt-3loop1PIaG32}
		\begin{fmfgraph}(37,37)
			\fmfset{dash_len}{1.2mm}
			\fmfset{wiggly_len}{1.1mm} \fmfset{dot_len}{0.5mm}
			\fmfpen{0.25mm}
			\fmfsurround{a,b,c,d}
			\fmf{phantom,fore=black,tension=1,curved}{a,b,c,d,a}
			\fmf{dashes,fore=black,tension=1,right}{a,b}
			\fmf{dashes,fore=black,tension=1,right}{c,d}
			\fmf{dashes,fore=black,tension=1,right}{b,c}
			\fmf{dashes,fore=black,tension=1,right}{d,a}
			\fmf{dashes,fore=black,tension=1,straight}{a,b}
			\fmf{dashes,fore=black,tension=1,straight}{c,d}
		\end{fmfgraph}
	\end{fmffile}
\end{gathered}\,\,
+\tfrac{1}{6}\!\!\!\!\!
\begin{gathered}
	\begin{fmffile}{dpeaceG32}
		\begin{fmfgraph}(105,105)
			\fmfset{dash_len}{1.2mm}
			\fmfset{wiggly_len}{1.1mm} \fmfset{dot_len}{0.5mm}
			\fmfpen{0.25mm}
			\fmfsurroundn{i}{3}
			\fmf{phantom,fore=black}{i1,v,i2}
			\fmf{phantom,fore=black}{i2,u,i3}
			\fmf{phantom,fore=black}{i3,s,i1}
			\fmfi{dashes,fore=black}{fullcircle scaled .4w shifted (.5w,.5h)}
			\fmf{dashes,fore=black}{v,c}
			\fmf{dashes,fore=black}{u,c}
			\fmf{dashes,fore=black}{s,c}
		\end{fmfgraph}
	\end{fmffile}
\end{gathered}\!\!\!\!\!
\Big)
-g^2\lambda
\Big(\tfrac{1}{8}\,\,
\begin{gathered}
	\begin{fmffile}{dg2L-bubbleG32}
		\begin{fmfgraph}(70,70)
			\fmfset{dash_len}{1.2mm}
			\fmfset{wiggly_len}{1.1mm} \fmfset{dot_len}{0.5mm}
			\fmfpen{0.25mm}
			\fmftop{t1,t2,t3}
			\fmfbottom{b1,b2,b3}
			\fmf{phantom}{t1,v1,b1}
			\fmf{phantom}{t2,v2,b2}
			\fmf{phantom}{t3,v3,b3}
			\fmffreeze
			\fmf{dashes,fore=black,right}{v1,v2,v1}
			\fmf{dashes,fore=black,right}{v2,v3,v2}
			\fmf{dashes,fore=black,tension=1}{t1,b1}
			\fmfforce{(0.25w,0.7h)}{t1}
			\fmfforce{(0.25w,0.3h)}{b1}
		\end{fmfgraph}
	\end{fmffile}
\end{gathered}
-\tfrac{3}{8}
\,\,
\begin{gathered}
	\begin{fmffile}{dg2L-3loopbubble-1PIG32}
		\begin{fmfgraph}(35,35)
			\fmfset{dash_len}{1.2mm}
			\fmfset{wiggly_len}{1.1mm} \fmfset{dot_len}{0.5mm}
			\fmfpen{0.25mm}
			\fmfsurroundn{i}{3}
			\fmf{dashes,fore=black,tension=1,right=1}{i1,i2}
			\fmf{dashes,fore=black,tension=1,right=1}{i2,i3}
			\fmf{dashes,fore=black,tension=1,right=0.8}{i3,i1}
			\fmf{dashes,fore=black,tension=1}{i1,i2}
			\fmf{dashes,fore=black,tension=1}{i2,i3}
		\end{fmfgraph}
	\end{fmffile}
\end{gathered}\,\,
\Big)
+g\kappa\Big(\tfrac{1}{12}\!
\begin{gathered}
	\begin{fmffile}{dgkappa-bubble-1PIG32}
		\begin{fmfgraph}(60,60)
			\fmfset{dash_len}{1.2mm}
			\fmfset{wiggly_len}{1.1mm} \fmfset{dot_len}{0.5mm}
			\fmfpen{0.25mm}
			\fmfleft{i}
			\fmfright{o}
			\fmf{phantom,tension=5}{i,v1}
			\fmf{phantom,tension=5}{v2,o}
			\fmf{dashes,fore=black,left,tension=0.5}{v1,v2,v1}
			\fmf{dashes,fore=black}{v1,v2}
			\fmffreeze
			\fmfi{dashes,fore=black}{fullcircle scaled .55w shifted (1.1w,.5h)}
		\end{fmfgraph}
	\end{fmffile}
\end{gathered}\,\,\,\,\Big)\\
&\hspace{0.4cm}
-\lambda
\Big(\tfrac{1}{8}
\begin{gathered}
	\begin{fmffile}{dlambdabubbled2}
		\begin{fmfgraph}(80,80)
			\fmfset{dash_len}{1.2mm}
			\fmfset{wiggly_len}{1.1mm} \fmfset{dot_len}{0.5mm}
			\fmfpen{0.25mm}
			\fmftop{t1,t2,t3}
			\fmfbottom{b1,b2,b3}
			\fmf{phantom}{t1,v1,b1}
			\fmf{phantom}{t2,v2,b2}
			\fmf{phantom}{t3,v3,b3}
			\fmffreeze
			\fmf{dashes,fore=black,right}{v1,v2,v1}
			\fmf{dashes,fore=black,right}{v2,v3,v2}
		\end{fmfgraph}
	\end{fmffile}
\end{gathered}
\Big)
-\lambda^2\Big(\tfrac{1}{24}\!
\begin{gathered}
	\begin{fmffile}{dL2-3loop-bubbleG42}
		\begin{fmfgraph}(75,75)
			\fmfset{dash_len}{1.2mm}
			\fmfset{wiggly_len}{1.1mm} \fmfset{dot_len}{0.5mm}
			\fmfpen{0.25mm}
			\fmfleft{i}
			\fmfright{o}
			\fmf{phantom,tension=10}{i,v1}
			\fmf{phantom,tension=10}{v2,o}
			\fmf{dashes,left,tension=0.4}{v1,v2,v1}
			\fmf{dashes,left=0.5}{v1,v2}
			\fmf{dashes,right=0.5}{v1,v2}
    		\end{fmfgraph}
	\end{fmffile}
\end{gathered}\!
\Big)
+\gamma
\Big(\tfrac{1}{48}\,
\begin{gathered}
	\begin{fmffile}{dgamma-vacuum-1PIkk}
		\begin{fmfgraph}(80,80)
			\fmfset{dash_len}{1.2mm}
			\fmfset{wiggly_len}{1.1mm} \fmfset{dot_len}{0.5mm}
			\fmfpen{0.25mm}
			\fmfsurroundn{x}{3}
			\fmf{phantom,fore=black}{x1,v}
			\fmf{phantom,fore=black}{x2,v}
			\fmf{phantom,fore=black}{x3,v}
			\fmf{dashes,fore=black,tension=0.7}{v,v}
			\fmf{dashes,fore=black,tension=0.7,right}{v,v}
			\fmf{dashes,fore=black,tension=0.7,left}{v,v}
			\fmffreeze
			\fmfbottom{b}
			\fmfforce{(0w,0.5h)}{b}
			\end{fmfgraph}
	\end{fmffile}
\end{gathered}\!\Big)
+\mathcal{O}(\ell^6) \, . \phantom{\Bigg\}}
\end{aligned}
\end{equation*}
Now let us turn to $Q_3$, the explicit vacuum diagrams that appear in $W(J)$. Substituting the mass counterterms of (\ref{eq:delta g etc b}) into the dressed propagators of the explicit vacuum diagrams in (\ref{eq:W(Jwiggly)}) leads to:
\begin{equation*}
\begin{aligned}
Q_3&
=
g^2\Big(\tfrac{1}{12}
\hspace{-.5cm}
\begin{gathered}
	\begin{fmffile}{dg3tavh}
		\begin{fmfgraph}(130,130)
			\fmfset{dash_len}{1.2mm}
			\fmfset{wiggly_len}{1.1mm} \fmfset{dot_len}{0.5mm}
			\fmfpen{0.25mm}
			\fmftop{t}
			\fmfbottom{b}
			\fmfleft{l}
			\fmfright{r}
			\fmf{phantom,fore=black,tension=9}{t,x,v,b}
			\fmf{phantom,fore=black,tension=9}{l,s,u,r}
			\fmf{dashes,fore=black,tension=.01,left}{x,v,x}
			\fmf{phantom,fore=black,tension=0.01}{s,x,s}
			\fmf{dashes,fore=black,tension=1}{x,v}
			\fmf{phantom,fore=black,tension=1}{u,r}
			\fmffreeze
			\fmfforce{(0.66w,0.5h)}{u}
		\end{fmfgraph}
	\end{fmffile}
\end{gathered}\hspace{-.5cm}\Big)
-\lambda\Big(\tfrac{1}{8}
\begin{gathered}
	\begin{fmffile}{dlambdabubbledh}
		\begin{fmfgraph}(80,80)
			\fmfset{dash_len}{1.2mm}
			\fmfset{wiggly_len}{1.1mm} \fmfset{dot_len}{0.5mm}
			\fmfpen{0.25mm}
			\fmftop{t1,t2,t3}
			\fmfbottom{b1,b2,b3}
			\fmf{phantom}{t1,v1,b1}
			\fmf{phantom}{t2,v2,b2}
			\fmf{phantom}{t3,v3,b3}
			\fmffreeze
			\fmf{dashes,fore=black,right}{v1,v2,v1}
			\fmf{dashes,fore=black,right}{v2,v3,v2}
		\end{fmfgraph}
	\end{fmffile}
\end{gathered}\Big)
+g^4
\Big(\tfrac{1}{16}\,
\begin{gathered}
	\begin{fmffile}{dg4-0pt-3loop1PIah}
		\begin{fmfgraph}(40,40)
			\fmfset{dash_len}{1.2mm}
			\fmfset{wiggly_len}{1.1mm} \fmfset{dot_len}{0.5mm}
			\fmfpen{0.25mm}
			\fmfsurround{a,b,c,d}
			\fmf{phantom,fore=black,tension=1,curved}{a,b,c,d,a}
			\fmf{dashes,fore=black,tension=1,right}{a,b}
			\fmf{dashes,fore=black,tension=1,right}{c,d}
			\fmf{dashes,fore=black,tension=1,right}{b,c}
			\fmf{dashes,fore=black,tension=1,right}{d,a}
			\fmf{dashes,fore=black,tension=1,straight}{a,b}
			\fmf{dashes,fore=black,tension=1,straight}{c,d}
		\end{fmfgraph}
	\end{fmffile}
\end{gathered}\,
+\tfrac{1}{4!}\!\!\!\!
\begin{gathered}
	\begin{fmffile}{dpeaceh}
		\begin{fmfgraph}(100,100)
			\fmfset{dash_len}{1.2mm}
			\fmfset{wiggly_len}{1.1mm} \fmfset{dot_len}{0.5mm}
			\fmfpen{0.25mm}
			\fmfsurroundn{i}{3}
			\fmf{phantom,fore=black}{i1,v,i2}
			\fmf{phantom,fore=black}{i2,u,i3}
			\fmf{phantom,fore=black}{i3,s,i1}
			\fmfi{dashes,fore=black}{fullcircle scaled .4w shifted (.5w,.5h)}
			\fmf{dashes,fore=black}{v,c}
			\fmf{dashes,fore=black}{u,c}
			\fmf{dashes,fore=black}{s,c}
		\end{fmfgraph}
	\end{fmffile}
\end{gathered}\!\!\!\!\Big)
-g^2\lambda
\Big(
\tfrac{1}{8}\,\,
\begin{gathered}
	\begin{fmffile}{dg2L-3loopbubble-1PIh}
		\begin{fmfgraph}(40,40)
			\fmfset{dash_len}{1.2mm}
			\fmfset{wiggly_len}{1.1mm} \fmfset{dot_len}{0.5mm}
			\fmfpen{0.25mm}
			\fmfsurroundn{i}{3}
			\fmf{dashes,fore=black,tension=1,right=1}{i1,i2}
			\fmf{dashes,fore=black,tension=1,right=1}{i2,i3}
			\fmf{dashes,fore=black,tension=1,right=0.8}{i3,i1}
			\fmf{dashes,fore=black,tension=1}{i1,i2}
			\fmf{dashes,fore=black,tension=1}{i2,i3}
		\end{fmfgraph}
	\end{fmffile}
\end{gathered}\,\,
\Big)\\
&\,\,\,\,\,\,\,\,+\lambda^2
\Big(
\tfrac{1}{48}\!
\begin{gathered}
	\begin{fmffile}{dL2-3loop-bubbleh}
		\begin{fmfgraph}(80,80)
			\fmfset{dash_len}{1.2mm}
			\fmfset{wiggly_len}{1.1mm} \fmfset{dot_len}{0.5mm}
			\fmfpen{0.25mm}
			\fmfleft{i}
			\fmfright{o}
			\fmf{phantom,tension=10}{i,v1}
			\fmf{phantom,tension=10}{v2,o}
			\fmf{dashes,left,tension=0.4}{v1,v2,v1}
			\fmf{dashes,left=0.5}{v1,v2}
			\fmf{dashes,right=0.5}{v1,v2}
    		\end{fmfgraph}
	\end{fmffile}
\end{gathered}\!\!
-\tfrac{1}{16}
\begin{gathered}
	\begin{fmffile}{dL2-bubbleh}
		\begin{fmfgraph}(70,70)
			\fmfset{dash_len}{1.2mm}
			\fmfset{wiggly_len}{1.1mm} \fmfset{dot_len}{0.5mm}
			\fmfpen{0.25mm}
			\fmftop{t1,t2,t3}
			\fmfbottom{b1,b2,b3}
			\fmf{phantom}{t1,v1,b1}
			\fmf{phantom}{t2,v2,b2}
			\fmf{phantom}{t3,v3,b3}
			\fmffreeze
			\fmf{dashes,fore=black,right}{v1,v2,v1}
			\fmf{dashes,fore=black,right}{v2,v3,v2}
			\fmfi{dashes,fore=black}{fullcircle scaled .5w shifted (1.25w,.5h)}
		\end{fmfgraph}
	\end{fmffile}
\end{gathered}\,\,\,\,\,\,\Big)
+\gamma\Big(
\tfrac{1}{24}
\begin{gathered}
	\begin{fmffile}{dgamma-vacuum-1PIh}
		\begin{fmfgraph}(70,70)
			\fmfset{dash_len}{1.2mm}
			\fmfset{wiggly_len}{1.1mm} \fmfset{dot_len}{0.5mm}
			\fmfpen{0.25mm}
			\fmfsurroundn{x}{3}
			\fmf{phantom,fore=black}{x1,v}
			\fmf{phantom,fore=black}{x2,v}
			\fmf{phantom,fore=black}{x3,v}
			\fmf{dashes,fore=black,tension=0.7}{v,v}
			\fmf{dashes,fore=black,tension=0.7,right}{v,v}
			\fmf{dashes,fore=black,tension=0.7,left}{v,v}
		\end{fmfgraph}
	\end{fmffile}
\end{gathered}\!\Big)+\mathcal{O}(\ell^6),
\end{aligned}
\end{equation*}
\end{spreadlines}
after various cancellations. Therefore, adding the three contributions to the vacuum, $Q_j$, $j=1,2,3$, it is seen that all diagrams involving self-contractions precisely cancel, the remaining terms being the 1PI diagrams:
\begin{spreadlines}{-0.3\baselineskip}
\begin{equation}\label{eq:Q}
\begin{aligned}
Q_{\rm W}&=-g^2\Big(\tfrac{1}{12}
\hspace{-.5cm}
\begin{gathered}
	\begin{fmffile}{dg3tavh}
		\begin{fmfgraph}(130,130)
			\fmfset{dash_len}{1.2mm}
			\fmfset{wiggly_len}{1.1mm} \fmfset{dot_len}{0.5mm}
			\fmfpen{0.25mm}
			\fmftop{t}
			\fmfbottom{b}
			\fmfleft{l}
			\fmfright{r}
			\fmf{phantom,fore=black,tension=9}{t,x,v,b}
			\fmf{phantom,fore=black,tension=9}{l,s,u,r}
			\fmf{dashes,fore=black,tension=.01,left}{x,v,x}
			\fmf{phantom,fore=black,tension=0.01}{s,x,s}
			\fmf{dashes,fore=black,tension=1}{x,v}
			\fmf{phantom,fore=black,tension=1}{u,r}
			\fmffreeze
			\fmfforce{(0.66w,0.5h)}{u}
		\end{fmfgraph}
	\end{fmffile}
\end{gathered}\hspace{-.5cm}\Big)
-g^4\Big(
\tfrac{3}{16}\,\,
\begin{gathered}
	\begin{fmffile}{dg4-0pt-3loop1PIaG32}
		\begin{fmfgraph}(37,37)
			\fmfset{dash_len}{1.2mm}
			\fmfset{wiggly_len}{1.1mm} \fmfset{dot_len}{0.5mm}
			\fmfpen{0.25mm}
			\fmfsurround{a,b,c,d}
			\fmf{phantom,fore=black,tension=1,curved}{a,b,c,d,a}
			\fmf{dashes,fore=black,tension=1,right}{a,b}
			\fmf{dashes,fore=black,tension=1,right}{c,d}
			\fmf{dashes,fore=black,tension=1,right}{b,c}
			\fmf{dashes,fore=black,tension=1,right}{d,a}
			\fmf{dashes,fore=black,tension=1,straight}{a,b}
			\fmf{dashes,fore=black,tension=1,straight}{c,d}
		\end{fmfgraph}
	\end{fmffile}
\end{gathered}\,\,+
\tfrac{1}{8}\!\!\!\!
\begin{gathered}
	\begin{fmffile}{dpeaceh}
		\begin{fmfgraph}(100,100)
			\fmfset{dash_len}{1.2mm}
			\fmfset{wiggly_len}{1.1mm} \fmfset{dot_len}{0.5mm}
			\fmfpen{0.25mm}
			\fmfsurroundn{i}{3}
			\fmf{phantom,fore=black}{i1,v,i2}
			\fmf{phantom,fore=black}{i2,u,i3}
			\fmf{phantom,fore=black}{i3,s,i1}
			\fmfi{dashes,fore=black}{fullcircle scaled .4w shifted (.5w,.5h)}
			\fmf{dashes,fore=black}{v,c}
			\fmf{dashes,fore=black}{u,c}
			\fmf{dashes,fore=black}{s,c}
		\end{fmfgraph}
	\end{fmffile}
\end{gathered}\!\!\!\!\Big)
+g^2\lambda
\Big(
\tfrac{1}{4}\,\,
\begin{gathered}
	\begin{fmffile}{dg2L-3loopbubble-1PIh}
		\begin{fmfgraph}(40,40)
			\fmfset{dash_len}{1.2mm}
			\fmfset{wiggly_len}{1.1mm} \fmfset{dot_len}{0.5mm}
			\fmfpen{0.25mm}
			\fmfsurroundn{i}{3}
			\fmf{dashes,fore=black,tension=1,right=1}{i1,i2}
			\fmf{dashes,fore=black,tension=1,right=1}{i2,i3}
			\fmf{dashes,fore=black,tension=1,right=0.8}{i3,i1}
			\fmf{dashes,fore=black,tension=1}{i1,i2}
			\fmf{dashes,fore=black,tension=1}{i2,i3}
		\end{fmfgraph}
	\end{fmffile}
\end{gathered}\,\,
\Big)\\
&\hspace{0.4cm}-\lambda^2\Big(\tfrac{1}{48}\!
\begin{gathered}
	\begin{fmffile}{dL2-3loop-bubbleG42}
		\begin{fmfgraph}(75,75)
			\fmfset{dash_len}{1.2mm}
			\fmfset{wiggly_len}{1.1mm} \fmfset{dot_len}{0.5mm}
			\fmfpen{0.25mm}
			\fmfleft{i}
			\fmfright{o}
			\fmf{phantom,tension=10}{i,v1}
			\fmf{phantom,tension=10}{v2,o}
			\fmf{dashes,left,tension=0.4}{v1,v2,v1}
			\fmf{dashes,left=0.5}{v1,v2}
			\fmf{dashes,right=0.5}{v1,v2}
    		\end{fmfgraph}
	\end{fmffile}
\end{gathered}\!
\Big)
+\tfrac{1}{2}\Big(
\hspace{0.5cm}
\begin{gathered}
	\begin{fmffile}{bubble1b}
		\begin{fmfgraph}(40,40)
			\fmfset{dash_len}{1.2mm}
			\fmfset{wiggly_len}{1.1mm} \fmfset{dot_len}{0.5mm}
			\fmfpen{0.25mm}
			\fmfvn{decor.shape=square,decor.filled=shaded, decor.size=5thin}{u}{1}
			\fmfleft{i}
			\fmfright{o}
			\fmf{plain,fore=black,tension=5,left}{i,u1,i}
			\fmffreeze
			\fmfforce{(-w,0.35h)}{i}
			\fmfforce{(0w,0.35h)}{u1}
			\fmfforce{(1.1w,0.35h)}{o}
		\end{fmfgraph}\!\!\!\!
	\end{fmffile}
\end{gathered}
+
\tfrac{1}{2}\hspace{0.6cm}
\begin{gathered}
	\begin{fmffile}{bubble2b}
		\begin{fmfgraph}(40,40)
			\fmfset{dash_len}{1.2mm}
			\fmfset{wiggly_len}{1.1mm} \fmfset{dot_len}{0.5mm}
			\fmfpen{0.25mm}
			\fmfvn{decor.shape=square,decor.filled=shaded, decor.size=5thin}{u}{2}
			\fmfleft{i}
			\fmfright{o}
			\fmf{plain,fore=black,tension=5,left}{i,u1,u2,i}
			\fmffreeze
			\fmfforce{(-w,0.35h)}{i}
			\fmfforce{(0w,0.35h)}{u1}
			\fmfforce{(-1w,0.35h)}{u2}
			\fmfforce{(1.1w,0.35h)}{o}
		\end{fmfgraph}\!\!\!\!
	\end{fmffile}
\end{gathered}
+\dots\Big)
+\mathcal{O}(\ell^6).
\end{aligned}
\end{equation}
\end{spreadlines}
It will also be useful to display $Q_{\rm W}$ with the wave-function counterterms made explicit. Substituting the defining expression for the dashed propagator (\ref{eq:dressed_prop}) into (\ref{eq:Q}) leads to:
\begin{spreadlines}{-0.3\baselineskip}
\begin{equation}\label{eq:QW}
\begin{aligned}
Q_{\rm W}&=-g^2\Big(\tfrac{1}{12}
\hspace{-.5cm}
\begin{gathered}
	\begin{fmffile}{pg3tavh}
		\begin{fmfgraph}(130,130)
			\fmfset{dash_len}{1.2mm}
			\fmfset{wiggly_len}{1.1mm} \fmfset{dot_len}{0.5mm}
			\fmfpen{0.25mm}
			\fmftop{t}
			\fmfbottom{b}
			\fmfleft{l}
			\fmfright{r}
			\fmf{phantom,fore=black,tension=9}{t,x,v,b}
			\fmf{phantom,fore=black,tension=9}{l,s,u,r}
			\fmf{plain,fore=black,tension=.01,left}{x,v,x}
			\fmf{phantom,fore=black,tension=0.01}{s,x,s}
			\fmf{plain,fore=black,tension=1}{x,v}
			\fmf{phantom,fore=black,tension=1}{u,r}
			\fmffreeze
			\fmfforce{(0.66w,0.5h)}{u}
		\end{fmfgraph}
	\end{fmffile}
\end{gathered}\hspace{-.5cm}+\tfrac{3}{12}\hspace{-.5cm}
\begin{gathered}
	\begin{fmffile}{pg3tavh2}
		\begin{fmfgraph}(130,130)
			\fmfset{dash_len}{1.2mm}
			\fmfset{wiggly_len}{1.1mm} \fmfset{dot_len}{0.5mm}
			\fmfpen{0.25mm}
			\fmftop{t}
			\fmfbottom{b}
			\fmfleft{l}
			\fmfright{r}
			\fmf{phantom,fore=black,tension=9}{t,x,v,b}
			\fmf{phantom,fore=black,tension=9}{l,s,u,r}
			\fmf{plain,fore=black,tension=.01,left}{x,v,x}
			\fmf{phantom,fore=black,tension=0.01}{s,x,s}
			\fmf{plain,fore=black,tension=1}{x,v}
			\fmf{phantom,fore=black,tension=1}{u,r}
			\fmffreeze
			\fmfforce{(0.66w,0.5h)}{u}
			\fmfv{decor.shape=square,decor.filled=shaded, decor.size=5thin}{c}
			\fmfforce{(0.5w,0.5h)}{c}
		\end{fmfgraph}
	\end{fmffile}
\end{gathered}\hspace{-.5cm}\Big)
-g^4\Big(
\tfrac{3}{16}\,\,
\begin{gathered}
	\begin{fmffile}{pg4-0pt-3loop1PIaG32}
		\begin{fmfgraph}(37,37)
			\fmfset{dash_len}{1.2mm}
			\fmfset{wiggly_len}{1.1mm} \fmfset{dot_len}{0.5mm}
			\fmfpen{0.25mm}
			\fmfsurround{a,b,c,d}
			\fmf{phantom,fore=black,tension=1,curved}{a,b,c,d,a}
			\fmf{plain,fore=black,tension=1,right}{a,b}
			\fmf{plain,fore=black,tension=1,right}{c,d}
			\fmf{plain,fore=black,tension=1,right}{b,c}
			\fmf{plain,fore=black,tension=1,right}{d,a}
			\fmf{plain,fore=black,tension=1,straight}{a,b}
			\fmf{plain,fore=black,tension=1,straight}{c,d}
		\end{fmfgraph}
	\end{fmffile}
\end{gathered}\,\,+
\tfrac{1}{8}\!\!\!\!
\begin{gathered}
	\begin{fmffile}{ppeaceh}
		\begin{fmfgraph}(100,100)
			\fmfset{dash_len}{1.2mm}
			\fmfset{wiggly_len}{1.1mm} \fmfset{dot_len}{0.5mm}
			\fmfpen{0.25mm}
			\fmfsurroundn{i}{3}
			\fmf{phantom,fore=black}{i1,v,i2}
			\fmf{phantom,fore=black}{i2,u,i3}
			\fmf{phantom,fore=black}{i3,s,i1}
			\fmfi{plain,fore=black}{fullcircle scaled .4w shifted (.5w,.5h)}
			\fmf{plain,fore=black}{v,c}
			\fmf{plain,fore=black}{u,c}
			\fmf{plain,fore=black}{s,c}
		\end{fmfgraph}
	\end{fmffile}
\end{gathered}\!\!\!\!\Big)
+g^2\lambda
\Big(
\tfrac{1}{4}\,\,
\begin{gathered}
	\begin{fmffile}{pg2L-3loopbubble-1PIh}
		\begin{fmfgraph}(40,40)
			\fmfset{dash_len}{1.2mm}
			\fmfset{wiggly_len}{1.1mm} \fmfset{dot_len}{0.5mm}
			\fmfpen{0.25mm}
			\fmfsurroundn{i}{3}
			\fmf{plain,fore=black,tension=1,right=1}{i1,i2}
			\fmf{plain,fore=black,tension=1,right=1}{i2,i3}
			\fmf{plain,fore=black,tension=1,right=0.8}{i3,i1}
			\fmf{plain,fore=black,tension=1}{i1,i2}
			\fmf{plain,fore=black,tension=1}{i2,i3}
		\end{fmfgraph}
	\end{fmffile}
\end{gathered}\,\,
\Big)\\
&\hspace{0.4cm}-\lambda^2\Big(\tfrac{1}{48}\!
\begin{gathered}
	\begin{fmffile}{pL2-3loop-bubbleG42}
		\begin{fmfgraph}(75,75)
			\fmfset{dash_len}{1.2mm}
			\fmfset{wiggly_len}{1.1mm} \fmfset{dot_len}{0.5mm}
			\fmfpen{0.25mm}
			\fmfleft{i}
			\fmfright{o}
			\fmf{phantom,tension=10}{i,v1}
			\fmf{phantom,tension=10}{v2,o}
			\fmf{plain,left,tension=0.4}{v1,v2,v1}
			\fmf{plain,left=0.5}{v1,v2}
			\fmf{plain,right=0.5}{v1,v2}
    		\end{fmfgraph}
	\end{fmffile}
\end{gathered}\!
\Big)
+\tfrac{1}{2}\Big(
\hspace{0.5cm}
\begin{gathered}
	\begin{fmffile}{bubble1b}
		\begin{fmfgraph}(40,40)
			\fmfset{dash_len}{1.2mm}
			\fmfset{wiggly_len}{1.1mm} \fmfset{dot_len}{0.5mm}
			\fmfpen{0.25mm}
			\fmfvn{decor.shape=square,decor.filled=shaded, decor.size=5thin}{u}{1}
			\fmfleft{i}
			\fmfright{o}
			\fmf{plain,fore=black,tension=5,left}{i,u1,i}
			\fmffreeze
			\fmfforce{(-w,0.35h)}{i}
			\fmfforce{(0w,0.35h)}{u1}
			\fmfforce{(1.1w,0.35h)}{o}
		\end{fmfgraph}\!\!\!\!
	\end{fmffile}
\end{gathered}
+
\tfrac{1}{2}\hspace{0.6cm}
\begin{gathered}
	\begin{fmffile}{bubble2b}
		\begin{fmfgraph}(40,40)
			\fmfset{dash_len}{1.2mm}
			\fmfset{wiggly_len}{1.1mm} \fmfset{dot_len}{0.5mm}
			\fmfpen{0.25mm}
			\fmfvn{decor.shape=square,decor.filled=shaded, decor.size=5thin}{u}{2}
			\fmfleft{i}
			\fmfright{o}
			\fmf{plain,fore=black,tension=5,left}{i,u1,u2,i}
			\fmffreeze
			\fmfforce{(-w,0.35h)}{i}
			\fmfforce{(0w,0.35h)}{u1}
			\fmfforce{(-1w,0.35h)}{u2}
			\fmfforce{(1.1w,0.35h)}{o}
		\end{fmfgraph}\!\!\!\!
	\end{fmffile}
\end{gathered}
+\dots\Big)
+\mathcal{O}(\ell^6),
\end{aligned}
\end{equation}
\end{spreadlines}
where we have assumed that $\wf\sim \mathcal{O}(\ell^2)$. 

This short calculation shows that weak complete normal ordering is somewhat unnatural, given that the combinatorial coefficients of the various diagrams in $Q_{\rm W}$ are not those that one would have expected from a naive application of the Feynman rules; the latter can be read off from the relevant vacuum diagrams in (\ref{eq:U(Jwiggly)}). This is in fact related to the comment above, that although weak complete normal ordering cancels all cephalopods from the full generating function, to get a more natural-looking expression (when the vacuum terms are physical) it is best to adopt the strong form of complete normal ordering, and we discuss this next. 

We mentioned above that adopting the strong form of complete normal ordering amounts to replacing $Q_{\rm W}$ by $Q_{\rm S}$ so that, according to (\ref{eq:QWQS}), we also need $Q_4$, namely the contribution coming from the complete normal ordering of the kinetic term in the action, which is determined as follows:
\begin{equation*}
\begin{aligned}
&\int \mathcal{D}\phi\,e^{-\n\int \frac{1}{2}[(\hbar\nabla\phi)^2
-\,\phi \! \begin{gathered}
	\begin{fmffile}{square-shaded}
		\begin{fmfgraph}(40,40)
			\fmfset{dash_len}{1.2mm}
			\fmfset{wiggly_len}{1.1mm} \fmfset{dot_len}{0.5mm}
			\fmfpen{0.25mm}
			\fmfvn{decor.shape=square,decor.filled=shaded, decor.size=5thin}{u}{1}
			\fmfleft{i}
			\fmfright{o}
			\fmf{phantom,fore=black,tension=5}{i,u1,o}
			\fmffreeze
			\fmfforce{(0.5w,0.35h)}{i}
			\fmfforce{(0.5w,0.35h)}{u1}
			\fmfforce{(0.5w,0.35h)}{o}
		\end{fmfgraph}
	\end{fmffile}
\end{gathered}\!\phi+m^2\phi^2]\n+\dots}\\
&\qquad= e^{\int \frac{1}{2}(\Delta+m^2)G_2-\int\frac{1}{2}\!\wf \!G_2}\int \mathcal{D}\phi\,e^{-\int \frac{1}{2}[(\hbar\nabla\phi)^2-\,\phi \! \begin{gathered}
	\begin{fmffile}{square-shaded}
		\begin{fmfgraph}(40,40)
			\fmfset{dash_len}{1.2mm}
			\fmfset{wiggly_len}{1.1mm} \fmfset{dot_len}{0.5mm}
			\fmfpen{0.25mm}
			\fmfvn{decor.shape=square,decor.filled=shaded, decor.size=5thin}{u}{1}
			\fmfleft{i}
			\fmfright{o}
			\fmf{phantom,fore=black,tension=5}{i,u1,o}
			\fmffreeze
			\fmfforce{(0.5w,0.35h)}{i}
			\fmfforce{(0.5w,0.35h)}{u1}
			\fmfforce{(0.5w,0.35h)}{o}
		\end{fmfgraph}
	\end{fmffile}
\end{gathered}\!\phi+m^2\phi^2]+\dots},
\end{aligned}
\end{equation*}
where the interaction terms are denoted by dots. The vacuum contribution, $Q_4$, is therefore given by:
\begin{equation}\label{eq:Q4a}
Q_4=\int \frac{1}{2}(\Delta+m^2)G_2-\int\frac{1}{2}\!\wf \!G_2,
\end{equation}
where it is to be understood that the Laplacian, $\Delta$, acts on \emph{one} of the two legs of the two-point function, and then the coincident limit is taken, $\Delta G_2\dfn \lim_{1,2\rightarrow z}\Delta_1G_2(1,2)$. 

Let us compute $Q_4$, starting from the first of the two terms in (\ref{eq:Q4a}). From p.~\!\pageref{eq:G_N wiggly} we have an expression for the full two-point function at coincident points, $G_2$, up to and including $\mathcal{O}(\ell^4)$. This is given in terms of the dressed propagator, defined in (\ref{eq:dressed_prop}), see also (\ref{eq:deltam2 deltaZ}). Keeping $\delta Z$ fixed for now, we can rewrite $G_2$ while making manifest the mass renormalisation counterterms, $\delta_{\n}  m^2$, as well as all coupling counterterms, see (\ref{eq:delta g etc b}). Substituting all counterterms (\ref{eq:delta g etc b}) into $G_2$ of p.~\!\pageref{eq:G_N wiggly}, various cancellations take place and we are left with
\begin{equation}\label{eq:G_2 dashes}
\begin{aligned}
G_2=&
\hspace{0.5cm}
\begin{gathered}
	\begin{fmffile}{dbubblexdc}
		\begin{fmfgraph}(40,40)
			\fmfset{dash_len}{1.2mm}
			\fmfset{wiggly_len}{1.1mm} \fmfset{dot_len}{0.5mm}
			\fmfpen{0.25mm}
			\fmfvn{decor.shape=circle,decor.filled=full, decor.size=3thin}{u}{1}
			\fmfleft{i}
			\fmfright{o}
			\fmf{dashes,fore=black,tension=5,left}{i,u1,i}
			\fmffreeze
			\fmfforce{(-w,0.35h)}{i}
			\fmfforce{(0w,0.35h)}{u1}
			\fmfforce{(1.1w,0.35h)}{o}
		\end{fmfgraph}\!\!\!\!\!\!
	\end{fmffile}
\end{gathered}+
g^2\Big(\tfrac{1}{2}
\hspace{-.5cm}
\begin{gathered}
	\begin{fmffile}{ddg3tav}
		\begin{fmfgraph}(130,130)
			\fmfset{dash_len}{1.2mm}
			\fmfset{wiggly_len}{1.1mm} \fmfset{dot_len}{0.5mm}
			\fmfpen{0.25mm}
			\fmftop{t}
			\fmfbottom{b}
			\fmfleft{l}
			\fmfright{r}
			\fmfv{decor.shape=circle,decor.filled=full, decor.size=3thin}{u}
			\fmf{phantom,fore=black,tension=9}{t,x,v,b}
			\fmf{phantom,fore=black,tension=9}{l,s,u,r}
			\fmf{dashes,fore=black,tension=.01,left}{x,v,x}
			\fmf{phantom,fore=black,tension=0.01}{s,x,s}
			\fmf{dashes,fore=black,tension=1}{x,v}
			\fmf{phantom,fore=black,tension=1}{u,r}
			\fmffreeze
			\fmfforce{(0.66w,0.5h)}{u}
		\end{fmfgraph}
	\end{fmffile}
\end{gathered}\hspace{-.5cm}
\Big)
+g^4\Big(
\tfrac{1}{4}\,\,
\begin{gathered}
	\begin{fmffile}{dg4-0pt-3loop1PIaG32a}
		\begin{fmfgraph}(37,37)
			\fmfset{dash_len}{1.2mm}
			\fmfset{wiggly_len}{1.1mm} \fmfset{dot_len}{0.5mm}
			\fmfpen{0.25mm}
			\fmfsurround{a,b,c,d}
			\fmf{phantom,fore=black,tension=1,curved}{a,b,c,d,a}
			\fmf{dashes,fore=black,tension=1,right}{a,b}
			\fmf{dashes,fore=black,tension=1,right}{c,d}
			\fmf{dashes,fore=black,tension=1,right}{b,c}
			\fmf{dashes,fore=black,tension=1,right}{d,a}
			\fmf{dashes,fore=black,tension=1,straight}{a,b}
			\fmf{dashes,fore=black,tension=1,straight}{c,d}
			\fmffreeze
			\fmfv{decor.shape=circle,decor.filled=full, decor.size=3thin}{u}
			\fmfforce{(0.97w,0.02h)}{u}
		\end{fmfgraph}
	\end{fmffile}
\end{gathered}\,\,
+\tfrac{1}{2}\!\!\!\!
\begin{gathered}
	\begin{fmffile}{dpeacehc}
		\begin{fmfgraph}(100,100)
			\fmfset{dash_len}{1.2mm}
			\fmfset{wiggly_len}{1.1mm} \fmfset{dot_len}{0.5mm}
			\fmfpen{0.25mm}
			\fmfsurroundn{i}{3}
			\fmf{phantom,fore=black}{i1,v,i2}
			\fmf{phantom,fore=black}{i2,u,i3}
			\fmf{phantom,fore=black}{i3,s,i1}
			\fmfi{dashes,fore=black}{fullcircle scaled .4w shifted (.5w,.5h)}
			\fmf{dashes,fore=black}{v,c}
			\fmf{dashes,fore=black}{u,c}
			\fmf{dashes,fore=black}{s,c}
			\fmffreeze
			\fmfv{decor.shape=circle,decor.filled=full, decor.size=3thin}{h}
			\fmfforce{(0.7w,0.5h)}{h}
		\end{fmfgraph}
	\end{fmffile}
\end{gathered}\!\!\!\!
+\tfrac{1}{2}\,\,
\begin{gathered}
	\begin{fmffile}{dg4-0pt-3loop1PIaG32b}
		\begin{fmfgraph}(37,37)
			\fmfset{dash_len}{1.2mm}
			\fmfset{wiggly_len}{1.1mm} \fmfset{dot_len}{0.5mm}
			\fmfpen{0.25mm}
			\fmfsurround{a,b,c,d}
			\fmf{phantom,fore=black,tension=1,curved}{a,b,c,d,a}
			\fmf{dashes,fore=black,tension=1,right}{a,b}
			\fmf{dashes,fore=black,tension=1,right}{c,d}
			\fmf{dashes,fore=black,tension=1,right}{b,c}
			\fmf{dashes,fore=black,tension=1,right}{d,a}
			\fmf{dashes,fore=black,tension=1,straight}{a,b}
			\fmf{dashes,fore=black,tension=1,straight}{c,d}
			\fmffreeze
			\fmfv{decor.shape=circle,decor.filled=full, decor.size=3thin}{u}
			\fmfforce{(0.99w,0.98h)}{u}
		\end{fmfgraph}
	\end{fmffile}
\end{gathered}\,\,
\Big)\\[-0.3cm]
&
-g^2\lambda\Big(
\,\,
\begin{gathered}
	\begin{fmffile}{dg2L-3loopbubble-1PIhc}
		\begin{fmfgraph}(40,40)
			\fmfset{dash_len}{1.2mm}
			\fmfset{wiggly_len}{1.1mm} \fmfset{dot_len}{0.5mm}
			\fmfpen{0.25mm}
			\fmfsurroundn{i}{3}
			\fmf{dashes,fore=black,tension=1,right=1}{i1,i2}
			\fmf{dashes,fore=black,tension=1,right=1}{i2,i3}
			\fmf{dashes,fore=black,tension=1,right=0.8}{i3,i1}
			\fmf{dashes,fore=black,tension=1}{i1,i2}
			\fmf{dashes,fore=black,tension=1}{i2,i3}
			\fmffreeze
			\fmfv{decor.shape=circle,decor.filled=full, decor.size=3thin}{u}
			\fmfforce{(0.9w,1.05h)}{u}
		\end{fmfgraph}
	\end{fmffile}
\end{gathered}\,\,
+\tfrac{1}{4}\,\,
\begin{gathered}
	\begin{fmffile}{dg2L-3loopbubble-1PIhd}
		\begin{fmfgraph}(40,40)
			\fmfset{dash_len}{1.2mm}
			\fmfset{wiggly_len}{1.1mm} \fmfset{dot_len}{0.5mm}
			\fmfpen{0.25mm}
			\fmfsurroundn{i}{3}
			\fmf{dashes,fore=black,tension=1,right=1}{i1,i2}
			\fmf{dashes,fore=black,tension=1,right=1}{i2,i3}
			\fmf{dashes,fore=black,tension=1,right=0.8}{i3,i1}
			\fmf{dashes,fore=black,tension=1}{i1,i2}
			\fmf{dashes,fore=black,tension=1}{i2,i3}
			\fmffreeze
			\fmfv{decor.shape=circle,decor.filled=full, decor.size=3thin}{u}
			\fmfforce{(0.8w,0.0h)}{u}
		\end{fmfgraph}
	\end{fmffile}
\end{gathered}\,\,
\Big)
+\lambda^2
\Big(\tfrac{1}{6}\!
\begin{gathered}
	\begin{fmffile}{dL2-3loop-bubbleG42b}
		\begin{fmfgraph}(75,75)
			\fmfset{dash_len}{1.2mm}
			\fmfset{wiggly_len}{1.1mm} \fmfset{dot_len}{0.5mm}
			\fmfpen{0.25mm}
			\fmfleft{i}
			\fmfright{o}
			\fmf{phantom,tension=10}{i,v1}
			\fmf{phantom,tension=10}{v2,o}
			\fmf{dashes,left,tension=0.4}{v1,v2,v1}
			\fmf{dashes,left=0.5}{v1,v2}
			\fmf{dashes,right=0.5}{v1,v2}
			\fmffreeze
			\fmfv{decor.shape=circle,decor.filled=full, decor.size=3thin}{u}
			\fmfforce{(0.52w,0.17h)}{u}
    		\end{fmfgraph}
	\end{fmffile}
\end{gathered}\!
\Big)+\mathcal{O}(\ell^6).
\end{aligned}
\end{equation}
We need to apply the operator $\Delta+m^2$ to this expression, and in doing so it will be useful to recall the definition of the plain propagator (\ref{eq:vanillapropagator}) and its relation to the dashed propagator in (\ref{eq:dressed_prop}). Taking these relations into account and substituting them into (\ref{eq:G_2 dashes}), which is in turn substituted into (\ref{eq:Q4a}), leads to:
\begin{spreadlines}{-0.3\baselineskip}
\begin{equation}\label{eq:Q4a1}
\begin{aligned}
\int \frac{1}{2}(\Delta+m^2)G_2&=\tfrac{1}{2}+g^2\Big(\tfrac{1}{4}
\hspace{-.5cm}
\begin{gathered}
	\begin{fmffile}{pg3tavh}
		\begin{fmfgraph}(130,130)
			\fmfset{dash_len}{1.2mm}
			\fmfset{wiggly_len}{1.1mm} \fmfset{dot_len}{0.5mm}
			\fmfpen{0.25mm}
			\fmftop{t}
			\fmfbottom{b}
			\fmfleft{l}
			\fmfright{r}
			\fmf{phantom,fore=black,tension=9}{t,x,v,b}
			\fmf{phantom,fore=black,tension=9}{l,s,u,r}
			\fmf{plain,fore=black,tension=.01,left}{x,v,x}
			\fmf{phantom,fore=black,tension=0.01}{s,x,s}
			\fmf{plain,fore=black,tension=1}{x,v}
			\fmf{phantom,fore=black,tension=1}{u,r}
			\fmffreeze
			\fmfforce{(0.66w,0.5h)}{u}
		\end{fmfgraph}
	\end{fmffile}
\end{gathered}\hspace{-.5cm}+\hspace{-.5cm}
\begin{gathered}
	\begin{fmffile}{pg3tavh2}
		\begin{fmfgraph}(130,130)
			\fmfset{dash_len}{1.2mm}
			\fmfset{wiggly_len}{1.1mm} \fmfset{dot_len}{0.5mm}
			\fmfpen{0.25mm}
			\fmftop{t}
			\fmfbottom{b}
			\fmfleft{l}
			\fmfright{r}
			\fmf{phantom,fore=black,tension=9}{t,x,v,b}
			\fmf{phantom,fore=black,tension=9}{l,s,u,r}
			\fmf{plain,fore=black,tension=.01,left}{x,v,x}
			\fmf{phantom,fore=black,tension=0.01}{s,x,s}
			\fmf{plain,fore=black,tension=1}{x,v}
			\fmf{phantom,fore=black,tension=1}{u,r}
			\fmffreeze
			\fmfforce{(0.66w,0.5h)}{u}
			\fmfv{decor.shape=square,decor.filled=shaded, decor.size=5thin}{c}
			\fmfforce{(0.5w,0.5h)}{c}
		\end{fmfgraph}
	\end{fmffile}
\end{gathered}\hspace{-.5cm}\Big)
+g^4\Big(
\tfrac{3}{8}\,\,
\begin{gathered}
	\begin{fmffile}{pg4-0pt-3loop1PIaG32}
		\begin{fmfgraph}(37,37)
			\fmfset{dash_len}{1.2mm}
			\fmfset{wiggly_len}{1.1mm} \fmfset{dot_len}{0.5mm}
			\fmfpen{0.25mm}
			\fmfsurround{a,b,c,d}
			\fmf{phantom,fore=black,tension=1,curved}{a,b,c,d,a}
			\fmf{plain,fore=black,tension=1,right}{a,b}
			\fmf{plain,fore=black,tension=1,right}{c,d}
			\fmf{plain,fore=black,tension=1,right}{b,c}
			\fmf{plain,fore=black,tension=1,right}{d,a}
			\fmf{plain,fore=black,tension=1,straight}{a,b}
			\fmf{plain,fore=black,tension=1,straight}{c,d}
		\end{fmfgraph}
	\end{fmffile}
\end{gathered}\,\,+
\tfrac{1}{4}\!\!\!\!
\begin{gathered}
	\begin{fmffile}{ppeaceh}
		\begin{fmfgraph}(100,100)
			\fmfset{dash_len}{1.2mm}
			\fmfset{wiggly_len}{1.1mm} \fmfset{dot_len}{0.5mm}
			\fmfpen{0.25mm}
			\fmfsurroundn{i}{3}
			\fmf{phantom,fore=black}{i1,v,i2}
			\fmf{phantom,fore=black}{i2,u,i3}
			\fmf{phantom,fore=black}{i3,s,i1}
			\fmfi{plain,fore=black}{fullcircle scaled .4w shifted (.5w,.5h)}
			\fmf{plain,fore=black}{v,c}
			\fmf{plain,fore=black}{u,c}
			\fmf{plain,fore=black}{s,c}
		\end{fmfgraph}
	\end{fmffile}
\end{gathered}\!\!\!\!\Big)\\
&\hspace{0.4cm}-g^2\lambda
\Big(
\tfrac{5}{8}\,\,
\begin{gathered}
	\begin{fmffile}{pg2L-3loopbubble-1PIh}
		\begin{fmfgraph}(40,40)
			\fmfset{dash_len}{1.2mm}
			\fmfset{wiggly_len}{1.1mm} \fmfset{dot_len}{0.5mm}
			\fmfpen{0.25mm}
			\fmfsurroundn{i}{3}
			\fmf{plain,fore=black,tension=1,right=1}{i1,i2}
			\fmf{plain,fore=black,tension=1,right=1}{i2,i3}
			\fmf{plain,fore=black,tension=1,right=0.8}{i3,i1}
			\fmf{plain,fore=black,tension=1}{i1,i2}
			\fmf{plain,fore=black,tension=1}{i2,i3}
		\end{fmfgraph}
	\end{fmffile}
\end{gathered}\,\,
\Big)+\lambda^2\Big(\tfrac{1}{12}\!
\begin{gathered}
	\begin{fmffile}{pL2-3loop-bubbleG42}
		\begin{fmfgraph}(75,75)
			\fmfset{dash_len}{1.2mm}
			\fmfset{wiggly_len}{1.1mm} \fmfset{dot_len}{0.5mm}
			\fmfpen{0.25mm}
			\fmfleft{i}
			\fmfright{o}
			\fmf{phantom,tension=10}{i,v1}
			\fmf{phantom,tension=10}{v2,o}
			\fmf{plain,left,tension=0.4}{v1,v2,v1}
			\fmf{plain,left=0.5}{v1,v2}
			\fmf{plain,right=0.5}{v1,v2}
    		\end{fmfgraph}
	\end{fmffile}
\end{gathered}\!
\Big)
+\mathcal{O}(\ell^6),
\end{aligned}
\end{equation}
\end{spreadlines}
where we have assumed that $\wf\sim \mathcal{O}(\ell^2)$ and dropped terms of $\mathcal{O}(\ell^6)$. Notice that the operator $\Delta+m^2$ and the subsequent integral has completed the vertices and hence there is no ``dot'' in the diagrams of (\ref{eq:Q4a1}).

Let us now determine the second term in (\ref{eq:Q4a}). Here we again make use of (\ref{eq:G_2 dashes}) and also the definition of the dashed propagator (\ref{eq:dressed_prop}). Assuming again $\wf\sim \mathcal{O}(\ell^2)$ we find
\begin{equation*}
-\int\tfrac{1}{2}\!\wf \!G_2=-\tfrac{1}{2}\Big(\hspace{0.5cm}
\begin{gathered}
	\begin{fmffile}{bubble1b}
		\begin{fmfgraph}(40,40)
			\fmfset{dash_len}{1.2mm}
			\fmfset{wiggly_len}{1.1mm} \fmfset{dot_len}{0.5mm}
			\fmfpen{0.25mm}
			\fmfvn{decor.shape=square,decor.filled=shaded, decor.size=5thin}{u}{1}
			\fmfleft{i}
			\fmfright{o}
			\fmf{plain,fore=black,tension=5,left}{i,u1,i}
			\fmffreeze
			\fmfforce{(-w,0.35h)}{i}
			\fmfforce{(0w,0.35h)}{u1}
			\fmfforce{(1.1w,0.35h)}{o}
		\end{fmfgraph}\!\!\!\!
	\end{fmffile}
\end{gathered}
+
\hspace{0.6cm}
\begin{gathered}
	\begin{fmffile}{bubble2b}
		\begin{fmfgraph}(40,40)
			\fmfset{dash_len}{1.2mm}
			\fmfset{wiggly_len}{1.1mm} \fmfset{dot_len}{0.5mm}
			\fmfpen{0.25mm}
			\fmfvn{decor.shape=square,decor.filled=shaded, decor.size=5thin}{u}{2}
			\fmfleft{i}
			\fmfright{o}
			\fmf{plain,fore=black,tension=5,left}{i,u1,u2,i}
			\fmffreeze
			\fmfforce{(-w,0.35h)}{i}
			\fmfforce{(0w,0.35h)}{u1}
			\fmfforce{(-1w,0.35h)}{u2}
			\fmfforce{(1.1w,0.35h)}{o}
		\end{fmfgraph}\!\!\!\!
	\end{fmffile}
\end{gathered}\Big)
-g^2\Big(\tfrac{3}{12}\hspace{-.5cm}
\begin{gathered}
	\begin{fmffile}{pg3tavh2}
		\begin{fmfgraph}(130,130)
			\fmfset{dash_len}{1.2mm}
			\fmfset{wiggly_len}{1.1mm} \fmfset{dot_len}{0.5mm}
			\fmfpen{0.25mm}
			\fmftop{t}
			\fmfbottom{b}
			\fmfleft{l}
			\fmfright{r}
			\fmf{phantom,fore=black,tension=9}{t,x,v,b}
			\fmf{phantom,fore=black,tension=9}{l,s,u,r}
			\fmf{plain,fore=black,tension=.01,left}{x,v,x}
			\fmf{phantom,fore=black,tension=0.01}{s,x,s}
			\fmf{plain,fore=black,tension=1}{x,v}
			\fmf{phantom,fore=black,tension=1}{u,r}
			\fmffreeze
			\fmfforce{(0.66w,0.5h)}{u}
			\fmfv{decor.shape=square,decor.filled=shaded, decor.size=5thin}{c}
			\fmfforce{(0.5w,0.5h)}{c}
		\end{fmfgraph}
	\end{fmffile}
\end{gathered}\hspace{-.5cm}\Big)+\mathcal{O}(\ell^6).
\end{equation*}
Collecting the above results, we learn that $Q_4$ takes the form:
\vspace{-0.3cm}
\begin{spreadlines}{-0.3\baselineskip}
\begin{equation*}\label{eq:Q4a3}
\begin{aligned}
Q_4&=\tfrac{1}{2}+g^2\Big(\tfrac{1}{4}
\hspace{-.5cm}
\begin{gathered}
	\begin{fmffile}{pg3tavh}
		\begin{fmfgraph}(130,130)
			\fmfset{dash_len}{1.2mm}
			\fmfset{wiggly_len}{1.1mm} \fmfset{dot_len}{0.5mm}
			\fmfpen{0.25mm}
			\fmftop{t}
			\fmfbottom{b}
			\fmfleft{l}
			\fmfright{r}
			\fmf{phantom,fore=black,tension=9}{t,x,v,b}
			\fmf{phantom,fore=black,tension=9}{l,s,u,r}
			\fmf{plain,fore=black,tension=.01,left}{x,v,x}
			\fmf{phantom,fore=black,tension=0.01}{s,x,s}
			\fmf{plain,fore=black,tension=1}{x,v}
			\fmf{phantom,fore=black,tension=1}{u,r}
			\fmffreeze
			\fmfforce{(0.66w,0.5h)}{u}
		\end{fmfgraph}
	\end{fmffile}
\end{gathered}\hspace{-.5cm}+\tfrac{3}{4}\hspace{-.5cm}
\begin{gathered}
	\begin{fmffile}{pg3tavh2}
		\begin{fmfgraph}(130,130)
			\fmfset{dash_len}{1.2mm}
			\fmfset{wiggly_len}{1.1mm} \fmfset{dot_len}{0.5mm}
			\fmfpen{0.25mm}
			\fmftop{t}
			\fmfbottom{b}
			\fmfleft{l}
			\fmfright{r}
			\fmf{phantom,fore=black,tension=9}{t,x,v,b}
			\fmf{phantom,fore=black,tension=9}{l,s,u,r}
			\fmf{plain,fore=black,tension=.01,left}{x,v,x}
			\fmf{phantom,fore=black,tension=0.01}{s,x,s}
			\fmf{plain,fore=black,tension=1}{x,v}
			\fmf{phantom,fore=black,tension=1}{u,r}
			\fmffreeze
			\fmfforce{(0.66w,0.5h)}{u}
			\fmfv{decor.shape=square,decor.filled=shaded, decor.size=5thin}{c}
			\fmfforce{(0.5w,0.5h)}{c}
		\end{fmfgraph}
	\end{fmffile}
\end{gathered}\hspace{-.5cm}\Big)
+g^4\Big(
\tfrac{3}{8}\,\,
\begin{gathered}
	\begin{fmffile}{pg4-0pt-3loop1PIaG32}
		\begin{fmfgraph}(37,37)
			\fmfset{dash_len}{1.2mm}
			\fmfset{wiggly_len}{1.1mm} \fmfset{dot_len}{0.5mm}
			\fmfpen{0.25mm}
			\fmfsurround{a,b,c,d}
			\fmf{phantom,fore=black,tension=1,curved}{a,b,c,d,a}
			\fmf{plain,fore=black,tension=1,right}{a,b}
			\fmf{plain,fore=black,tension=1,right}{c,d}
			\fmf{plain,fore=black,tension=1,right}{b,c}
			\fmf{plain,fore=black,tension=1,right}{d,a}
			\fmf{plain,fore=black,tension=1,straight}{a,b}
			\fmf{plain,fore=black,tension=1,straight}{c,d}
		\end{fmfgraph}
	\end{fmffile}
\end{gathered}\,\,+
\tfrac{1}{4}\!\!\!\!
\begin{gathered}
	\begin{fmffile}{ppeaceh}
		\begin{fmfgraph}(100,100)
			\fmfset{dash_len}{1.2mm}
			\fmfset{wiggly_len}{1.1mm} \fmfset{dot_len}{0.5mm}
			\fmfpen{0.25mm}
			\fmfsurroundn{i}{3}
			\fmf{phantom,fore=black}{i1,v,i2}
			\fmf{phantom,fore=black}{i2,u,i3}
			\fmf{phantom,fore=black}{i3,s,i1}
			\fmfi{plain,fore=black}{fullcircle scaled .4w shifted (.5w,.5h)}
			\fmf{plain,fore=black}{v,c}
			\fmf{plain,fore=black}{u,c}
			\fmf{plain,fore=black}{s,c}
		\end{fmfgraph}
	\end{fmffile}
\end{gathered}\!\!\!\!\Big)\\
&\hspace{0.4cm}-g^2\lambda
\Big(
\tfrac{5}{8}\,\,
\begin{gathered}
	\begin{fmffile}{pg2L-3loopbubble-1PIh}
		\begin{fmfgraph}(40,40)
			\fmfset{dash_len}{1.2mm}
			\fmfset{wiggly_len}{1.1mm} \fmfset{dot_len}{0.5mm}
			\fmfpen{0.25mm}
			\fmfsurroundn{i}{3}
			\fmf{plain,fore=black,tension=1,right=1}{i1,i2}
			\fmf{plain,fore=black,tension=1,right=1}{i2,i3}
			\fmf{plain,fore=black,tension=1,right=0.8}{i3,i1}
			\fmf{plain,fore=black,tension=1}{i1,i2}
			\fmf{plain,fore=black,tension=1}{i2,i3}
		\end{fmfgraph}
	\end{fmffile}
\end{gathered}\,\,
\Big)+\lambda^2\Big(\tfrac{1}{12}\!
\begin{gathered}
	\begin{fmffile}{pL2-3loop-bubbleG42}
		\begin{fmfgraph}(75,75)
			\fmfset{dash_len}{1.2mm}
			\fmfset{wiggly_len}{1.1mm} \fmfset{dot_len}{0.5mm}
			\fmfpen{0.25mm}
			\fmfleft{i}
			\fmfright{o}
			\fmf{phantom,tension=10}{i,v1}
			\fmf{phantom,tension=10}{v2,o}
			\fmf{plain,left,tension=0.4}{v1,v2,v1}
			\fmf{plain,left=0.5}{v1,v2}
			\fmf{plain,right=0.5}{v1,v2}
    		\end{fmfgraph}
	\end{fmffile}
\end{gathered}\!
\Big)
-\tfrac{1}{2}\Big(\hspace{0.5cm}
\begin{gathered}
	\begin{fmffile}{bubble1b}
		\begin{fmfgraph}(40,40)
			\fmfset{dash_len}{1.2mm}
			\fmfset{wiggly_len}{1.1mm} \fmfset{dot_len}{0.5mm}
			\fmfpen{0.25mm}
			\fmfvn{decor.shape=square,decor.filled=shaded, decor.size=5thin}{u}{1}
			\fmfleft{i}
			\fmfright{o}
			\fmf{plain,fore=black,tension=5,left}{i,u1,i}
			\fmffreeze
			\fmfforce{(-w,0.35h)}{i}
			\fmfforce{(0w,0.35h)}{u1}
			\fmfforce{(1.1w,0.35h)}{o}
		\end{fmfgraph}\!\!\!\!
	\end{fmffile}
\end{gathered}
+
\hspace{0.6cm}
\begin{gathered}
	\begin{fmffile}{bubble2b}
		\begin{fmfgraph}(40,40)
			\fmfset{dash_len}{1.2mm}
			\fmfset{wiggly_len}{1.1mm} \fmfset{dot_len}{0.5mm}
			\fmfpen{0.25mm}
			\fmfvn{decor.shape=square,decor.filled=shaded, decor.size=5thin}{u}{2}
			\fmfleft{i}
			\fmfright{o}
			\fmf{plain,fore=black,tension=5,left}{i,u1,u2,i}
			\fmffreeze
			\fmfforce{(-w,0.35h)}{i}
			\fmfforce{(0w,0.35h)}{u1}
			\fmfforce{(-1w,0.35h)}{u2}
			\fmfforce{(1.1w,0.35h)}{o}
		\end{fmfgraph}\!\!\!\!
	\end{fmffile}
\end{gathered}\Big)
+\mathcal{O}(\ell^6).
\end{aligned}
\end{equation*}
\end{spreadlines}
We can now state the result for $Q_{\rm S}$. According to (\ref{eq:QWQS}), we should add $Q_4$ to $Q_{\rm W}$ in order to extract $Q_{\rm S}$, leading to the full vacuum contribution, $Q_{\rm S}$, in the context of strong complete normal ordering:\footnote{It might be worth noting that this can also be written as:
\vspace{-0.3cm}
\begin{equation}\label{eq:QSred}
\begin{aligned}
Q_{\rm S}&=\tfrac{1}{2}+2\bigg[g^2\Big(\tfrac{1}{12}
\hspace{-.5cm}
\begin{gathered}
	\begin{fmffile}{pg3tavh}
		\begin{fmfgraph}(130,130)
			\fmfset{dash_len}{1.2mm}
			\fmfset{wiggly_len}{1.1mm} \fmfset{dot_len}{0.5mm}
			\fmfpen{0.25mm}
			\fmftop{t}
			\fmfbottom{b}
			\fmfleft{l}
			\fmfright{r}
			\fmf{phantom,fore=black,tension=9}{t,x,v,b}
			\fmf{phantom,fore=black,tension=9}{l,s,u,r}
			\fmf{plain,fore=black,tension=.01,left}{x,v,x}
			\fmf{phantom,fore=black,tension=0.01}{s,x,s}
			\fmf{plain,fore=black,tension=1}{x,v}
			\fmf{phantom,fore=black,tension=1}{u,r}
			\fmffreeze
			\fmfforce{(0.66w,0.5h)}{u}
		\end{fmfgraph}
	\end{fmffile}
\end{gathered}\hspace{-.5cm}\Big)\bigg]
+3\bigg[g^2\Big(\tfrac{3}{12}\hspace{-.5cm}
\begin{gathered}
	\begin{fmffile}{pg3tavh2}
		\begin{fmfgraph}(130,130)
			\fmfset{dash_len}{1.2mm}
			\fmfset{wiggly_len}{1.1mm} \fmfset{dot_len}{0.5mm}
			\fmfpen{0.25mm}
			\fmftop{t}
			\fmfbottom{b}
			\fmfleft{l}
			\fmfright{r}
			\fmf{phantom,fore=black,tension=9}{t,x,v,b}
			\fmf{phantom,fore=black,tension=9}{l,s,u,r}
			\fmf{plain,fore=black,tension=.01,left}{x,v,x}
			\fmf{phantom,fore=black,tension=0.01}{s,x,s}
			\fmf{plain,fore=black,tension=1}{x,v}
			\fmf{phantom,fore=black,tension=1}{u,r}
			\fmffreeze
			\fmfforce{(0.66w,0.5h)}{u}
			\fmfv{decor.shape=square,decor.filled=shaded, decor.size=5thin}{c}
			\fmfforce{(0.5w,0.5h)}{c}
		\end{fmfgraph}
	\end{fmffile}
\end{gathered}\hspace{-.5cm}\Big)+g^4\Big(
\tfrac{1}{16}\,\,
\begin{gathered}
	\begin{fmffile}{pg4-0pt-3loop1PIaG32}
		\begin{fmfgraph}(37,37)
			\fmfset{dash_len}{1.2mm}
			\fmfset{wiggly_len}{1.1mm} \fmfset{dot_len}{0.5mm}
			\fmfpen{0.25mm}
			\fmfsurround{a,b,c,d}
			\fmf{phantom,fore=black,tension=1,curved}{a,b,c,d,a}
			\fmf{plain,fore=black,tension=1,right}{a,b}
			\fmf{plain,fore=black,tension=1,right}{c,d}
			\fmf{plain,fore=black,tension=1,right}{b,c}
			\fmf{plain,fore=black,tension=1,right}{d,a}
			\fmf{plain,fore=black,tension=1,straight}{a,b}
			\fmf{plain,fore=black,tension=1,straight}{c,d}
		\end{fmfgraph}
	\end{fmffile}
\end{gathered}\,\,+
\tfrac{1}{4!}\!\!\!\!
\begin{gathered}
	\begin{fmffile}{ppeaceh}
		\begin{fmfgraph}(100,100)
			\fmfset{dash_len}{1.2mm}
			\fmfset{wiggly_len}{1.1mm} \fmfset{dot_len}{0.5mm}
			\fmfpen{0.25mm}
			\fmfsurroundn{i}{3}
			\fmf{phantom,fore=black}{i1,v,i2}
			\fmf{phantom,fore=black}{i2,u,i3}
			\fmf{phantom,fore=black}{i3,s,i1}
			\fmfi{plain,fore=black}{fullcircle scaled .4w shifted (.5w,.5h)}
			\fmf{plain,fore=black}{v,c}
			\fmf{plain,fore=black}{u,c}
			\fmf{plain,fore=black}{s,c}
		\end{fmfgraph}
	\end{fmffile}
\end{gathered}\!\!\!\!\Big)\\[-0.3cm]
&\hspace{0.4cm}
-g^2\lambda
\Big(
\tfrac{1}{8}\,\,
\begin{gathered}
	\begin{fmffile}{pg2L-3loopbubble-1PIh}
		\begin{fmfgraph}(40,40)
			\fmfset{dash_len}{1.2mm}
			\fmfset{wiggly_len}{1.1mm} \fmfset{dot_len}{0.5mm}
			\fmfpen{0.25mm}
			\fmfsurroundn{i}{3}
			\fmf{plain,fore=black,tension=1,right=1}{i1,i2}
			\fmf{plain,fore=black,tension=1,right=1}{i2,i3}
			\fmf{plain,fore=black,tension=1,right=0.8}{i3,i1}
			\fmf{plain,fore=black,tension=1}{i1,i2}
			\fmf{plain,fore=black,tension=1}{i2,i3}
		\end{fmfgraph}
	\end{fmffile}
\end{gathered}\,\,
\Big)
+\lambda^2\Big(\tfrac{1}{48}\!
\begin{gathered}
	\begin{fmffile}{pL2-3loop-bubbleG42}
		\begin{fmfgraph}(75,75)
			\fmfset{dash_len}{1.2mm}
			\fmfset{wiggly_len}{1.1mm} \fmfset{dot_len}{0.5mm}
			\fmfpen{0.25mm}
			\fmfleft{i}
			\fmfright{o}
			\fmf{phantom,tension=10}{i,v1}
			\fmf{phantom,tension=10}{v2,o}
			\fmf{plain,left,tension=0.4}{v1,v2,v1}
			\fmf{plain,left=0.5}{v1,v2}
			\fmf{plain,right=0.5}{v1,v2}
    		\end{fmfgraph}
	\end{fmffile}
\end{gathered}\!
\Big)\bigg]
+\tfrac{1}{2}\Big(\hspace{0.5cm}
\begin{gathered}
	\begin{fmffile}{bubble1b}
		\begin{fmfgraph}(40,40)
			\fmfset{dash_len}{1.2mm}
			\fmfset{wiggly_len}{1.1mm} \fmfset{dot_len}{0.5mm}
			\fmfpen{0.25mm}
			\fmfvn{decor.shape=square,decor.filled=shaded, decor.size=5thin}{u}{1}
			\fmfleft{i}
			\fmfright{o}
			\fmf{plain,fore=black,tension=5,left}{i,u1,i}
			\fmffreeze
			\fmfforce{(-w,0.35h)}{i}
			\fmfforce{(0w,0.35h)}{u1}
			\fmfforce{(1.1w,0.35h)}{o}
		\end{fmfgraph}\!\!\!\!
	\end{fmffile}
\end{gathered}
+\tfrac{1}{2}
\hspace{0.6cm}
\begin{gathered}
	\begin{fmffile}{bubble2b}
		\begin{fmfgraph}(40,40)
			\fmfset{dash_len}{1.2mm}
			\fmfset{wiggly_len}{1.1mm} \fmfset{dot_len}{0.5mm}
			\fmfpen{0.25mm}
			\fmfvn{decor.shape=square,decor.filled=shaded, decor.size=5thin}{u}{2}
			\fmfleft{i}
			\fmfright{o}
			\fmf{plain,fore=black,tension=5,left}{i,u1,u2,i}
			\fmffreeze
			\fmfforce{(-w,0.35h)}{i}
			\fmfforce{(0w,0.35h)}{u1}
			\fmfforce{(-1w,0.35h)}{u2}
			\fmfforce{(1.1w,0.35h)}{o}
		\end{fmfgraph}\!\!\!\!
	\end{fmffile}
\end{gathered}\Big)-\int\frac{1}{2}\!\wf \!G_2
+\mathcal{O}(\ell^6),
\end{aligned}
\end{equation}
which makes it clear what we would have found had we not completely normal ordered the wave-function renormalisation contribution to the kinetic term, i.e., if we had dropped $-\int\frac{1}{2}\!\wf \!G_2$ from (\ref{eq:Q4a}).
}
\vspace{-0.3cm}
\begin{spreadlines}{-0.3\baselineskip}
\begin{equation}\label{eq:QS}
\begin{aligned}
Q_{\rm S}&=\tfrac{1}{2}+2\bigg[g^2\Big(\tfrac{1}{12}
\hspace{-.5cm}
\begin{gathered}
	\begin{fmffile}{pg3tavh}
		\begin{fmfgraph}(130,130)
			\fmfset{dash_len}{1.2mm}
			\fmfset{wiggly_len}{1.1mm} \fmfset{dot_len}{0.5mm}
			\fmfpen{0.25mm}
			\fmftop{t}
			\fmfbottom{b}
			\fmfleft{l}
			\fmfright{r}
			\fmf{phantom,fore=black,tension=9}{t,x,v,b}
			\fmf{phantom,fore=black,tension=9}{l,s,u,r}
			\fmf{plain,fore=black,tension=.01,left}{x,v,x}
			\fmf{phantom,fore=black,tension=0.01}{s,x,s}
			\fmf{plain,fore=black,tension=1}{x,v}
			\fmf{phantom,fore=black,tension=1}{u,r}
			\fmffreeze
			\fmfforce{(0.66w,0.5h)}{u}
		\end{fmfgraph}
	\end{fmffile}
\end{gathered}\hspace{-.5cm}+\tfrac{3}{12}\hspace{-.5cm}
\begin{gathered}
	\begin{fmffile}{pg3tavh2}
		\begin{fmfgraph}(130,130)
			\fmfset{dash_len}{1.2mm}
			\fmfset{wiggly_len}{1.1mm} \fmfset{dot_len}{0.5mm}
			\fmfpen{0.25mm}
			\fmftop{t}
			\fmfbottom{b}
			\fmfleft{l}
			\fmfright{r}
			\fmf{phantom,fore=black,tension=9}{t,x,v,b}
			\fmf{phantom,fore=black,tension=9}{l,s,u,r}
			\fmf{plain,fore=black,tension=.01,left}{x,v,x}
			\fmf{phantom,fore=black,tension=0.01}{s,x,s}
			\fmf{plain,fore=black,tension=1}{x,v}
			\fmf{phantom,fore=black,tension=1}{u,r}
			\fmffreeze
			\fmfforce{(0.66w,0.5h)}{u}
			\fmfv{decor.shape=square,decor.filled=shaded, decor.size=5thin}{c}
			\fmfforce{(0.5w,0.5h)}{c}
		\end{fmfgraph}
	\end{fmffile}
\end{gathered}\hspace{-.5cm}\Big)\bigg]
+3\bigg[g^4\Big(
\tfrac{1}{16}\,\,
\begin{gathered}
	\begin{fmffile}{pg4-0pt-3loop1PIaG32}
		\begin{fmfgraph}(37,37)
			\fmfset{dash_len}{1.2mm}
			\fmfset{wiggly_len}{1.1mm} \fmfset{dot_len}{0.5mm}
			\fmfpen{0.25mm}
			\fmfsurround{a,b,c,d}
			\fmf{phantom,fore=black,tension=1,curved}{a,b,c,d,a}
			\fmf{plain,fore=black,tension=1,right}{a,b}
			\fmf{plain,fore=black,tension=1,right}{c,d}
			\fmf{plain,fore=black,tension=1,right}{b,c}
			\fmf{plain,fore=black,tension=1,right}{d,a}
			\fmf{plain,fore=black,tension=1,straight}{a,b}
			\fmf{plain,fore=black,tension=1,straight}{c,d}
		\end{fmfgraph}
	\end{fmffile}
\end{gathered}\,\,+
\tfrac{1}{4!}\!\!\!\!
\begin{gathered}
	\begin{fmffile}{ppeaceh}
		\begin{fmfgraph}(100,100)
			\fmfset{dash_len}{1.2mm}
			\fmfset{wiggly_len}{1.1mm} \fmfset{dot_len}{0.5mm}
			\fmfpen{0.25mm}
			\fmfsurroundn{i}{3}
			\fmf{phantom,fore=black}{i1,v,i2}
			\fmf{phantom,fore=black}{i2,u,i3}
			\fmf{phantom,fore=black}{i3,s,i1}
			\fmfi{plain,fore=black}{fullcircle scaled .4w shifted (.5w,.5h)}
			\fmf{plain,fore=black}{v,c}
			\fmf{plain,fore=black}{u,c}
			\fmf{plain,fore=black}{s,c}
		\end{fmfgraph}
	\end{fmffile}
\end{gathered}\!\!\!\!\Big)\\
&\hspace{0.4cm}
-g^2\lambda
\Big(
\tfrac{1}{8}\,\,
\begin{gathered}
	\begin{fmffile}{pg2L-3loopbubble-1PIh}
		\begin{fmfgraph}(40,40)
			\fmfset{dash_len}{1.2mm}
			\fmfset{wiggly_len}{1.1mm} \fmfset{dot_len}{0.5mm}
			\fmfpen{0.25mm}
			\fmfsurroundn{i}{3}
			\fmf{plain,fore=black,tension=1,right=1}{i1,i2}
			\fmf{plain,fore=black,tension=1,right=1}{i2,i3}
			\fmf{plain,fore=black,tension=1,right=0.8}{i3,i1}
			\fmf{plain,fore=black,tension=1}{i1,i2}
			\fmf{plain,fore=black,tension=1}{i2,i3}
		\end{fmfgraph}
	\end{fmffile}
\end{gathered}\,\,
\Big)
+\lambda^2\Big(\tfrac{1}{48}\!
\begin{gathered}
	\begin{fmffile}{pL2-3loop-bubbleG42}
		\begin{fmfgraph}(75,75)
			\fmfset{dash_len}{1.2mm}
			\fmfset{wiggly_len}{1.1mm} \fmfset{dot_len}{0.5mm}
			\fmfpen{0.25mm}
			\fmfleft{i}
			\fmfright{o}
			\fmf{phantom,tension=10}{i,v1}
			\fmf{phantom,tension=10}{v2,o}
			\fmf{plain,left,tension=0.4}{v1,v2,v1}
			\fmf{plain,left=0.5}{v1,v2}
			\fmf{plain,right=0.5}{v1,v2}
    		\end{fmfgraph}
	\end{fmffile}
\end{gathered}\!
\Big)\bigg]
-\tfrac{1}{4}\Big(
\hspace{0.6cm}
\begin{gathered}
	\begin{fmffile}{bubble2b}
		\begin{fmfgraph}(40,40)
			\fmfset{dash_len}{1.2mm}
			\fmfset{wiggly_len}{1.1mm} \fmfset{dot_len}{0.5mm}
			\fmfpen{0.25mm}
			\fmfvn{decor.shape=square,decor.filled=shaded, decor.size=5thin}{u}{2}
			\fmfleft{i}
			\fmfright{o}
			\fmf{plain,fore=black,tension=5,left}{i,u1,u2,i}
			\fmffreeze
			\fmfforce{(-w,0.35h)}{i}
			\fmfforce{(0w,0.35h)}{u1}
			\fmfforce{(-1w,0.35h)}{u2}
			\fmfforce{(1.1w,0.35h)}{o}
		\end{fmfgraph}\!\!\!\!
	\end{fmffile}
\end{gathered}\Big)
+\mathcal{O}(\ell^6).
\end{aligned}
\end{equation}
\end{spreadlines}
Clearly, both $Q_{\rm W}$ and $Q_{\rm S}$ are free from cephalopods. 
Curiously however, all the coefficients appearing within the square brackets have an obvious combinatorial interpretation and agree with those expected from (\ref{eq:W(Jwiggly)}) (and with the correct signs) \emph{up to} the overall factor of 2 for the 2-loop diagrams and a factor of 3 for the 3-loop diagrams. This suggests that, upon strong complete normal ordering, in addition to cancelling all cephalopods, the subtractions are reshuffling the various terms such that the $L$-loop vacuum diagrams are multiplied by a factor of $L$.  It would be interesting to check to what extent this holds true at higher loop orders. Note also that the first term in $Q_{\rm S}$, namely $1/2$, is metric-independent and can be absorbed into a redefinition of $N'$ displayed below (\ref{eq:U(J)expo defn}), $N'\rightarrow N_0=\sqrt{e}N'$.

The wave-function renormalisation, $\delta Z$, is scheme-dependent and can, for example, be fixed by requiring that the kinetic term of the quantum effective action has canonical normalisation, $\frac{1}{\hbar}\Gamma(\varphi)=\int \frac{\hbar^2}{2}(\nabla\varphi)^2+\dots$

\subsection{Completely Normal Ordered Generating Function}\label{sec:CNOGF}
\vspace{0.4cm}
Let us now return to the full renormalised generating function of connected Green functions, $W(J)$, see (\ref{eq:W(Jwiggly)}). The objective is now to substitute the mass counterterms into the dressed propagators that appear in (\ref{eq:W(Jwiggly)}), and the coupling counterterms (\ref{eq:delta g etc b}) that appear in the explicit couplings $\hat{g}$, $\hat{\lambda}$, etc., in (\ref{eq:W(Jwiggly)}). 

In particular, making all counterterms (except for $\delta Z$) explicit in (\ref{eq:W(Jwiggly)}) and dropping terms of $\mathcal{O}(\ell^5)$ leads to
\begin{equation}\label{eq:W(Jdashes)a}
\begin{aligned}
\tfrac{1}{\hbar}&W(J) = \tfrac{1}{2}
\Big(\hspace{0.4cm}
\begin{gathered}
	\begin{fmffile}{dressed-plain0r}
		\begin{fmfgraph}(30,30)
			\fmfset{dash_len}{1.2mm}
			\fmfset{wiggly_len}{1.1mm} \fmfset{dot_len}{0.5mm}
			\fmfpen{0.25mm}
			\fmfleft{i}
			\fmfright{o}
			\fmf{dashes,fore=black}{i,o}
			\fmffreeze
			\fmfforce{(-1w,0.35h)}{i}
			\fmfforce{(1w,0.35h)}{o}
		\end{fmfgraph}
	\end{fmffile}
\end{gathered}\,\,
+\hspace{0.4cm}
\begin{gathered}
	\begin{fmffile}{dressed-plain1r}
		\begin{fmfgraph}(30,30)
			\fmfset{dash_len}{1.2mm}
			\fmfset{wiggly_len}{1.1mm} \fmfset{dot_len}{0.5mm}
			\fmfpen{0.25mm}
			\fmfvn{decor.shape=circle,decor.filled=shaded, decor.size=3.5thin}{u}{1}
			\fmfleft{i}
			\fmfright{o}
			\fmf{dashes,fore=black,tension=5}{i,u1,o}
			\fmffreeze
			\fmfforce{(-w,0.35h)}{i}
			\fmfforce{(0w,0.35h)}{u1}
			\fmfforce{(1.1w,0.35h)}{o}
		\end{fmfgraph}
	\end{fmffile}
\end{gathered}\hspace{0.1cm}
+\hspace{0.4cm}
\begin{gathered}
	\begin{fmffile}{dressed-plain2r}
		\begin{fmfgraph}(30,30)
			\fmfset{dash_len}{1.2mm}
			\fmfset{wiggly_len}{1.1mm} \fmfset{dot_len}{0.5mm}
			\fmfpen{0.25mm}
			\fmfvn{decor.shape=circle,decor.filled=shaded, decor.size=3.5thin}{u}{2}
			\fmfleft{i}
			\fmfright{o}
			\fmf{dashes,fore=black,tension=5}{i,u1,u2,o}
			\fmffreeze
			\fmfforce{(-w,0.35h)}{i}
			\fmfforce{(0w,0.35h)}{u1}
			\fmfforce{(1w,0.35h)}{u2}
			\fmfforce{(2w,0.35h)}{o}
		\end{fmfgraph}
	\end{fmffile}
\end{gathered}\hspace{0.4cm}\Big)
-\Big[g-\kappa\big(\tfrac{1}{2}\hspace{0.4cm}
\begin{gathered}
	\begin{fmffile}{dbubblex2h}
		\begin{fmfgraph}(30,30)
			\fmfset{dash_len}{1.2mm}
			\fmfset{wiggly_len}{1.1mm} \fmfset{dot_len}{0.5mm}
			\fmfpen{0.25mm}
			\fmfvn{decor.shape=hexagram,decor.filled=full, decor.size=3thin}{u}{1}
			\fmfleft{i}
			\fmfright{o}
			\fmf{dashes,fore=black,tension=5,left}{i,u1,i}
			\fmffreeze
			\fmfforce{(-w,0.35h)}{i}
			\fmfforce{(0w,0.35h)}{u1}
			\fmfforce{(1.1w,0.35h)}{o}
		\end{fmfgraph}\!\!\!\!\!\!
	\end{fmffile}
\end{gathered}\,\,\big)\Big]
\Big(
\tfrac{1}{3!}
\begin{gathered}
	\begin{fmffile}{dg-3p-0dm2}
		\begin{fmfgraph}(40,40)
			\fmfset{dash_len}{1.2mm}
			\fmfset{wiggly_len}{1.1mm} \fmfset{dot_len}{0.5mm}
			\fmfpen{0.25mm}
			\fmfleft{i}
			\fmfright{o1,o2}
			\fmf{dashes,fore=black,tension=5}{i,v1}
			\fmf{dashes,fore=black,tension=5}{v1,o1}
			\fmf{dashes,fore=black,tension=5}{v1,o2}
			\fmfv{decor.shape=hexagram,decor.filled=full, decor.size=3thin}{r}
			\fmfforce{(0.6w,0.5h)}{r}
		\end{fmfgraph}
	\end{fmffile}
\end{gathered}
+\tfrac{1}{2}\,\,\,
\begin{gathered}
	\begin{fmffile}{dg-3p-1dm2}
		\begin{fmfgraph}(40,40)
			\fmfset{dash_len}{1.2mm}
			\fmfset{wiggly_len}{1.1mm} \fmfset{dot_len}{0.5mm}
			\fmfpen{0.25mm}
			\fmfleft{i}
			\fmfright{o1,o2}
			\fmf{dashes,fore=black,tension=5}{i,v1}
			\fmf{dashes,fore=black,tension=5}{v1,o1}
			\fmf{dashes,fore=black,tension=5}{v1,o2}
			\fmffreeze
			\fmfforce{(-0.3w,0.5h)}{i}
			\fmfv{decor.shape=circle,decor.filled=shaded, decor.size=3.5thin}{u}
			\fmfforce{(0.17w,0.5h)}{u}
			\fmfv{decor.shape=hexagram,decor.filled=full, decor.size=3thin}{r}
			\fmfforce{(0.6w,0.5h)}{r}
		\end{fmfgraph}
	\end{fmffile}
\end{gathered}
\Big)\\
&+\Big[g-\kappa\big(\tfrac{1}{2}\hspace{0.4cm}
\begin{gathered}
	\begin{fmffile}{dbubblex2h}
		\begin{fmfgraph}(30,30)
			\fmfset{dash_len}{1.2mm}
			\fmfset{wiggly_len}{1.1mm} \fmfset{dot_len}{0.5mm}
			\fmfpen{0.25mm}
			\fmfvn{decor.shape=hexagram,decor.filled=full, decor.size=3thin}{u}{1}
			\fmfleft{i}
			\fmfright{o}
			\fmf{dashes,fore=black,tension=5,left}{i,u1,i}
			\fmffreeze
			\fmfforce{(-w,0.35h)}{i}
			\fmfforce{(0w,0.35h)}{u1}
			\fmfforce{(1.1w,0.35h)}{o}
		\end{fmfgraph}\!\!\!\!\!\!
	\end{fmffile}
\end{gathered}\,\,\big)\Big]^2
\Big(
\tfrac{1}{12}\!
\begin{gathered}
	\begin{fmffile}{dg2-2loopbubble-1PI-0dm2}
		\begin{fmfgraph}(50,50)
			\fmfset{dash_len}{1.2mm}
			\fmfset{wiggly_len}{1.1mm} \fmfset{dot_len}{0.5mm}
			\fmfpen{0.25mm}
			\fmfleft{i}
			\fmfright{o}
			\fmf{phantom,tension=5}{i,v1}
			\fmf{phantom,tension=5}{v2,o}
			\fmf{dashes,fore=black,left,tension=0.4}{v1,v2,v1}
			\fmf{dashes,fore=black}{v1,v2}
			\fmffreeze
			\fmfv{decor.shape=hexagram,decor.filled=full, decor.size=3thin}{r}
			\fmfforce{(0.2w,0.5h)}{r}
		\end{fmfgraph}
	\end{fmffile}
\end{gathered}\!
+\tfrac{1}{4}\!
\begin{gathered}
	\begin{fmffile}{dg2-2loopbubble-1PI-1dm2}
		\begin{fmfgraph}(60,60)
			\fmfset{dash_len}{1.2mm}
			\fmfset{wiggly_len}{1.1mm} \fmfset{dot_len}{0.5mm}
			\fmfpen{0.25mm}
			\fmfleft{i}
			\fmfright{o}
			\fmf{phantom,tension=5}{i,v1}
			\fmf{phantom,tension=5}{v2,o}
			\fmf{dashes,fore=black,left,tension=0.4}{v1,v2,v1}
			\fmf{dashes,fore=black}{v1,v2}
			\fmffreeze
			\fmfv{decor.shape=circle,decor.filled=shaded, decor.size=3.5thin}{u}
			\fmfforce{(0.5w,0.5h)}{u}
			\fmffreeze
			\fmfv{decor.shape=hexagram,decor.filled=full, decor.size=3thin}{r}
			\fmfforce{(0.2w,0.5h)}{r}
		\end{fmfgraph}
	\end{fmffile}
\end{gathered}\!
+\tfrac{1}{4}\,   
\begin{gathered}
	\begin{fmffile}{dg2-2pt-0dm2}
		\begin{fmfgraph}(65,65)
			\fmfset{dash_len}{1.2mm}
			\fmfset{wiggly_len}{1.1mm} \fmfset{dot_len}{0.5mm}
			\fmfpen{0.25mm}
			\fmfleft{i}
			\fmfright{o}
			\fmf{dashes,fore=black,tension=1}{i,v1}
			\fmf{dashes,fore=black,tension=1}{v2,o}
			\fmf{dashes,fore=black,left,tension=0.4}{v1,v2,v1}
			\fmffreeze
			\fmfforce{(-.12w,0.5h)}{i}
			\fmfforce{(1.1w,0.5h)}{o}
			\fmfv{decor.shape=hexagram,decor.filled=full, decor.size=3thin}{r}
			\fmfforce{(0.3w,0.5h)}{r}
		\end{fmfgraph}
	\end{fmffile}
\end{gathered}\,
+\tfrac{1}{2}\,   
\begin{gathered}
	\begin{fmffile}{dg2-2pt-1dm2a}
		\begin{fmfgraph}(65,65)
			\fmfset{dash_len}{1.2mm}
			\fmfset{wiggly_len}{1.1mm} \fmfset{dot_len}{0.5mm}
			\fmfpen{0.25mm}
			\fmfleft{i}
			\fmfright{o}
			\fmf{dashes,fore=black,tension=1}{i,v1}
			\fmf{dashes,fore=black,tension=1}{v2,o}
			\fmf{dashes,fore=black,left,tension=0.4}{v1,v2,v1}
			\fmffreeze
			\fmfforce{(-.12w,0.5h)}{i}
			\fmfforce{(1.1w,0.5h)}{o}
			\fmfv{decor.shape=hexagram,decor.filled=full, decor.size=3thin}{r}
			\fmfforce{(0.3w,0.5h)}{r}
			\fmfv{decor.shape=circle,decor.filled=shaded, decor.size=3.5thin}{m}
			\fmfforce{(0.5w,0.7h)}{m}
		\end{fmfgraph}
	\end{fmffile}
\end{gathered}\,
+\tfrac{1}{4}\,\,\,\,\,
\begin{gathered}
	\begin{fmffile}{dg2-2pt-1dm2b}
		\begin{fmfgraph}(60,60)
			\fmfset{dash_len}{1.2mm}
			\fmfset{wiggly_len}{1.1mm} \fmfset{dot_len}{0.5mm}
			\fmfpen{0.25mm}
			\fmfleft{i}
			\fmfright{o}
			\fmf{dashes,fore=black,tension=1}{i,v1}
			\fmf{dashes,fore=black,tension=1}{v2,o}
			\fmf{dashes,fore=black,left,tension=0.4}{v1,v2,v1}
			\fmffreeze
			\fmfforce{(-.5w,0.5h)}{i}
			\fmfforce{(1.1w,0.5h)}{o}
			\fmfv{decor.shape=circle,decor.filled=shaded, decor.size=3.5thin}{m}
			\fmfforce{(-0.12w,0.5h)}{m}
			\fmfv{decor.shape=hexagram,decor.filled=full, decor.size=3thin}{r}
			\fmfforce{(0.33w,0.5h)}{r}
		\end{fmfgraph}
	\end{fmffile}
\end{gathered}\,
+\tfrac{1}{4}\,\,\,\,\,
\begin{gathered}
	\begin{fmffile}{dg2-2pt-1dm2bb}
		\begin{fmfgraph}(60,60)
			\fmfset{dash_len}{1.2mm}
			\fmfset{wiggly_len}{1.1mm} \fmfset{dot_len}{0.5mm}
			\fmfpen{0.25mm}
			\fmfleft{i}
			\fmfright{o}
			\fmf{dashes,fore=black,tension=1}{i,v1}
			\fmf{dashes,fore=black,tension=1}{v2,o}
			\fmf{dashes,fore=black,left,tension=0.4}{v1,v2,v1}
			\fmffreeze
			\fmfforce{(-.5w,0.5h)}{i}
			\fmfforce{(1.1w,0.5h)}{o}
			\fmfv{decor.shape=circle,decor.filled=shaded, decor.size=3.5thin}{m}
			\fmfforce{(-0.12w,0.5h)}{m}
			\fmfv{decor.shape=hexagram,decor.filled=full, decor.size=3thin}{r}
			\fmfforce{(0.66w,0.5h)}{r}
		\end{fmfgraph}
	\end{fmffile}
\end{gathered}\,\\
&+\tfrac{1}{8}
\begin{gathered}
	\begin{fmffile}{d2-2_g2-0dm2}
		\begin{fmfgraph}(65,65)
			\fmfset{dash_len}{1.2mm}
			\fmfset{wiggly_len}{1.1mm} \fmfset{dot_len}{0.5mm}
			\fmfpen{0.25mm}
			\fmfsurroundn{i}{4}
			\fmf{dashes,fore=black}{i1,n,m,i4}
			\fmf{dashes,fore=black}{i2,n}
			\fmf{dashes,fore=black}{m,i3}
			\fmfv{decor.shape=hexagram,decor.filled=full, decor.size=3thin}{r}
			\fmfforce{(0.64w,0.61h)}{r}
		\end{fmfgraph}
	\end{fmffile}
\end{gathered}
+\tfrac{1}{8}
\begin{gathered}
	\begin{fmffile}{d2-2_g2-1dm2a}
		\begin{fmfgraph}(70,70)
			\fmfset{dash_len}{1.2mm}
			\fmfset{wiggly_len}{1.1mm} \fmfset{dot_len}{0.5mm}
			\fmfpen{0.25mm}
			\fmfsurroundn{i}{4}
			\fmf{dashes,fore=black}{i1,n,m,i4}
			\fmf{dashes,fore=black}{i2,n}
			\fmf{dashes,fore=black}{m,i3}
			\fmffreeze
			\fmfv{decor.shape=hexagram,decor.filled=full, decor.size=3thin}{r}
			\fmfforce{(0.62w,0.61h)}{r}
			\fmfv{decor.shape=circle,decor.filled=shaded, decor.size=3.5thin}{g}
			\fmfforce{(0.5w,0.5h)}{g}
		\end{fmfgraph}
	\end{fmffile}
\end{gathered}
+\tfrac{1}{4}\,\,\,
\begin{gathered}
	\begin{fmffile}{d2-2_g2-1dm2b}
		\begin{fmfgraph}(70,70)
			\fmfset{dash_len}{1.2mm}
			\fmfset{wiggly_len}{1.1mm} \fmfset{dot_len}{0.5mm}
			\fmfpen{0.25mm}
			\fmfsurroundn{i}{4}
			\fmf{dashes,fore=black}{i1,n,m,i4}
			\fmf{dashes,fore=black}{i2,n}
			\fmf{dashes,fore=black}{m,i3}
			\fmffreeze
			\fmfv{decor.shape=hexagram,decor.filled=full, decor.size=3thin}{r}
			\fmfforce{(0.64w,0.61h)}{r}
			\fmfv{decor.shape=circle,decor.filled=shaded, decor.size=3.5thin}{g}
			\fmfforce{(0.1w,0.43h)}{g}
			\fmfforce{(-0.3w,0.5h)}{i3}
		\end{fmfgraph}
	\end{fmffile}
\end{gathered}
+\tfrac{1}{4}\,\,\,
\begin{gathered}
	\begin{fmffile}{d2-2_g2-1dm2c}
		\begin{fmfgraph}(70,70)
			\fmfset{dash_len}{1.2mm}
			\fmfset{wiggly_len}{1.1mm} \fmfset{dot_len}{0.5mm}
			\fmfpen{0.25mm}
			\fmfsurroundn{i}{4}
			\fmf{dashes,fore=black}{i1,n,m,i4}
			\fmf{dashes,fore=black}{i2,n}
			\fmf{dashes,fore=black}{m,i3}
			\fmffreeze
			\fmfv{decor.shape=hexagram,decor.filled=full, decor.size=3thin}{m}
			\fmfv{decor.shape=circle,decor.filled=shaded, decor.size=3.5thin}{g}
			\fmfforce{(0.1w,0.43h)}{g}
			\fmfforce{(-0.3w,0.5h)}{i3}
		\end{fmfgraph}
	\end{fmffile}
\end{gathered}
\Big)
-\Big[\lambda-\gamma\big(\tfrac{1}{2}\hspace{0.4cm}
\begin{gathered}
	\begin{fmffile}{dbubblex2h}
		\begin{fmfgraph}(30,30)
			\fmfset{dash_len}{1.2mm}
			\fmfset{wiggly_len}{1.1mm} \fmfset{dot_len}{0.5mm}
			\fmfpen{0.25mm}
			\fmfvn{decor.shape=hexagram,decor.filled=full, decor.size=3thin}{u}{1}
			\fmfleft{i}
			\fmfright{o}
			\fmf{dashes,fore=black,tension=5,left}{i,u1,i}
			\fmffreeze
			\fmfforce{(-w,0.35h)}{i}
			\fmfforce{(0w,0.35h)}{u1}
			\fmfforce{(1.1w,0.35h)}{o}
		\end{fmfgraph}\!\!\!\!\!\!
	\end{fmffile}
\end{gathered}\,\,\big)\Big]
\Big(
\tfrac{1}{8}
\begin{gathered}
	\begin{fmffile}{dlambdabubble-0dm2}
		\begin{fmfgraph}(45,45)
			\fmfset{dash_len}{1.2mm}
			\fmfset{wiggly_len}{1.1mm} \fmfset{dot_len}{0.5mm}
			\fmfpen{0.25mm}
			\fmftop{t1,t2,t3}
			\fmfbottom{b1,b2,b3}
			\fmf{phantom}{t1,v1,b1}
			\fmf{phantom}{t2,v2,b2}
			\fmf{phantom}{t3,v3,b3}
			\fmffreeze
			\fmf{dashes,fore=black,right}{v1,v2,v1}
			\fmf{dashes,fore=black,right}{v2,v3,v2}
			\fmfv{decor.shape=hexagram,decor.filled=full, decor.size=3thin}{v2}
		\end{fmfgraph}
	\end{fmffile}
\end{gathered}
+\tfrac{1}{4}
\begin{gathered}
	\begin{fmffile}{dlambdabubble-1dm2}
		\begin{fmfgraph}(55,55)
			\fmfset{dash_len}{1.2mm}
			\fmfset{wiggly_len}{1.1mm} \fmfset{dot_len}{0.5mm}
			\fmfpen{0.25mm}
			\fmftop{t1,t2,t3}
			\fmfbottom{b1,b2,b3}
			\fmf{phantom}{t1,v1,b1}
			\fmf{phantom}{t2,v2,b2}
			\fmf{phantom}{t3,v3,b3}
			\fmffreeze
			\fmf{dashes,fore=black,right}{v1,v2,v1}
			\fmf{dashes,fore=black,right}{v2,v3,v2}
			\fmfv{decor.shape=circle,decor.filled=shaded, decor.size=3.5thin}{m}
			\fmfforce{(0w,0.5h)}{m}
			\fmfv{decor.shape=hexagram,decor.filled=full, decor.size=3thin}{v2}
		\end{fmfgraph}
	\end{fmffile}
\end{gathered}
\\
&+\tfrac{1}{4}
\begin{gathered}
	\begin{fmffile}{dlambdaself-0dm2}
		\begin{fmfgraph}(45,45)
			\fmfset{dash_len}{1.2mm}
			\fmfset{wiggly_len}{1.1mm} \fmfset{dot_len}{0.5mm}
			\fmfpen{0.25mm}
			\fmftop{s}
			\fmfleft{a}
			\fmfright{b}
			\fmf{dashes,fore=black}{a,v}
			\fmf{dashes,fore=black}{b,v}
			\fmf{dashes,fore=black,right,tension=.7}{v,v}
			\fmffreeze
			\fmfforce{(0w,0.2h)}{a}
			\fmfforce{(w,0.2h)}{b}
			\fmfv{decor.shape=hexagram,decor.filled=full, decor.size=3thin}{v}
		\end{fmfgraph}
	\end{fmffile}
\end{gathered}
+\tfrac{1}{2}
\begin{gathered}
	\begin{fmffile}{dlambdaself-1dm2a}
		\begin{fmfgraph}(45,45)
			\fmfset{dash_len}{1.2mm}
			\fmfset{wiggly_len}{1.1mm} \fmfset{dot_len}{0.5mm}
			\fmfpen{0.25mm}
			\fmftop{s}
			\fmfleft{a}
			\fmfright{b}
			\fmf{dashes,fore=black}{a,v}
			\fmf{dashes,fore=black}{b,v}
			\fmf{dashes,fore=black,right,tension=.7}{v,v}
			\fmffreeze
			\fmfforce{(0w,0.2h)}{a}
			\fmfforce{(w,0.2h)}{b}
			\fmfv{decor.shape=circle,decor.filled=shaded, decor.size=3.5thin}{m}
			\fmfforce{(0.23w,0.32h)}{m}
			\fmfv{decor.shape=hexagram,decor.filled=full, decor.size=3thin}{v}
		\end{fmfgraph}
	\end{fmffile}
\end{gathered}
+\tfrac{1}{4}
\begin{gathered}
	\begin{fmffile}{dlambdaself-1dm2b}
		\begin{fmfgraph}(45,45)
			\fmfset{dash_len}{1.2mm}
			\fmfset{wiggly_len}{1.1mm} \fmfset{dot_len}{0.5mm}
			\fmfpen{0.25mm}
			\fmftop{s}
			\fmfleft{a}
			\fmfright{b}
			\fmf{dashes,fore=black}{a,v}
			\fmf{dashes,fore=black}{b,v}
			\fmf{dashes,fore=black,right,tension=.7}{v,v}
			\fmffreeze
			\fmfforce{(0w,0.2h)}{a}
			\fmfforce{(w,0.2h)}{b}
			\fmfv{decor.shape=circle,decor.filled=shaded, decor.size=3.5thin}{m}
			\fmfforce{(0.5w,h)}{m}
			\fmfv{decor.shape=hexagram,decor.filled=full, decor.size=3thin}{v}
		\end{fmfgraph}
	\end{fmffile}
\end{gathered}
+\tfrac{1}{4!}
\begin{gathered}
	\begin{fmffile}{dlambdax-0dm2}
		\begin{fmfgraph}(55,55)
			\fmfset{dash_len}{1.2mm}
			\fmfset{wiggly_len}{1.1mm} \fmfset{dot_len}{0.5mm}
			\fmfpen{0.25mm}
			\fmfleft{i1,i2}
			\fmfright{o1,o2}
			\fmf{dashes,fore=black}{i1,v,o2}
			\fmf{dashes,fore=black}{i2,v,o1}
			\fmfv{decor.shape=hexagram,decor.filled=full, decor.size=3thin}{v}
		\end{fmfgraph}
	\end{fmffile}
\end{gathered}
+\tfrac{1}{3!}
\begin{gathered}
	\begin{fmffile}{dlambdax-1dm2}
		\begin{fmfgraph}(55,55)
			\fmfset{dash_len}{1.2mm}
			\fmfset{wiggly_len}{1.1mm} \fmfset{dot_len}{0.5mm}
			\fmfpen{0.25mm}
			\fmfleft{i1,i2}
			\fmfright{o1,o2}
			\fmf{dashes,fore=black}{i1,v,o2}
			\fmf{dashes,fore=black}{i2,v,o1}
			\fmfv{decor.shape=hexagram,decor.filled=full, decor.size=3thin}{v}
			\fmfv{decor.shape=circle,decor.filled=shaded, decor.size=3.5thin}{m}
			\fmfforce{(0.27w,0.77h)}{m}
		\end{fmfgraph}
	\end{fmffile}
\end{gathered}
\Big)
-g^3
\Big(
\tfrac{1}{4}\,\,
\begin{gathered}
	\begin{fmffile}{dfishz}
		\begin{fmfgraph}(50,50)
			\fmfset{dash_len}{1.2mm}
			\fmfset{wiggly_len}{1.1mm} \fmfset{dot_len}{0.5mm}
			\fmfpen{0.25mm}
			\fmfleft{i,j}
			\fmfright{o}
			\fmf{dashes,fore=black,tension=1}{i,v1}
			\fmf{dashes,fore=black,tension=1}{j,v1}
			\fmf{dashes,fore=black}{v1,v2}
			\fmf{dashes,fore=black,tension=1}{v3,o}
			\fmf{dashes,fore=black,left,tension=0.3}{v2,v3,v2}
			\fmffreeze
			\fmfforce{(-.2w,0.8h)}{i}
			\fmfforce{(-.2w,0.2h)}{j}
			\fmfforce{(0.05w,0.5h)}{v1}
			\fmfforce{(1.2w,0.5h)}{o}
		\end{fmfgraph}
	\end{fmffile}
\end{gathered}
+\tfrac{1}{8}
\begin{gathered}
	\begin{fmffile}{dg3-crystalz}
		\begin{fmfgraph}(55,55)
			\fmfset{dash_len}{1.2mm}
			\fmfset{wiggly_len}{1.1mm} \fmfset{dot_len}{0.5mm}
			\fmfpen{0.25mm}
			\fmfsurround{i1,i2,i3,i4,i5,i6}
			\fmf{dashes,fore=black,tension=1}{i6,v}
			\fmf{dashes,fore=black,tension=1}{i1,v}
			\fmf{dashes,fore=black,tension=1}{v,c}
		 	\fmf{dashes,fore=black,tension=1}{c,u}
			\fmf{phantom,tension=1}{u,i2}
			\fmf{phantom,fore=black,tension=1}{u,i3}
			\fmf{dashes,fore=black,tension=1}{c,s,i4}
			\fmf{dashes,fore=black,tension=1}{s,i5}
		\end{fmfgraph}
	\end{fmffile}
\end{gathered}
+\tfrac{1}{6}
\begin{gathered}
	\begin{fmffile}{dg3-3ptlogz}
		\begin{fmfgraph}(40,40)
			\fmfset{dash_len}{1.2mm}
			\fmfset{wiggly_len}{1.1mm} \fmfset{dot_len}{0.5mm}
			\fmfpen{0.25mm}
			\fmfsurroundn{i}{6}
			\fmf{phantom,fore=black}{i1,v,u,i4}
			\fmf{dashes,fore=black}{v,i1}
			\fmf{phantom,fore=black}{i2,s,t,i5}
			\fmf{dashes,fore=black}{i5,t}
			\fmf{phantom,fore=black}{i3,w,x,i6}
			\fmf{dashes,fore=black}{i3,w}
			\fmfi{dashes,fore=black}{fullcircle scaled .55w shifted (.51w,.5h)}
			\fmffreeze
			\fmfforce{(1.25w,0.5h)}{i1}
			\fmfforce{(0.16w,1.1h)}{i3}
			\fmfforce{(0.16w,-.1h)}{i5}
		\end{fmfgraph}
	\end{fmffile}
\end{gathered}\,\,
\Big)\\
&+g\lambda\Big(
\tfrac{1}{4}\,
\begin{gathered}
	\begin{fmffile}{dglamba-swimz}
		\begin{fmfgraph}(50,50)
			\fmfset{dash_len}{1.2mm}
			\fmfset{wiggly_len}{1.1mm} \fmfset{dot_len}{0.5mm}
			\fmfpen{0.25mm}
			\fmfleft{a}
			\fmfright{f,g}
			\fmf{dashes,fore=black}{a,v,b}
			\fmf{dashes,fore=black,tension=.6}{v,v}
			\fmf{dashes,fore=black,tension=1}{b,f}
			\fmf{dashes,fore=black,tension=1}{b,g}
			\fmffreeze
			\fmfforce{(1.1w,0.8h)}{f}
			\fmfforce{(1.1w,0.2h)}{g}
			\fmffreeze
			\fmfforce{(-.1w,0.5h)}{a}
		\end{fmfgraph}
	\end{fmffile}
\end{gathered}\,
+\tfrac{1}{4}\,\,
\begin{gathered}
	\begin{fmffile}{dglambda-dartz}
		\begin{fmfgraph}(50,50)
			\fmfset{dash_len}{1.2mm}
			\fmfset{wiggly_len}{1.1mm} \fmfset{dot_len}{0.5mm}
			\fmfpen{0.25mm}
			\fmfleft{i,j}
			\fmfright{o}
			\fmf{dashes,fore=black,tension=5}{i,v1}
			\fmf{dashes,fore=black,tension=5}{j,v1}
			\fmf{dashes,fore=black,tension=0.8}{v2,o}
			\fmf{dashes,fore=black,left,tension=0.08}{v1,v2,v1}
			\fmf{phantom}{v1,v2}
			\fmffreeze
			\fmfforce{(-.2w,0.8h)}{i}
			\fmfforce{(-.2w,0.2h)}{j}
		\end{fmfgraph}
	\end{fmffile}
\end{gathered}
+\tfrac{1}{12}
\begin{gathered}
	\begin{fmffile}{dglambda-5ptreez}
		\begin{fmfgraph}(40,40)
			\fmfset{dash_len}{1.2mm}
			\fmfset{wiggly_len}{1.1mm} \fmfset{dot_len}{0.5mm}
			\fmfpen{0.25mm}
			\fmftop{t1,t2}
			\fmfbottom{a,b,c}
			\fmf{dashes,fore=black,tension=1}{a,v}
			\fmf{dashes,fore=black,tension=1}{b,v}
			\fmf{dashes,fore=black,tension=1}{c,v}
			\fmf{dashes,fore=black,tension=2}{v,t}
			\fmf{dashes,fore=black,tension=0.7,left,straight}{t,t1}
			\fmf{dashes,fore=black,tension=0.7,left,straight}{t,t2}
			\fmffreeze
			\fmfforce{(0.5w,-0.2h)}{b}
		\end{fmfgraph}
	\end{fmffile}
\end{gathered}
\Big)
-\kappa
\Big(
\tfrac{1}{12}
\begin{gathered}
	\begin{fmffile}{dkappa-3ptself}
		\begin{fmfgraph}(40,40)
			\fmfset{dash_len}{1.2mm}
			\fmfset{wiggly_len}{1.1mm} \fmfset{dot_len}{0.5mm}
			\fmfpen{0.25mm}
			\fmftop{t}
			\fmfbottom{a,b,c}
			\fmf{dashes,fore=black,tension=1}{a,v}
			\fmf{dashes,fore=black,tension=1}{b,v}
			\fmf{dashes,fore=black,tension=1}{c,v}
			\fmf{dashes,fore=black,tension=1.4,left}{v,t,v}
		\end{fmfgraph}
	\end{fmffile}
\end{gathered}+
\tfrac{1}{5!}
\begin{gathered}
	\begin{fmffile}{dkappa-5pt}
		\begin{fmfgraph}(40,40)
			\fmfset{dash_len}{1.2mm}
			\fmfset{wiggly_len}{1.1mm} \fmfset{dot_len}{0.5mm}
			\fmfpen{0.25mm}
			\fmfsurround{u1,u2,u3,u4,u5}
			\fmf{dashes,fore=black,tension=1}{u1,v}
			\fmf{dashes,fore=black,tension=1}{u2,v}
			\fmf{dashes,fore=black,tension=1}{u3,v}
			\fmf{dashes,fore=black,tension=1}{u4,v}
			\fmf{dashes,fore=black,tension=1}{u5,v}
		\end{fmfgraph}
	\end{fmffile}
\end{gathered}
\Big)
+g^4
\Big(
\tfrac{1}{48}
\begin{gathered}
	\begin{fmffile}{dg4-6pttree}
		\begin{fmfgraph}(45,45)
			\fmfset{dash_len}{1.2mm}
			\fmfset{wiggly_len}{1.1mm} \fmfset{dot_len}{0.5mm}
			\fmfpen{0.25mm}
			\fmfsurround{u1,u2,u3,u4,u5,u6}
			\fmf{dashes,fore=black,tension=1}{u1,v}
			\fmf{dashes,fore=black,tension=1}{u2,v}
			\fmf{dashes,fore=black,tension=1}{u3,u}
			\fmf{dashes,fore=black,tension=1}{u4,u}
			\fmf{dashes,fore=black,tension=1}{u5,s}
			\fmf{dashes,fore=black,tension=1}{u6,s}
			\fmf{dashes,fore=black,tension=1,left,straight}{s,c}
			\fmf{dashes,fore=black,tension=1,left,straight}{u,c}
			\fmf{dashes,fore=black,tension=1,left,straight}{v,c}
		\end{fmfgraph}
	\end{fmffile}
\end{gathered}
+\tfrac{1}{16}\,
\begin{gathered}
	\begin{fmffile}{dg4-0pt-3loop1PIa}
		\begin{fmfgraph}(27,27)
			\fmfset{dash_len}{1.2mm}
			\fmfset{wiggly_len}{1.1mm} \fmfset{dot_len}{0.5mm}
			\fmfpen{0.25mm}
			\fmfsurround{a,b,c,d}
			\fmf{phantom,fore=black,tension=1,curved}{a,b,c,d,a}
			\fmf{dashes,fore=black,tension=1,right}{a,b}
			\fmf{dashes,fore=black,tension=1,right}{c,d}
			\fmf{dashes,fore=black,tension=1,right}{b,c}
			\fmf{dashes,fore=black,tension=1,right}{d,a}
			\fmf{dashes,fore=black,tension=1,straight}{a,b}
			\fmf{dashes,fore=black,tension=1,straight}{c,d}
		\end{fmfgraph}
	\end{fmffile}
\end{gathered}\,\\
&+\tfrac{1}{16}\,\,\,
\begin{gathered}
	\begin{fmffile}{dg4-4pt-1loop}
		\begin{fmfgraph}(40,40)
			\fmfset{dash_len}{1.2mm}
			\fmfset{wiggly_len}{1.1mm} \fmfset{dot_len}{0.5mm}
			\fmfpen{0.25mm}
			\fmfleft{a,b}
			\fmfright{c,d}
			\fmf{dashes,fore=black,tension=1}{a,v}
			\fmf{dashes,fore=black,tension=1}{b,v}
			\fmf{dashes,fore=black,tension=1}{v,s}
			\fmf{dashes,fore=black,tension=1}{t,u}
			\fmf{dashes,fore=black,tension=0.3,left}{s,t,s}
			\fmf{dashes,fore=black,tension=1}{u,c}
			\fmf{dashes,fore=black,tension=1}{u,d}	
			\fmffreeze
			\fmfforce{(1.3w,0.8h)}{c}
			\fmfforce{(1.3w,0.2h)}{d}
			\fmfforce{(1w,0.5h)}{u}
			\fmfforce{(-0.06w,0.5h)}{v}
			\fmffreeze
			\fmfforce{(-0.3w,0.8h)}{a}
			\fmfforce{(-0.3w,0.2h)}{b}
		\end{fmfgraph}
	\end{fmffile}
\end{gathered}\,\,\,
+\tfrac{1}{8}\,\,
\begin{gathered}
	\begin{fmffile}{dg4-2pt-1PR-twoloop}
		\begin{fmfgraph}(50,50)
			\fmfset{dash_len}{1.2mm}
			\fmfset{wiggly_len}{1.1mm} \fmfset{dot_len}{0.5mm}
			\fmfpen{0.25mm}
			\fmfleft{i}
			\fmfright{o}
			\fmf{dashes,fore=black,tension=5}{i,v1}
			\fmf{dashes,fore=black,tension=5}{v2,o}
			\fmf{dashes,fore=black,left,tension=0.4}{v1,v3,v1}
			\fmf{dashes,fore=black,right,tension=0.4}{v2,v4,v2}
			\fmf{dashes,fore=black}{v3,v4}
			\fmffreeze
			\fmfforce{(-.25w,0.5h)}{i}
			\fmfforce{(1.25w,0.5h)}{o}
			\end{fmfgraph}
	\end{fmffile}
\end{gathered}\,\,
+\tfrac{1}{8}\,\,
\begin{gathered}
	\begin{fmffile}{dg4-4ptlog}
		\begin{fmfgraph}(40,40)
			\fmfset{dash_len}{1.2mm}
			\fmfset{wiggly_len}{1.1mm} \fmfset{dot_len}{0.5mm}
			\fmfpen{0.25mm}
			\fmfsurroundn{i}{4}
			\fmf{phantom,fore=black}{i1,v,u,i3}
			\fmf{dashes,fore=black}{v,i1}
			\fmf{dashes,fore=black}{i3,u}
			\fmf{phantom,fore=black}{i2,s,t,i4}
			\fmf{dashes,fore=black}{i4,t}
			\fmf{dashes,fore=black}{i2,s}
			\fmfi{dashes,fore=black}{fullcircle scaled .5w shifted (.51w,.5h)}
			\fmffreeze
			\fmfforce{(1.15w,0.5h)}{i1}
			\fmfforce{(0.5w,1.1h)}{i2}
			\fmfforce{(-.1w,0.5h)}{i3}
			\fmfforce{(0.5w,-.1h)}{i4}
		\end{fmfgraph}
	\end{fmffile}
\end{gathered}\,
+\tfrac{1}{8}\begin{gathered}\,\,
	\begin{fmffile}{dg4-6pttreelong}
		\begin{fmfgraph}(55,55)
			\fmfset{dash_len}{1.2mm}
			\fmfset{wiggly_len}{1.1mm} \fmfset{dot_len}{0.5mm}
			\fmfpen{0.25mm}
			\fmfsurroundn{i}{6}
			\fmf{dashes,fore=black}{i1,v,u,s,t,i4}
			\fmf{dashes,fore=black}{i6,v}
			\fmf{dashes,fore=black}{i4,t}
			\fmf{dashes,fore=black}{i5,t}
			\fmf{dashes,fore=black}{i2,u}
			\fmf{dashes,fore=black}{i3,s}
			\fmffreeze
			\fmfforce{(1.12w,0.4h)}{i1}
			\fmfforce{(-.15w,0.423h)}{i4}
			\fmfforce{(.26w,0.16h)}{i5}
			\fmfforce{(0.67w,0.15h)}{i6}
		\end{fmfgraph}
	\end{fmffile}
\end{gathered}\,\,
+\tfrac{1}{4!}\!\!
\begin{gathered}
	\begin{fmffile}{dpeace}
		\begin{fmfgraph}(75,75)
			\fmfset{dash_len}{1.2mm}
			\fmfset{wiggly_len}{1.1mm} \fmfset{dot_len}{0.5mm}
			\fmfpen{0.25mm}
			\fmfsurroundn{i}{3}
			\fmf{phantom,fore=black}{i1,v,i2}
			\fmf{phantom,fore=black}{i2,u,i3}
			\fmf{phantom,fore=black}{i3,s,i1}
			\fmfi{dashes,fore=black}{fullcircle scaled .4w shifted (.5w,.5h)}
			\fmf{dashes,fore=black}{v,c}
			\fmf{dashes,fore=black}{u,c}
			\fmf{dashes,fore=black}{s,c}
		\end{fmfgraph}
	\end{fmffile}
\end{gathered}\!\!
+\tfrac{1}{4}
\begin{gathered}
	\begin{fmffile}{dg4-2pt2loop-1PIa}
		\begin{fmfgraph}(60,60)
			\fmfset{dash_len}{1.2mm}
			\fmfset{wiggly_len}{1.1mm} \fmfset{dot_len}{0.5mm}
			\fmfpen{0.25mm}
			\fmftop{t}
			\fmfbottom{b}
			\fmfleft{l}
			\fmfright{r}
			\fmf{phantom,fore=black,tension=9}{t,u,v,b}
			\fmf{phantom,fore=black,tension=9}{l,s,x,r}
			\fmf{dashes,fore=black,tension=.01,left}{u,v,u}
			\fmf{phantom,fore=black,tension=0.01}{s,x,s}
			\fmf{dashes,fore=black,tension=1}{u,v}
			\fmf{dashes,fore=black,tension=1}{x,r}
			\fmf{dashes,fore=black,tension=1}{l,s}
		\end{fmfgraph}
	\end{fmffile}
\end{gathered}
+\tfrac{1}{4}
\begin{gathered}
	\begin{fmffile}{dg4-2pt2loop-1PIb}
		\begin{fmfgraph}(35,35)
			\fmfset{dash_len}{1.2mm}
			\fmfset{wiggly_len}{1.1mm} \fmfset{dot_len}{0.5mm}
			\fmfpen{0.25mm}
			\fmfleft{i}
			\fmfright{o}
			\fmf{phantom,tension=5}{i,v1}
			\fmf{phantom,tension=5}{v2,o}
			\fmf{dashes,fore=black,left,tension=0.4}{v1,v2,v1}
			\fmf{dashes,fore=black}{v1,v2}
			\fmfsurroundn{i}{6}
			\fmf{phantom}{i2,a,c,x1,i5}
			\fmf{phantom}{i3,b,c,x2,i6}
			\fmf{dashes,fore=black}{x1,i5}
			\fmf{dashes,fore=black}{x2,i6}
			\fmffreeze
			\fmfforce{(0.1w,-.2h)}{i5}
			\fmfforce{(0.9w,-.2h)}{i6}
			\end{fmfgraph}
	\end{fmffile}
\end{gathered}\,
+\tfrac{1}{4}
\begin{gathered}
	\begin{fmffile}{dg4-4ptlog1PR}
		\begin{fmfgraph}(35,35)
			\fmfset{dash_len}{1.2mm}
			\fmfset{wiggly_len}{1.1mm} \fmfset{dot_len}{0.5mm}
			\fmfpen{0.25mm}
			\fmfsurroundn{i}{6}
			\fmf{phantom,fore=black}{i1,v,u,i4}
			\fmf{dashes,fore=black}{v,i1}
			\fmf{phantom,fore=black}{i2,s,t,i5}
			\fmf{dashes,fore=black}{i5,t}
			\fmf{phantom,fore=black}{i3,w,x,i6}
			\fmf{dashes,fore=black}{i3,w}
			\fmfi{dashes,fore=black}{fullcircle scaled .55w shifted (.51w,.5h)}
			\fmffreeze
			\fmfforce{(1.25w,0.5h)}{i1}
			\fmfforce{(0.16w,1.1h)}{i3}
			\fmfforce{(0.16w,-.1h)}{i5}
			\fmffreeze
			\fmfright{x,z}
			\fmf{dashes,fore=black,tension=1}{i1,x}
			\fmf{dashes,fore=black,tension=1}{i1,z}
			\fmfforce{(1.5w,0.9h)}{z}
			\fmfforce{(1.5w,0.1h)}{x}
		\end{fmfgraph}
	\end{fmffile}
\end{gathered}\,\,\,\,
+\tfrac{1}{4}\,\,\,\,\,
\begin{gathered}
	\begin{fmffile}{dg4-4pt-1loopb}
		\begin{fmfgraph}(40,40)
			\fmfset{dash_len}{1.2mm}
			\fmfset{wiggly_len}{1.1mm} \fmfset{dot_len}{0.5mm}
			\fmfpen{0.25mm}
			\fmfsurroundn{i}{4}
			\fmf{dashes,fore=black}{i1,n,m,i4}
			\fmf{dashes,fore=black}{i2,n}
			\fmf{dashes,fore=black}{m,i3}
			\fmffreeze
			\fmfleft{x}
			\fmf{dashes,fore=black,tension=10,left}{i3,u,i3}
			\fmf{dashes,fore=black}{u,x}
			\fmfforce{(-.4w,0.6h)}{u}
			\fmfforce{(-.75w,0.75h)}{x}
		\end{fmfgraph}
	\end{fmffile}
\end{gathered}
\Big)\\
&-g^2\lambda\Big(
\tfrac{1}{8}
\begin{gathered}
	\begin{fmffile}{dg2L-bubble}
		\begin{fmfgraph}(45,45)
			\fmfset{dash_len}{1.2mm}
			\fmfset{wiggly_len}{1.1mm} \fmfset{dot_len}{0.5mm}
			\fmfpen{0.25mm}
			\fmftop{t1,t2,t3}
			\fmfbottom{b1,b2,b3}
			\fmf{phantom}{t1,v1,b1}
			\fmf{phantom}{t2,v2,b2}
			\fmf{phantom}{t3,v3,b3}
			\fmffreeze
			\fmf{dashes,fore=black,right}{v1,v2,v1}
			\fmf{dashes,fore=black,right}{v2,v3,v2}
			\fmf{dashes,fore=black,tension=1}{t1,b1}
			\fmfforce{(0.25w,0.7h)}{t1}
			\fmfforce{(0.25w,0.3h)}{b1}
		\end{fmfgraph}
	\end{fmffile}
\end{gathered}
+\tfrac{1}{8}
\begin{gathered}
	\begin{fmffile}{dg2L-bubble2pt}
		\begin{fmfgraph}(45,45)
			\fmfset{dash_len}{1.2mm}
			\fmfset{wiggly_len}{1.1mm} \fmfset{dot_len}{0.5mm}
			\fmfpen{0.25mm}
			\fmftop{t1,t2,t3}
			\fmfbottom{b1,b2,b3}
			\fmf{phantom}{t1,v1,b1}
			\fmf{phantom}{t2,v2,b2}
			\fmf{phantom}{t3,v3,b3}
			\fmffreeze
			\fmf{dashes,fore=black,right}{v1,v2,v1}
			\fmf{dashes,fore=black}{v2,t3}
			\fmf{dashes,fore=black}{v2,b3}
			\fmf{dashes,fore=black,tension=1}{t1,b1}
			\fmfforce{(0.25w,0.7h)}{t1}
			\fmfforce{(0.25w,0.3h)}{b1}
			\fmfforce{(0.9w,0.9h)}{t3}
			\fmfforce{(0.9w,0.1h)}{b3}
		\end{fmfgraph}
	\end{fmffile}
\end{gathered}
+\tfrac{1}{4}\,\,\,
\begin{gathered}
	\begin{fmffile}{dg2L-fishself}
		\begin{fmfgraph}(45,45)
			\fmfset{dash_len}{1.2mm}
			\fmfset{wiggly_len}{1.1mm} \fmfset{dot_len}{0.5mm}
			\fmfpen{0.25mm}
			\fmfleft{j}
			\fmfright{o}
			\fmf{dashes,fore=black,tension=1}{j,v1}
			\fmf{dashes,fore=black}{v1,v2}
			\fmf{dashes,fore=black,tension=1}{v3,o}
			\fmf{dashes,fore=black,left,tension=0.3}{v2,v3,v2}
			\fmf{dashes,fore=black,tension=0.6}{v1,v1}
			\fmffreeze
			\fmfforce{(-.2w,0.2h)}{j}
			\fmfforce{(0.05w,0.5h)}{v1}
			\fmfforce{(1.2w,0.5h)}{o}
			\end{fmfgraph}
	\end{fmffile}
\end{gathered}\,
+\tfrac{1}{12}\,\,\,
\begin{gathered}
	\begin{fmffile}{dg2L-4ptfish}
		\begin{fmfgraph}(40,40)
			\fmfset{dash_len}{1.2mm}
			\fmfset{wiggly_len}{1.1mm} \fmfset{dot_len}{0.5mm}
			\fmfpen{0.25mm}
			\fmfleftn{i}{3}
			\fmfright{o}
			\fmf{dashes,fore=black,tension=1}{i1,v}
			\fmf{dashes,fore=black,tension=1}{i2,v}
			\fmf{dashes,fore=black,tension=1}{i3,v}
			\fmf{dashes,fore=black,tension=0.5}{v,u}
			\fmf{dashes,fore=black,tension=0.2,left}{u,x,u}
			\fmf{dashes,fore=black,tension=1}{x,o}
			\fmffreeze
			\fmfforce{(-.1w,0.9h)}{i1}
			\fmfforce{(-.1w,0.1h)}{i3}
			\fmfforce{(-.35w,0.5h)}{i2}
			\fmfforce{(1.3w,0.5h)}{o}
		\end{fmfgraph}\,\,
	\end{fmffile}
\end{gathered}
+\tfrac{1}{8}\,\,
\begin{gathered}
	\begin{fmffile}{dg2L-3loopbubble-1PI}
		\begin{fmfgraph}(27,27)
			\fmfset{dash_len}{1.2mm}
			\fmfset{wiggly_len}{1.1mm} \fmfset{dot_len}{0.5mm}
			\fmfpen{0.25mm}
			\fmfsurroundn{i}{3}
			\fmf{dashes,fore=black,tension=1,right=1}{i1,i2}
			\fmf{dashes,fore=black,tension=1,right=1}{i2,i3}
			\fmf{dashes,fore=black,tension=1,right=0.8}{i3,i1}
			\fmf{dashes,fore=black,tension=1}{i1,i2}
			\fmf{dashes,fore=black,tension=1}{i2,i3}
		\end{fmfgraph}
	\end{fmffile}
\end{gathered}\,\,
+\tfrac{1}{4}\,\,\,\,\,
\begin{gathered}
	\begin{fmffile}{dg2L-2ptdoublebubble-1PI}
		\begin{fmfgraph}(17,17)
			\fmfset{dash_len}{1.2mm}
			\fmfset{wiggly_len}{1.1mm} \fmfset{dot_len}{0.5mm}
			\fmfpen{0.25mm}
			\fmfright{o1,o2}
			\fmfleft{i}
			\fmf{dashes,fore=black,tension=1,right=0.63}{i,o1}
			\fmf{dashes,fore=black,tension=1,right=0.6}{o1,o2}
			\fmf{dashes,fore=black,tension=1,right=0.63}{o2,i}
			\fmf{dashes,fore=black,tension=5,left}{i,u,i}
			\fmffreeze
			\fmfforce{(-1.2w,0.5h)}{u}
			\fmffreeze
			\fmfright{x1,x2}
			\fmf{dashes,fore=black,tension=1}{o1,x2}
			\fmf{dashes,fore=black,tension=1}{o2,x1}
			\fmfforce{(1.5w,1.8h)}{x1}
			\fmfforce{(1.5w,-0.8h)}{x2}
		\end{fmfgraph}
	\end{fmffile}
\end{gathered}\,\,\,
+\tfrac{1}{4}\,\,\,\,\,
\begin{gathered}
	\begin{fmffile}{dg2L-4pt1loop-1PI}
		\begin{fmfgraph}(17,17)
			\fmfset{dash_len}{1.2mm}
			\fmfset{wiggly_len}{1.1mm} \fmfset{dot_len}{0.5mm}
			\fmfpen{0.25mm}
			\fmfright{o1,o2}
			\fmfleft{i}
			\fmf{dashes,fore=black,tension=1,right=0.63}{i,o1}
			\fmf{dashes,fore=black,tension=1,right=0.6}{o1,o2}
			\fmf{dashes,fore=black,tension=1,right=0.63}{o2,i}
			\fmffreeze
			\fmfleft{s1,s2}
			\fmf{dashes,fore=black,tension=1}{i,s2}
			\fmf{dashes,fore=black,tension=1}{i,s1}
			\fmfforce{(-0.8w,1.6h)}{s1}
			\fmfforce{(-0.8w,-0.6h)}{s2}
			\fmffreeze
			\fmfright{x1,x2}
			\fmf{dashes,fore=black,tension=1}{o1,x2}
			\fmf{dashes,fore=black,tension=1}{o2,x1}
			\fmfforce{(1.4w,2h)}{x1}
			\fmfforce{(1.4w,-h)}{x2}
		\end{fmfgraph}
	\end{fmffile}
\end{gathered}\,\,\,
+\tfrac{1}{2}
\begin{gathered}
	\begin{fmffile}{dg2L-2pt2loop-1PI}
		\begin{fmfgraph}(40,40)
			\fmfset{dash_len}{1.2mm}
			\fmfset{wiggly_len}{1.1mm} \fmfset{dot_len}{0.5mm}
			\fmfpen{0.25mm}
			\fmfsurroundn{i}{4}
			\fmfi{dashes,fore=black}{fullcircle scaled .63w shifted (0.5w,.5h)}
			\fmf{phantom,fore=black}{i1,u,v,i3}
			\fmf{dashes,fore=black}{i1,v}
			\fmf{phantom,fore=black}{i2,s,t,i4}
			\fmf{dashes,fore=black}{i2,s}
			\fmffreeze
			\fmfforce{(0.5w,1.2h)}{i2}
			\fmfforce{(0.2w,0.5h)}{v}
			\fmfforce{(1.3w,0.5h)}{i1}
		\end{fmfgraph}
	\end{fmffile}
\end{gathered}\,\,\\
&+\tfrac{1}{4}\,\,\,\,\,
\begin{gathered}
	\begin{fmffile}{dg2L-4pt-1loopcz}
		\begin{fmfgraph}(50,50)
			\fmfset{dash_len}{1.2mm}
			\fmfset{wiggly_len}{1.1mm} \fmfset{dot_len}{0.5mm}
			\fmfpen{0.25mm}
			\fmfsurroundn{i}{4}
			\fmf{dashes,fore=black}{i1,n,m,i4}
			\fmf{dashes,fore=black}{i2,n}
			\fmf{dashes,fore=black}{m,i3}
			\fmffreeze
			\fmfleft{x}
			\fmf{dashes,fore=black,tension=0.6,right}{i3,i3}
			\fmf{dashes,fore=black}{i3,x}
			\fmfforce{(-.4w,0.63h)}{x}
		\end{fmfgraph}
	\end{fmffile}
\end{gathered}
+\tfrac{1}{8}\,\,\,
\begin{gathered}
	\begin{fmffile}{dg2L-4pt1loopwigglyz}
		\begin{fmfgraph}(40,40)
			\fmfset{dash_len}{1.2mm}
			\fmfset{wiggly_len}{1.1mm} \fmfset{dot_len}{0.5mm}
			\fmfpen{0.25mm}
			\fmfleft{i1,i2}
			\fmfright{o1,o2}
			\fmf{dashes,fore=black,tension=1}{i1,n}
			\fmf{dashes,fore=black,tension=1}{i2,n}
			\fmf{dashes,fore=black,tension=1}{n,m}
			\fmf{dashes,fore=black,tension=0.15,left}{m,s,m}
			\fmf{dashes,fore=black,tension=1}{s,o1}
			\fmf{dashes,fore=black,tension=1}{s,o2}
			\fmffreeze
			\fmfforce{(-0.15w,0.5h)}{n}
			\fmfforce{(-0.45w,h)}{i1}
			\fmfforce{(-0.45w,0h)}{i2}
			\fmfforce{(1.15w,h)}{o1}
			\fmfforce{(1.15w,0h)}{o2}
		\end{fmfgraph}
	\end{fmffile}
\end{gathered}\,\,\,
+\tfrac{1}{12}
\begin{gathered}
	\begin{fmffile}{dg2L-crystalz}
		\begin{fmfgraph}(50,50)
			\fmfset{dash_len}{1.2mm}
			\fmfset{wiggly_len}{1.1mm} \fmfset{dot_len}{0.5mm}
			\fmfpen{0.25mm}
			\fmfsurround{i1,i2,i3,i4,i5,i6}
			\fmf{dashes,fore=black,tension=1}{i6,v}
			\fmf{dashes,fore=black,tension=1}{i1,v}
			\fmf{dashes,fore=black,tension=1}{v,c}
		 	\fmf{dashes,fore=black,tension=1}{c,u}
			\fmf{phantom,tension=1}{u,i2}
			\fmf{phantom,fore=black,tension=1}{u,i3}
			\fmf{dashes,fore=black,tension=1}{c,s,i4}
			\fmf{dashes,fore=black,tension=1}{s,i5}
			\fmffreeze
			\fmfright{x}
			\fmf{dashes,fore=black}{v,x}
			\fmfforce{(1.1w,0.2h)}{x}
			\fmfforce{(1.05w,0.54h)}{i1}
			\fmfforce{(-.05w,0.54h)}{i4}
		\end{fmfgraph}
	\end{fmffile}
\end{gathered}
+\tfrac{1}{8}\,\,\,\,
\begin{gathered}
	\begin{fmffile}{dg2L-2pt2loop-1PIbz}
		\begin{fmfgraph}(55,55)
			\fmfset{dash_len}{1.2mm}
			\fmfset{wiggly_len}{1.1mm} \fmfset{dot_len}{0.5mm}
			\fmfpen{0.25mm}
			\fmftop{t1,t2,t3,t4}
        			\fmfbottom{b1,b2,b3,b4}
        			\fmf{phantom}{t1,v1,b1}
        			\fmf{phantom}{t2,v2,b2}
			\fmf{phantom}{t3,v3,b3}
			\fmf{phantom}{t4,v4,b4}
        			\fmffreeze
			\fmf{dashes,fore=black,right,tension=0.7}{v1,v2,v1}
        			\fmf{dashes,fore=black,right,tension=0.7}{v2,v3,v2}
        			\fmf{dashes,fore=black,tension=3}{v3,v4}
			\fmffreeze
			\fmfforce{(1.1w,0.5h)}{v4}
			\fmfleft{l}
			\fmf{dashes,fore=black}{v1,l}
			\fmfforce{(-.4w,0.5h)}{l}
		\end{fmfgraph}
	\end{fmffile}
\end{gathered}
+\tfrac{1}{4}\,\,\,\,
\begin{gathered}
	\begin{fmffile}{dg2L-4ptdartz}
		\begin{fmfgraph}(45,45)
			\fmfset{dash_len}{1.2mm}
			\fmfset{wiggly_len}{1.1mm} \fmfset{dot_len}{0.5mm}
			\fmfpen{0.25mm}
			\fmfleft{i,j}
			\fmfright{o}
			\fmf{dashes,fore=black,tension=5}{i,v1}
			\fmf{dashes,fore=black,tension=5}{j,v1}
			\fmf{dashes,fore=black,tension=0.8}{v2,o}
			\fmf{dashes,fore=black,left,tension=0.08}{v1,v2,v1}
			\fmf{phantom}{v1,v2}
			\fmffreeze
			\fmfforce{(-.2w,0.8h)}{i}
			\fmfforce{(-.2w,0.2h)}{j}
			\fmfleft{n,m}
			\fmf{dashes,fore=black,tension=1}{i,n}
			\fmf{dashes,fore=black,tension=1}{i,m}
			\fmfforce{(-.5w,1.1h)}{n}
			\fmfforce{(-.55w,0.5h)}{m}			
		\end{fmfgraph}
	\end{fmffile}
\end{gathered}
+\tfrac{1}{16}\,
\begin{gathered}
	\begin{fmffile}{dg2L-4pt-1PR}
		\begin{fmfgraph}(35,35)
			\fmfset{dash_len}{1.2mm}
			\fmfset{wiggly_len}{1.1mm} \fmfset{dot_len}{0.5mm}
			\fmfpen{0.25mm}
			\fmfsurround{a,b,c}
			\fmf{dashes,fore=black,tension=1}{c,v,a}
			\fmf{dashes,fore=black,left,tension=0.8}{v,b,v}
			\fmffreeze
			\fmfright{n,m}
			\fmf{dashes,fore=black,tension=1.3}{a,n}
			\fmf{dashes,fore=black,tension=1.3}{a,m}
			\fmfforce{(1.3w,0.8h)}{n}
			\fmfforce{(1.1w,0.1h)}{m}
			\fmffreeze
			\fmfbottom{x,z}
			\fmf{dashes,fore=black,tension=1.3}{c,x}
			\fmf{dashes,fore=black,tension=1.3}{c,z}
			\fmfforce{(-.12w,-0.12h)}{x}
			\fmfforce{(.55w,-0.2h)}{z}
		\end{fmfgraph}
	\end{fmffile}
\end{gathered}\,\,
+\tfrac{1}{16}\,
\begin{gathered}
	\begin{fmffile}{dg2L-6pt-1PR}
		\begin{fmfgraph}(35,35)
			\fmfset{dash_len}{1.2mm}
			\fmfset{wiggly_len}{1.1mm} \fmfset{dot_len}{0.5mm}
			\fmfpen{0.25mm}
			\fmfsurround{a,b,c}
			\fmf{dashes,fore=black,tension=1}{c,v,a}
			\fmf{phantom,fore=black,left,tension=1.2}{v,b,v}
			\fmffreeze
			\fmftop{n,m}
			\fmf{dashes,fore=black,tension=1}{n,v}
			\fmf{dashes,fore=black,tension=1}{m,v}
			\fmfforce{(-0.05w,0.77h)}{n}
			\fmfforce{(0.57w,1.15h)}{m}
			\fmffreeze
			\fmfbottom{x,z}
			\fmf{dashes,fore=black,tension=1.3}{c,x}
			\fmf{dashes,fore=black,tension=1.3}{c,z}
			\fmfforce{(-.12w,-0.12h)}{x}
			\fmfforce{(.55w,-0.2h)}{z}
			\fmffreeze
			\fmfright{s,t}
			\fmf{dashes,fore=black,tension=1.3}{a,s}
			\fmf{dashes,fore=black,tension=1.3}{a,t}
			\fmfforce{(1.3w,0.8h)}{s}
			\fmfforce{(1.1w,0.1h)}{t}
		\end{fmfgraph}
	\end{fmffile}
\end{gathered}\,\,
\Big)\\
&+\lambda^2
\Big(
\tfrac{1}{48}
\begin{gathered}
	\begin{fmffile}{dL2-3loop-bubble}
		\begin{fmfgraph}(50,50)
			\fmfset{dash_len}{1.2mm}
			\fmfset{wiggly_len}{1.1mm} \fmfset{dot_len}{0.5mm}
			\fmfpen{0.25mm}
			\fmfleft{i}
			\fmfright{o}
			\fmf{phantom,tension=10}{i,v1}
			\fmf{phantom,tension=10}{v2,o}
			\fmf{dashes,left,tension=0.4}{v1,v2,v1}
			\fmf{dashes,left=0.5}{v1,v2}
			\fmf{dashes,right=0.5}{v1,v2}
    		\end{fmfgraph}
	\end{fmffile}
\end{gathered}
+\tfrac{1}{16}
\begin{gathered}
	\begin{fmffile}{dL2-bubblez}
		\begin{fmfgraph}(35,35)
			\fmfset{dash_len}{1.2mm}
			\fmfset{wiggly_len}{1.1mm} \fmfset{dot_len}{0.5mm}
			\fmfpen{0.25mm}
			\fmftop{t1,t2,t3}
			\fmfbottom{b1,b2,b3}
			\fmf{phantom}{t1,v1,b1}
			\fmf{phantom}{t2,v2,b2}
			\fmf{phantom}{t3,v3,b3}
			\fmffreeze
			\fmf{dashes,fore=black,right}{v1,v2,v1}
			\fmf{dashes,fore=black,right}{v2,v3,v2}
			\fmfi{dashes,fore=black}{fullcircle scaled .5w shifted (1.25w,.5h)}
		\end{fmfgraph}
	\end{fmffile}
\end{gathered}\,\,\,
+\tfrac{1}{12}\,\,
\begin{gathered}
	\begin{fmffile}{dL2-2pt2loop-1PI}
		\begin{fmfgraph}(45,45)
			\fmfset{dash_len}{1.2mm}
			\fmfset{wiggly_len}{1.1mm} \fmfset{dot_len}{0.5mm}
			\fmfpen{0.25mm}
			\fmfleft{i}
			\fmfright{o}
			\fmf{dashes,fore=black,tension=5}{i,v1}
			\fmf{dashes,fore=black,tension=5}{v2,o}
			\fmf{dashes,fore=black,left,tension=0.4}{v1,v2,v1}
			\fmf{dashes,fore=black}{v1,v2}
			\fmffreeze
			\fmfforce{(-0.2w,0.5h)}{i}
			\fmfforce{(1.2w,0.5h)}{o}
		\end{fmfgraph}
	\end{fmffile}
\end{gathered}\,\,
+\tfrac{1}{8}
\begin{gathered}
	\begin{fmffile}{dL2-2pt2loop-1PIb}
		\begin{fmfgraph}(35,35)
			\fmfset{dash_len}{1.2mm}
			\fmfset{wiggly_len}{1.1mm} \fmfset{dot_len}{0.5mm}
			\fmfpen{0.25mm}
			\fmftop{t1,t2,t3}
			\fmfbottom{b1,b2,b3}
			\fmf{phantom}{t1,v1,b1}
			\fmf{phantom}{t2,v2,b2}
			\fmf{phantom}{t3,v3,b3}
			\fmffreeze
			\fmf{dashes,fore=black,right}{v1,v2,v1}
			\fmf{dashes,fore=black,right}{v2,v3,v2}
			\fmffreeze
			\fmfright{x,z}
			\fmf{dashes,fore=black}{v3,x}
			\fmf{dashes,fore=black}{v3,z}
			\fmfforce{(1.3w,0.9h)}{x}
			\fmfforce{(1.3w,0.1h)}{z}
		\end{fmfgraph}
	\end{fmffile}
\end{gathered}\,\,\, 
+\tfrac{1}{8}\,
\begin{gathered}
	\begin{fmffile}{dL2-2pt2loop-1PR}
		\begin{fmfgraph}(50,50)
			\fmfset{dash_len}{1.2mm}
			\fmfset{wiggly_len}{1.1mm} \fmfset{dot_len}{0.5mm}
			\fmfpen{0.25mm}
			\fmfleft{i}
			\fmfright{o}
			\fmf{dashes,fore=black}{i,u,v,o}
			\fmf{dashes,fore=black,tension=0.5,left}{u,u}
			\fmf{dashes,fore=black,tension=0.5,right}{v,v}
		\end{fmfgraph}
	\end{fmffile}
\end{gathered}\,
+\tfrac{1}{16}\,
\begin{gathered}
	\begin{fmffile}{dL2-4pt1loop-1PI}
		\begin{fmfgraph}(50,50)
			\fmfset{dash_len}{1.2mm}
			\fmfset{wiggly_len}{1.1mm} \fmfset{dot_len}{0.5mm}
			\fmfpen{0.25mm}
			\fmfleft{i1,i2}
			\fmfright{o1,o2}
			\fmf{dashes,fore=black,tension=1}{i1,u}
			\fmf{dashes,fore=black,tension=1}{i2,u}
			\fmf{dashes,fore=black,tension=0.4,left}{u,v,u}
			\fmf{dashes,fore=black}{v,o1}
			\fmf{dashes,fore=black}{v,o2}
			\fmffreeze
			\fmfforce{(w,0.8h)}{o1}
			\fmfforce{(w,0.2h)}{o2}
			\fmfforce{(0w,0.8h)}{i1}
			\fmfforce{(0w,0.2h)}{i2}
		\end{fmfgraph}
	\end{fmffile}
\end{gathered}\,
+\tfrac{1}{12}\,\,
\begin{gathered}
	\begin{fmffile}{dL2-4pt1loop31-1PR}
		\begin{fmfgraph}(50,50)
			\fmfset{dash_len}{1.2mm}
			\fmfset{wiggly_len}{1.1mm} \fmfset{dot_len}{0.5mm}
			\fmfpen{0.25mm}
			\fmfleftn{i}{3}
			\fmfright{o}
			\fmf{dashes,fore=black}{i1,u}
			\fmf{dashes,fore=black}{i2,u}
			\fmf{dashes,fore=black}{i3,u}
			\fmf{dashes,fore=black}{u,v,o}
			\fmf{dashes,fore=black,tension=0.6,right}{v,v}
			\fmffreeze
			\fmfforce{(0w,0.9h)}{i1}
			\fmfforce{(-.2w,0.5h)}{i2}
			\fmfforce{(0w,0.1h)}{i3}
		\end{fmfgraph}
	\end{fmffile}
\end{gathered}\,\\
&+\tfrac{1}{72}\,\,
\begin{gathered}
	\begin{fmffile}{dL2-6pttree33-1PR}
		\begin{fmfgraph}(45,45)
			\fmfset{dash_len}{1.2mm}
			\fmfset{wiggly_len}{1.1mm} \fmfset{dot_len}{0.5mm}
			\fmfpen{0.25mm}
			\fmfleftn{i}{3}
			\fmfrightn{o}{3}
			\fmf{dashes,fore=black}{i1,u}
			\fmf{dashes,fore=black}{i2,u}
			\fmf{dashes,fore=black}{i3,u}
			\fmf{dashes,fore=black}{o1,v}
			\fmf{dashes,fore=black}{o2,v}
			\fmf{dashes,fore=black}{o3,v}
			\fmf{dashes,fore=black}{u,v}
			\fmffreeze
			\fmfforce{(0w,0.9h)}{i1}
			\fmfforce{(-.2w,0.5h)}{i2}
			\fmfforce{(0w,0.1h)}{i3}
			\fmfforce{(1w,0.9h)}{o1}
			\fmfforce{(1.2w,0.5h)}{o2}
			\fmfforce{(1w,0.1h)}{o3}
		\end{fmfgraph}
	\end{fmffile}
\end{gathered}\,
\Big)
+g\kappa\Big(
\tfrac{1}{12}\!
\begin{gathered}
	\begin{fmffile}{dgkappa-bubble-1PIz}
		\begin{fmfgraph}(40,40)
			\fmfset{dash_len}{1.2mm}
			\fmfset{wiggly_len}{1.1mm} \fmfset{dot_len}{0.5mm}
			\fmfpen{0.25mm}
			\fmfleft{i}
			\fmfright{o}
			\fmf{phantom,tension=5}{i,v1}
			\fmf{phantom,tension=5}{v2,o}
			\fmf{dashes,fore=black,left,tension=0.5}{v1,v2,v1}
			\fmf{dashes,fore=black}{v1,v2}
			\fmffreeze
			\fmfi{dashes,fore=black}{fullcircle scaled .55w shifted (1.1w,.5h)}
		\end{fmfgraph}
	\end{fmffile}
\end{gathered}\,\,
+\tfrac{1}{12}\!
\begin{gathered}
	\begin{fmffile}{dgkappa-bubble-2pt-1PIz}
		\begin{fmfgraph}(45,45)
			\fmfset{dash_len}{1.2mm}
			\fmfset{wiggly_len}{1.1mm} \fmfset{dot_len}{0.5mm}
			\fmfpen{0.25mm}
			\fmfleft{i}
			\fmfright{o}
			\fmf{phantom,tension=5}{i,v1}
			\fmf{phantom,tension=5}{v2,o}
			\fmf{dashes,fore=black,left,tension=0.4}{v1,v2,v1}
			\fmf{dashes,fore=black}{v1,v2}
			\fmffreeze
			\fmfright{o1,o2}
			\fmf{dashes,fore=black,tension=1}{v2,o1}
			\fmf{dashes,fore=black,tension=1}{v2,o2}
			\fmfforce{(1.1w,0.9h)}{o1}
			\fmfforce{(1.1w,0.1h)}{o2}
		\end{fmfgraph}
	\end{fmffile}
\end{gathered}\,
+\tfrac{1}{4}\,
\begin{gathered}
	\begin{fmffile}{dgkappa-2pt2loop-1PIbz}
		\begin{fmfgraph}(70,70)
			\fmfset{dash_len}{1.2mm}
			\fmfset{wiggly_len}{1.1mm} \fmfset{dot_len}{0.5mm}
			\fmfpen{0.25mm}
			\fmftop{t1,t2,t3}
			\fmfbottom{b1,b2,b3}
			\fmf{phantom}{t1,v1,b1}
			\fmf{phantom}{t2,v2,b2}
			\fmf{phantom}{t3,v3,b3}
			\fmf{dashes,fore=black,right,tension=1}{v1,v2,v1}
			\fmf{dashes,fore=black,right,tension=1}{v2,v3,v2}
			\fmf{dashes,fore=black,tension=1}{v2,b2}
			\fmffreeze
			\fmfleft{i}
			\fmf{dashes,fore=black}{i,v1}
			\fmfforce{(-.1w,0.44h)}{i}
		\end{fmfgraph}
	\end{fmffile}
\end{gathered}\!\!
+\tfrac{1}{12}
\begin{gathered}
	\begin{fmffile}{dgkappa-4pt-1PIz}
		\begin{fmfgraph}(42,42)
			\fmfset{dash_len}{1.2mm}
			\fmfset{wiggly_len}{1.1mm} \fmfset{dot_len}{0.5mm}
			\fmfpen{0.25mm}
			\fmftop{t}
			\fmfbottom{a,b,c}
			\fmf{dashes,fore=black,tension=1}{a,v}
			\fmf{dashes,fore=black,tension=1}{b,v}
			\fmf{dashes,fore=black,tension=1}{c,v}
			\fmf{dashes,fore=black,tension=1.4,left}{v,t,v}
			\fmffreeze
			\fmftop{x}
			\fmf{dashes,fore=black}{t,x}
			\fmfforce{(0.5w,1.45h)}{x}
		\end{fmfgraph}
	\end{fmffile}
\end{gathered}
+\tfrac{1}{8}
\begin{gathered}
	\begin{fmffile}{dgkappa-4ptself1PRz}
		\begin{fmfgraph}(40,40)
			\fmfset{dash_len}{1.2mm}
			\fmfset{wiggly_len}{1.1mm} \fmfset{dot_len}{0.5mm}
			\fmfpen{0.25mm}
			\fmftop{t}
			\fmfbottom{a,b,c}
			\fmf{dashes,fore=black,tension=1}{a,v}
			\fmf{dashes,fore=black,tension=1}{b,v}
			\fmf{dashes,fore=black,tension=1}{c,v}
			\fmf{dashes,fore=black,tension=1.4,left}{v,t,v}
			\fmffreeze
			\fmfbottom{x,z}
			\fmf{dashes,fore=black}{c,x}
			\fmf{dashes,fore=black}{c,z}
			\fmfforce{(w,-.4h)}{x}
			\fmfforce{(1.45w,.4h)}{z}
		\end{fmfgraph}
	\end{fmffile}
\end{gathered}\,\,
+\tfrac{1}{48}
\begin{gathered}
	\begin{fmffile}{dgkappa-6pt-1PRz}
		\begin{fmfgraph}(45,45)
			\fmfset{dash_len}{1.2mm}
			\fmfset{wiggly_len}{1.1mm} \fmfset{dot_len}{0.5mm}
			\fmfpen{0.25mm}
			\fmfsurroundn{i}{5}
			\fmfright{s,t}
			\fmf{dashes,fore=black}{i1,c}
			\fmf{dashes,fore=black}{i2,c}
			\fmf{dashes,fore=black}{i3,c}
			\fmf{dashes,fore=black}{i4,c}
			\fmf{dashes,fore=black}{i5,c}
			\fmf{dashes,fore=black}{i1,s}
			\fmf{dashes,fore=black}{i1,t}
			\fmffreeze
			\fmfforce{(1.2w,0.9h)}{s}
			\fmfforce{(1.2w,0.1h)}{t}
		\end{fmfgraph}
	\end{fmffile}
\end{gathered}\,\,
\Big)\\
&-\gamma\Big(
\tfrac{1}{48}
\begin{gathered}
	\begin{fmffile}{dgamma-vacuum-1PIz}
		\begin{fmfgraph}(55,55)
			\fmfset{dash_len}{1.2mm}
			\fmfset{wiggly_len}{1.1mm} \fmfset{dot_len}{0.5mm}
			\fmfpen{0.25mm}
			\fmfsurroundn{x}{3}
			\fmf{phantom,fore=black}{x1,v}
			\fmf{phantom,fore=black}{x2,v}
			\fmf{phantom,fore=black}{x3,v}
			\fmf{dashes,fore=black,tension=0.7}{v,v}
			\fmf{dashes,fore=black,tension=0.7,right}{v,v}
			\fmf{dashes,fore=black,tension=0.7,left}{v,v}
		\end{fmfgraph}
	\end{fmffile}
\end{gathered}\!
+\tfrac{1}{16}
\begin{gathered}
	\begin{fmffile}{dgamma-2pt1loop-1PIz}
		\begin{fmfgraph}(50,50)
			\fmfset{dash_len}{1.2mm}
			\fmfset{wiggly_len}{1.1mm} \fmfset{dot_len}{0.5mm}
			\fmfpen{0.25mm}
			\fmfleft{i}
			\fmfright{o}
			\fmf{dashes,fore=black,tension=0.7}{i,v,v,o}
			\fmf{dashes,fore=black,left=90,tension=0.7}{v,v}
		\end{fmfgraph}
	\end{fmffile}
\end{gathered}\!
+\tfrac{1}{48}
\begin{gathered}
	\begin{fmffile}{dgamma-4pt1loop-1PIz}
		\begin{fmfgraph}(50,50)
			\fmfset{dash_len}{1.2mm}
			\fmfset{wiggly_len}{1.1mm} \fmfset{dot_len}{0.5mm}
			\fmfpen{0.25mm}
			\fmfsurroundn{x}{8}
			\fmf{phantom,fore=black}{x1,c,x5}
			\fmf{phantom,fore=black}{x2,c,x6}
			\fmf{phantom,fore=black}{x3,c,x7}
			\fmf{phantom,fore=black}{x4,c,x8}
			\fmf{dashes,fore=black}{x1,c}
			\fmf{dashes,fore=black}{x8,c}
			\fmf{dashes,fore=black}{x7,c}
			\fmf{dashes,fore=black}{x6,c}
			\fmfi{dashes,fore=black}{fullcircle scaled .38w shifted (0.46w,.58h)}
		\end{fmfgraph}
	\end{fmffile}
\end{gathered}
+\tfrac{1}{6!}
\begin{gathered}
	\begin{fmffile}{dgamma-6pttree-1PIz}
		\begin{fmfgraph}(45,45)
			\fmfset{dash_len}{1.2mm}
			\fmfset{wiggly_len}{1.1mm} \fmfset{dot_len}{0.5mm}
			\fmfpen{0.25mm}
			\fmfsurroundn{x}{6}
			\fmf{dashes,fore=black}{x1,c,x4}
			\fmf{dashes,fore=black}{x2,c,x5}
			\fmf{dashes,fore=black}{x3,c,x6}
		\end{fmfgraph}
	\end{fmffile}
\end{gathered}
\Big)
+\mathcal{O}(\ell^5)+\ln N
-\smallint \Lambda-\smallint\delta_{\n}\Lambda+Q_4\\
&+\tfrac{1}{2}\Big(\hspace{0.4cm}
\begin{gathered}
	\begin{fmffile}{bubble1wer}
		\begin{fmfgraph}(30,30)
			\fmfset{dash_len}{1.2mm}
			\fmfset{wiggly_len}{1.1mm} \fmfset{dot_len}{0.5mm}
			\fmfpen{0.25mm}
			\fmfvn{decor.shape=circle,decor.filled=shaded, decor.size=3.5thin}{u}{1}
			\fmfleft{i}
			\fmfright{o}
			\fmf{dashes,fore=black,tension=5,left}{i,u1,i}
			\fmffreeze
			\fmfforce{(-w,0.35h)}{i}
			\fmfforce{(0w,0.35h)}{u1}
			\fmfforce{(1.1w,0.35h)}{o}
		\end{fmfgraph}\!\!\!\!
	\end{fmffile}
\end{gathered}\hspace{0cm}
+
\tfrac{1}{2}\hspace{0.45cm}
\begin{gathered}
	\begin{fmffile}{bubble2wer}
		\begin{fmfgraph}(30,30)
			\fmfset{dash_len}{1.2mm}
			\fmfset{wiggly_len}{1.1mm} \fmfset{dot_len}{0.5mm}
			\fmfpen{0.25mm}
			\fmfvn{decor.shape=circle,decor.filled=shaded, decor.size=3.5thin}{u}{2}
			\fmfleft{i}
			\fmfright{o}
			\fmf{dashes,fore=black,tension=5,left}{i,u1,u2,i}
			\fmffreeze
			\fmfforce{(-w,0.35h)}{i}
			\fmfforce{(0w,0.35h)}{u1}
			\fmfforce{(-1w,0.35h)}{u2}
			\fmfforce{(1.1w,0.35h)}{o}
		\end{fmfgraph}\!\!\!\!
	\end{fmffile}
\end{gathered}\hspace{0cm}
+
\tfrac{1}{3}\!\!
\begin{gathered}
	\begin{fmffile}{bubble3ert}
		\begin{fmfgraph}(60,60)
			\fmfset{dash_len}{1.2mm}
			\fmfset{wiggly_len}{1.1mm} \fmfset{dot_len}{0.5mm}
			\fmfpen{0.25mm}
			\fmfvn{decor.shape=circle,decor.filled=shaded, decor.size=3.5thin}{x}{3}
			\fmfsurroundn{u}{6}
			\fmf{phantom,fore=black,tension=1}{u1,x1,c,v,u4}
			\fmf{phantom,fore=black,tension=1}{u2,u,c,x3,u5}
			\fmf{phantom,fore=black,tension=1}{u3,x2,c,t,u6}
			\fmffreeze
			\fmf{dashes,fore=black,tension=1,right=.7}{x1,x2}
			\fmf{dashes,fore=black,tension=1,right=.7}{x2,x3}
			\fmf{dashes,fore=black,tension=1,right=.7}{x3,x1}
		\end{fmfgraph}\!\!
	\end{fmffile}
\end{gathered}
+\dots+
\hspace{0.35cm}
\begin{gathered}
	\begin{fmffile}{bubble1byu}
		\begin{fmfgraph}(30,30)
			\fmfset{dash_len}{1.2mm}
			\fmfset{wiggly_len}{1.1mm} \fmfset{dot_len}{0.5mm}
			\fmfpen{0.25mm}
			\fmfvn{decor.shape=square,decor.filled=shaded, decor.size=3.5thin}{u}{1}
			\fmfleft{i}
			\fmfright{o}
			\fmf{plain,fore=black,tension=5,left}{i,u1,i}
			\fmffreeze
			\fmfforce{(-w,0.35h)}{i}
			\fmfforce{(0w,0.35h)}{u1}
			\fmfforce{(1.1w,0.35h)}{o}
		\end{fmfgraph}\!\!\!\!
	\end{fmffile}
\end{gathered}
+
\tfrac{1}{2}\hspace{0.45cm}
\begin{gathered}
	\begin{fmffile}{bubble2bhj}
		\begin{fmfgraph}(30,30)
			\fmfset{dash_len}{1.2mm}
			\fmfset{wiggly_len}{1.1mm} \fmfset{dot_len}{0.5mm}
			\fmfpen{0.25mm}
			\fmfvn{decor.shape=square,decor.filled=shaded, decor.size=3.5thin}{u}{2}
			\fmfleft{i}
			\fmfright{o}
			\fmf{plain,fore=black,tension=5,left}{i,u1,u2,i}
			\fmffreeze
			\fmfforce{(-w,0.35h)}{i}
			\fmfforce{(0w,0.35h)}{u1}
			\fmfforce{(-1w,0.35h)}{u2}
			\fmfforce{(1.1w,0.35h)}{o}
		\end{fmfgraph}\!\!\!\!
	\end{fmffile}
\end{gathered}
+
\tfrac{1}{3}\!\!
\begin{gathered}
	\begin{fmffile}{bubble3bhj}
		\begin{fmfgraph}(60,60)
			\fmfset{dash_len}{1.2mm}
			\fmfset{wiggly_len}{1.1mm} \fmfset{dot_len}{0.5mm}
			\fmfpen{0.25mm}
			\fmfvn{decor.shape=square,decor.filled=shaded, decor.size=3.5thin}{x}{3}
			\fmfsurroundn{u}{6}
			\fmf{phantom,fore=black,tension=1}{u1,x1,c,v,u4}
			\fmf{phantom,fore=black,tension=1}{u2,u,c,x3,u5}
			\fmf{phantom,fore=black,tension=1}{u3,x2,c,t,u6}
			\fmffreeze
			\fmf{plain,fore=black,tension=1,right=.7}{x1,x2}
			\fmf{plain,fore=black,tension=1,right=.7}{x2,x3}
			\fmf{plain,fore=black,tension=1,right=.7}{x3,x1}
		\end{fmfgraph}\!\!
	\end{fmffile}
\end{gathered}
+\dots\Big),
\end{aligned}
\end{equation}
where we have now also included the counterterm $Q_4$ associated to the strong version of complete normal ordering, see above. 

The vacuum diagram contributions have already been determined, see (\ref{eq:QS}), and these can also be read off from (\ref{eq:W(Jdashes)a}). Also, it should be clear by now how the cephalopod vacuum diagrams cancel: as an example, we focus on the vacuum cephalopod diagram in the third to last line, namely $g\kappa\big(
\tfrac{1}{12}\!
\begin{gathered}
	\begin{fmffile}{dgkappa-bubble-1PI}
		\begin{fmfgraph}(30,30)
			\fmfset{dash_len}{1.2mm}
			\fmfset{wiggly_len}{1.1mm} \fmfset{dot_len}{0.5mm}
			\fmfpen{0.25mm}
			\fmfleft{i}
			\fmfright{o}
			\fmf{phantom,tension=5}{i,v1}
			\fmf{phantom,tension=5}{v2,o}
			\fmf{dashes,fore=black,left,tension=0.5}{v1,v2,v1}
			\fmf{dashes,fore=black}{v1,v2}
			\fmffreeze
			\fmfi{dashes,fore=black}{fullcircle scaled .55w shifted (1.1w,.5h)}
		\end{fmfgraph}
	\end{fmffile}
\end{gathered}\,\,\big)$. This is clearly cancelled by the $g\kappa$ vacuum diagram in the second line in (\ref{eq:W(Jdashes)a}) once it has been expanded out and the two-loop components have been sewn together at the common vertex, denoted by `
$\!\!\!\!\!
\begin{fmffile}{hexagram}
\begin{fmfgraph}(30,30)
\fmfsurroundn{v}{1}
\fmfv{decor.shape=hexagram,decor.filled=full, decor.size=3.5thin}{v1}
\end{fmfgraph}
\end{fmffile}
$\,\,'. Similar remarks hold for all the remaining vacuum cephalopods. 

The procedure is similar for \emph{all} cephalopod counterterm diagrams: we simply substitute into $W(J)$ in (\ref{eq:W(Jdashes)a}) the mass counterterm (\ref{eq:deltam2b}), taking account of (\ref{eq:deltam2 deltaZ}), and then follow through the algebra, the relevant manipulations being similar to those spelled out in (\ref{eq:wigglyexample}), (\ref{eq:lambdasewingexample}), or (\ref{eq:g2sewingexample}). 
We list a couple of examples to be completely explicit; notice that the following combination of terms appears in (\ref{eq:W(Jdashes)a}):
\begin{equation*}
\begin{aligned}
&\tfrac{1}{2}\hspace{0.4cm}
\begin{gathered}
	\begin{fmffile}{dressed-plain2r}
		\begin{fmfgraph}(30,30)
			\fmfset{dash_len}{1.2mm}
			\fmfset{wiggly_len}{1.1mm} \fmfset{dot_len}{0.5mm}
			\fmfpen{0.25mm}
			\fmfvn{decor.shape=circle,decor.filled=shaded, decor.size=3.5thin}{u}{2}
			\fmfleft{i}
			\fmfright{o}
			\fmf{dashes,fore=black,tension=5}{i,u1,u2,o}
			\fmffreeze
			\fmfforce{(-w,0.35h)}{i}
			\fmfforce{(0w,0.35h)}{u1}
			\fmfforce{(1w,0.35h)}{u2}
			\fmfforce{(2w,0.35h)}{o}
		\end{fmfgraph}
	\end{fmffile}
\end{gathered}\hspace{0.4cm}
-\lambda \Big(\tfrac{1}{2}
\begin{gathered}
	\begin{fmffile}{dlambdaself-1dm2ac}
		\begin{fmfgraph}(45,45)
			\fmfset{dash_len}{1.2mm}
			\fmfset{wiggly_len}{1.1mm} \fmfset{dot_len}{0.5mm}
			\fmfpen{0.25mm}
			\fmftop{s}
			\fmfleft{a}
			\fmfright{b}
			\fmf{dashes,fore=black}{a,v}
			\fmf{dashes,fore=black}{b,v}
			\fmf{dashes,fore=black,right,tension=.7}{v,v}
			\fmffreeze
			\fmfforce{(0w,0.2h)}{a}
			\fmfforce{(w,0.2h)}{b}
			\fmfv{decor.shape=circle,decor.filled=shaded, decor.size=3.5thin}{m}
			\fmfforce{(0.23w,0.32h)}{m}
		\end{fmfgraph}
	\end{fmffile}
\end{gathered}\Big)
+\lambda^2\Big(\tfrac{1}{8}\,
\begin{gathered}
	\begin{fmffile}{dL2-2pt2loop-1PRz}
		\begin{fmfgraph}(55,55)
			\fmfset{dash_len}{1.2mm}
			\fmfset{wiggly_len}{1.1mm} \fmfset{dot_len}{0.5mm}
			\fmfpen{0.25mm}
			\fmfleft{i}
			\fmfright{o}
			\fmf{dashes,fore=black}{i,u,v,o}
			\fmf{dashes,fore=black,tension=0.5,left}{u,u}
			\fmf{dashes,fore=black,tension=0.5,right}{v,v}
		\end{fmfgraph}
	\end{fmffile}
\end{gathered}\,\Big)=\mathcal{O}(\ell^6) \, , \\
&\tfrac{1}{2}\hspace{0.4cm}
\begin{gathered}
	\begin{fmffile}{dressed-plain1r}
		\begin{fmfgraph}(30,30)
			\fmfset{dash_len}{1.2mm}
			\fmfset{wiggly_len}{1.1mm} \fmfset{dot_len}{0.5mm}
			\fmfpen{0.25mm}
			\fmfvn{decor.shape=circle,decor.filled=shaded, decor.size=3.5thin}{u}{1}
			\fmfleft{i}
			\fmfright{o}
			\fmf{dashes,fore=black,tension=5}{i,u1,o}
			\fmffreeze
			\fmfforce{(-w,0.35h)}{i}
			\fmfforce{(0w,0.35h)}{u1}
			\fmfforce{(1.1w,0.35h)}{o}
		\end{fmfgraph}
	\end{fmffile}
\end{gathered}\hspace{0.1cm}
-\lambda\Big(\tfrac{1}{4}
\begin{gathered}
	\begin{fmffile}{dlambdaself-0dm2c}
		\begin{fmfgraph}(45,45)
			\fmfset{dash_len}{1.2mm}
			\fmfset{wiggly_len}{1.1mm} \fmfset{dot_len}{0.5mm}
			\fmfpen{0.25mm}
			\fmftop{s}
			\fmfleft{a}
			\fmfright{b}
			\fmf{dashes,fore=black}{a,v}
			\fmf{dashes,fore=black}{b,v}
			\fmf{dashes,fore=black,right,tension=.7}{v,v}
			\fmffreeze
			\fmfforce{(0w,0.2h)}{a}
			\fmfforce{(w,0.2h)}{b}
		\end{fmfgraph}
	\end{fmffile}
\end{gathered}\Big)
+-g^2\lambda\Big(\tfrac{1}{8}
\begin{gathered}
	\begin{fmffile}{dg2L-bubble2pt}
		\begin{fmfgraph}(45,45)
			\fmfset{dash_len}{1.2mm}
			\fmfset{wiggly_len}{1.1mm} \fmfset{dot_len}{0.5mm}
			\fmfpen{0.25mm}
			\fmftop{t1,t2,t3}
			\fmfbottom{b1,b2,b3}
			\fmf{phantom}{t1,v1,b1}
			\fmf{phantom}{t2,v2,b2}
			\fmf{phantom}{t3,v3,b3}
			\fmffreeze
			\fmf{dashes,fore=black,right}{v1,v2,v1}
			\fmf{dashes,fore=black}{v2,t3}
			\fmf{dashes,fore=black}{v2,b3}
			\fmf{dashes,fore=black,tension=1}{t1,b1}
			\fmfforce{(0.25w,0.7h)}{t1}
			\fmfforce{(0.25w,0.3h)}{b1}
			\fmfforce{(0.9w,0.9h)}{t3}
			\fmfforce{(0.9w,0.1h)}{b3}
		\end{fmfgraph}
	\end{fmffile}
\end{gathered}\Big)
+g\kappa\Big(\tfrac{1}{12}\!
\begin{gathered}
	\begin{fmffile}{dgkappa-bubble-2pt-1PIz}
		\begin{fmfgraph}(45,45)
			\fmfset{dash_len}{1.2mm}
			\fmfset{wiggly_len}{1.1mm} \fmfset{dot_len}{0.5mm}
			\fmfpen{0.25mm}
			\fmfleft{i}
			\fmfright{o}
			\fmf{phantom,tension=5}{i,v1}
			\fmf{phantom,tension=5}{v2,o}
			\fmf{dashes,fore=black,left,tension=0.4}{v1,v2,v1}
			\fmf{dashes,fore=black}{v1,v2}
			\fmffreeze
			\fmfright{o1,o2}
			\fmf{dashes,fore=black,tension=1}{v2,o1}
			\fmf{dashes,fore=black,tension=1}{v2,o2}
			\fmfforce{(1.1w,0.9h)}{o1}
			\fmfforce{(1.1w,0.1h)}{o2}
		\end{fmfgraph}
	\end{fmffile}
\end{gathered}\,\Big)
-\gamma\Big(\tfrac{1}{16}
\begin{gathered}
	\begin{fmffile}{dgamma-2pt1loop-1PIz}
		\begin{fmfgraph}(50,50)
			\fmfset{dash_len}{1.2mm}
			\fmfset{wiggly_len}{1.1mm} \fmfset{dot_len}{0.5mm}
			\fmfpen{0.25mm}
			\fmfleft{i}
			\fmfright{o}
			\fmf{dashes,fore=black,tension=0.7}{i,v,v,o}
			\fmf{dashes,fore=black,left=90,tension=0.7}{v,v}
		\end{fmfgraph}
	\end{fmffile}
\end{gathered}\!\Big)=\mathcal{O}(\ell^6) \, , \\
&\hspace{5cm} \vdots
\end{aligned}
\end{equation*}
and these can clearly  be dropped on account of (\ref{eq:deltam2b}), given that we are working up to and including $\mathcal{O}(\ell^4)$. Similar cancellations occur for \emph{all} cephalopods, as can easily be checked. Taking into account that the all vacuum diagram contributions are contained in $Q_{\rm S}$ (or in the weak version, $Q_{\rm W}$) in (\ref{eq:QS}), carrying out this procedure explicitly for all counterterms (other than $\delta Z$), the result for the full generating function is:
\begin{equation}\label{eq:W(Jwiggly)d}
\begin{aligned}
\tfrac{1}{\hbar}&W(J) = \tfrac{1}{2}
\hspace{0.6cm}
\begin{gathered}
	\begin{fmffile}{dressed-dashes0}
		\begin{fmfgraph}(40,40)
			\fmfset{dash_len}{1.2mm}
			\fmfset{wiggly_len}{1.1mm} \fmfset{dot_len}{0.5mm}
			\fmfpen{0.25mm}
			\fmfleft{i}
			\fmfright{o}
			\fmf{dashes,fore=black}{i,o}
			\fmffreeze
			\fmfforce{(-1w,0.35h)}{i}
			\fmfforce{(1w,0.35h)}{o}
		\end{fmfgraph}
	\end{fmffile}
\end{gathered}\,\,
-g
\Big(
\tfrac{1}{3!}
\begin{gathered}
	\begin{fmffile}{dg-3p}
		\begin{fmfgraph}(40,40)
			\fmfset{dash_len}{1.2mm}
			\fmfset{wiggly_len}{1.1mm} \fmfset{dot_len}{0.5mm}
			\fmfpen{0.25mm}
			\fmfleft{i}
			\fmfright{o1,o2}
			\fmf{dashes,fore=black,tension=5}{i,v1}
			\fmf{dashes,fore=black,tension=5}{v1,o1}
			\fmf{dashes,fore=black,tension=5}{v1,o2}
		\end{fmfgraph}
	\end{fmffile}
\end{gathered}
\Big)
+g^2
\Big(
\tfrac{1}{4}\,   
\begin{gathered}
	\begin{fmffile}{dg2-2pt}
		\begin{fmfgraph}(40,40)
			\fmfset{dash_len}{1.2mm}
			\fmfset{wiggly_len}{1.1mm} \fmfset{dot_len}{0.5mm}
			\fmfpen{0.25mm}
			\fmfleft{i}
			\fmfright{o}
			\fmf{dashes,fore=black,tension=1}{i,v1}
			\fmf{dashes,fore=black,tension=1}{v2,o}
			\fmf{dashes,fore=black,left,tension=0.4}{v1,v2,v1}
			\fmffreeze
			\fmfforce{(-.12w,0.5h)}{i}
			\fmfforce{(1.1w,0.5h)}{o}
		\end{fmfgraph}
	\end{fmffile}
\end{gathered}\,
+\tfrac{1}{8}
\begin{gathered}
	\begin{fmffile}{d2-2_g2}
		\begin{fmfgraph}(45,45)
			\fmfset{dash_len}{1.2mm}
			\fmfset{wiggly_len}{1.1mm} \fmfset{dot_len}{0.5mm}
			\fmfpen{0.25mm}
			\fmfsurroundn{i}{4}
			\fmf{dashes,fore=black}{i1,n,m,i4}
			\fmf{dashes,fore=black}{i2,n}
			\fmf{dashes,fore=black}{m,i3}
		\end{fmfgraph}
	\end{fmffile}
\end{gathered}
\Big)
-\lambda
\Big(
\tfrac{1}{4!}
\begin{gathered}
	\begin{fmffile}{dlambdax}
		\begin{fmfgraph}(30,30)
			\fmfset{dash_len}{1.2mm}
			\fmfset{wiggly_len}{1.1mm} \fmfset{dot_len}{0.5mm}
			\fmfpen{0.25mm}
			\fmfleft{i1,i2}
			\fmfright{o1,o2}
			\fmf{dashes,fore=black}{i1,v,o2}
			\fmf{dashes,fore=black}{i2,v,o1}
		\end{fmfgraph}
	\end{fmffile}
\end{gathered}
\Big)\\
&-g^3
\Big(
\tfrac{1}{4}\,\,
\begin{gathered}
	\begin{fmffile}{dfish}
		\begin{fmfgraph}(40,40)
			\fmfset{dash_len}{1.2mm}
			\fmfset{wiggly_len}{1.1mm} \fmfset{dot_len}{0.5mm}
			\fmfpen{0.25mm}
			\fmfleft{i,j}
			\fmfright{o}
			\fmf{dashes,fore=black,tension=1}{i,v1}
			\fmf{dashes,fore=black,tension=1}{j,v1}
			\fmf{dashes,fore=black}{v1,v2}
			\fmf{dashes,fore=black,tension=1}{v3,o}
			\fmf{dashes,fore=black,left,tension=0.3}{v2,v3,v2}
			\fmffreeze
			\fmfforce{(-.2w,0.8h)}{i}
			\fmfforce{(-.2w,0.2h)}{j}
			\fmfforce{(0.05w,0.5h)}{v1}
			\fmfforce{(1.2w,0.5h)}{o}
		\end{fmfgraph}
	\end{fmffile}
\end{gathered}
+\tfrac{1}{8}
\begin{gathered}
	\begin{fmffile}{dg3-crystal}
		\begin{fmfgraph}(45,45)
			\fmfset{dash_len}{1.2mm}
			\fmfset{wiggly_len}{1.1mm} \fmfset{dot_len}{0.5mm}
			\fmfpen{0.25mm}
			\fmfsurround{i1,i2,i3,i4,i5,i6}
			\fmf{dashes,fore=black,tension=1}{i6,v}
			\fmf{dashes,fore=black,tension=1}{i1,v}
			\fmf{dashes,fore=black,tension=1}{v,c}
		 	\fmf{dashes,fore=black,tension=1}{c,u}
			\fmf{phantom,tension=1}{u,i2}
			\fmf{phantom,fore=black,tension=1}{u,i3}
			\fmf{dashes,fore=black,tension=1}{c,s,i4}
			\fmf{dashes,fore=black,tension=1}{s,i5}
		\end{fmfgraph}
	\end{fmffile}
\end{gathered}
+\tfrac{1}{6}
\begin{gathered}
	\begin{fmffile}{dg3-3ptlog}
		\begin{fmfgraph}(35,35)
			\fmfset{dash_len}{1.2mm}
			\fmfset{wiggly_len}{1.1mm} \fmfset{dot_len}{0.5mm}
			\fmfpen{0.25mm}
			\fmfsurroundn{i}{6}
			\fmf{phantom,fore=black}{i1,v,u,i4}
			\fmf{dashes,fore=black}{v,i1}
			\fmf{phantom,fore=black}{i2,s,t,i5}
			\fmf{dashes,fore=black}{i5,t}
			\fmf{phantom,fore=black}{i3,w,x,i6}
			\fmf{dashes,fore=black}{i3,w}
			\fmfi{dashes,fore=black}{fullcircle scaled .55w shifted (.51w,.5h)}
			\fmffreeze
			\fmfforce{(1.25w,0.5h)}{i1}
			\fmfforce{(0.16w,1.1h)}{i3}
			\fmfforce{(0.16w,-.1h)}{i5}
		\end{fmfgraph}
	\end{fmffile}
\end{gathered}\,\,
\Big)
+g\lambda\Big(
\tfrac{1}{4}\,\,
\begin{gathered}
	\begin{fmffile}{dglambda-dart}
		\begin{fmfgraph}(40,40)
			\fmfset{dash_len}{1.2mm}
			\fmfset{wiggly_len}{1.1mm} \fmfset{dot_len}{0.5mm}
			\fmfpen{0.25mm}
			\fmfleft{i,j}
			\fmfright{o}
			\fmf{dashes,fore=black,tension=5}{i,v1}
			\fmf{dashes,fore=black,tension=5}{j,v1}
			\fmf{dashes,fore=black,tension=0.8}{v2,o}
			\fmf{dashes,fore=black,left,tension=0.08}{v1,v2,v1}
			\fmf{phantom}{v1,v2}
			\fmffreeze
			\fmfforce{(-.2w,0.8h)}{i}
			\fmfforce{(-.2w,0.2h)}{j}
		\end{fmfgraph}
	\end{fmffile}
\end{gathered}
+\tfrac{1}{12}
\begin{gathered}
	\begin{fmffile}{dglambda-5ptree}
		\begin{fmfgraph}(30,30)
			\fmfset{dash_len}{1.2mm}
			\fmfset{wiggly_len}{1.1mm} \fmfset{dot_len}{0.5mm}
			\fmfpen{0.25mm}
			\fmftop{t1,t2}
			\fmfbottom{a,b,c}
			\fmf{dashes,fore=black,tension=1}{a,v}
			\fmf{dashes,fore=black,tension=1}{b,v}
			\fmf{dashes,fore=black,tension=1}{c,v}
			\fmf{dashes,fore=black,tension=2}{v,t}
			\fmf{dashes,fore=black,tension=0.7,left,straight}{t,t1}
			\fmf{dashes,fore=black,tension=0.7,left,straight}{t,t2}
			\fmffreeze
			\fmfforce{(0.5w,-0.2h)}{b}
		\end{fmfgraph}
	\end{fmffile}
\end{gathered}
\Big)
-\kappa
\Big(
\tfrac{1}{5!}
\begin{gathered}
	\begin{fmffile}{dkappa-5pt}
		\begin{fmfgraph}(40,40)
			\fmfset{dash_len}{1.2mm}
			\fmfset{wiggly_len}{1.1mm} \fmfset{dot_len}{0.5mm}
			\fmfpen{0.25mm}
			\fmfsurround{u1,u2,u3,u4,u5}
			\fmf{dashes,fore=black,tension=1}{u1,v}
			\fmf{dashes,fore=black,tension=1}{u2,v}
			\fmf{dashes,fore=black,tension=1}{u3,v}
			\fmf{dashes,fore=black,tension=1}{u4,v}
			\fmf{dashes,fore=black,tension=1}{u5,v}
		\end{fmfgraph}
	\end{fmffile}
\end{gathered}
\Big)\\
&+g^4
\Big(
\tfrac{1}{48}
\begin{gathered}
	\begin{fmffile}{dg4-6pttree}
		\begin{fmfgraph}(45,45)
			\fmfset{dash_len}{1.2mm}
			\fmfset{wiggly_len}{1.1mm} \fmfset{dot_len}{0.5mm}
			\fmfpen{0.25mm}
			\fmfsurround{u1,u2,u3,u4,u5,u6}
			\fmf{dashes,fore=black,tension=1}{u1,v}
			\fmf{dashes,fore=black,tension=1}{u2,v}
			\fmf{dashes,fore=black,tension=1}{u3,u}
			\fmf{dashes,fore=black,tension=1}{u4,u}
			\fmf{dashes,fore=black,tension=1}{u5,s}
			\fmf{dashes,fore=black,tension=1}{u6,s}
			\fmf{dashes,fore=black,tension=1,left,straight}{s,c}
			\fmf{dashes,fore=black,tension=1,left,straight}{u,c}
			\fmf{dashes,fore=black,tension=1,left,straight}{v,c}
		\end{fmfgraph}
	\end{fmffile}
\end{gathered}
+\tfrac{1}{16}\,\,\,
\begin{gathered}
	\begin{fmffile}{dg4-4pt-1loopn}
		\begin{fmfgraph}(50,50)
			\fmfset{dash_len}{1.2mm}
			\fmfset{wiggly_len}{1.1mm} \fmfset{dot_len}{0.5mm}
			\fmfpen{0.25mm}
			\fmfleft{a,b}
			\fmfright{c,d}
			\fmf{dashes,fore=black,tension=1}{a,v}
			\fmf{dashes,fore=black,tension=1}{b,v}
			\fmf{dashes,fore=black,tension=1}{v,s}
			\fmf{dashes,fore=black,tension=1}{t,u}
			\fmf{dashes,fore=black,tension=0.3,left}{s,t,s}
			\fmf{dashes,fore=black,tension=1}{u,c}
			\fmf{dashes,fore=black,tension=1}{u,d}	
			\fmffreeze
			\fmfforce{(1.3w,0.8h)}{c}
			\fmfforce{(1.3w,0.2h)}{d}
			\fmfforce{(1w,0.5h)}{u}
			\fmfforce{(-0.06w,0.5h)}{v}
			\fmffreeze
			\fmfforce{(-0.3w,0.8h)}{a}
			\fmfforce{(-0.3w,0.2h)}{b}
		\end{fmfgraph}
	\end{fmffile}
\end{gathered}\,\,\,
+\tfrac{1}{8}\,\,
\begin{gathered}
	\begin{fmffile}{dg4-2pt-1PR-twoloop}
		\begin{fmfgraph}(50,50)
			\fmfset{dash_len}{1.2mm}
			\fmfset{wiggly_len}{1.1mm} \fmfset{dot_len}{0.5mm}
			\fmfpen{0.25mm}
			\fmfleft{i}
			\fmfright{o}
			\fmf{dashes,fore=black,tension=5}{i,v1}
			\fmf{dashes,fore=black,tension=5}{v2,o}
			\fmf{dashes,fore=black,left,tension=0.4}{v1,v3,v1}
			\fmf{dashes,fore=black,right,tension=0.4}{v2,v4,v2}
			\fmf{dashes,fore=black}{v3,v4}
			\fmffreeze
			\fmfforce{(-.25w,0.5h)}{i}
			\fmfforce{(1.25w,0.5h)}{o}
			\end{fmfgraph}
	\end{fmffile}
\end{gathered}\,\,
+\tfrac{1}{8}\,\,
\begin{gathered}
	\begin{fmffile}{dg4-4ptlog}
		\begin{fmfgraph}(40,40)
			\fmfset{dash_len}{1.2mm}
			\fmfset{wiggly_len}{1.1mm} \fmfset{dot_len}{0.5mm}
			\fmfpen{0.25mm}
			\fmfsurroundn{i}{4}
			\fmf{phantom,fore=black}{i1,v,u,i3}
			\fmf{dashes,fore=black}{v,i1}
			\fmf{dashes,fore=black}{i3,u}
			\fmf{phantom,fore=black}{i2,s,t,i4}
			\fmf{dashes,fore=black}{i4,t}
			\fmf{dashes,fore=black}{i2,s}
			\fmfi{dashes,fore=black}{fullcircle scaled .5w shifted (.51w,.5h)}
			\fmffreeze
			\fmfforce{(1.15w,0.5h)}{i1}
			\fmfforce{(0.5w,1.1h)}{i2}
			\fmfforce{(-.1w,0.5h)}{i3}
			\fmfforce{(0.5w,-.1h)}{i4}
		\end{fmfgraph}
	\end{fmffile}
\end{gathered}\,
+\tfrac{1}{8}\begin{gathered}\,\,
	\begin{fmffile}{dg4-6pttreelong}
		\begin{fmfgraph}(55,55)
			\fmfset{dash_len}{1.2mm}
			\fmfset{wiggly_len}{1.1mm} \fmfset{dot_len}{0.5mm}
			\fmfpen{0.25mm}
			\fmfsurroundn{i}{6}
			\fmf{dashes,fore=black}{i1,v,u,s,t,i4}
			\fmf{dashes,fore=black}{i6,v}
			\fmf{dashes,fore=black}{i4,t}
			\fmf{dashes,fore=black}{i5,t}
			\fmf{dashes,fore=black}{i2,u}
			\fmf{dashes,fore=black}{i3,s}
			\fmffreeze
			\fmfforce{(1.12w,0.4h)}{i1}
			\fmfforce{(-.15w,0.423h)}{i4}
			\fmfforce{(.26w,0.16h)}{i5}
			\fmfforce{(0.67w,0.15h)}{i6}
		\end{fmfgraph}
	\end{fmffile}
\end{gathered}\,\,
+\tfrac{1}{4}
\begin{gathered}
	\begin{fmffile}{dg4-2pt2loop-1PIa}
		\begin{fmfgraph}(60,60)
			\fmfset{dash_len}{1.2mm}
			\fmfset{wiggly_len}{1.1mm} \fmfset{dot_len}{0.5mm}
			\fmfpen{0.25mm}
			\fmftop{t}
			\fmfbottom{b}
			\fmfleft{l}
			\fmfright{r}
			\fmf{phantom,fore=black,tension=9}{t,u,v,b}
			\fmf{phantom,fore=black,tension=9}{l,s,x,r}
			\fmf{dashes,fore=black,tension=.01,left}{u,v,u}
			\fmf{phantom,fore=black,tension=0.01}{s,x,s}
			\fmf{dashes,fore=black,tension=1}{u,v}
			\fmf{dashes,fore=black,tension=1}{x,r}
			\fmf{dashes,fore=black,tension=1}{l,s}
		\end{fmfgraph}
	\end{fmffile}
\end{gathered}
+\tfrac{1}{4}
\begin{gathered}
	\begin{fmffile}{dg4-2pt2loop-1PIb}
		\begin{fmfgraph}(35,35)
			\fmfset{dash_len}{1.2mm}
			\fmfset{wiggly_len}{1.1mm} \fmfset{dot_len}{0.5mm}
			\fmfpen{0.25mm}
			\fmfleft{i}
			\fmfright{o}
			\fmf{phantom,tension=5}{i,v1}
			\fmf{phantom,tension=5}{v2,o}
			\fmf{dashes,fore=black,left,tension=0.4}{v1,v2,v1}
			\fmf{dashes,fore=black}{v1,v2}
			\fmfsurroundn{i}{6}
			\fmf{phantom}{i2,a,c,x1,i5}
			\fmf{phantom}{i3,b,c,x2,i6}
			\fmf{dashes,fore=black}{x1,i5}
			\fmf{dashes,fore=black}{x2,i6}
			\fmffreeze
			\fmfforce{(0.1w,-.2h)}{i5}
			\fmfforce{(0.9w,-.2h)}{i6}
			\end{fmfgraph}
	\end{fmffile}
\end{gathered}\,\\
&+\tfrac{1}{4}
\begin{gathered}
	\begin{fmffile}{dg4-4ptlog1PR}
		\begin{fmfgraph}(35,35)
			\fmfset{dash_len}{1.2mm}
			\fmfset{wiggly_len}{1.1mm} \fmfset{dot_len}{0.5mm}
			\fmfpen{0.25mm}
			\fmfsurroundn{i}{6}
			\fmf{phantom,fore=black}{i1,v,u,i4}
			\fmf{dashes,fore=black}{v,i1}
			\fmf{phantom,fore=black}{i2,s,t,i5}
			\fmf{dashes,fore=black}{i5,t}
			\fmf{phantom,fore=black}{i3,w,x,i6}
			\fmf{dashes,fore=black}{i3,w}
			\fmfi{dashes,fore=black}{fullcircle scaled .55w shifted (.51w,.5h)}
			\fmffreeze
			\fmfforce{(1.25w,0.5h)}{i1}
			\fmfforce{(0.16w,1.1h)}{i3}
			\fmfforce{(0.16w,-.1h)}{i5}
			\fmffreeze
			\fmfright{x,z}
			\fmf{dashes,fore=black,tension=1}{i1,x}
			\fmf{dashes,fore=black,tension=1}{i1,z}
			\fmfforce{(1.5w,0.9h)}{z}
			\fmfforce{(1.5w,0.1h)}{x}
		\end{fmfgraph}
	\end{fmffile}
\end{gathered}\,\,\,\,
+\tfrac{1}{4}\,\,\,\,\,
\begin{gathered}
	\begin{fmffile}{dg4-4pt-1loopb}
		\begin{fmfgraph}(40,40)
			\fmfset{dash_len}{1.2mm}
			\fmfset{wiggly_len}{1.1mm} \fmfset{dot_len}{0.5mm}
			\fmfpen{0.25mm}
			\fmfsurroundn{i}{4}
			\fmf{dashes,fore=black}{i1,n,m,i4}
			\fmf{dashes,fore=black}{i2,n}
			\fmf{dashes,fore=black}{m,i3}
			\fmffreeze
			\fmfleft{x}
			\fmf{dashes,fore=black,tension=10,left}{i3,u,i3}
			\fmf{dashes,fore=black}{u,x}
			\fmfforce{(-.4w,0.6h)}{u}
			\fmfforce{(-.75w,0.75h)}{x}
		\end{fmfgraph}
	\end{fmffile}
\end{gathered}
\Big)
-g^2\lambda\Big(
\tfrac{1}{12}\,\,\,
\begin{gathered}
	\begin{fmffile}{dg2L-4ptfish}
		\begin{fmfgraph}(40,40)
			\fmfset{dash_len}{1.2mm}
			\fmfset{wiggly_len}{1.1mm} \fmfset{dot_len}{0.5mm}
			\fmfpen{0.25mm}
			\fmfleftn{i}{3}
			\fmfright{o}
			\fmf{dashes,fore=black,tension=1}{i1,v}
			\fmf{dashes,fore=black,tension=1}{i2,v}
			\fmf{dashes,fore=black,tension=1}{i3,v}
			\fmf{dashes,fore=black,tension=0.5}{v,u}
			\fmf{dashes,fore=black,tension=0.2,left}{u,x,u}
			\fmf{dashes,fore=black,tension=1}{x,o}
			\fmffreeze
			\fmfforce{(-.1w,0.9h)}{i1}
			\fmfforce{(-.1w,0.1h)}{i3}
			\fmfforce{(-.35w,0.5h)}{i2}
			\fmfforce{(1.3w,0.5h)}{o}
		\end{fmfgraph}\,\,
	\end{fmffile}
\end{gathered}
+\tfrac{1}{4}\,\,\,\,\,
\begin{gathered}
	\begin{fmffile}{dg2L-4pt1loop-1PI}
		\begin{fmfgraph}(17,17)
			\fmfset{dash_len}{1.2mm}
			\fmfset{wiggly_len}{1.1mm} \fmfset{dot_len}{0.5mm}
			\fmfpen{0.25mm}
			\fmfright{o1,o2}
			\fmfleft{i}
			\fmf{dashes,fore=black,tension=1,right=0.63}{i,o1}
			\fmf{dashes,fore=black,tension=1,right=0.6}{o1,o2}
			\fmf{dashes,fore=black,tension=1,right=0.63}{o2,i}
			\fmffreeze
			\fmfleft{s1,s2}
			\fmf{dashes,fore=black,tension=1}{i,s2}
			\fmf{dashes,fore=black,tension=1}{i,s1}
			\fmfforce{(-0.8w,1.6h)}{s1}
			\fmfforce{(-0.8w,-0.6h)}{s2}
			\fmffreeze
			\fmfright{x1,x2}
			\fmf{dashes,fore=black,tension=1}{o1,x2}
			\fmf{dashes,fore=black,tension=1}{o2,x1}
			\fmfforce{(1.4w,2h)}{x1}
			\fmfforce{(1.4w,-h)}{x2}
		\end{fmfgraph}
	\end{fmffile}
\end{gathered}\,\,\,
+\tfrac{1}{2}
\begin{gathered}
	\begin{fmffile}{dg2L-2pt2loop-1PI}
		\begin{fmfgraph}(40,40)
			\fmfset{dash_len}{1.2mm}
			\fmfset{wiggly_len}{1.1mm} \fmfset{dot_len}{0.5mm}
			\fmfpen{0.25mm}
			\fmfsurroundn{i}{4}
			\fmfi{dashes,fore=black}{fullcircle scaled .63w shifted (0.5w,.5h)}
			\fmf{phantom,fore=black}{i1,u,v,i3}
			\fmf{dashes,fore=black}{i1,v}
			\fmf{phantom,fore=black}{i2,s,t,i4}
			\fmf{dashes,fore=black}{i2,s}
			\fmffreeze
			\fmfforce{(0.5w,1.2h)}{i2}
			\fmfforce{(0.2w,0.5h)}{v}
			\fmfforce{(1.3w,0.5h)}{i1}
		\end{fmfgraph}
	\end{fmffile}
\end{gathered}\,\,
+\tfrac{1}{8}\,\,\,
\begin{gathered}
	\begin{fmffile}{dg2L-4pt1loopwiggly}
		\begin{fmfgraph}(35,35)
			\fmfset{dash_len}{1.2mm}
			\fmfset{wiggly_len}{1.1mm} \fmfset{dot_len}{0.5mm}
			\fmfpen{0.25mm}
			\fmfleft{i1,i2}
			\fmfright{o1,o2}
			\fmf{dashes,fore=black,tension=1}{i1,n}
			\fmf{dashes,fore=black,tension=1}{i2,n}
			\fmf{dashes,fore=black,tension=1}{n,m}
			\fmf{dashes,fore=black,tension=0.15,left}{m,s,m}
			\fmf{dashes,fore=black,tension=1}{s,o1}
			\fmf{dashes,fore=black,tension=1}{s,o2}
			\fmffreeze
			\fmfforce{(-0.15w,0.5h)}{n}
			\fmfforce{(-0.45w,h)}{i1}
			\fmfforce{(-0.45w,0h)}{i2}
			\fmfforce{(1.15w,h)}{o1}
			\fmfforce{(1.15w,0h)}{o2}
		\end{fmfgraph}
	\end{fmffile}
\end{gathered}\,\,\,
+\tfrac{1}{12}
\begin{gathered}
	\begin{fmffile}{dg2L-crystal}
		\begin{fmfgraph}(45,45)
			\fmfset{dash_len}{1.2mm}
			\fmfset{wiggly_len}{1.1mm} \fmfset{dot_len}{0.5mm}
			\fmfpen{0.25mm}
			\fmfsurround{i1,i2,i3,i4,i5,i6}
			\fmf{dashes,fore=black,tension=1}{i6,v}
			\fmf{dashes,fore=black,tension=1}{i1,v}
			\fmf{dashes,fore=black,tension=1}{v,c}
		 	\fmf{dashes,fore=black,tension=1}{c,u}
			\fmf{phantom,tension=1}{u,i2}
			\fmf{phantom,fore=black,tension=1}{u,i3}
			\fmf{dashes,fore=black,tension=1}{c,s,i4}
			\fmf{dashes,fore=black,tension=1}{s,i5}
			\fmffreeze
			\fmfright{x}
			\fmf{dashes,fore=black}{v,x}
			\fmfforce{(1.1w,0.2h)}{x}
			\fmfforce{(1.05w,0.54h)}{i1}
			\fmfforce{(-.05w,0.54h)}{i4}
		\end{fmfgraph}
	\end{fmffile}
\end{gathered}\\
&+\tfrac{1}{8}\,\,\,\,
\begin{gathered}
	\begin{fmffile}{dg2L-2pt2loop-1PIb}
		\begin{fmfgraph}(50,50)
			\fmfset{dash_len}{1.2mm}
			\fmfset{wiggly_len}{1.1mm} \fmfset{dot_len}{0.5mm}
			\fmfpen{0.25mm}
			\fmftop{t1,t2,t3,t4}
        			\fmfbottom{b1,b2,b3,b4}
        			\fmf{phantom}{t1,v1,b1}
        			\fmf{phantom}{t2,v2,b2}
			\fmf{phantom}{t3,v3,b3}
			\fmf{phantom}{t4,v4,b4}
        			\fmffreeze
			\fmf{dashes,fore=black,right,tension=0.7}{v1,v2,v1}
        			\fmf{dashes,fore=black,right,tension=0.7}{v2,v3,v2}
        			\fmf{dashes,fore=black,tension=3}{v3,v4}
			\fmffreeze
			\fmfforce{(1.1w,0.5h)}{v4}
			\fmfleft{l}
			\fmf{dashes,fore=black}{v1,l}
			\fmfforce{(-.4w,0.5h)}{l}
		\end{fmfgraph}
	\end{fmffile}
\end{gathered}
+\tfrac{1}{4}\,\,\,\,
\begin{gathered}
	\begin{fmffile}{dg2L-4ptdart}
		\begin{fmfgraph}(40,40)
			\fmfset{dash_len}{1.2mm}
			\fmfset{wiggly_len}{1.1mm} \fmfset{dot_len}{0.5mm}
			\fmfpen{0.25mm}
			\fmfleft{i,j}
			\fmfright{o}
			\fmf{dashes,fore=black,tension=5}{i,v1}
			\fmf{dashes,fore=black,tension=5}{j,v1}
			\fmf{dashes,fore=black,tension=0.8}{v2,o}
			\fmf{dashes,fore=black,left,tension=0.08}{v1,v2,v1}
			\fmf{phantom}{v1,v2}
			\fmffreeze
			\fmfforce{(-.2w,0.8h)}{i}
			\fmfforce{(-.2w,0.2h)}{j}
			\fmfleft{n,m}
			\fmf{dashes,fore=black,tension=1}{i,n}
			\fmf{dashes,fore=black,tension=1}{i,m}
			\fmfforce{(-.5w,1.1h)}{n}
			\fmfforce{(-.55w,0.5h)}{m}			
		\end{fmfgraph}
	\end{fmffile}
\end{gathered}
+\tfrac{1}{16}\,
\begin{gathered}
	\begin{fmffile}{dg2L-6pt-1PR}
		\begin{fmfgraph}(35,35)
			\fmfset{dash_len}{1.2mm}
			\fmfset{wiggly_len}{1.1mm} \fmfset{dot_len}{0.5mm}
			\fmfpen{0.25mm}
			\fmfsurround{a,b,c}
			\fmf{dashes,fore=black,tension=1}{c,v,a}
			\fmf{phantom,fore=black,left,tension=1.2}{v,b,v}
			\fmffreeze
			\fmftop{n,m}
			\fmf{dashes,fore=black,tension=1}{n,v}
			\fmf{dashes,fore=black,tension=1}{m,v}
			\fmfforce{(-0.05w,0.77h)}{n}
			\fmfforce{(0.57w,1.15h)}{m}
			\fmffreeze
			\fmfbottom{x,z}
			\fmf{dashes,fore=black,tension=1.3}{c,x}
			\fmf{dashes,fore=black,tension=1.3}{c,z}
			\fmfforce{(-.12w,-0.12h)}{x}
			\fmfforce{(.55w,-0.2h)}{z}
			\fmffreeze
			\fmfright{s,t}
			\fmf{dashes,fore=black,tension=1.3}{a,s}
			\fmf{dashes,fore=black,tension=1.3}{a,t}
			\fmfforce{(1.3w,0.8h)}{s}
			\fmfforce{(1.1w,0.1h)}{t}
		\end{fmfgraph}
	\end{fmffile}
\end{gathered}\,\,
\Big)
+\lambda^2
\Big(
\tfrac{1}{12}\,\,
\begin{gathered}
	\begin{fmffile}{dL2-2pt2loop-1PI}
		\begin{fmfgraph}(45,45)
			\fmfset{dash_len}{1.2mm}
			\fmfset{wiggly_len}{1.1mm} \fmfset{dot_len}{0.5mm}
			\fmfpen{0.25mm}
			\fmfleft{i}
			\fmfright{o}
			\fmf{dashes,fore=black,tension=5}{i,v1}
			\fmf{dashes,fore=black,tension=5}{v2,o}
			\fmf{dashes,fore=black,left,tension=0.4}{v1,v2,v1}
			\fmf{dashes,fore=black}{v1,v2}
			\fmffreeze
			\fmfforce{(-0.2w,0.5h)}{i}
			\fmfforce{(1.2w,0.5h)}{o}
		\end{fmfgraph}
	\end{fmffile}
\end{gathered}\,\,
+\tfrac{1}{16}\,
\begin{gathered}
	\begin{fmffile}{dL2-4pt1loop-1PI}
		\begin{fmfgraph}(50,50)
			\fmfset{dash_len}{1.2mm}
			\fmfset{wiggly_len}{1.1mm} \fmfset{dot_len}{0.5mm}
			\fmfpen{0.25mm}
			\fmfleft{i1,i2}
			\fmfright{o1,o2}
			\fmf{dashes,fore=black,tension=1}{i1,u}
			\fmf{dashes,fore=black,tension=1}{i2,u}
			\fmf{dashes,fore=black,tension=0.4,left}{u,v,u}
			\fmf{dashes,fore=black}{v,o1}
			\fmf{dashes,fore=black}{v,o2}
			\fmffreeze
			\fmfforce{(w,0.8h)}{o1}
			\fmfforce{(w,0.2h)}{o2}
			\fmfforce{(0w,0.8h)}{i1}
			\fmfforce{(0w,0.2h)}{i2}
		\end{fmfgraph}
	\end{fmffile}
\end{gathered}\,
+\tfrac{1}{72}\,\,
\begin{gathered}
	\begin{fmffile}{dL2-6pttree33-1PR}
		\begin{fmfgraph}(45,45)
			\fmfset{dash_len}{1.2mm}
			\fmfset{wiggly_len}{1.1mm} \fmfset{dot_len}{0.5mm}
			\fmfpen{0.25mm}
			\fmfleftn{i}{3}
			\fmfrightn{o}{3}
			\fmf{dashes,fore=black}{i1,u}
			\fmf{dashes,fore=black}{i2,u}
			\fmf{dashes,fore=black}{i3,u}
			\fmf{dashes,fore=black}{o1,v}
			\fmf{dashes,fore=black}{o2,v}
			\fmf{dashes,fore=black}{o3,v}
			\fmf{dashes,fore=black}{u,v}
			\fmffreeze
			\fmfforce{(0w,0.9h)}{i1}
			\fmfforce{(-.2w,0.5h)}{i2}
			\fmfforce{(0w,0.1h)}{i3}
			\fmfforce{(1w,0.9h)}{o1}
			\fmfforce{(1.2w,0.5h)}{o2}
			\fmfforce{(1w,0.1h)}{o3}
		\end{fmfgraph}
	\end{fmffile}
\end{gathered}\,
\Big)\\
&+g\kappa\Big(
\tfrac{1}{12}
\begin{gathered}
	\begin{fmffile}{dgkappa-4pt-1PI}
		\begin{fmfgraph}(32,32)
			\fmfset{dash_len}{1.2mm}
			\fmfset{wiggly_len}{1.1mm} \fmfset{dot_len}{0.5mm}
			\fmfpen{0.25mm}
			\fmftop{t}
			\fmfbottom{a,b,c}
			\fmf{dashes,fore=black,tension=1}{a,v}
			\fmf{dashes,fore=black,tension=1}{b,v}
			\fmf{dashes,fore=black,tension=1}{c,v}
			\fmf{dashes,fore=black,tension=1.4,left}{v,t,v}
			\fmffreeze
			\fmftop{x}
			\fmf{dashes,fore=black}{t,x}
			\fmfforce{(0.5w,1.45h)}{x}
		\end{fmfgraph}
	\end{fmffile}
\end{gathered}
+\tfrac{1}{48}
\begin{gathered}
	\begin{fmffile}{dgkappa-6pt-1PR}
		\begin{fmfgraph}(40,40)
			\fmfset{dash_len}{1.2mm}
			\fmfset{wiggly_len}{1.1mm} \fmfset{dot_len}{0.5mm}
			\fmfpen{0.25mm}
			\fmfsurroundn{i}{5}
			\fmfright{s,t}
			\fmf{dashes,fore=black}{i1,c}
			\fmf{dashes,fore=black}{i2,c}
			\fmf{dashes,fore=black}{i3,c}
			\fmf{dashes,fore=black}{i4,c}
			\fmf{dashes,fore=black}{i5,c}
			\fmf{dashes,fore=black}{i1,s}
			\fmf{dashes,fore=black}{i1,t}
			\fmffreeze
			\fmfforce{(1.2w,0.9h)}{s}
			\fmfforce{(1.2w,0.1h)}{t}
		\end{fmfgraph}
	\end{fmffile}
\end{gathered}\,\,
\Big)
-\gamma\Big(
\tfrac{1}{6!}
\begin{gathered}
	\begin{fmffile}{dgamma-6pttree-1PI}
		\begin{fmfgraph}(40,40)
			\fmfset{dash_len}{1.2mm}
			\fmfset{wiggly_len}{1.1mm} \fmfset{dot_len}{0.5mm}
			\fmfpen{0.25mm}
			\fmfsurroundn{x}{6}
			\fmf{dashes,fore=black}{x1,c,x4}
			\fmf{dashes,fore=black}{x2,c,x5}
			\fmf{dashes,fore=black}{x3,c,x6}
		\end{fmfgraph}
	\end{fmffile}
\end{gathered}
\Big)
+\mathcal{O}(\ell^5)+\ln N+Q_{\rm S}
-\smallint \Lambda.
\end{aligned}
\end{equation}
Thus complete normal ordering of all interaction terms in the action  has completely cancelled all cephalopods (including all tadpole diagrams). It is manifest the trivial vacuum, $\bar{\varphi}=0$, around which we have expanded is a true minimum (up to this order in $\ell$) of the full quantum effective action --recall the comments related to (\ref{eq:dGdv=0}) and (\ref{eq:dWdJ=v}). 
In $n=2$ dimensions this generating function automatically yields UV-finite Green functions.

Comparing (\ref{eq:W(Jwiggly)d}) with the equivalent bare expression (\ref{eq:W(J)inter}) and taking (\ref{eq:U(Jwiggly)}) into account, it is seen that the complete normal ordered generating function can be obtained from the following \emph{effective rules}; starting from the bare theory generating function we are to:
\begin{itemize}\label{effective rules}
\item[{\bf i.}] \emph{drop all cephalopod Feynman diagrams;}
\item[{\bf ii.}] \emph{replace all counterterms $\delta g_N$ by $\delta_{\perp\n}g_N$;}
\item[{\bf iii.}] \emph{shift the renormalised couplings, $g_N$, by the finite part of $\delta_{\n}g_N$;}
\item[{\bf iv.}] \emph{multiply $L$-loop vacuum contributions by a factor of $L$.}
\end{itemize}
This last point has only been verified up to $L=3$, and it would be very interesting to understand to what extent this remains true at higher loop orders. Furthermore, there is no unique way of carrying out step 3.~as it is renormalisation scheme-dependent: different subtraction schemes will give rise to different finite parts in $\delta_{\n }g_N$, and ultimately the finite parts are fixed at some energy scale of interest by ``experiment''. 

These effective rules clearly simplify computations considerably compared to the traditional approach, because the latter does not give a closed form expression for the counterterms, $\delta_{\n}g_N$, associated to cephalopod cancellation. These are in turn determined from the \emph{cephalopod-free} generating function, and so everything is self-consistent.

\subsection{Derivative Interactions}\label{sec:PS-CNOn}

We next show that the cephalopod cancellation associated to complete normal ordering that resulted in the generating function (\ref{eq:W(Jwiggly)d}) is actually expected to hold for generic (local) derivative interaction theories too, at least within perturbation theory (where the notion of cephalopod makes sense). 
In order to proceed, we observe that in the absence of anomalies it is formally possible (at least within perturbation theory) to extract information about a theory with derivative interactions from a theory without. This is achieved by formally promoting the local couplings of the non-derivative interaction theory to certain (strictly speaking non-linear) operators using a point-splitting trick. We discuss this next.

Suppose the non-derivative interaction theory has interaction terms in the action of the form:
\begin{equation}\label{eq:gNphiN}
\sum_{N=0}^{\infty}\frac{1}{N!}\hat{g}_N\phi(z)^N,
\end{equation}
where we recall that $\hat{g}_N=g_N+\delta_{\n}g_N$, that we are as usual absorbing all non-cephalopod counterterms, $\delta_{\perp \n}g_N$, into $g_N$, and that we are not displaying explicitly the coefficients $\alpha^{N-2}$ (as discussed above). 


Let us choose $N$ arbitrary distinct points in a small neighbourhood of the spacetime point $z^{\mu}$, for $\mu=0,1,\dots,n-1$, and suppose we are able to define the \emph{formal} point-splitting operator, $\hat{S}$, whose action on a polynomial~\footnote{I.e., polynomial such as $\phi^N(z)$, or an $N$-point Green function, some external legs of which may be coincident, such as $G_N(z,\dots,z)$ or $G_{N+m}(z,\dots,z,w_1,\dots,w_m)$, or some polynomial combination thereof, such as $\phi^{N-2}(z)G_3(z,z,w)$, etc.} is to split the point $z$ and sum over all distinct permutations of the points, say 1 to $N$, e.g.,
\begin{equation}\label{eq:SphiNetc}
\begin{aligned}
&\hat{S}_z\phi^N(z)=\phi(1)\phi(2)\dots\phi(N),\\
&\hat{S}_z\,\big(3G_2(z,z)\phi(z)\big)=G_2(1,2)\phi(3)+G_2(1,3)\phi(2)+G_2(2,3)\phi(1),\\
&\hat{S}_z3G_3(z,z,w)^2=G_3(1,2,w)G_3(3,4,w)+G_3(1,3,w)G_3(2,4,w)+\\
&\qquad\qquad\qquad\qquad\qquad\qquad+G_3(2,3,w)G_3(1,4,w),\\
&\qquad\vdots
\end{aligned}
\end{equation}
where it will sometimes be convenient to put a subscript on $\hat{S}$ to denote the space-time point at which it acts, $\hat{S}_z $. 
The inverse of this operator, denoted by $\hat{S}^{-1}$, is the corresponding (formal) `point-merging' operator:
\begin{equation*}\label{eq:SphiNetc-1}
\begin{aligned}
&\hat{S}^{-1}_z\big(\phi(1)\phi(2)\dots\phi(N)\big)=\phi^N(z),\\
&\hat{S}^{-1}_z\big(G_2(1,2)\phi(3)+G_2(1,3)\phi(2)+G_2(2,3)\phi(1)\big)=3G_2(z,z)\phi(z),\\
&\hat{S}^{-1}_z\big(G_3(1,2,w)G_3(3,4,w)+G_3(1,3,w)G_3(2,4,w)+\\
&\qquad\qquad+G_3(2,3,w)G_3(1,4,w)\big)=3G_3(z,z,w)^2,\\
&\qquad \vdots
\end{aligned}
\end{equation*}
and similarly for other monomials (or polynomials). Here the subscript $z$ on $\hat{S}^{-1}_z$ denotes the space-time point to which points merge. This clearly will not be a well-defined operator in general, and requires particular care if it is to be used in the quantum theory, in particular. The range of $\hat{S}^{-1}$ will generically be singular, seemingly making it impossible to preserve $\hat{S}^{-1}\hat{S}=1$. However, the divergence associated to acting with $\hat{S}^{-1}$ is already familiar and used often (sometimes implicitly) in practically all interacting quantum field theories. In particular, this point-merging gives rise to Green functions evaluated at coincident points, and so any resulting UV divergences can be subtracted in the usual manner by introducing some regularisation procedure. Many of the resulting divergences will actually be associated to cephalopods and will hence be absent upon complete normal ordering.

Using these effective rules for the operator $\hat{S}$ and its inverse, $\hat{S}^{-1}$, let us now consider more general operators of the form:
\begin{equation}\label{eq:S-1OS}
\hat{S}^{-1}\hat{O}\hat{S},
\end{equation}
where $\hat{O}$ is some (scalar) combination of derivative operators that act on point-split expressions. For example, we may take $\hat{O}=\nabla_1\nabla_2$, so that using the above effective rules for $\hat{S}$:
\begin{equation*}
\begin{aligned}
\hat{S}_z^{-1}\hat{O}\hat{S}_z\phi(z)^N&=\hat{S}_z^{-1}\nabla_1\nabla_2\Big(\phi(1)\phi(2)\dots\phi(N)\Big)\\
&=\hat{S}_z^{-1}\Big(\nabla_1\phi(1)\nabla_2\phi(2)\dots\phi(N)\Big)\\
&=(\nabla\phi)^2\phi^{N-2},
\end{aligned}
\end{equation*}
with spacetime index contractions implicit. This example makes it explicit that if we were to shift the coupling $\hat{g}_N$ in (\ref{eq:gNphiN}) by an operator of the form (\ref{eq:S-1OS}), we can generate derivative interaction theories from non-derivative interaction theories, provided the two theories contain monomials $\phi^N$ with the same powers of $N$. 

The same argument applies to complete normal ordered actions as well, and to make this explicit let us again consider actions with interaction terms of the form (\ref{eq:gNphiN}). Upon complete normal ordering this is of the form:
\begin{equation}\label{eq:*gNphiN*n}
\sum_{N=0}^{\infty}\frac{1}{N!}g_N\n\phi(z)^N\n,
\end{equation}
where notice that $\hat{g}_N$ has been replaced by $g_N$, see (\ref{eq:counterterms gen}). And so we want to now consider the effect of shifting the couplings $g_N$ in (\ref{eq:*gNphiN*n}) by operators of the form (\ref{eq:S-1OS}). In order to do so we will first need to establish that $\hat{S}^{-1}\hat{S}\n\phi^N\n=\n\phi^N\n$.

We start with an observation: the \emph{non}-coincident point limit of complete normal ordered monomials,\footnote{The arguments, $1,\dots,N$, of this expression again labelling some generic distinct points in the neighbourhood of $z$.} $\n\phi(1)\phi(2)\dots\phi(N)\n$, can be extracted from the corresponding coincident point limit expressions, $\n\phi(z)^N\n$. 
A good illustrative example is the $N=4$ case (\ref{eq:*phi4*}),
\begin{equation}\label{eq:*phi4*2}
\begin{aligned}
\n\phi(z)^4&\n=\phi(z)^4-6G_2(z,z)\phi(z)^2-4G_3(z,z,z)\phi(z)+3G_2(z,z)^2-G_4(z,z,z,z),
\end{aligned}
\end{equation}
For elementary fields that are not at coincident points, the \emph{degeneracy} that gives rise to the multiplicity factors 1, 6, 4, 3 and 1 on the right-hand side of (\ref{eq:*phi4*2}) is broken, and we end up instead with $1+6+4+3+1=15$ distinct terms:
\begin{equation}\label{eq:*phi4*nc}
\begin{aligned}
\n\phi&(1)\phi(2)\phi(3)\phi(4)\n\,=\phi(1)\phi(2)\phi(3)\phi(4)-\Big(G_2(1,2)\phi(3)\phi(4)+G_2(1,3)\phi(2)\phi(4)\\
&\qquad+G_2(1,4)\phi(3)\phi(2)+G_2(3,2)\phi(1)\phi(4)+G_2(4,2)\phi(3)\phi(1)+G_2(3,4)\phi(1)\phi(2)\Big)\\
&\qquad -\Big(G_3(1,2,3)\phi(4)+G_3(1,2,4)\phi(3)+G_3(1,4,3)\phi(2)+G_3(4,2,3)\phi(1)\Big)\\
&\qquad +\Big(G_2(1,2)G_2(3,4)+G_2(1,3)G_2(2,4)+G_2(1,4)G_2(3,2)\Big)-G_4(1,2,3,4)\, .
\end{aligned}
\end{equation}
We may depict this pictorially by attaching labels to each of the four dots in the combinatorial diagram associated to (\ref{eq:*phi4*}), making each dot distinct. More generally, given a complete normal-ordered expression at coincident points, see Sec.~\ref{sec:CI-CNO}, the multiplicities appearing on the right-hand sides of each of the terms tell us to how many terms the point-splitting procedure will give rise to. 

We made no use of the point-splitting operator to arrive at (\ref{eq:*phi4*nc}). 
However, this result (\ref{eq:*phi4*nc}) is precisely what we would find had we acted with $\hat{S}_z$ on $\n\phi(z)^4\n$; using the rules that led to (\ref{eq:SphiNetc}),
\begin{equation}\label{eq:Snphi4n}
\begin{aligned}
&\hat{S}\big(\phi^4\big)=\phi(1)\phi(2)\dots\phi(4) \, ,\\
&\hat{S}\big(6\,G_2\phi^2\big)=G_2(1,2)\phi(3)\phi(4)+5\,\,{\rm perms} \, ,\\
&\hat{S}\big(4G_3\phi\big)=G_3(1,2,3)\phi(4)+3\,\,{\rm perms} \, ,\\
&\hat{S}\big(3\,G_2^2\big)=G_2(1,2)G_2(3,4)+2\,\,{\rm perms} \, ,\\
&\hat{S}\big(G_4\big)=G_4(1,2,3,4),
\end{aligned}
\end{equation}
where we drop the subscript $z$ on $\hat{S}_z$ when there is no potential ambiguity. 
From (\ref{eq:*phi4*2}), (\ref{eq:*phi4*nc}) and (\ref{eq:Snphi4n}) we learn that indeed:
$$
\hat{S}\big(\!\n\!\phi(z)^4\n\!\big)=\,\n\,\phi(1)\phi(2)\dots\phi(4)\,\n,
$$
the right-hand side of this, according to (\ref{eq:nmathcalFn2}), being given by (\ref{eq:*phi4*nc}). The inverse operator, $\hat{S}^{-1}$, correspondingly point merges:
\begin{equation*}\label{eq:hatS-1onphi4*}
\begin{aligned}
&\hat{S}^{-1}\big(\phi(1)\phi(2)\dots\phi(4)\big)=\phi^4 \, ,\\
&\hat{S}^{-1}\big(G_2(1,2)\phi(3)\phi(4)+5\,\,{\rm perms}\big)=6\, G_2\phi^2 \, ,\\
&\hat{S}^{-1}\big(G_3(1,2,3)\phi(4)+3\,\,{\rm perms}\big)=4\,G_3\phi \, ,\\
&\hat{S}^{-1}\big(G_2(1,2)G_2(3,4)+2\,\,{\rm perms}\big)=3\,G_2^2\phi \, ,\\
&\hat{S}^{-1}\big(G_4(1,2,3,4)\big)=G_4 \, ,
\end{aligned}
\end{equation*}
all quantities on the right-hand sides being evaluated at the coincident point $z$. Clearly therefore, 
$
\hat{S}^{-1}\hat{S}\n\phi^4\n\,=\,\n\phi^4\n,
$ 
and this procedure can evidently be defined for any of the monomials $\n\phi^N\n$ in (\ref{eq:nphiNn}), with the formal result
\begin{equation}\label{eq:S-1SphiN}
\hat{S}^{-1}\hat{S}\n\phi^N\n\,=\,\n\,\phi^N\n.
\end{equation}
The arbitrariness in the definition of $\hat{S}$ (such as the choice of points $1,2,\dots$) will be immaterial provided these points are ``sufficiently close'' to the space-time point at which $\hat{S}$ acts, and when this operator is used in the manner we suggest. 

Having established (\ref{eq:S-1SphiN}), let us now consider terms of the form:
\begin{equation*}\label{eq:nS-1OSn}
\hat{S}^{-1}\hat{O}\hat{S}\,\n\phi^N\n,
\end{equation*}
where $\hat{O}$ is (as specified above) some scalar operator (such as a collection of derivatives) that acts on the \emph{point-split} expression, and $\hat{S}^{-1}$ point merges the result. 

We again proceed by example, considering the case $\hat{O}=\nabla_1\nabla_2$ and, say $N=3$. From the expression for $\n\phi^3\n$ given in Sec.~\ref{eq:*phi3*},
\begin{equation}\label{eq:S-1DDSphi3}
\begin{aligned}
&\hat{S}^{-1}\nabla_1\nabla_2\hat{S}\big(\!\n\phi^3\n\!\big)=\hat{S}^{-1}\nabla_1\nabla_2\hat{S}\,\big(\phi^3-3G_2\phi-G_3\big)\\
&=\hat{S}^{-1}\nabla_1\nabla_2\,\big(\phi(1)\phi(2)\phi(3)- G_2(1,2)\phi(3)-G_2(1,3)\phi(2)-G_2(2,3)\phi(1)-G_3(1,2,3)\big)\\
&=\hat{S}^{-1}\big(\nabla_1\phi(1)\nabla_2\phi(2)\phi(3)- \nabla_1\nabla_2G_2(1,2)\phi(3)\\
&\qquad\quad-\nabla_1G_2(1,3)\nabla_2\phi(2)-\nabla_2G_2(2,3)\nabla_1\phi(1)-\nabla_1\nabla_2G_3(1,2,3)\big)\\
&=(\nabla\phi)^2\phi- (\nabla^2G_2)\phi-2(\nabla G_2)\nabla \phi-\nabla^2G_3,
\end{aligned}
\end{equation}
where in the last line we used the abbreviation 
$
\nabla^2G_2(z)\dfn \lim_{1,2\rightarrow z}\nabla_1\nabla_2G_2(1,2),
$ and also 
 $2(\nabla  G_2)\nabla\phi(z)=\lim_{1,2,3\rightarrow z}\nabla_1G_2(1,3)\nabla_2\phi(2)+\nabla_2G_2(2,3)\nabla_1\phi(1)$ and\\ $\nabla^2G_3\dfn \lim_{1,2,3\rightarrow z}\nabla_1\nabla_2G_3(1,2,3)$. Now, the result on the last line in (\ref{eq:S-1DDSphi3}) is precisely what we would have found using the independent pictorial approach (such as that leading to (\ref{eq:*phi2dphi2*})) for the case $\n(\nabla\phi)^2\phi\,\n\,$, and so we have shown that:
$$
\n(\nabla\phi)^2\phi\,\n\,=\hat{S}^{-1}\nabla_1\nabla_2\hat{S}\big(\!\n\phi^3\n\!\big).
$$
Similarly, for $N=2$ one finds:
\begin{equation*}\label{eq:(dphi)2phi2b}
\begin{aligned}
\hat{S}^{-1}\nabla_1\nabla_2\hat{S}\big(\!\n\phi^4\n\!\big)&=\phi^2(\nabla\phi)^2-(\nabla^2G_2)\phi^2-4(\nabla G_2)\phi\nabla\phi-G_2(\nabla\phi)^2\\
&\quad-2\,(\nabla^2G_3)\,\phi-2\,(\nabla G_3)\nabla \phi+G_2\nabla^2 G_2+2(\nabla G_2)^2 -\nabla^2G_4,
\end{aligned}
\end{equation*}
where we took into account the result (\ref{eq:*phi4*2}) and the last five relations in (\ref{eq:Snphi4n}) before finally point merging. This is precisely equivalent to an independent computation of the quantity $\n(\nabla\phi)^2\phi^2\n\,$, see (\ref{eq:*phi2dphi2*}), and so learn that:
$$
\n(\nabla\phi)^2\phi^2\n\,=\hat{S}^{-1}\nabla_1\nabla_2\hat{S}\big(\!\n\phi^4\n\!\big).
$$

This procedure can clearly be generalised to any well-defined and local scalar combination of derivative operators, and so these examples lead us to conjecture that \emph{all} local derivative interaction (including complete normal ordered) theories can be generated from generic non-derivative interaction theories with interaction terms $\sum_{N=0}^{\infty}\frac{1}{N!}g_N\n\phi^N\n$, by an appropriate shift in the couplings,
\begin{equation}\label{eq:gNshift}
g_N\rightarrow g_N+\hat{S}^{-1}\hat{O}_N\hat{S},
\end{equation}
where we have added a subscript `$N$', $\hat{O}\rightarrow \hat{O}_N$, to denote the fact that different values of $N$ can acquire different derivative interactions, including of course different scalar combinations of the Riemann tensor, Ricci scalar, any other fields in the theory, etc.

For most purposes (and in the absence of anomalies \cite{NovotnySchnabl00}) we do not need to worry about the uniqueness of the operator $\hat{S}$ and its inverse $\hat{S}^{-1}$, \emph{or} about whether it is well-defined, \emph{when} it is used in just the manner in which we have described.

What remains now is to explain how shifts of the form (\ref{eq:gNshift}) are implemented in the generating function of connected Green functions, $W(J)$, and this will make it clear that complete normal ordering cancels cephalopods in generic scalar field theories. 

Suppose we return to the computation of the generating function (\ref{eq:W(J)01}) within perturbation theory with shifted couplings (\ref{eq:gNshift}). There may exist a term in the action such as, say, $\frac{1}{2}\hat{\zeta} (\nabla\phi)^2\phi$, so from the above this can be extracted by shifting the $\phi^3$ coupling:
$$
\hat{g}\rightarrow \hat{g}+3\hat{\zeta}\hat{S}^{-1}\nabla_1\nabla_2\hat{S}.
$$
The various diagrams will again be of the form (\ref{eq:U(Jwiggly)}), but now the coupling $\hat{g}$ appearing is replaced as above. An instructive example of one such diagram is, 
\begin{equation*}
\begin{aligned}
\frac{1}{12}\hat{g}^2
\begin{gathered}
	\begin{fmffile}{wg2-2loopbubble-1PI}
		\begin{fmfgraph}(40,40)
			\fmfset{dash_len}{1.2mm}
			\fmfset{wiggly_len}{1.1mm} \fmfset{dot_len}{0.5mm}
			\fmfpen{0.25mm}
			\fmfleft{i}
			\fmfright{o}
			\fmf{phantom,tension=5}{i,v1}
			\fmf{phantom,tension=5}{v2,o}
			\fmf{wiggly,fore=black,left,tension=0.4}{v1,v2,v1}
			\fmf{wiggly,fore=black}{v1,v2}
		\end{fmfgraph}
	\end{fmffile}
\end{gathered}
&\rightarrow
\frac{1}{12}(\hat{g}+3\hat{\zeta}\hat{S}^{-1}\nabla_1\nabla_2\hat{S})^2
\begin{gathered}
	\begin{fmffile}{wg2-2loopbubble-1PI}
		\begin{fmfgraph}(40,40)
			\fmfset{dash_len}{1.2mm}
			\fmfset{wiggly_len}{1.1mm} \fmfset{dot_len}{0.5mm}
			\fmfpen{0.25mm}
			\fmfleft{i}
			\fmfright{o}
			\fmf{phantom,tension=5}{i,v1}
			\fmf{phantom,tension=5}{v2,o}
			\fmf{wiggly,fore=black,left,tension=0.4}{v1,v2,v1}
			\fmf{wiggly,fore=black}{v1,v2}
		\end{fmfgraph}
	\end{fmffile}
\end{gathered}\!=\\
&= \frac{1}{12}\int_{z,w}(\hat{g}(z)+3\hat{\zeta}(z)\hat{S}^{-1}_z\nabla_1\nabla_2\hat{S}_z)(\hat{g}(w)+3\hat{\zeta}(w)\hat{S}^{-1}_w\nabla_1\nabla_2\hat{S}_w)\mathcal{G}(z,w)^3\\
&=\frac{1}{12}\hat{g}^2
\begin{gathered}
	\begin{fmffile}{wg2-2loopbubble-1PI}
		\begin{fmfgraph}(40,40)
			\fmfset{dash_len}{1.2mm}
			\fmfset{wiggly_len}{1.1mm} \fmfset{dot_len}{0.5mm}
			\fmfpen{0.25mm}
			\fmfleft{i}
			\fmfright{o}
			\fmf{phantom,tension=5}{i,v1}
			\fmf{phantom,tension=5}{v2,o}
			\fmf{wiggly,fore=black,left,tension=0.4}{v1,v2,v1}
			\fmf{wiggly,fore=black}{v1,v2}
		\end{fmfgraph}
	\end{fmffile}
\end{gathered}
+\frac{1}{2}\int_{z,w}\hat{g}(z)\hat{\zeta}(z)\hat{S}^{-1}_w\nabla_1\nabla_2\hat{S}_w\mathcal{G}(z,w)^3\\
&\qquad+\frac{3}{4}\int_{z,w}(\hat{\zeta}(z)\hat{S}^{-1}_z\nabla_1\nabla_2\hat{S}_z)(\hat{\zeta}(w)\hat{S}^{-1}_w\nabla_1\nabla_2\hat{S}_w)\mathcal{G}(z,w)^3.
\end{aligned}
\end{equation*}
The second term is explicitly, according to the above rules:
\begin{equation*}
\begin{aligned}
\frac{1}{2}\int_{z,w}&\hat{g}(z)\hat{\zeta}(w)\big(\nabla_w\mathcal{G}\big)^2\mathcal{G},
\end{aligned}
\end{equation*}
with the argument of $\mathcal{G}(z,w)$ and spacetime index contractions implicit, whereas the third reads:
\begin{equation*}
\begin{aligned}
&\frac{1}{4}\int_{z,w}\hat{\zeta}(z)\hat{\zeta}(w)(\nabla_z\nabla_w\mathcal{G})^2\mathcal{G}+\frac{1}{2}\int_{z,w}\hat{\zeta}(z)\hat{\zeta}(w)\nabla_z\mathcal{G}\big(\nabla_z\nabla_w\mathcal{G}\big)\nabla_w\mathcal{G}.
\end{aligned}
\end{equation*}
Therefore, the above shift has produced four diagrams out of the single diagram, $\frac{1}{12}\hat{g}^2
\begin{gathered}
	\begin{fmffile}{wg2-2loopbubble-1PI}
		\begin{fmfgraph}(40,40)
			\fmfset{dash_len}{1.2mm}
			\fmfset{wiggly_len}{1.1mm} \fmfset{dot_len}{0.5mm}
			\fmfpen{0.25mm}
			\fmfleft{i}
			\fmfright{o}
			\fmf{phantom,tension=5}{i,v1}
			\fmf{phantom,tension=5}{v2,o}
			\fmf{wiggly,fore=black,left,tension=0.4}{v1,v2,v1}
			\fmf{wiggly,fore=black}{v1,v2}
		\end{fmfgraph}
	\end{fmffile}
\end{gathered}$, with the correct combinatoric coefficients, and one can clearly carry out the same procedure for all diagrams appearing in (\ref{eq:U(Jwiggly)}). However, the key point now is that in fact we need-not make the operator nature of the couplings explicit at that stage of the computation. Rather, we can carry out the complete normal ordering, write down expressions for the counterterms just as we did in the non-derivative interaction case above, keeping in mind that the couplings appearing are really operators. We can go through the entire calculation with the operator nature of the couplings implicit, leading all the way up to our final result for the cephalopod-free generating function (\ref{eq:W(Jwiggly)d}). To derive the final expression for the renormalised generating function of connected Green functions in the presence of derivative interactions thus amounts to interpreting the couplings appearing in the final answer as operators produced by shifts (\ref{eq:gNshift}), and then acting with these operators on the internal vertices in the manner described here. This procedure can be carried out (in principle) for any number of derivatives in the interaction terms, provided these do not spoil diffeomorphism invariance, and \emph{provided} the path integral measure does not introduce additional terms that cannot be absorbed into redefinitions of the various couplings appearing\footnote{\label{foot:coleman}See, e.g., Coleman \cite{Coleman} for a concise description of how to define the path integral measure in the presence of derivative interactions.}

Therefore, we conclude that complete normal ordering cancels all cephalopod Feynman diagrams for generic derivative interaction scalar field theories (when the true vacuum of the full quantum effective action is at vanishing field value). 

\section{Discussion}\label{sec:D}
We have generalised the standard field theory notion of `normal ordering' to what we call `complete normal ordering'. 
Although normal ordering the bare action of a given field theory eliminates all diagrams with simple internal propagators that begin and end on the same internal vertex \cite{Coleman75}, 
there still remain an infinite number of tadpoles. We have shown that `complete normal ordering' a bare (scalar) field theory action 
(whose full quantum effective action has a trivial vacuum at zero field value) cancels \emph{all} cephalopod Feynman diagrams, including all tadpoles and all other
graphs that are removed by conventional normal ordering. This cancellation is automatic and can be implemented in practice by a set of \emph{effective rules}, spelled out on p.~\pageref{effective rules}. 
When it is of interest to compute the renormalisation group flow (RG) of the various couplings, for instance, the resulting truncated generating function is sufficient, 
and at no point is it necessary to consider Green functions with cephalopod Feynman diagrams present. The number of diagrams that one has to compute is thus vastly reduced (by a factor of 2 or 3 or more, depending on the theory), whereas closed-form expressions for the necessary counterterms are automatically produced. Therefore, the cancellation of cephalopods via complete normal ordering is much more efficient than the traditional ``brute-force'' method. 

Complete normal ordering works in any number of spacetime dimensions and is to a large extent independent of the background 
spacetime on which the field theory is formulated (under the usual assumptions, such as the requirement of global hyperbolicity). By using a point-splitting trick\footnote{The use of point splitting in the current paper is certainly naive, and should only be viewed as a ``trick'' that enables 
one to extract expressions for the generating functions of connected Green functions for theories with derivative interactions from the corresponding 
quantities in the absence of derivative interactions. For uses of point-splitting as a regularisation tool we refer the reader to \cite{NovotnySchnabl00}.} we have shown that these results hold for generic interacting (local) scalar field theories, including derivative interactions. 

This simple result is expected to be of use in a number of contexts, for example where it is crucial to carry out perturbation theory around a 
minimum of the full quantum effective action \cite{ColemanWeinberg73}, recently emphasised in \cite{GarbrechtMillington15}. 
It would be of great interest to study complete normal ordering in theories with local and global internal symmetries, 
and theories of gravity. We suspect it will also lead to additional insight to understand how to implement complete normal ordering in
theories with spontaneously broken supersymmetry (and in particular in a string theory context) where typically \cite{DudasPradisiNicolosiSagnotti05} 
the presence of massless tadpoles destabilises the vacuum forcing one to carry out perturbation theory around a new vacuum \cite{PiusRudraSen14}, 
see also \cite{FischlerSusskind86a,FischlerSusskind86b}: there is a flurry of somewhat independent recent developments here, 
see \cite{AngelantonjFlorakisTsulaia14}, \cite{Witten15}, and \cite{Sen15}, and references therein, and it would be extremely interesting 
to understand how to implement complete normal ordering at the level of the worldsheet string theory, 
keeping in mind that the tadpoles one wishes to cancel here are target space notions. In particular, understanding how to implement 
complete normal ordering in such a string worldsheet context would ensure that one is doing superstring perturbation theory around a ``correct'' vacuum, 
i.e., a vacuum that is a true minimum of the full (e.g.~1PI) quantum effective action.

This is the first of a series of papers on complete normal ordering. In the follow-up article \cite{EllisMavromatosSkliros15b} we discuss complete normal ordering of theories whose full quantum effective actions have minima at non-vanishing field values (the case of interest for theories with spontaneous symmetry breaking), 
or theories with instantons or solitons, for example. We also discuss complete normal ordering in the context of Liouville theory \cite{Zams} in Euclidean ${\rm AdS}_{2}$. In \cite{EllisMavromatosSkliros15c} we apply these results to non-linear sigma models. 

\section*{Acknowledgements}
DPS would like to thank Apostolos Pilaftsis and Arkady Tseytlin for useful comments, and Luis Alvarez-Gaume and especially Jean Alexandre, Jo\~ao Penedones and Peter Millington for valuable discussions. DPS would also like to acknowledge the kind hospitality of the CERN theory group and the Institut des Hautes \'Etudes Scientifiques (IH\'ES) where part of this work was carried out. 
This work was supported in part by the London Centre for Terauniverse Studies (LCTS), using funding from the European Research Council (ERC) via the Advanced Investigator Grant 267352. 


\end{document}